%% file: aanda.tex
\documentclass{aa}  

\usepackage{graphicx}
\usepackage{txfonts}
\usepackage{booktabs}
\usepackage{subcaption}
\usepackage{placeins}
\usepackage{ulem}
\usepackage{textgreek}

\usepackage[colorlinks=true]{hyperref}
\hypersetup{citecolor=blue}
\newcommand{\bdsf}{\texttt{PyBDSF}}
\newcommand{\wsclean}{\texttt{WSClean}}
\newcommand{\casa}{\texttt{CASA}}
\newcommand{\spinnaker}{\texttt{SPINNAKER}}
\newcommand{\changes}[0]{CHANG-ES}
\usepackage{upgreek}
\newcommand{\muJyBeam}{\textmu Jy/Beam}
\newcommand{\porfileargone}{\frac{\sigma^2 - z z_0}{\sqrt{2}\sigma z_0}}
\newcommand{\porfileargtwo}{\frac{\sigma^2 + z z_0}{\sqrt{2}\sigma z_0}}
\newcommand{\porfileargthree}{\frac{\sigma^2 - z z_1}{\sqrt{2}\sigma z_1}}
\newcommand{\porfileargfour}{\frac{\sigma^2 + z z_1}{\sqrt{2}\sigma z_1}}

\begin{document}

   \title{CHANG-ES XXVI: Insights into cosmic-ray transport from radio halos in edge-on galaxies}


   \author{M. Stein
          \inst{1}
          \and
          V. Heesen
          \inst{2}
          \and
          R.-J. Dettmar
          \inst{1}
          \and
          Y. Stein
          \inst{1}
          \and
          M. Brüggen
          \inst{2}
          \and
          R. Beck
          \inst{3}
          \and
          B. Adebahr
          \inst{1}
          \and
          T. Wiegert
          \inst{4}
          \and
          C.\,J. Vargas
          \inst{5}\\
          D.\,J. Bomans
          \inst{1,6}
          \and
          J. Li
          \inst{7,8}
          \and
          J. English
          \inst{9}
          \and
          K.\,T.~Chy\.zy
          \inst{10}
          \and
          R. Paladino
          \inst{11}
          \and
          F.\,S. Tabatabaei
          \inst{12}
          \and
          A. Strong
          \inst{13}
          }

   \institute{Ruhr University Bochum, Faculty of Physics and Astronomy, Astronomical Institute (AIRUB), 44780 Bochum, Germany\\
              \email{mstein@astro.ruhr-uni-bochum.de}
         \and
             Hamburger Sternwarte, University of Hamburg, Gojenbergsweg 112, 21029 Hamburg, Germany
        \and
        Max-Planck-Institut für Radioastronomie, Auf dem Hügel 69, 53121 Bonn, Germany
        \and
        Instituto de Astrof\'isica de Andaluc\'ia - CSIC, Glorieta de la Astronom\'ia s/n, 18008 Granada, Spain
        \and
        University of Arizona, Steward Observatory, 933 N Cherry Ave., Tucson, AZ 85721, USA
        \and
        Ruhr-Universität Bochum, Research Department, Plasmas with Complex Interactions, 44780 Bochum, Germany
        \and
        Department of Astronomy, University of Michigan, 311 West Hall, 1085 South University Avenue, Ann Arbor, MI 48109, USA
        \and
        Purple Mountain Observatory, Chinese Academy of Sciences, 10 Yuanhua Road, Nanjing 210023, China
        \and
        University of Manitoba, Dept of Physics and Astronomy, Winnipeg, Manitoba, Canada, R3T 2N2
        \and
        Astronomical Observatory, Jagiellonian University, ul. Orla 171, 30-244 Krak\' ow, Poland
        \and
        INAF - Istituto di Radioastronomia, Via P. Gobetti 101, 40129 Bologna, Italy
        \and
        School of Astronomy, Institute for Research in Fundamental Sciences, 19395-5531 Tehran, Iran
        \and
        Max-Planck-Institut für extraterrestrische Physik, Gießenbachstraße 1, 85748 Garching, Germany
}

   \date{October 17, 2022}

 
  \abstract
   {Galactic winds play a key role in regulating the evolution of galaxies over cosmic time. In recent years, the role of cosmic rays (CR) in the formation of the galactic wind has increasingly gained attention. Therefore, we use radio continuum data to analyse the cosmic ray transport in edge-on galaxies.

   }
   {With newly reduced radio continuum data of five edge on galaxies (NGC~891, NGC~3432, NGC~4013, NGC~4157, and NGC~4631), we plan to set new constraints on the morphology of radio halos and the physical properties of galactic winds driven by stellar feedback. By distinguishing between the central and outer regions of the galaxies, our study setup allows us to search for variations in the radio halo profile or CR transport along the Galactic disk. 
   }
   {Data from the LOFAR Two-metre Sky Survey (LoTSS) data release 2 at 144\,MHz (HBA) and reprocessed VLA data at 1.6\,GHz (L-band) from the Continuum Halos in Nearby Galaxies – an EVLA Survey (CHANG-ES) enable us to increase the extent of the analysed radio continuum profile significantly (up to a factor of 2) compared to previous studies. We compute thermal emission maps using a mixture approach of H$\upalpha$ and near infrared data, which is then subtracted to yield radio synchrotron emission maps. Then we compile non-thermal spectral index maps and compute intensity profiles using a box integration approach. Lastly, we perform 1D cosmic ray transport modelling.}
   {The non-thermal spectral index maps show evidence that the LoTSS maps are affected by thermal absorption, in star forming regions. The scale height analysis reveals that most of the galaxies are equally well fitted with an one-component instead of a two-component exponential profile. We find a bi-modality within our sample. While NGC~3432 and NGC~4013 have similar scale heights in the L-band and HBA, the low-frequency scale heights of NGC~891, NGC~4157, and NGC~4631 exceed their high-frequency counterpart significantly. The 1D CR transport modelling shows agreement of the predicted magnetic field strength and the magnetic field strength estimates of equipartition measurements. Additionally we find an increasing difference of wind velocities (with increasing height over the galactic disk) between central and outer regions of the analysed galaxies.}
   {}


   \maketitle
%

\section{Introduction}
The evolution of galaxies is a highly complex and not yet fully understood process. The circumgalactic medium (CGM) builds the venue where most of the driving factors interplay. There are multiple processes that directly influence the CGM. Gas is introduced by cosmic accretion of the intergalactic medium (IGM), satellite galaxies and their winds add gas from outside of the galaxy halo \citep[cf.][]{2019MNRAS.488.1248H}, thermal instabilities convert hot to cold gas \citep[cf.][]{2020ApJ...903...77B}, and galactic winds of the host galaxy itself also transport gas from the disk into its halo. The constituents of this interplay need to be analysed separately and discussed together in order to derive a holistic view. Analyses of the CGM show that the CGM fuels the galaxy's star formation and moderates feedback processes \citep{2017ARA&A..55..389T}. Therefore, analysing details of the CGM (e.g. multiphase structure, gas dynamics, metallicity) is key to a better understanding of  the evolution of galaxies. Besides all these dynamical and thermal processes, simulations show that there are also non-thermal processes like magnetic fields and cosmic rays (CR) that directly influence the CGM \citep[cf.][]{2020MNRAS.496.4221J,2021MNRAS.501.4888V}. In this article, we focus on galactic winds of the host galaxy that drive the processed interstellar medium (ISM) into the galactic halo. 

In recent years, cosmic rays gained increasing attention from observers as well as theorists, as their influence on the ISM and galactic evolution in general might have been overlooked before. The interactions of CRs with the ISM are manifold and complex. CRs propagate from their origin, supernova (SN) remnants, through the galactic disk and into the galactic halo. Several propagation mechanisms can operate in this environment, namely advection, diffusion, and streaming \citep{2007ARNPS..57..285S,2018ApJ...868..108B,2022MNRAS.510.1184Q,2022MNRAS.510..920Q}. CRs interact mainly with the galactic magnetic field and can therefore transfer some of their momentum to the thermal gas through scattering in magnetic field irregularities with the force directed down the CR pressure gradient. This then leads to a fluid description of the interaction between CRs and thermal gas and is sometimes referred to as `cosmic ray hydrodynamics' \citep{Zweibel2017}. In this context, the CR electrons (CREs) are used as proxies for the whole CR population (CREs and CR protons (CRPs)), which peaks at a few GeV. This is motivated by the similar shape of the CRE and CRP spectrum\footnote{The CRPs have a much higher energy density than CREs but due to the higher rest mass of the protons, CRPs with energies of a few GeV are not observable because they emit most of their synchrotron radiation at very low and therefore non-observable frequencies \citep{2013PhPl...20e5501Z}.}. This hydrodynamic description has been implemented in galactic wind models which show that the mass-loss rate, gaseous distribution and even the existence of winds are dependent on CR transport \citep[e.g.][]{1991A&A...245...79B,2008ApJ...674..258E,2012MNRAS.423.2374U,2014MNRAS.437.3312S,2016MNRAS.462.4227R,2016ApJ...816L..19G}. It is the properties (i.e. wind speed, implied magnetic field strength in the galactic halo) of these winds, driven by stellar feedback, that we would like to explore further in this work.

Radio continuum observations offer us the possibility to study the transport of CRs using the electrons as a tracer \citep[see][for an overview]{2021Ap&SS.366..117H}. With the non-thermal component of the radio continuum emission, one can follow CREs from their origin, SNe in star forming regions, into the galactic halo. On their way through the galaxy, these charged particles gyrate around the magnetic field lines and therefore emit synchrotron radiation. Among the questions we would like to answer are whether these winds exist in the first place, and if so, whether they are able to overcome the galactic gravitational potential and therefore enrich the CGM. Edge-on spiral galaxies are good laboratories to study galactic winds because only here, the path of the galactic wind into the halo can be examined. \citet{2015AJ....150...81W} show that such radio halos can be detected using deep radio continuum data, they even detect radio halos with vertical extents larger than the diameter of the disk. \cite{2017ARA&A..55..389T} note that the CGM reaches several hundreds of kilo parsecs (kpc) (based on absorption studies). Such a scale cannot be probed with today's radio continuum observations. Here, the emission reaches about ten kpc \citep[e.g.][]{2022MNRAS.509..658H}. However, as pointed out before, the analysis of these galactic winds is very important for a better understanding of the CGM properties, since the material of the galactic disk can only reach the CGM if and only if a bona fide wind develops.  Since synchrotron radiation preferably depletes the highest energy electrons, the non-thermal spectral index (SPIX) 
of the emission is expected to steepen on its way from the disk into the galaxy halo. This steepening of the radio spectrum is usually referred to as spectral ageing. The spectral index $\alpha$  $(\text{I}_\nu \propto \nu^\alpha)$ in this work is defined as:
\begin{equation}
    \label{eq:spec_index}
    \alpha\ = \frac{\log\left(S_{\nu_1}/S_{\nu_2} \right)}{\log\left( \nu_1/\nu_2 \right)}. 
\end{equation}
Such a steepening is observed in radio galaxies \citep[e.g.][]{1996A&ARv...7....1C,heesen_2018a} as well as in star-forming galaxies \citep[e.g.][]{2019A&A...632A..12S,2019A&A...632A..10M,2019A&A...622A...9M,2019A&A...632A..13S,2022MNRAS.509..658H}.
For star forming galaxies, the steepening of the SPIX can be used to further analyse the transport of CRs in galactic winds. As CRs are accelerated by SNe in the galactic disk, there is a gradient in the CR pressure pointing from the  galactic disk into the halo. As pointed out by \citet{2016MNRAS.462.4227R} and \citet{2022MNRAS.509..658H}, this gradient influences the background plasma by adding an additional pressure term, P\textsubscript{CR}, to the overall pressure, which can exceed the gravitational pull and therefore launch a CR-driven wind.

The use of low-frequency radio data to analyse non-thermal galactic winds via CR ageing offers a twofold advantage. First, for such an analysis, a reliable spectral index measurement is essential. From Eq.~\eqref{eq:spec_index}, one can deduce that the accuracy of the SPIX measurement grows with the relative distance of the two analysed spectral bands. Therefore combining low- and high-frequency data sets increases the quality of the SPIX measurement \citep{1978A&A....66..205S,2019A&A...622A...9M, 2019A&A...632A..13S}.  Second, with low-frequency data one can trace the older CR population that has already travelled far into the galactic halo, and therefore increase the extent to which the halo is observable. On basis of such data sets (high- and low-frequency radio continuum maps), one can reliably distinguish whether advection or diffusion is the dominant transport mechanism. Advection-dominated transport seems to produce large radio halos, while diffusion-dominated transport results in much smaller halos \citep[cf.][]{2016MNRAS.458..332H,2019A&A...623A..33S}.

To conclude this section, we note that the importance of CRs for the galactic winds and for the evolution of galaxies as a whole is now widely recognised. However, the interaction of CRs with the ISM and the galactic magnetic field are not yet understood in detail. In particular, the potential of CRs to drive galactic winds is the subject of ongoing discussion.  A crucial observational experiment would be to measure both the outflow velocity of the ionised gas and the CR advection speed. With CR-driven winds, we would expect the CRs to be transported faster than the gas, which is the basic mechanism that can offset the adiabatic cooling of the gas in the hydrodynamic model of CR-driven winds \citep{Zweibel2017}. Such a differential velocity could be explained for instance by CR streaming with the Alfv\'en velocity or by CR diffusion down the density gradient \citep{2017MNRAS.467..906W}. Observationally, this is a difficult task as it is hard to measure both velocities due to geometry. While edge-on galaxies allow us to study the radio halo and the cosmic-ray advection speed, face-on galaxies are better suited to study outflow speeds of the ionised gas with optical or ultra-violet spectroscopy \citep{2015ApJ...809..147H}.
Therefore, as an alternative, we aim to build up a coherent picture of both CR and gaseous transport in galactic halos in a sample of galaxies with similar properties over a range of fundamental parameters to better understand the influence of CR transport mechanisms on stellar feedback-driven winds \citep[see][for a review]{2021Ap&SS.366..117H}. To expand this picture with new deep radio observations is the scope of this work.

This paper is structured as follows: In Sect. \ref{sec:sample}, we describe the galaxies that are analysed in this work and put them into context by comparing them to other nearby galaxies. Section \ref{sec:data} describes the data that are used and Sect. \ref{sec:method} describes the processing of the data. We then present the first results in Sect. \ref{sec:results} before we further analyse the radio halo profiles in Sect. \ref{sec:profiles} and derive physical properties of the non-thermal galactic wind by applying a 1D cosmic ray transport model in Sect. \ref{sec:CRwinds}. Lastly, we conclude the paper in Sects. \ref{sec:discussion} and \ref{sec:SandO}.

\section{Sample Description}
\label{sec:sample}
\input{tab_basic_info}
\begin{figure}
 \centering
\includegraphics[width=\linewidth]{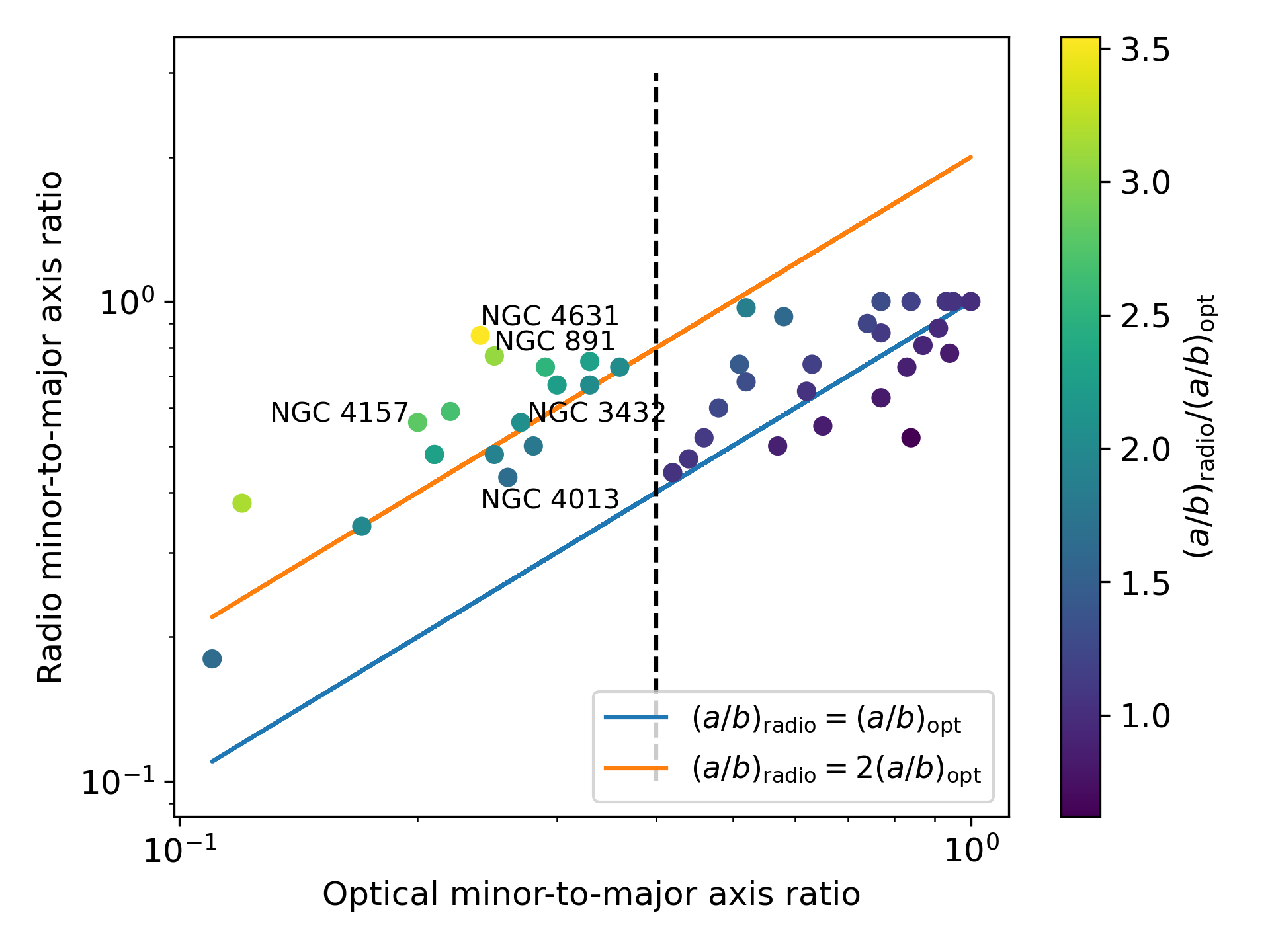}
\caption{Minor-to-major axis ratios in the radio continuum and optical light. Data points are from LoTSS-DR2, where CHANG-ES galaxies are located to the left of the vertical dashed line. Data points are coloured according to their radio-to-optical axis ratios. Radio halos are particularly prominent if $(a/b)_{\rm radio}\geq 2(a/b)_{\rm opt}$ as indicated by the orange line.}
\label{fig:axis_ratio}
\end{figure}
In order to analyse the interaction of the galactic disk with its halo, the orientation of the galaxy is an important factor. Edge-on galaxies allow us to trace the CREs from their origin in the disk over several kpc into the galaxy halo. Therefore, the data acquired for the Continuum Halos in Nearby Galaxies – an EVLA Survey (CHANG-ES), is the starting point for our analysis. The CHANG-ES survey consists of 35 nearby (d $\leq$ 42\,Mpc)
edge-on (76$^\circ\leq\text{i}\leq\text{90}^\circ$) galaxies \citep{2012AJ....144...43I,2015AJ....150...81W}. As previously outlined, adding low-frequency data to the information drawn from the VLA maps highly improves the quality of SPIX maps, which is a crucial for the planned analysis. Therefore, we only choose the 17 galaxies that lie in the footprint of the LoTSS-DR2.

Radio halos of edge-on galaxies can be identified by comparing the minor-to-major axis ratios in the radio continuum and optical emission. If a radio halo is present, we expect the ratio in the radio continuum to be significantly larger than the optical one. In Fig.~\ref{fig:axis_ratio}, we show such a comparison for the LoTSS-DR2 sample. As the CHANG-ES galaxies have high inclination angles, they lie on the left side of the plot, clearly separated from the other galaxies with moderate inclination angles. We find that of the 17 CHANG-ES galaxies in LoTSS-DR2, nine galaxies have axis ratios with $(a/b)_{\rm radio}\gtrsim 2(a/b)_{\rm opt}$. These galaxies are NGC~2820, NGC~4157, NGC~5297, NGC~3079, NGC~4631, NGC~891, NGC~3556, NGC~2683, and NGC~4217 (listed with increasing $(a/b)_{\rm opt}$). Hence, these galaxies are most promising to study galactic winds. Of these galaxies, two were studied previously with LOFAR, namely NGC~891 \citep{2018A&A...615A..98M} and NGC~3556 \citep{2019A&A...622A...9M}, where CR transport was specifically modelled only in NGC~3556. Of the remaining galaxies, NGC~ 4631, NGC~891, NGC~2820, and NGC~4157 have the highest radio-to-optical ratios. We chose to omit NGC~2820 as previous analyses (e.g.:\citet{1985A&A...150L...7V} (VLA 1.5\,GHz), \citet{2015AJ....150...81W} (CHANG-ES D-configuration),  \citet{2019AJ....158...21I} (CHANG-ES B-configuration), \citet{2005A&A...435..483K} (GMRT), and \cite{2022arXiv220400635H} (LOFAR LoTSS)) show a strong radio bridge, reaching from the galaxy to its closest interaction partner NGC~2814, which will most probably disturb the radio halo and therefore prevent the planned analysis. As NGC~891 and NGC~4157 are fairly fast rotating, compared to NGC~4631, we decided to also include NGC~3432, which has a less prominent radio halo, to have another slow rotating galaxy in our sample. We further include NGC~4013, which was already studied by \citet{2019A&A...632A..13S} who found a diffusion-dominated radio halo. We chose to re-analyse this galaxy with consistent methods, as we expect a more accurate measurement of the radio halo intensity profile due to the newly processed data sets. Hence, we chose five galaxies from LoTSS-DR2 for further analysis, namely NGC~891, NGC~3423, NGC~4013, NGC~4157, and NGC~4631. Table \ref{tab:basic_info} lists some fundamental parameters of each analysed galaxy. 



\input{total_int_figs}

\section{Data}
\label{sec:data}
\subsection{Radio-Continuum Data}
In order to avoid losses of diffuse emission in these extended galaxies and their halos, we combine C-configuration and D-configuration L-band data. All VLA data sets that are used in this project where observed for the CHANG-ES project (Project Code 10C-119). The observation and calibration strategy is fully described for the D-configuration data in 
 \citet{2017AJ....153..202W} \footnote{The data are publicly available: \url{https://www.queensu.ca/changes}}. In Table \ref{tab:obs_info}, we list the relevant information for the VLA C-configuration L-band observations. Phase and flux density calibration have been performed using the \texttt{Common Astronomy Software Applications package} (\casa) \citep{2007ASPC..376..127M}. Self calibration was incorporated in the calibration process if it increased the data quality. The frequency range for all targets spans from 1.247\,GHz to 1.503\,GHz and from 1.647\,GHz to 1.903\,GHz.
 \input{tab_obs_info}
 Cleaning is performed using \wsclean \ \citep{2014MNRAS.444..606O}. We incorporate multiscale cleaning, automatic masking and a Briggs robust weighting of 0.4. Additionally, a Gaussian taper of 6\,\arcsec\ is applied to the uv-weights and the restoring beam size is set to 20\,\arcsec. 

The LOFAR data are based on observations from the LOFAR Two-metre Sky Survey \citep[LoTSS,][]{2017A&A...598A.104S,2019A&A...622A...1S}, using data from LoTSS-DR2 \citep{2022A&A...659A...1S}. These data are taken with the high-band antenna (HBA) system with an effective frequency of 144~MHz. The data were presented in \citet{2022arXiv220400635H}, the data release paper of nearby galaxies in LoTSS-DR2. Besides following standard observing and data reduction procedures for LoTSS data as described by \citet{2017A&A...598A.104S,2019A&A...622A...1S}, we use the re-calibration pipeline described by \citet{2021A&A...651A.115V} which results in new $(u,v)$ data sets that are specifically tailored to our target galaxies. The maps are {\sc clean}ed using \wsclean \ with a Briggs weighting of $-0.25$ and a Gaussian taper of 10\,\arcsec\ with a fixed restoring beam size of 20\,\arcsec \citep[see][for further information]{2022arXiv220400635H}. The imaging process of NGC~891 was challenging due to its proximity to a $\sim 10$\,Jy radio source \citep[at 1.4\,GHz, ][]{1969ApJ...157....1K}, the radio galaxy 3C~66, and some fine tuning of parameters was necessary. Therefore, we use a slightly different strategy when creating a new 144\,MHz map for NGC~891 with \wsclean. Here, we incorporate a Briggs weighting of $-0.15$ together with multi-scale {\sc CLEAN}. In order to accurately account for the influence of 3C~66, which lies to the north of NGC~891, the map spans over $\sim1.9^{\circ} \times 1.9^{\circ}$. A Gaussian taper of 7\,\arcsec\ is applied in the ($uv$)-plane and the final beam size is set to 20\,\arcsec. This map has also been published by \citet{2022arXiv220400635H}. 
To estimate the noise levels of all radio maps, we run \bdsf\ \citep{2015ascl.soft02007M}, using standard parameters (source detection threshold 5$\sigma$; island boundary threshold 3$\sigma$). The resulting noise estimates are listed in Table \ref{tab:flux}.

\subsection{Additional Data Sets}

As in \citet{2018ApJ...853..128V}, we use resolution enhanced (12\farcs 4)  WISE 22\,\textmu m images to correct for thermal emission in our \changes\ galaxies \citep[][and T.~Jarrett, private communication]{2012AJ....144...68J}, combined with H$\upalpha$ data. Since the resulting WISE point spread function (PSF) may deviate significantly from a purely Gaussian PSF, we use an empirical PSF model when convolving the maps to the resolution of the radio maps \citep{2019ApJS..245...25J}. This model was compiled by stacking stars in a large field to extract the shape of the PSF \citep{2019ApJS..245...25J}.
H$\upalpha$ images of all CHANG-ES galaxies have been analysed in \citet{2019ApJ...881...26V}. For this analysis, 25 of the 35 \changes\ galaxies (including NGC~3432, NGC~4013, and NGC~4157) were observed with the 3.5\,m telescope of the Apache Point Observatory \citep[see][for further information]{2019ApJ...881...26V}. For NGC~4631, \citet{2019ApJ...881...26V} use an H$\upalpha$ map that was observed with the 2.1\,m telescope at the Kitt Peak National Observatory as ancillary data for the Spitzer Infrared Nearby Galaxies Survey \citep[SINGS;][]{2006MNRAS.367..469D,2008MNRAS.385..553D}. The H$\upalpha$ map of NGC~891 was obtained from the ancillary data set for the WSRT Hydrogen Accretion in LOcal GAlaxieS (HALOGAS) survey \citep{2011A&A...526A.118H,Heald2020}.


\section{Method}
\label{sec:method}

\subsection{Point Source Removal}

In order to extract the radio profile, background and foreground point-like sources need to be removed. We detect compact sources by using \bdsf with an adaptive\_rms\_box and a maximum detection island size. The detected sources are then fitted as Gaussian components and subtracted from the original image. Only large, extended sources remain in the residual image.

We further inspect the images manually to mask out previously undetected or nested sources by using the corresponding LoTSS 6\arcsec map, published in \citet{2022arXiv220400635H}, for an exact positioning of the masks. With this approach, the profile of the galaxy can be reliably followed, without being affected by back- or foreground sources.

\subsection{Thermal Correction}
\label{ss:thermal_correction}
In star forming galaxies, radio continuum emission is a superposition of thermal and non-thermal processes: 
\begin{equation}
    S_{\mathrm{tot}}(\nu) = S_{\mathrm{th}}(\nu_0)\left(\frac{\nu}{\nu_0}\right)^{-0.1}  +  S_{\mathrm{nth}}(\nu_0)\left(\frac{\nu}{\nu_0}\right)^{\alpha_{\mathrm{nth}}},
    \label{eq:th-non-th}
\end{equation}
where $S$ denotes the flux density at a given frequency $\nu_0$. The standard picture is that the thermal emission follows a power-law with a spectral index of $-0.1$. Overall, the measured combined spectral index is about $\alpha_{\mathrm{tot}} \approx -0.7$. Therefore, in order to analyse non-thermal processes such as winds driven by stellar feedback, the thermal emission needs to be estimated and subtracted. However, from Eq.~\eqref{eq:th-non-th}, one can deduce that the fraction of the thermal emission drops when reaching lower frequencies and hence that such a correction might not be needed for low-frequency data. It is also important to note that there are significant deviations from a constant non-thermal spectral index when reaching higher frequencies \citep{2018A&A...611A..55K}. In the literature, there are several approaches to separate the synchrotron emission from the thermal emission, e.g.
assuming a constant non-thermal SPIX \citep{1982A&A...108..176K}, estimating the thermal emission from H\,{\sc ii} regions \citep{1982A&A...105..192B}, or using de-reddened H$\upalpha$ emission \citep{2007A&A...475..133T,2018NatAs...2...83T}.

For this project, we re-implemented\footnote{The Python3 package can be downloaded via \url{https://github.com/msteinastro/thermal_maps_chang-es_xxvi}} the mixture-method technique developed by \citet{2018ApJ...853..128V}, which is specifically designed to estimate the thermal emission in edge-on galaxies. This technique is based on the combination of H$\upalpha$ and mid-infrared (WISE 22\,\textmu m) images to compile a star formation map and then convert this map to a thermal emission map. This technique also has been applied in the second \changes\ data release paper \citep{2019ApJ...881...26V}. For in depth details of the technique, we refer to \citet{2018ApJ...853..128V,2019ApJ...881...26V}.

Here, we shortly describe the overall workflow and point out where we deviate from the current implementation. In order to combine H$\upalpha$ with mid-infrared data and later apply them to a radio map, we first need to match resolutions. As the mid-infrared images have an resolution of $\sim$12\farcs 4  but deviate significantly from a Gaussian PSF, it is not advisable to convolve the original images with a purely Gaussian kernel as the result will not be a Gaussian PSF. This is especially important when the target resolution of the radio maps is only slightly above the FWHM of the mid-infrared images. Therefore, it is imperative to use individually tuned homogenisation kernels. The former implementation used convolution kernels provided by \citet{2011PASP..123.1218A}, where one could select from a variety of kernels to produce a Gaussian PSF with a predefined FWHM. As this approach limits analyses to certain resolutions, we now use the WISE PSF model described in Sect. \ref{sec:data} and the Python package PyPHER\footnote{\url{https://github.com/aboucaud/pypher}} \citep{2016A&A...596A..63B,alexandre_boucaud_2016_61392} to convolve the mid-infrared image to the target resolution. The benefit of such an approach is that one can select any type of target kernel, rather than having to choose from a couple of pre-selected kernels. As the H$\upalpha$ observations are mainly seeing limited, their resolution is of the order of $\sim$1\farcs 5. Since the target resolution of any study that uses the mid-infrared data needs to be above 12\farcs 4, the resulting PSF for the H$\upalpha$ image will be dominated by the convolution kernel, so that  we can convolve the H$\upalpha$ images with a purely Gaussian kernel. 

The resolution-matched and calibrated (erg\,s\textsuperscript{-1}\,cm\textsuperscript{-2}) images can then be further processed. As the following method relies on 24\,\textmu m data, the WISE band 4 data, which are centred on 22~\textmu m, need to be corrected. As shown by \citet{2015AJ....150...81W}, there is a tight linear relationship between 22\,\textmu m and 24\,\textmu m emission and therefore, the data can be scaled using a factor of 1.03. 
Following \cite{2007ApJ...666..870C}, we can use the mid-infrared data to correct the H$\upalpha$ maps for dust extinction:
\begin{equation}
    L_{\mathrm{H\alpha\_corr}}=L_{\mathrm{H\alpha\_obs}} + 0.042\cdot L_{24\mathrm{\mu m}}. 
\end{equation}
The factor 0.042 was derived in \citet{2018ApJ...853..128V} and is only valid for edge-on or very dusty galaxies and corrects the original factor by \citet{2007ApJ...666..870C} with an additional factor of $1.36$. The corrected H$\upalpha$ emission can then be related to a star formation rate and hence to the thermal radio emission at a frequency, \ $\nu$, \ \citep{2011ApJ...737...67M,2018ApJ...853..128V}:

\begin{equation}
    \mathrm{SFR_{mix}\,[M_\odot/yr]} = 5.37\cdot10^{-42}\cdot L_{\mathrm{H\alpha\_corr\,[erg\,s^{-1}]}},
\end{equation}
\begin{equation}
    L_\nu^{\mathrm{T}}\,\mathrm{[erg\,s^{-1}\,Hz^{-1}]} = 2.2\cdot10^{27}\left(  \frac{T_{e}\,[\mathrm{K}]}{10^{4}}  \right)^{0.45} \left(\nu\,[\mathrm{GHz}]^{-0.1} \right)\mathrm{SFR_{mix}}.
    \label{eq:L_therm}
\end{equation}

Here, we assume an average electron temperature of $T_e = 10^{4}$\,K \citep{2006agna.book.....O,2007ApJ...666..870C}. With Eq.~\eqref{eq:L_therm} we compute thermal emission maps for 144\,MHz as well as for 1.5\,GHz.  These thermal maps can then be subtracted from the radio maps to obtain the pure synchrotron emission.

As a concluding remark, we note that several assumptions and systematic uncertainties influence the estimate of the thermal emission (e.g. a varying H$\upalpha$/N\,[{\sc ii}] ratio  in the diffuse ionised gas \citep{1999A&A...343..705O}, which makes accounting for  N\,[{\sc ii}] contamination in the H$\upalpha$ images more complicated, or varying electron temperatures) that dominate the statistical uncertainties coming from the H$\upalpha$ and mid-infrared background noises. \citet{2018ApJ...853..128V} present a detailed analysis of all relevant factors and conclude that the thermal emission estimates have an overall uncertainty of approximately 14\%. Considering the fact that the contribution of thermal emission to the overall radio emission in L-band in the observed galaxies is of the order of approximately 30\% in star forming regions, the introduced uncertainties onto the radio data are expected to be approximately 4\% (even less than the HBA data) and are therefore smaller than standard calibration errors.
\input{si_fig}

\section{Results}
\label{sec:results}
\subsection{Total Intensity}
\label{sec:intensity_maps}
\input{tab_radio_param}

Table \ref{tab:flux} lists the integrated flux density measurements as well as the noise estimates for all analysed galaxies in LOFAR HBA and VLA (CHANG-ES) L-band. The uncertainties of the flux density measurements $\Delta_{\text{S}_{\nu}}$ with a flux density of $\text{S}_{\nu}$, are computed via:
\begin{equation}
    \label{eq:flux_density_error}
    \Delta_{\text{S}_{\nu}}  =  \sqrt{\left( \sigma \sqrt{N_{\text{beams}}} \right)^2 + \left( \epsilon S_{\nu} \right)^2}.
\end{equation}
Here, $\sigma$ denotes the background noise of the individual map and $N$\textsubscript{beams} the number of telescope beams that fit into the defined aperture. As an additional error term, one needs to account for calibration errors. This is implemented as a relative error term by setting $\epsilon=0.1$ for the HBA measurements \citep{2017A&A...598A.104S} and setting $\epsilon=0.05$ for the L-band measurements. Calibration errors are typically much more severe in the low-frequency regime, that is why we choose different weightings for the relative error terms. Since all the galaxies that are analysed in this work are relatively bright, the error terms are dominated by the calibration error rather than the background noise. In Fig. \ref{fig:int_maps_1} and \ref{fig:int_maps_2}, we present the radio maps after the subtraction and masking of background or foreground sources. All galaxies show a distinct extent orthogonal to the galactic disk. Comparing the extent of the outermost (3$\sigma$) contour, we detect larger extents in the HBA maps compared to the L-band contours (for NGC~3432, this comparison is more challenging as the overall structure of the contours is more complex compared to the other galaxies).  Of course, comparing the outermost contours is not a perfect way to determine the extent of a radio halo, as this measurement is limited only by the sensitivity of the particular data set.  However, one can already assume that this larger extent in the lower frequency data set reflects the ageing process of the CREs. Another interesting point to discuss is that the size of the individual radio halos varies strongly between the galaxies when compared to the galaxy's major axis. This is best observed in the low-frequency maps. The physical extents of the galactic disk and the halo based on the $3\sigma$-contours of the HBA maps are listed in Table \ref{tab:flux}.  While the radio halo of NGC~4013 shows only a small extent in $z$ direction, the radio halos of NGC~891, NGC~3432, and NGC~4157 reach already far up into the galactic halo. The most extreme case in this sample is certainly NGC~4631. Here, the extent in $z$-direction almost exceeds the galaxy's major axis. Such a comparison between the galaxies can be made because all LOFAR maps have a similar background noise (cf. Table \ref{tab:flux}). At this point, we can conclude that the dominating CRE transport mechanisms of the galaxies studied here are different. In a previous study, \cite{2019A&A...632A..13S} have shown that the intensity and SPIX profiles of NGC~4013 might best be modelled with a diffusion dominated CRE transport, which might result in a smaller vertical extent of the galactic halo.
  
\subsection{Thermal \& Non-Thermal Emission}
Here we present the result of the thermal emission correction, using the mixture method approach described in Sect. \ref{sec:method}. Table \ref{tab:therm flux} lists the integrated thermal flux densities integrated over defined aperture as well as the fraction of non-thermal emission (NTF), which is the remaining synchrotron radiation. As expected, the contribution of thermal emission to the total radio emission decreases when lower frequencies are reached.
\input{tab_thermal_emission}
As an integrated value, the thermal emission in the L-band is on the order of magnitude of the uncertainty due to the calibration error. However, the importance of the thermal emission changes when looking at the resolved fraction of thermal emission. To keep this work concise, we refer the reader to App. \ref{sec:app_tf_maps} for the thermal emission and thermal fraction maps.  In the thermal fraction maps, one can see that both the intensity of the thermal emission and the fraction of thermal emission in the overall radio emission decrease strongly when leaving the galactic disk. In L-band the thermal emission constitutes $\sim30\%$ of the overall observed radio emission in the galactic disk (for NGC~4631, the thermal contribution reaches even up to $\sim40\%$ in the galactic disk). However, the thermal fraction in the HBA data is much smaller. Here, the thermal emission normally does not exceed 10\% of the total emission (only NGC~4631 shows some regions where the thermal fraction reaches 15\%). Considering the overall calibration error of 10\% in the LoTSS data, the effect of thermal emission does not seem to be significant at the observed frequency. Therefore, we decided to correct only the L-band data for thermal emission in the further analysis.

\subsection{Spectral Index}
\label{sec:spectral_maps}
In Fig. \ref{fig:si_maps}, we present the non-thermal SPIX maps\footnote{Custom colour maps can be downloaded at: \url{https://github.com/mlarichardson/CosmosCanvas/}} for all analysed galaxies. As mentioned before, only L-band has been corrected for thermal emission, since the thermal contamination in the HBA is not significant when considering calibration errors. Overall, the galaxies show a common trend. In the galactic plane, we detect a non-thermal SPIX of approximately -0.5. Similar values have been reported in studies of NGC~3556 \citep{2019A&A...622A...9M}, NGC~5775 \citep{2022MNRAS.509..658H}, and previous studies of NGC~891 \citep{2019A&A...632A..12S}, NGC~4013 \citep{2019A&A...632A..13S} and NGC~4631\citep{2019A&A...632A..10M,2022MNRAS.511.3150V}. NGC~4631 shows in the outer part of the galactic disk similar SPIX values compared to the other galaxies of our sample, but deviates from the other galaxies in the centre. Here, we report a SPIX of approximately -0.4. This area coincides with the region of highest star formation and thermal radio emission (cf. bottom left map in Fig. \ref{fig:int_maps_2}). In Sec. \ref{sec:dis_si}, these results will be discussed in more detail. 

NGC~891, NGC~4157, and NGC~4631 show a strong and uniform gradient (declining SPIX) from the galactic disk into the halo. Such a gradient can be attributed to the ageing of CREs. In NGC~4013, we also detect a gradient in the non-thremal SPIX. However, we find an extended region in the centre of the galaxy with very flat SPIX values ($\alpha_{\mathrm{nth}} < -0.7$) and the gradient in the outer regions seems to be less prominent. The case of NGC~3432 is much less ordered. A gradient in the SPIX is not clearly visible and and only further analysis will show if CRE ageing is observable in this galaxy. However, one could have expected such a confused SPIX map, considering the fact that NGC~3432 is closely interacting with the dwarf galaxy UGC~5983\footnote{UGC~5983 is located at 10h52m16.749s +36d35m40.24s (coordinates taken from NASA NED) but does not show radio continuum emission.}. Therefore the dynamics of the galaxy might have disturbed the propagation of the CREs.

Another process that could explain the non-detection of a gradient in the spectral index is CR diffusion with a very strong energy dependence ($\mu=1$)\footnote{In this paper, the energy dependence of the CR diffusion is described as $D(E) = D_0\cdot(E/\mathrm{GeV})^\mu$}. The theoretical implications for a $\mu=1$ energy dependence of the CR diffusion is discussed in \cite{1974Ap&SS..29..305B}. This transport type would lead to a homogeneous non-thermal SPIX map, as the effects of CR transport and CR ageing would counterbalance each other. However, a $\mu > 0.7$ would require extreme conditions of the ISM turbulence or CR energies of about $10^4$\,GeV \citep{2021ApJ...923...53L}. By comparing their models to radio data, \citet{1974Ap&SS..29..305B} obtain a limit of $\mu < 0.4$ for the energy dependence of the CR diffusion. \citet{2012PhRvL.109f1101B} report an energy dependence of $D(E) \propto E^{0.7}$ for the energy range $10\,\mathrm{GeV} < E <200\,\mathrm{GeV}$\footnote{The energy dependence of the CR diffusion strongly depends on the observed energy range.}. For energies slightly below 10\,GeV, recent studies point to an even weaker ($\mu \sim 0.3$) energy dependence or no energy dependence of the CR diffusion at all \citep{2021ApJ...923...53L,2022arXiv220611670D}.
\section{Vertical Intensity Profiles}
\label{sec:profiles}
\subsection{Box Integration \& Scale Height Fitting}
\begin{figure}
    \centering
    \includegraphics[width=0.95\linewidth]{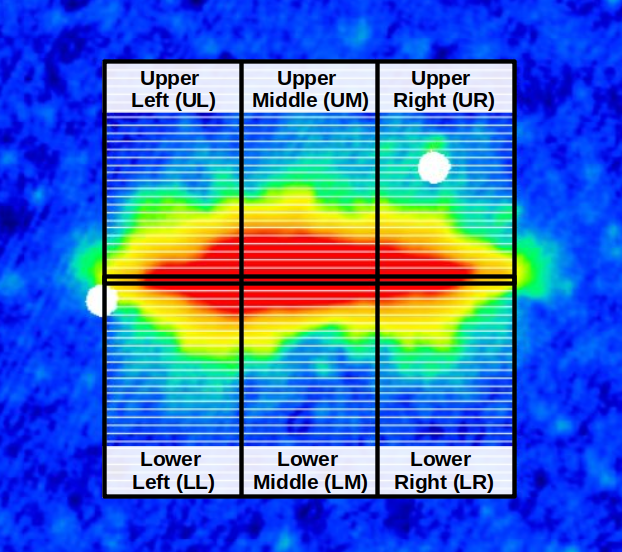}
    \caption{Strip setup on the HBA map of NGC~891. The map has been point source subtracted and masked. The six strips (UL, UM, UR, LW, LM, LR) are indicated with thick black lines. A similar setup is used on all galaxies during box integration with \texttt{NOD3}. Central boxes are part of the lower as well as the upper strips. Left strips have a positive offset, right strips have a negative one. Individual boxes (thin white lines) have a height of 10\arcsec and a width of 177\arcsec (as listed in Table \ref{tab:box_setup}).}
    \label{fig:stripe_setup}
\end{figure}

To trace the synchrotron emission further into the galactic halo, we average horizontally over large areas of the galaxy. In the literature, there is some variety in the number of strips into which the integration areas are divided. While \citet{2022MNRAS.509..658H} split up the galaxy into quadrants by choosing two vertical strips and splitting those again into northern and southern strips, there are also more detailed approaches (e.g.: \citet{2019A&A...632A..13S}: five strips, \citet{2019A&A...632A..12S}: five and seven strips). The number of strips used is obviously dependent on the spatial extend of the target compared to the spatial resolution of the observation, and less strips makes it possible to follow the galaxy's profile a bit further. We place three strips on each galaxy (left, middle, right) and analyse the upper and lower sections separately. Here, the idea is to average as much data as possible while accounting for the different conditions in the galaxy regions (e.g. stronger magnetic field in the centre of the galaxy; localised star formation regions, etc.). Separating the central part of the galaxy from the outer regions is very valuable for our study as we know from our Galaxy that AGN activity can heavily influence the galactic halo (e.g. Fermi-Bubbles \citep{2010ApJ...724.1044S}). Box integration is performed using the \texttt{BoxModels} method of \texttt{NOD3}\footnote{\url{https://gitlab.mpifr-bonn.mpg.de/peter/NOD3}} \citep{2017ascl.soft11024M,2017A&A...606A..41M}. We use the 3$\sigma$ contour of the HBA map as outline for defining the width of the boxes. The box setup for all galaxies is described in Table \ref{tab:box_setup}. Concerning the box height, we follow the approach of \citet{2018A&A...611A..72K} and \citet{2019A&A...632A..13S} and set it to half of the beam width. As an example, we show the box setup for NGC~891 on the HBA map in Fig. \ref{fig:stripe_setup}.
\input{tab_box_setup}

In order to derive reliable scale heights for each strip, we fit each strip with four models of decreasing complexity and evaluate the Akaike information criterion (AIC) \citep{akaike1998information} to select the best-fitting model while accounting for the different model complexities. As the number of data points are quite limited per strip, we apply the small sample correction to the AIC (AICc). For models that are generated by least square fitting routines, the AICc is computed in the following way:
\begin{equation}
    \mathrm{AICc} = 2k + n \cdot \mathrm{ln}(\mathrm{RSS})+\frac{2k^2+2k}{n-k-1}.
\end{equation}
Here $k$ denotes the number of model parameters, $n$ the number of data points and RSS the residual sum of squares.
Overall, we fit exponential profiles, however we need to account for the radio beam. Therefore, the models that are actually fitted are convolutions of exponential profiles with Gaussian kernels \citep[cf.][]{1995A&A...302..691D}.
Intrinsic exponential profiles with an intensity $w_0$ and a scale height  $z_0$:
\begin{equation}
    \mathrm{w}(z)=w_0 \mathrm{exp}\left(\frac{-z}{z_0}\right)
\end{equation}
need to be convoluted with a Gaussian kernel with a standard deviation $\sigma$:
\begin{equation}
    \mathrm{g}(z) = \frac{1}{\sqrt{2\pi \sigma^2}}\mathrm{exp}\left(\frac{-z^2}{2\sigma^2}\right)
\end{equation}
to account for the limited resolution of the images. We tried to fit different profiles, including Gaussian profiles, but the  analysed intensity profiles are overall better fitted with exponential functions. We use those fits to get the scale heights used in the following analysis. The most complex model we fit to the vertical profiles is a two-component exponential profile with intensities $w_0$, $w_1$ and scale heights $z_0$, $z_1$:
\begin{align}
\begin{split}
    \mathrm{w_{dual}}(z) =  \frac{w_0}{2}\mathrm{exp}\left(\frac{{-z}^2}{2\sigma^2}\right)   &\left[\mathrm{exp}\left(\left( \porfileargone\right)^2\right) \mathrm{erfc} \left( \porfileargone \right)\right.\\
       +&\left. \mathrm{exp}\left( \left(\porfileargtwo \right)^2\right) \mathrm{erfc} \left( \porfileargtwo \right) \right]
    \\
     + \frac{w_1}{2}\mathrm{exp}\left(\frac{{-z}^2}{2\sigma^2}\right) 
    & \left[\mathrm{exp}\left(\left( \porfileargthree\right)^2\right) \mathrm{erfc} \left( \porfileargthree \right)  \right.
    \\
      + & \left. \mathrm{exp}\left(\left( \porfileargfour \right)^2\right) \mathrm{erfc} \left( \porfileargfour \right) \right].
\end{split}
\end{align}
\label{eq:2comp}
Here, erfc is the complementary error function. With such a two-component model, one can try to decompose galactic disks and halos. However, such a distinction is not always possible. Furthermore, the intensities $w_0$ and $w_1$ as well as the scale heights $z_0$ and $z_1$ in the two-component model are degenerate. This can result in very large uncertainties in the model parameters, especially when fitting the model to only a limited number of noisy data points. When the estimated uncertainty of a fitted model parameter exceeds the parameter value, we reduce the complexity of the model to obtain well-constrained fitting parameters by using a one-component exponential profile.
\begin{align}
\begin{split}
     \mathrm{w}_{\mathrm{single}}(z) =  \frac{w_0}{2}\mathrm{exp}\left(\frac{{-z}^2}{2\sigma^2}\right)   &\left[\mathrm{exp}\left( \left( \porfileargone\right)^2\right) \mathrm{erfc} \left( \porfileargone \right)\right.\\
       +&\left. \mathrm{exp}\left(\left( \porfileargtwo \right)^2\right) \mathrm{erfc} \left( \porfileargtwo \right) \right].
\end{split}
\end{align}
\label{eq:1comp}
With these models, we can fit radio halo scale heights in each strip. We additionally allow for a shift in z-direction using the following coordinate transformation:
\begin{equation}
    z^\prime= z - Z_0.
\end{equation}
To summarise our scale height fitting process: Each strip is fitted with four models: A two-component exponential with shift in $z$-direction (two), a two-component exponential without shift in $z$-direction (two\_fo), a one-component exponential with shift in $z$-direction (one), and a one-component exponential without shift in $z$-direction (one\_fo). The model selection is based on the AICc and the model complexity is reduced if the uncertainty of the best fit parameter exceeds the parameter itself.

\subsection{Scale Heights}
\begin{figure*}
\centering
\begin{subfigure}{0.49\linewidth}
\centering
\includegraphics[width=1\linewidth]{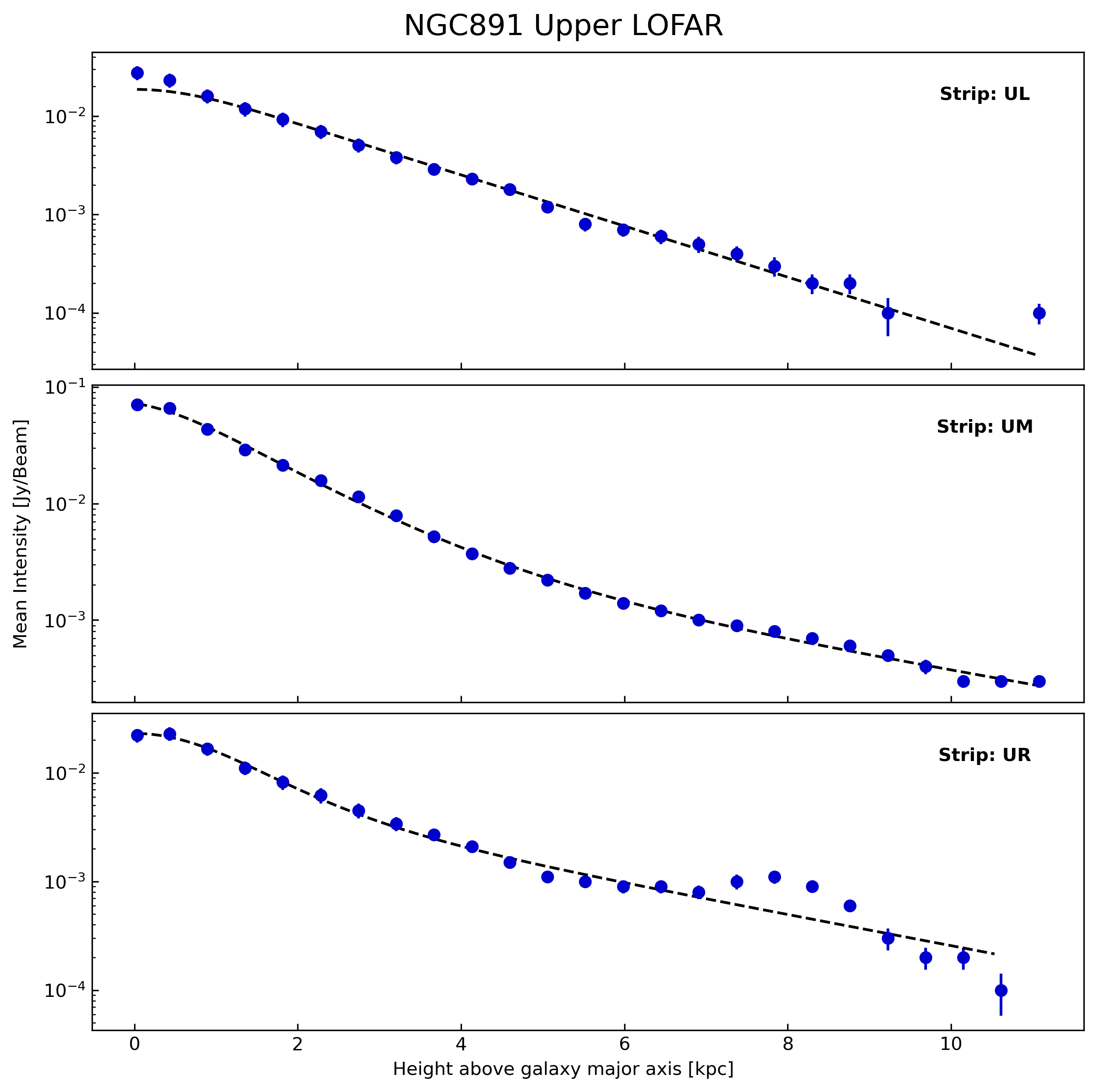}
\end{subfigure}
\hfill
\begin{subfigure}{0.49\linewidth}
\centering
\includegraphics[width=1\linewidth]{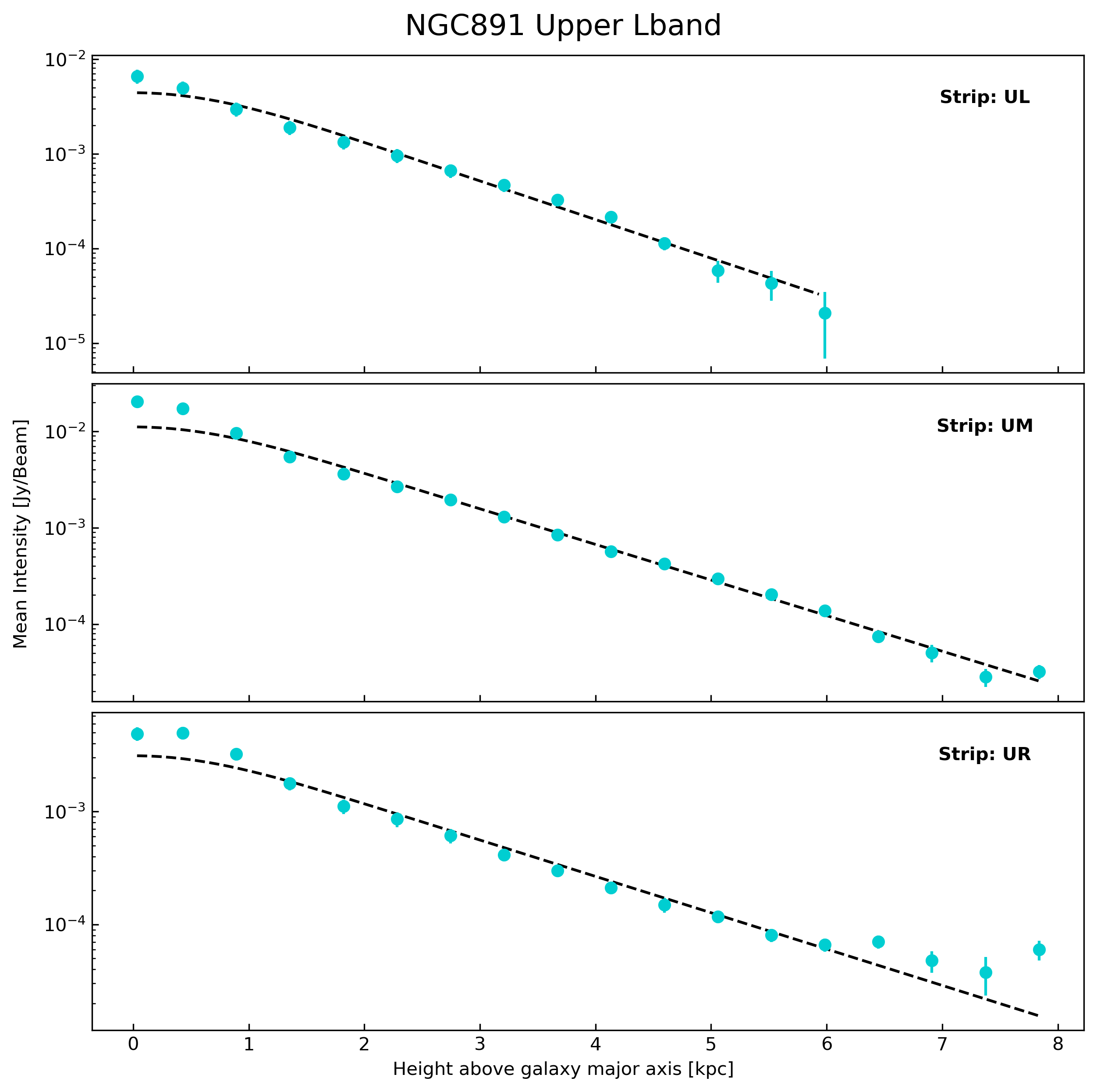}
\end{subfigure}
\\
\begin{subfigure}{0.49\linewidth}
\centering
\includegraphics[width=1\linewidth]{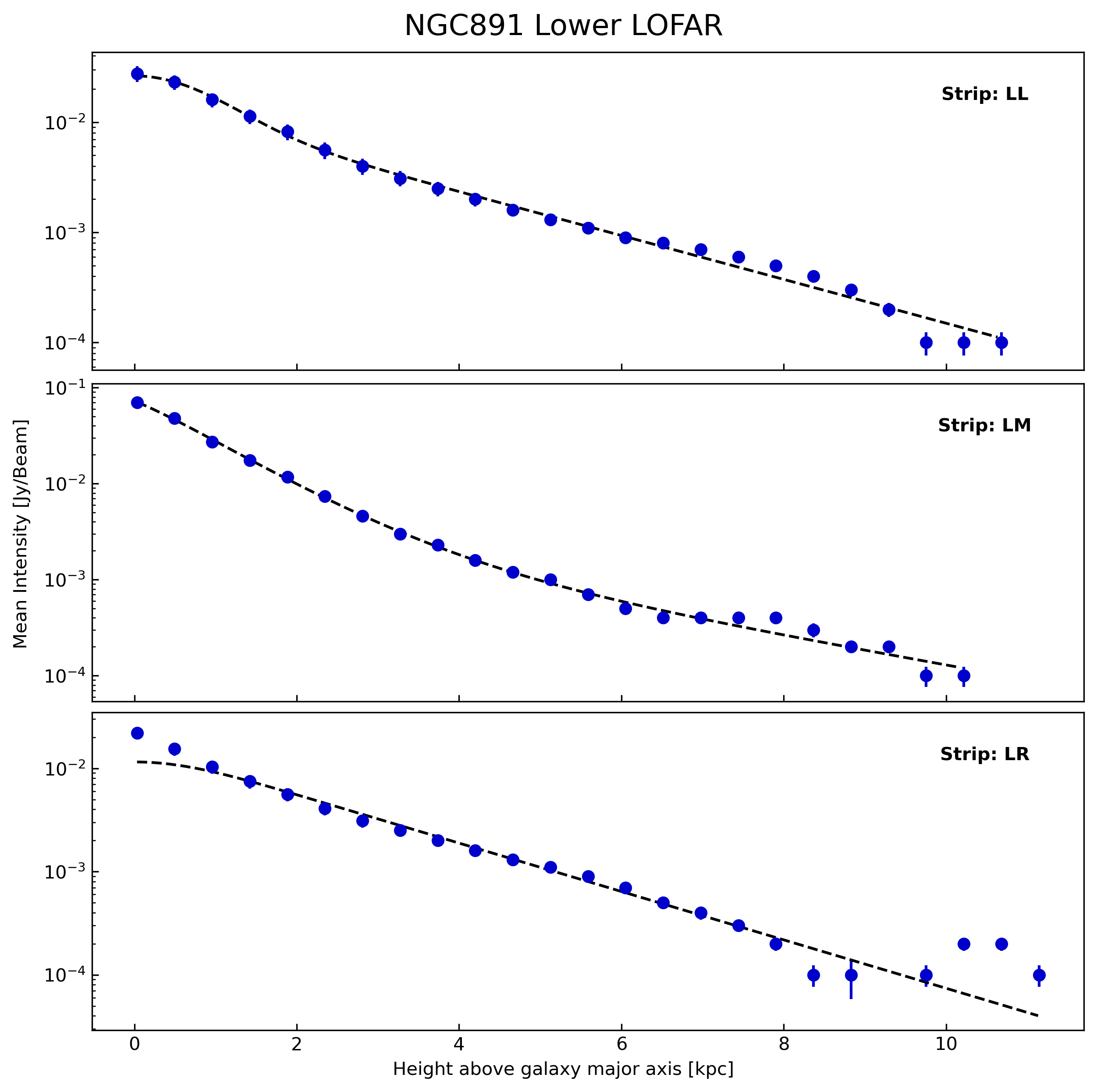}
\end{subfigure}
\hfill
\begin{subfigure}{0.49\linewidth}
\centering
\includegraphics[width=1\linewidth]{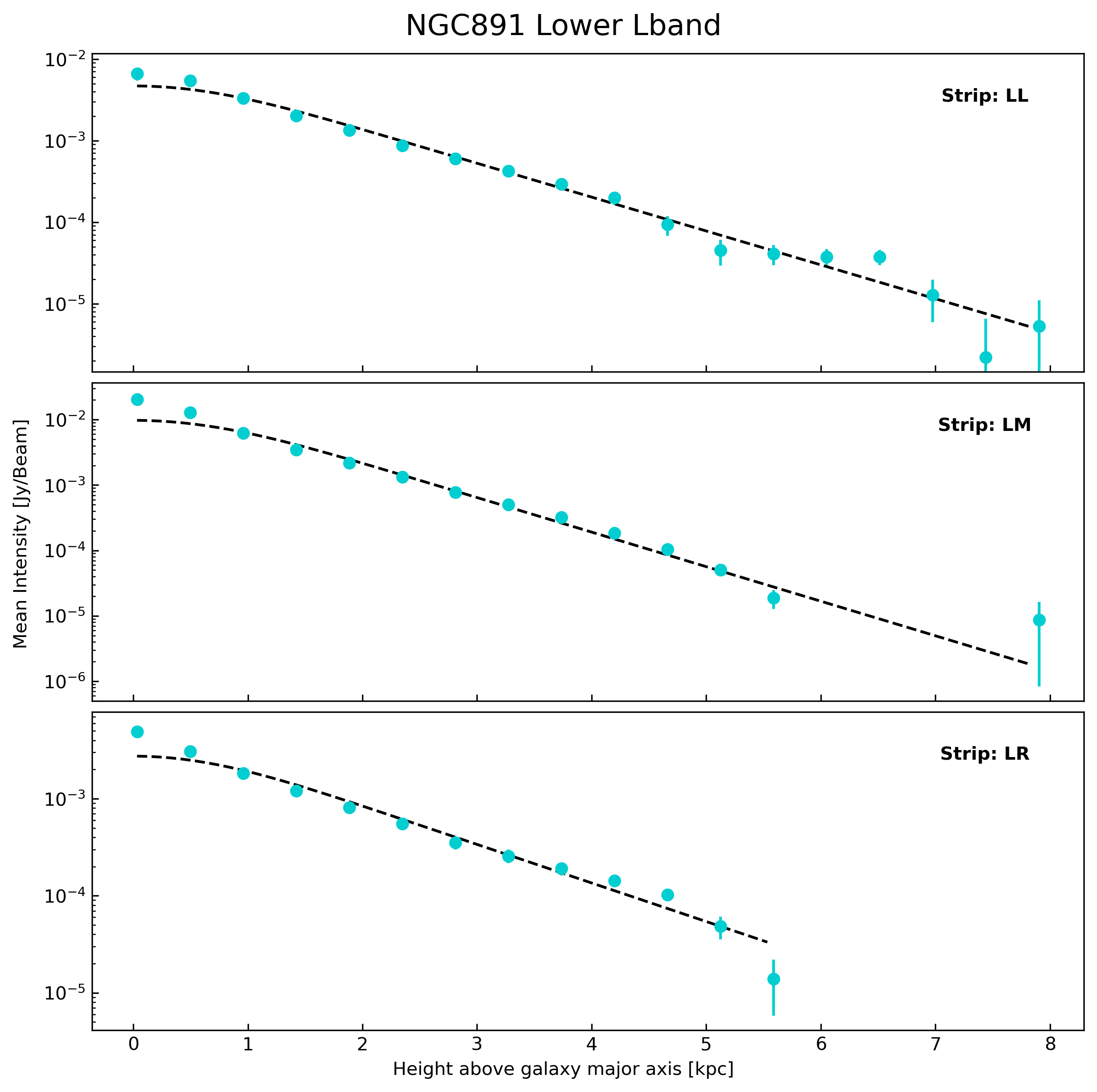}
\end{subfigure}
\caption{Vertical radio continuum intensity profiles in the three strips of NGC~891. Left panels show 144-MHz LOFAR data right panels show $1.5$-GHz JVLA data. Best-fitting double exponential profiles are shown as dashed lines. 
}
\label{fig:int_prof_891}
\end{figure*}

To keep this paper concise, we refer the reader to App. \ref{sec:app_int_prof}, where we list the fitted model parameters and present the fitted models for each strip. The scale heights averaged per galaxy are listed in Table \ref{tab:scale_summary}. As an example, we  present the fitted profiles of NGC~891 (Fig. \ref{fig:int_prof_891}) and the corresponding fitting parameters (Table \ref{tab:int_prof_891}).

\begin{table*}
\centering
\caption{Parameters from box integration model fitting of NGC~891: frequency of fitted data, model identifier, strip identifier, peak intensity and scale height of the first exponential component, peak intensity and scale height of the second exponential component (if applicable), offset in z-direction (if applicable), reduced χ-square. The corresponding profiles are shown in Fig. \ref{fig:int_prof_891}. A complete list of all galaxies can be found in Table \ref{tab:profile_param}.}
\label{tab:int_prof_891}
\begin{tabular}{lrlllllllr}
\hline
\hline
 $\nu$ & Model & ID & $w_0$ & $z_0$ & $w_1$ & $z_1$ & $Z_0$ & $\chi^2_\nu$\\
 & [MHz] & & & [\textmu Jy\,Beam\textsuperscript{-1}] & [kpc] & [\textmu Jy\,Beam\textsuperscript{-1}] & [kpc] & [kpc] &\\
\toprule
 1500 & one\_fo & LR &   4.3$\pm$0.3 &   1.09$\pm$0.04 &             - &             - &              - &  2.55 \\
 144 & one\_fo & LR &  15.2$\pm$1.0 &   1.85$\pm$0.04 &             - &             - &              - &  4.02 \\
 1500 & one\_fo & UR &   4.5$\pm$0.3 &   1.35$\pm$0.04 &             - &             - &              - &  3.50 \\
 144 & two\_fo & UR &      32$\pm$5 &   0.77$\pm$0.15 &   6.7$\pm$1.4 & 3.04$\pm$0.26 &              - &  3.41 \\
 1500 & one\_fo & LM &  17.2$\pm$0.8 & 0.822$\pm$0.016 &             - &             - &              - &  7.41 \\
 144 &     two & LM &   188$\pm$113 &   0.87$\pm$0.08 &   5.8$\pm$2.1 &   2.9$\pm$0.4 &   -1.0$\pm$0.7 &  1.01 \\
 1500 & one\_fo & UM &  16.9$\pm$0.6 & 1.176$\pm$0.014 &             - &             - &              - &  5.88 \\
 144 &     two & UM &    111$\pm$18 &   1.05$\pm$0.10 &   6.6$\pm$2.3 &   3.6$\pm$0.5 & -0.33$\pm$0.27 &  0.50 \\
 1500 & one\_fo & LL &   7.4$\pm$0.6 &   1.05$\pm$0.03 &             - &             - &              - &  1.21 \\
 144 & two\_fo & LL &     43$\pm$22 &   0.37$\pm$0.21 &  13.9$\pm$1.6 & 2.18$\pm$0.08 &              - &  1.22 \\
 1500 & one\_fo & UL &   6.9$\pm$0.6 &   1.07$\pm$0.04 &             - &             - &              - &  1.05 \\
 144 & one\_fo & UL &  25.4$\pm$1.9 &   1.67$\pm$0.05 &             - &             - &              - &  1.05 \\
\bottomrule
\end{tabular}
\end{table*}
\input{tab_scale_summary}
The intensity profile of NGC~891 has already been analysed in the previous works by \citet{2018A&A...615A..98M} (MU18) and \citet{2019A&A...632A..12S}. While \citet{2019A&A...632A..12S} analysed a combination of VLA C-band and L-band data, MU18 also analysed the galaxy with help of data from the LOFAR telescope. Both studies extract intensity profiles up to a height of $\sim$3.5\,kpc. With the data used in this work, we can follow the intensity profiles much farther into the halo. The newly cleaned \changes\ L-band allows us to extract accurate intensity profiles up to $\sim$8\,kpc for the upper half and to 6\,kpc for the lower half of NGC~891. Therefore, we can nearly double the extent of the analysed halo in the L-band data. The improvement in data quality is even more pronounced for the low-frequency data set. Here, we can follow the intensity profile to a height of $\sim$11\,kpc and therefore more than triple the extent of the analysed halo. When averaging over strips where the data allowed to fit two exponential components (four out of six strips for HBA), we derive mean scale heights $z_{\mathrm{disk}}=0.76\pm0.25$\,kpc and $z_{\mathrm{halo}}=2.9\pm0.5$\,kpc for the galactic disk and the halo of NGC~891 in the HBA data. Therefore, the derived scale height of the disk in this work is slightly larger than that derived by MU18  ($z_{\mathrm{disk}}=0.32\pm0.08$\,kpc; $z_{\mathrm{halo}}=2.3\pm0.7$\,kpc), while the halo scale heights are similar. Both studies correct for the effect of smoothing by the radio beam during the fitting routine. Nevertheless, especially the difference in the measurement of the disk scale height might result from different resolutions (12\arcsec\ in MU18, 20\arcsec\ in this work). Due to the larger extent of the strips analysed in this work, the derived halo scale heights might be more reliable compared to previous studies. Judging from the AICc model assessment, none of L-band data strips shows sufficient complexity to be fit with a two-component model. Averaging over all L-band strips that were fitted with one component, we derive a mean scale height of NGC~891 of $z_{\mathrm{1comp}}=1.09\pm0.16$\,kpc.

Scale heights of NGC~4631 have been analysed before, by \citet{2019A&A...632A..10M} (MP19) in the 6\,GHz and 1.5\,GHz bands. With the strips that were fitted with two components in our analysis, we find larger radio halo scale heights than MP19. However, both measurements have fairly large uncertainties ($z_{\mathrm{halo}}=2.1 \pm 0.3$\,kpc (this work),  $z_{\mathrm{halo}}=1.75 \pm 0.27$\,kpc (MP19)). In the HBA data, we report a mean scale height of the galactic halo of $z_{\mathrm{halo}}=4.9 \pm 0.4$\,kpc. This is more than double the scale height compared to the higher frequency data. While MP19 extracted profiles of $\sim6.6$\,kpc reaching from south to north of the galaxy, we follow the intensity profile $\sim8-10$\,kpc in either direction of the galactic disk, which most probably increases the reliability of our analysis. In the HBA data, the extent of the galactic halo can be analysed on unprecedented scales. Here, we can follow the intensity profile to a height of 14\,kpc\footnote{\citet{2022MNRAS.509..658H} reached similar extents in their analysis of NGC~5775 but had a lower physical resolution of 2.2\,kpc.}. 

Comparing our results of NGC~4013 to \citet{2019A&A...632A..13S} (ST19) is more difficult than in the case of NGC~891 or NGC~4631, since their profile fitting is based on Gaussian instead of exponential profiles. ST19 report disk (halo) scale heights of $0.36\pm0.05$\,kpc ($2.0\pm0.1$\,kpc) in VLA L-band and $0.47\pm0.10$\,kpc ($3.1\pm0.3$\,kpc) in LOFAR HBA. Our model comparison based on the AICc shows again that the strips are equally well fitted with just a single exponential fit rather than a two-component approach. The radio halo of NGC~4013 will be discussed in more detail in Sec. \ref{sec:discussion:N4013}.  

Scale heights of NGC~3432 and NGC~4157 have been fitted with two-component exponential functions in C-band and L-band by \citet{2018A&A...611A..72K} (L-band: NGC~3432:  $z_{\mathrm{halo}}=0.92\pm0.25$\,kpc, NGC~4157: $z_{\mathrm{halo}}=1.08\pm0.07$\,kpc)\footnote{The authors node that their derived disk scale heights are limited in their reliability, because of the resolution of the observations}. This is in agreement with the values we find, considering the fact that we only fit one-component models for NGC~3432.

\begin{figure*}
 \centering
\includegraphics[width=0.8\linewidth]{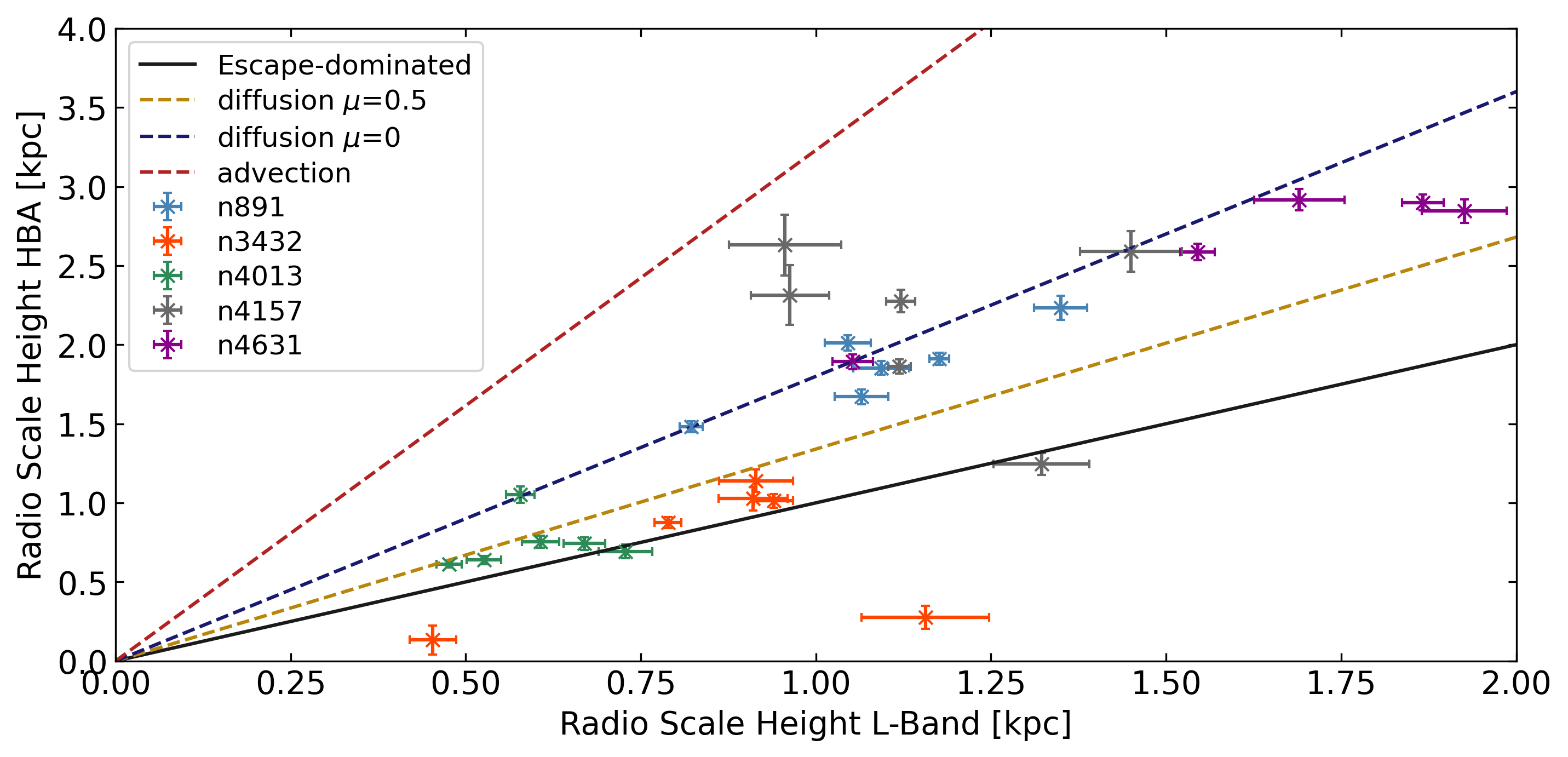}
\caption{Comparison of fitted non-thermal radio scale heights in VLA L-band and LOFAR HBA. For each strip of each galaxy the radio scale height in the HBA data is plotted against its L-band counterpart. Individual galaxies are colour coded as indicated in the legend. The solid black line represents the identity relation which one would expect in escape-dominated halos. Dashed lines indicate expected scale height ratios for loss-dominated halos: dark-blue: energy independent diffusion; gold: energy dependent diffusion; red: advection.}
\label{fig:scale_height_comp}
\end{figure*}

Now, we analyse the scale heights of all galaxies as a sample. The model comparison with the AICc shows that most strips can be fitted, with the chosen box integration setup (spatial resolution and box size), with one-component models. For the following analysis we use results from the one-component fits only, since mixing scale heights of one- and two-component models is not advisable at all (also strips that showed enough complexity to be fitted with two components are re-fitted with one-component models for this part of the analysis).  In Fig. \ref{fig:scale_height_comp} we compare the radio scale heights of each strip in the L-band and HBA data. In their analysis of 13 galaxies in L- and C-band, \citet{2018A&A...611A..72K} report overall similar scale heights in both spectral bands. This picture changes drastically when comparing L-band to HBA profiles. Here, one can see a bi-modality in the data. Galaxies that showed a strong and uniform gradient in the non-thermal SPIX maps (NGC~891, 4157, and 4631) have much larger scale heights in HBA than in L-band and galaxies without such a gradient (NGC~3432) or with a less prominent SPIX gradient (NGC~4013) lie much closer to the one-to-one line. The relative scattering of the individual strips of a galaxy is largest in NGC~4157. Comparing the HBA and L-band contour lines of NGC~4157 in Fig. \ref{fig:int_maps_2}, one can notice that not only the minor axis, but also the major axis is larger in HBA than in L-band. Therefore, the comparison of the box profiles might show a larger variety.Assuming a rather symmetric galactic magnetic field structure, one interpretation explaining the scatter in the measured scale-height ratios of a given galaxy could also be a space-dependent CR transport mechanism \citep[e.g. space-dependent diffusion coefficient][]{2018PhRvL.121b1102E}. In the literature, there are also studies that have observed larger scale heights in the low-frequency data. \citet{2018A&A...615A..98M} report a scale height ratio of $1.7\pm0.3$ for NGC~891 and \citet{2022MNRAS.509..658H} find a ratio of $1.2\pm0.3$ in NGC~5775 when comparing L-band and HBA scale heights.  We report the mean scale height ratio (average of the ratios of individual strips) for each galaxy in Table~\ref{tab:scale_ratio}. These ratios reflect the results from the non-thermal SPIX maps. Those galaxies that show a clear gradient in the SPIX maps also have significantly larger scale heights in HBA than in L-band, as expected.
\begin{table}
    \centering
    \caption{One component radio scale height ratios of HBA and L-band profiles. Uncertainties represent the spread of the different strips of a galaxy.}
    \label{tab:scale_ratio}
    \begin{tabular}{l r r}
    \hline\hline
    Galaxy  & HBA/L-band Ratio & SPIX-gradient\\
    \hline
    NGC~891  & 1.71$\pm$0.12  & strong, uniform\\
    NGC~3432 & 0.8$\pm$0.4  & no / disordered\\
    NGC~4013 & 1.27$\pm$0.27& less prominent \\
    NGC~4157 & 1.9$\pm$0.6 & strong, uniform\\
    NGC~4631 & 1.65$\pm$0.11 & strong, uniform\\
    \hline
    \end{tabular}
\end{table}

Following the argumentation of \citet{2018A&A...611A..72K}, the observed scale height ratio can predict the dominating CR transport mechanism in the galaxy. First of all, loss-dominated and escape dominated halos need to be distinguished. In a loss-dominated halo, the synchrotron emission falls below the detection limit because the CREs have lost so much energy through radiation that they can no longer be detected. In escape-dominated halos, the synchrotron radiation drops below the detection limit because of the decreasing number of CREs and lower magnetic field strengths. Of course, the observed expansion of the galaxy halo is actually limited by a combination of both effects, which makes it difficult to argue on the basis of scale heights alone. Nevertheless, for a loss-dominated halo, \citet{2018A&A...611A..72K} report the following frequency dependencies on the advection length l\textsubscript{con} and diffusion length l\textsubscript{diff} for diffusion with a non-energy dependent diffusion coefficient D ($\mu$=0), energy-dependent diffusion with D$\propto$E\textsuperscript{0.5} ($\mu$=0.5) and advection, which can than be translated into expected scale height ratios r for the frequencies analysed in this paper (144\,MHz \& 1.5\,GHz). We list the derived proportionality and expected ration in Table \ref{tab:transport}. 

\begin{table}
    \centering
    \caption{Derived proportionalities of the diffusion- / advection-lengths and the observed radio frequency (as derived by \citet{2018A&A...611A..72K} and expected scale height ratios for HBA and L-band.}
    \label{tab:transport}
    \begin{tabular}{lrr}
    \hline\hline
         Transport & Proportionality & Ratio\\
         \hline
         \multicolumn{3}{l}{\textbf{Loss-dominated:}}\\
         Diffusion ($\mu=0$)    & $\mathrm{l_{diff}\propto\nu^{-1/4}}$ & 1.80 \\
         Diffusion ($\mu=0.5$)  & $\mathrm{l_{diff}\propto\nu^{-1/8}}$ & 1.34 \\
         Advection              & $\mathrm{l_{con}\propto\nu^{-1/2}}$  & 3.23 \\
         \hline
         \multicolumn{3}{l}{\textbf{Escape-dominated:}}\\
         Advection              & $\mathrm{l_{con}=const}$              & 1.00\\
         \hline
    \end{tabular}
    
\end{table}

We show these relations in Fig. \ref{fig:scale_height_comp} to compare them to the measured scale heights. Most of the scale heights measured in NGC~3432 and NGC~4013 fall even below the energy-dependent diffusion line, while NGC~891 and NGC~4631 seem to be well represented by the $\mu$=0-diffusion transport. The aforementioned CRE escape process drives the measured scale ratios closer to the one-to-one relation. Because of that, \citet{2019A&A...622A...9M} find in the 1D CR transport modelling that advection models are the best-fitting transport mechanism in NGC~3556, although they find an HBA/L-band ratio of 1.8 in the halo, which would point to energy-independent diffusion if only the scale height ratio is considered. Therefore, it seems likely that the halo of NGC~3556 is not loss-dominated. 

\begin{figure*}
    \centering
    \begin{subfigure}{0.49\linewidth}
       \includegraphics[width=\hsize]{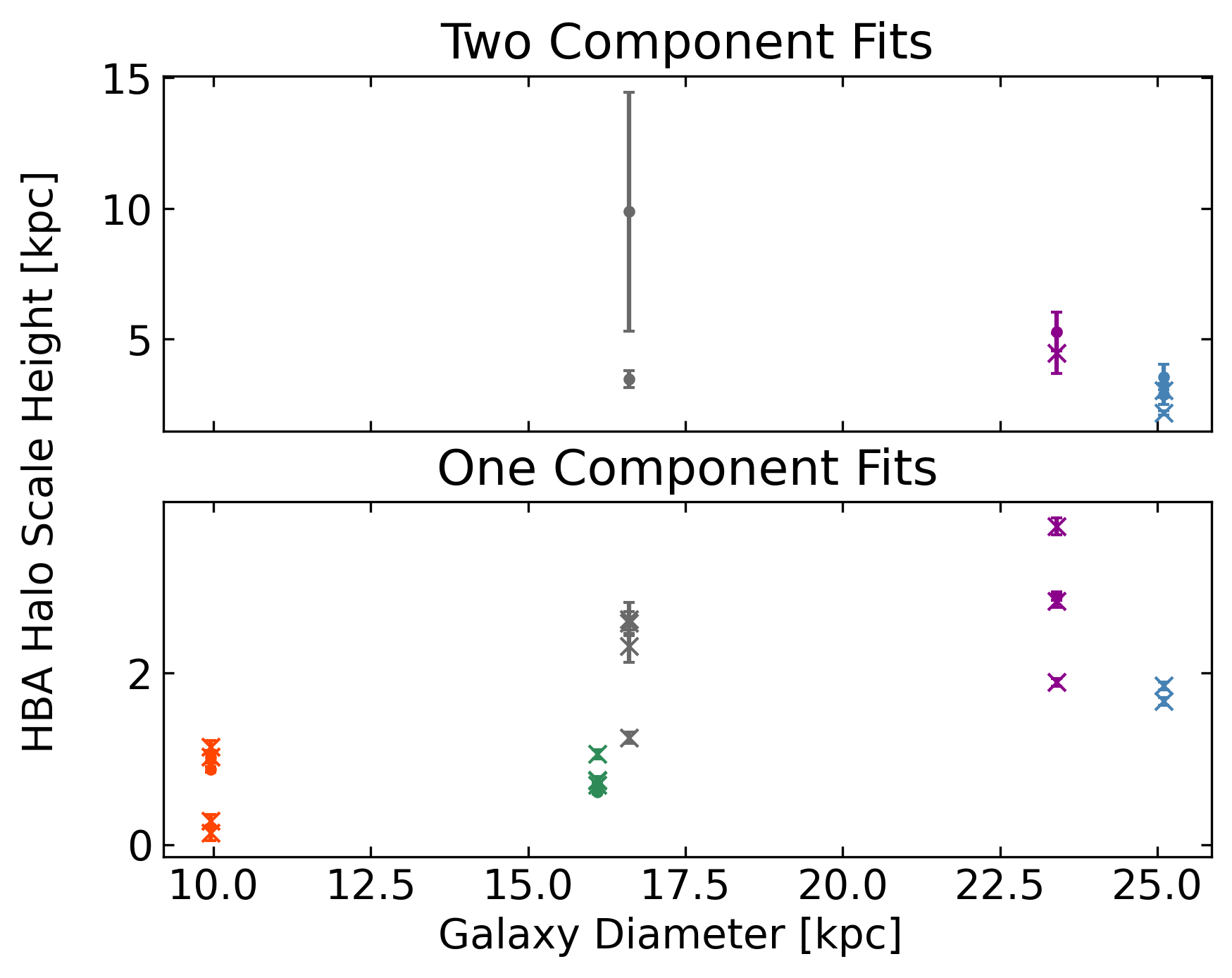}
    \end{subfigure}
   \begin{subfigure}{0.49\linewidth}
       \includegraphics[width=\hsize]{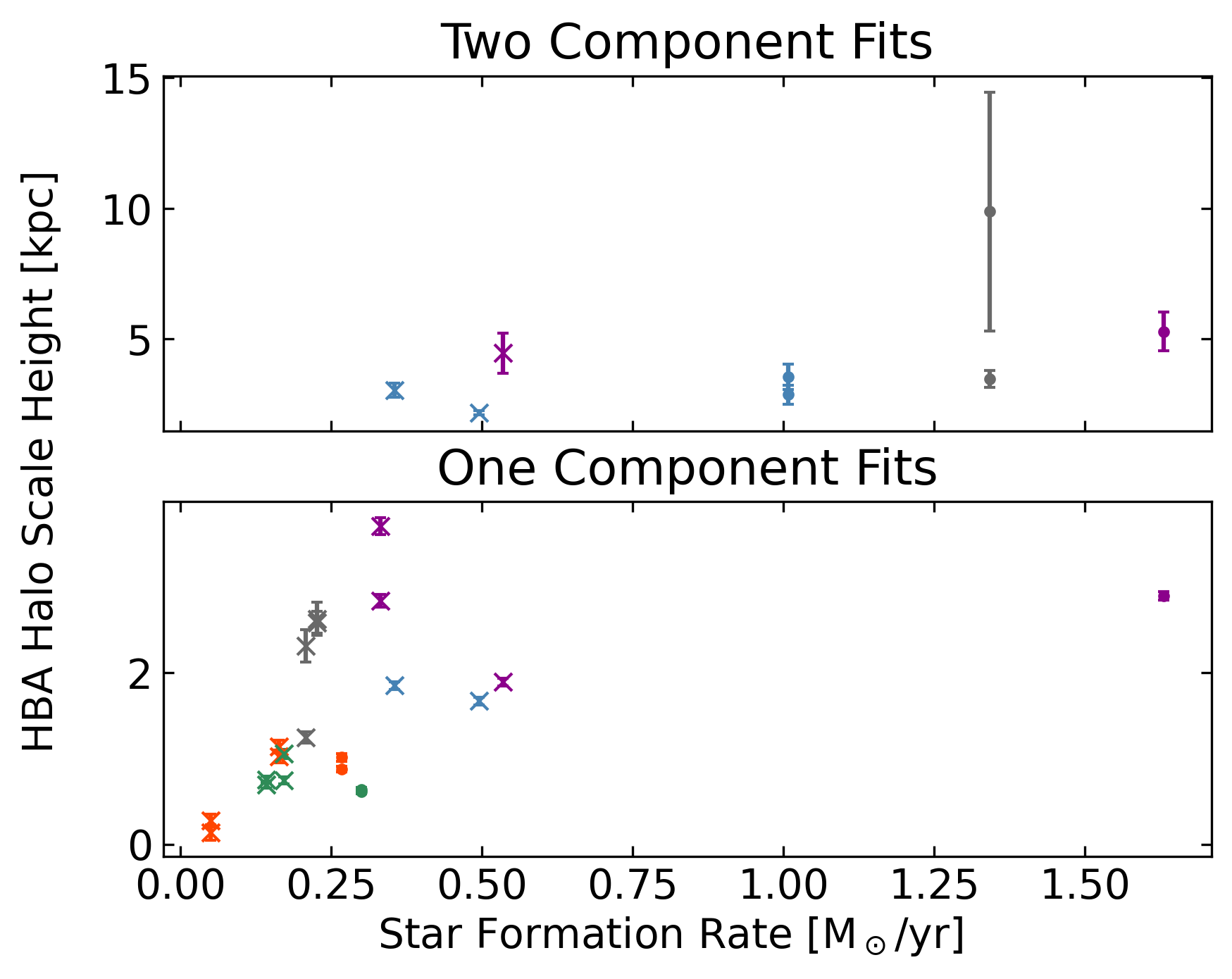}
    \end{subfigure}
   \caption{Individual radio halo scale heights plotted against galaxy diameter (left), and star formation within the individual strip (right). Two-component and one-component fits are displayed separately. In the plots of the two-component fits, only the halo scale heights are shown. Dots represent the central strips and x-symbols the outer strips of each galaxy.  Colour coding: NGC~891 blue; NGC~3432 orange; NGC~4013 green; NGC~4157 grey; NGC~4631 magenta (same as Fig. \ref{fig:scale_height_comp}).}
   \label{fig:scale_height_phys}
   \end{figure*}

\citet{2018MNRAS.476..158H} performed a correlation analysis to find possible relations between the fitted radio scale heights and other measured properties such as SFR, SFR surface density, and the rotation velocity. They did not find a strong correlation between the radio scale height and any of these properties. In Fig. \ref{fig:scale_height_phys}, we present the fitted HBA radio scale heights for one- and two-component models separately in comparison to the diameter of the galactic disk, measured on resolution enhanced WISE 22\,\textmu m images, and the SFR within the individual strips. In the two-component plots, only the halo scale heights are shown, which neither show a clear trend with the galaxy diameter nor the SFR.

\begin{figure*}
    \centering
    \includegraphics[width=1\linewidth]{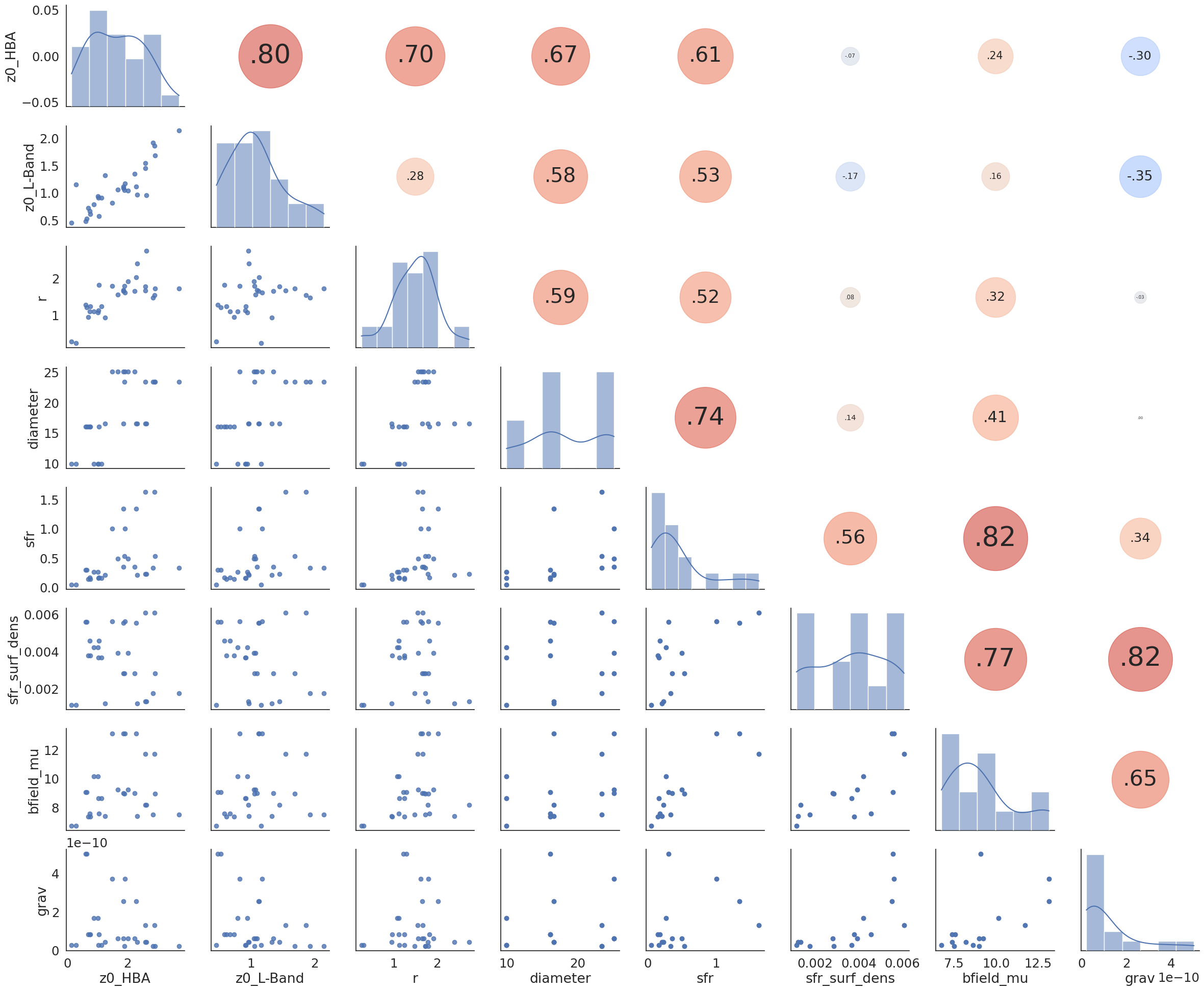}
    \caption{Correlation analysis of several physical properties in comparison to the fitted HBA and L-band radio scale heights (z\textsubscript{0, HBA}, z\textsubscript{0, L-band}) and the ratio of HBA and L-band (r). The main diagonal displays the distribution of z0 and r as well as the other analysed quantities (SFR, SFR surface density, and equipartition magnetic field strength) as histograms and kernel density estimates. Bottom: bivariate scatter plots for pairs of the analysed quantities with interpolation line. Top: Colour coded (blue: negative, red: positive) Spearman rank correlation coefficients.}
    \label{fig:corr_study}
\end{figure*}

For the one-component models, one might expect a positive trend for the galaxy diameter as well as the SFR. Therefore, we perform a more detailed correlation analysis where all strips have been fitted with one-component models in Fig. \ref{fig:corr_study}. Obviously there is a strong link between the HBA radio scale height, L-band scale height and the HBA/L-band ratio (r). However, here we are more interested in a link of the radio measurements of the halo with other quantities such as SFR or the magnetic field. The diameter of the galactic disk seems to be a quite good predictor for the measured HBA scale height (Spearman-r: 0.67, p-value: 5e-5)\footnote{It is important to note that the p-value for such small samples is not completely reliable. However, the measured correlation coefficients are large enough to be reliable and the trends are also visible in the scatter plots.} The SFR is equally strongly linked to $z_0$, however, the scatter plot does show outlier values with much larger SFRs than the average trend. We also checked if the specific star formation rate (sSFR) (based on the mass estimates in Table \ref{tab:basic_info}) is a better predictor but the sSFR does not perform better than the SFR. Therefore we just present the SFR values in Fig. \ref{fig:corr_study}.  We do not find significant correlations between z\textsubscript{0} and the SFR surface density nor the magnetic field strength. Overall, the L-band scale height (and therefore also the ratio r) is less strongly linked to the diameter and SFR compared to the HBA scale height.

The strongest correlations we find in this study are a) the link between the equipartition magnetic field strength and the SFR in the individual strips, and b) between the SFR surface density and the gravitational potential. A correlation between the SFR and the magnetic field strength has already been reported in studies of dwarf galaxies \citep{2011A&A...529A..94C}, the  KINGFISHER nearby galaxy sample \citep{2017ApJ...836..185T}, and nearby edge-on galaxies \citep{2018MNRAS.476..158H}.
However, the relation found in this study is most probably affected by the study setup. Meaning that we simply measure higher SFRs and magnetic field strengths in the centre of a galaxy, because of the radial dependence and not only because of a link between the SFR and the galactic magnetic field. The same is also most likely true for the link between the SFR surface density and the gravitational pull.

Concluding this section, we find that even the predictors that were found to be best in this study - measured SFR within a strip and the galaxy diameter - are not suited to reliably constrain the radio halo size. The formation of a galactic wind seems to be too complex to be predicted by a single parameter in such a small sample. We continue the analysis of galactic winds driven by stellar feedback by fitting the extracted intensity profiles with a 1D CR-transport model.

\section{Stellar Feedback Driven Winds}
\label{sec:CRwinds}
In this section, we extend the analysis of the extracted profiles in order to derive physical parameters of the galactic wind. The section is structured as follows: First, we introduce the models that are used to analyse the CR transport mechanisms in the galaxies and the input and output parameters for each galaxy. Further, we motivate the model choice for each galaxy and explain the exclusion of NGC~3432 from the further analysis. In Sec. \ref{sec:sub_cr_diff} the results of the diffusion model fitting of NGC~4013 are displayed. In the last subsection (Sec. \ref{sec:sub_cr_adv}), we present the results of the completed advection model fitting for NGC~891, NGC~4157, and NGC~4631.
\subsection{Cosmic Ray Transport Models}
The 1D cosmic ray transport modelling is performed with the SPectral INdex Numeral Analysis of K(c)osmic-ray Electron Radio-emission (\spinnaker) code \citep{2016MNRAS.458..332H,2018MNRAS.476..158H}. \spinnaker\ fits vertical intensity profiles of multiple frequencies and the resulting SPIX profiles to several wind models (diffusion or advection). The latest \spinnaker\ extension for advection-dominated galactic winds, the so called flux tube model, has been published by \citet{2022MNRAS.509..658H}. In the literature, several studies already made use of synchrotron radiation to analyse the cosmic ray transport of nearby edge-on galaxies \citep[cf.][]{2019A&A...632A..12S,2019A&A...622A...9M, 2019A&A...632A..13S}. Some of those studies analysed galaxies that are also part of the present work. However, as pointed out by \citet{2022MNRAS.509..658H}, the new flux tube model has several benefits compared to the previous approaches (e.g. magnetic and velocity scale heights are fitted simultaneously in the flux tube model, whereas they had to be fitted separately before). Therefore, we apply the advection flux tube model (accounting for advection losses) introduced by \citet{2022MNRAS.509..658H}. The model assumes an hyperboloidal tuneable flux tube, an isothermal wind, and exponentially declining gravitational acceleration for increasing heights above the galactic disk. Therefore, the cross sectional area $A$ of the flux tube above the disk is described as:
\begin{equation}
    A(z) = A_0 \left [1 + \left (\frac{z}{z_0}\right )^\beta \right ].
    \label{eq:flux_tube}
\end{equation}
The model results in a hyperboloid if the free parameter $\beta$ is set to $\beta = 0$. Further, \citet{2022MNRAS.509..658H} assume a magnetic field that declines with the midplane flow radius $r$ and the advection speed $v$:
\begin{equation}
    B(r,v) = B_0 \left (\frac{r}{r_0}\right )^{-1} \left ( \frac{v}{v_0}\right )^{-1}.
\end{equation}
Here, $B_0$\,[$\mu$G] describes the magnetic field strength in the galactic plane. For further details about the model, we refer to \citet[][Sec. 4.3 and App. A]{2022MNRAS.509..658H}. To describe the hyperbolodial flux tube, we introduce the base radius of the flux tube $R_0$, its scale height $z_0$ and the power law index that describes its opening angle $\beta$. With this model, we fit the vertical height $z_c$ of the critical point, the wind speed at the critical point $\varv_c$, and the power law index of the CRE injection spectrum $\gamma$. The critical point denotes the transition  of a subsonic to a supersonic galactic wind. As input parameters, we provide the $z=0$ radius of the flux tube $R_0$ (in this work defined as the strip width of each galaxy), rotational velocities, and the magnetic field strength $B_0$ within the galactic disk. $B_0$ is determined as the average magnetic field strength for each strip measured on equipartition magnetic field maps, published in \citet{2022arXiv220811068H} in 20\arcsec\ wide boxes (see Table \ref{tab:box_setup} for physical extents). \citet{2022arXiv220811068H} use the revised equipartition formula of \citet{2005AN....326..414B} to compile their magnetic field strength maps. In order to derive magnetic field strength values from radio continuum data, some assumptions have to be made. First of all, one assumes equipartition between the CRs and the total magnetic energy which is similar to a minimisation of the total energy density. Another important value that needs to be estimated to predict the magnetic field strength via the equipartition approach, is the proton-electron ratio $K_0$ . \citet{2022arXiv220811068H} use $K_0=100$. The applied value of the proton-electron ratio bases on measurements in our Galaxy \citep{2016ApJ...831...18C}. We fit for optimal solutions by incorporating a grid-search algorithm implemented in the interactive extension of \spinnaker, \texttt{SPINTERACTIVE} \citep{2019A&A...622A...9M}.

As shown by ST19, the CR transport of NGC~4013 is most probably dominated by diffusion. Therefore we recheck for all strips of NGC~4013 if the diffusion model (as described in \citet{2019A&A...628L...3H} and \citet{2019A&A...632A..13S}) still better fits the data, compared to the newly developed flux tube advection model. Here, we only consider energy independent ($\mu=0$) diffusion\footnote{We note that energy independent diffusion refers to the CR energy range that is probed with radio synchrotron observations (hundreds of MeV to a few GeV). Energy independent diffusion has found to be a good fitting model for CR transport probed with data that is similar to ours in M~51 \citep{2022arXiv220611670D}. For higher energies, the CR diffusion coefficient is expected to be energy dependent.}. All other galaxies (NGC~891, NGC~3432, NGC~4157, and NGC~4631) are only fitted with the flux tube advection model from \citet{2022MNRAS.509..658H}.

\begin{figure}
    \centering
    \begin{subfigure}{1\linewidth}
       \includegraphics[width=\hsize]{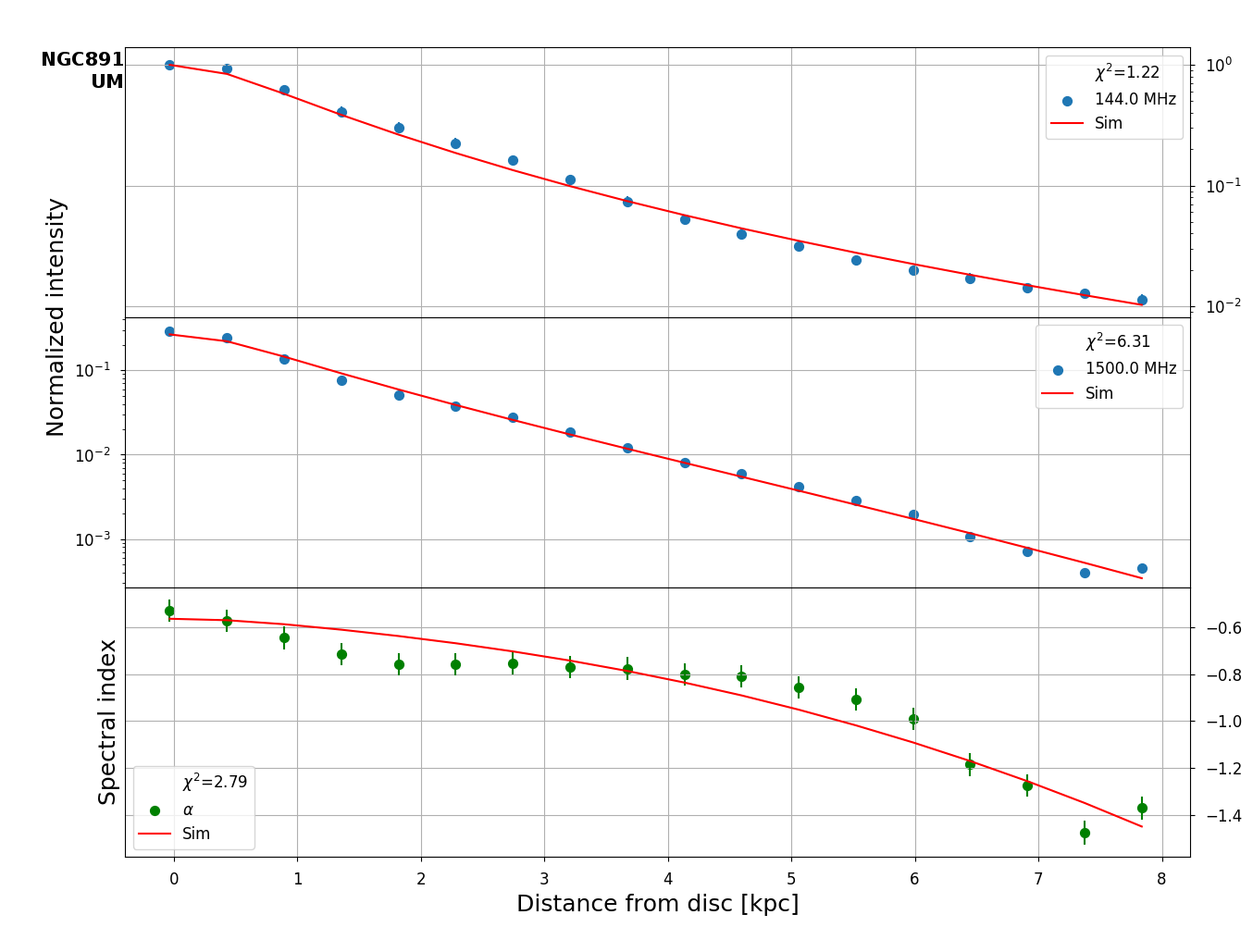}
    \end{subfigure}
    \\
   \begin{subfigure}{1\linewidth}
       \includegraphics[width=\hsize]{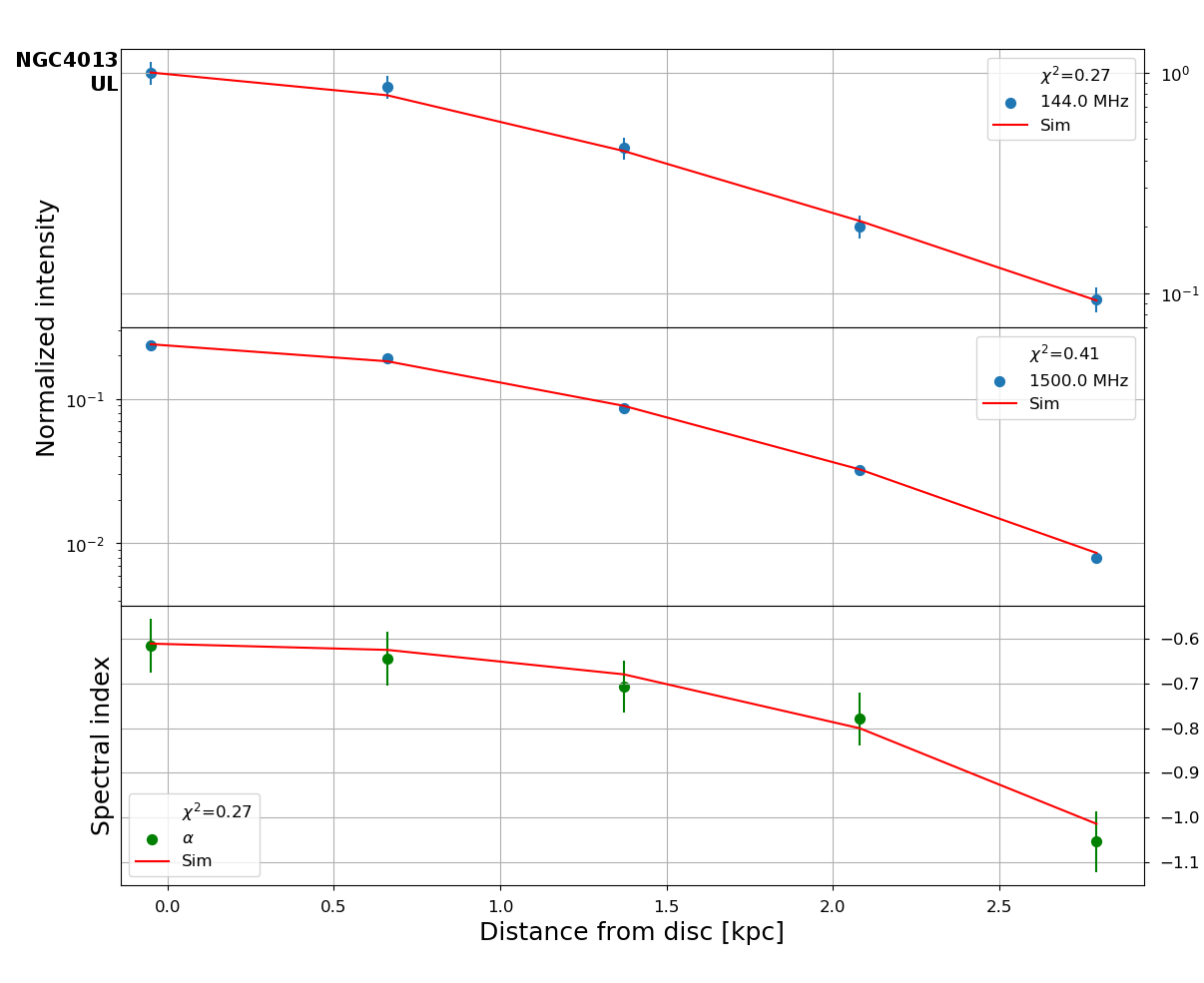}
    \end{subfigure}
    \\
    \begin{subfigure}{1\linewidth}
       \includegraphics[width=\hsize]{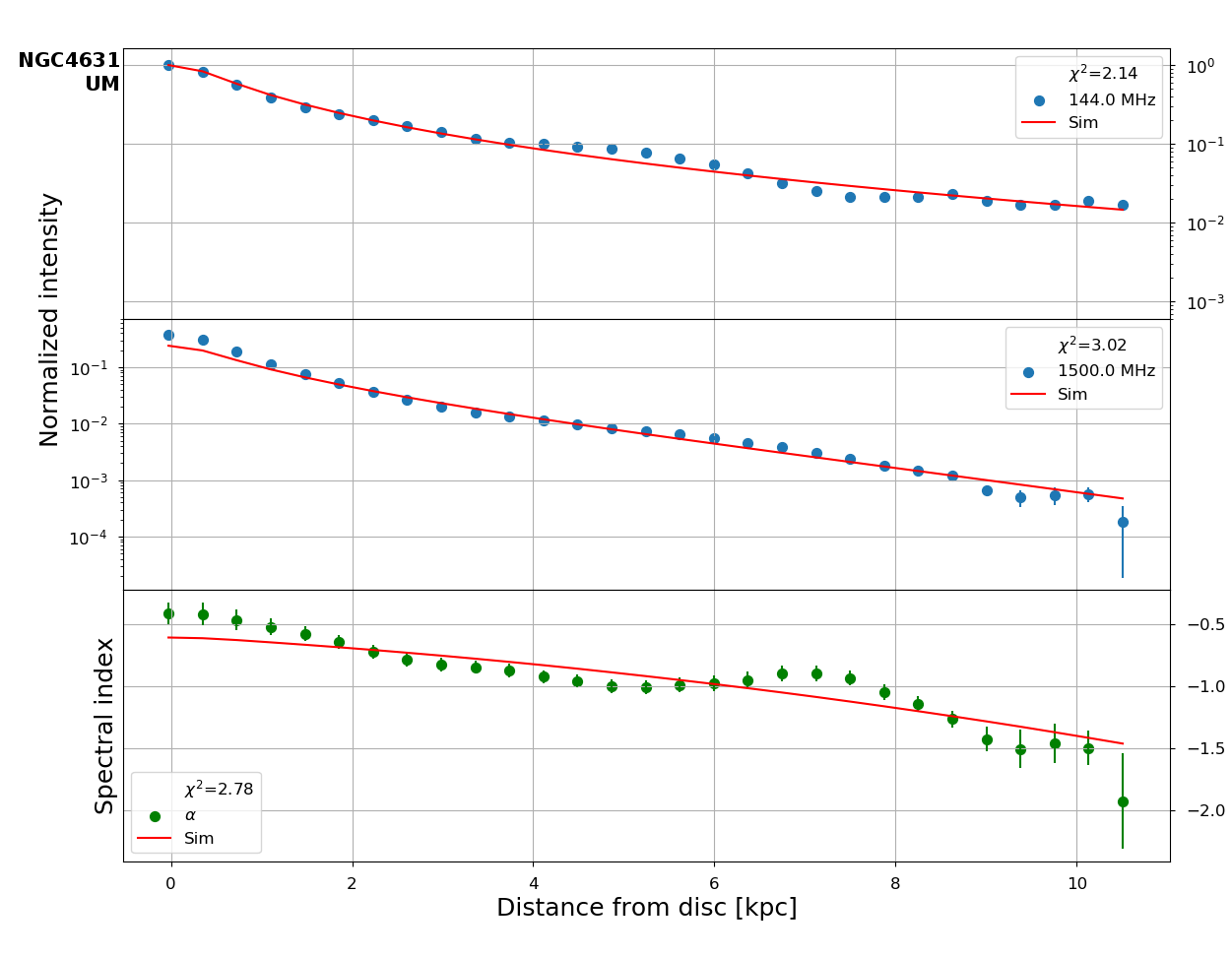}
    \end{subfigure}
     \\
    \caption{Best fitting profiles for NGC~891 UM (top panel), NGC~4013 UL (middle panel), and NGC~4631 UM (bottom panel). For each galaxy, \spinnaker\ fits the LOFAR HBA (top graph) and L-band (middle graph) intensity profiles as well as the spectral index profile (bottom graph) simultaneously. The model is shown as a red line. The advection model is applied to NGC~891 and NGC~4631, while the presented strip of NGC~4013 is fitted with a diffusion model.}
    \label{fig:spin_example}
\end{figure}

Input parameters for the galaxies can be found in Table \ref{tab:basic_info} ($\varv$\textsubscript{rot}) and Table \ref{tab:box_setup} ($B_0$, $R_0$, beam resolution). The resulting output parameters of the individual strips for all \spinnaker\ advection fits are listed in Table \ref{tab:spinnaker_param} and the results from the diffusion fits are summarised in Table \ref{tab:spin_diff}. As an example, in Fig. \ref{fig:spin_example} we show best fitting models for one strip each of NGC~891, NGC~4013, and NGC~4631. All other \spinnaker\ fits are displayed in Appendix \ref{sec:app_spinn}. To summarise the \spinnaker-fitting results, we report that all strips of NGC~891 and NGC~4631 have been successfully fitted with the flux tube advection model. For NGC~4157, we fitted four strips with the advection model and excluded the upper and lower right strips from the further analysis as the resulting SPIX profiles did not allow a proper fitting of the model. As expected from the SPIX map of NGC~3432, we do not find a clear gradient in the SPIX profiles of this galaxy (which is a needed requirement for every CR-propagation process). Nonetheless, to be consistent we present to best fitting advection model (for the two middle strips), but the wind velocity is basically unconstrained. We therefore exclude NGC~3432 from further analysis, as its interaction with UGC~5983 has most likely disrupted its radio halo. 
\begin{table*}
\caption{Output parameters from \spinnaker\ advection-model fitting: the maximal height of the fitting region $z_{\mathrm{max}}$; power law index of the cosmic ray injection spectrum $\gamma$; wind speed at the critical point $\varv_c$, scale height of the flux tube $z_0$; power law index of the flux tube $\beta$; height of the critical point $z_c$; reduced chi square of the best fitting model $\chi^2_\nu$ computed as the quadratic mean of the two intensity profiles and the SPIX profile.}
\input{tab_spin_param}
\label{tab:spinnaker_param}
\end{table*}
\begin{table}
    \centering
    \caption{Output parameters from \spinnaker\ energy-independent diffusion-model fitting: the maximal height of the fitting region $z_{\mathrm{max}}$; power law index of the cosmic ray injection spectrum $\gamma$; energy-independent diffusion coefficient $D_0$; exponential scale height of the magnetic field $h_z$; reduced chi square of best the fitting model $\chi^2_\nu$ computed as the quadratic mean of the two intensity profiles and the SPIX profile.}
    \label{tab:spin_diff}
    \begingroup
\renewcommand{\arraystretch}{1.3} 
    \begin{tabular}{lrrrrr}
    \hline\hline
    strip & $z_{\mathrm{max}}$ & $\gamma$ & $D_0$ & $h_z$ &$\chi^2_\nu$\\
     & [kpc] & & $[\mathrm{cm^2\,s^{-1}}]\cdot10^{28}$& [kpc] & \\
     \hline
     \multicolumn{6}{c}{NGC~4013}\\
     \hline
     UR    & 3.5 & 2.30$^{+0.2}_{-0.1}$ & 8$^{+30}_{-5}$ & 1.40$^{+0.1}_{-0.2}$ & 1.0\\
     UM   & 3.5 & 2.20$^{+0.2}_{-0.1}$ & 1.4$^{+2.8}_{-0.7}$ & 1.20$^{+0.1}_{-0.1}$ & 1.0\\
     UL     & 3.5 & 2.20$^{+0.2}_{-0.1}$ & 1.4$^{+2.8}_{-0.7}$ & 1.20$^{+0.1}_{-0.1}$ & 1.0										\\
    \hline
    \end{tabular}
    \endgroup
\end{table}
\subsection{Cosmic Ray Diffusion}
\label{sec:sub_cr_diff}
NGC~4013 shows mixed results. We find that the upper half of the halo fits better with the energy independent diffusion model while the lower halo seems to be dominated by advection. Due to the limited extend of the halo, each profile of NGC~4013 is sampled with only five data points (see middle panel of Fig. \ref{fig:spin_example}). Therefore, we reduce the complexity of the advection model by excluding $\beta$ from the fitting process. Only two of the three strips (UM and UL) of NGC~4013 that have been fitted with a diffusion model, result in a reasonably constrained diffusion coefficient $D_0$.
\begin{figure*}
    \centering
    \begin{subfigure}{0.45\linewidth}
       \includegraphics[width=\hsize]{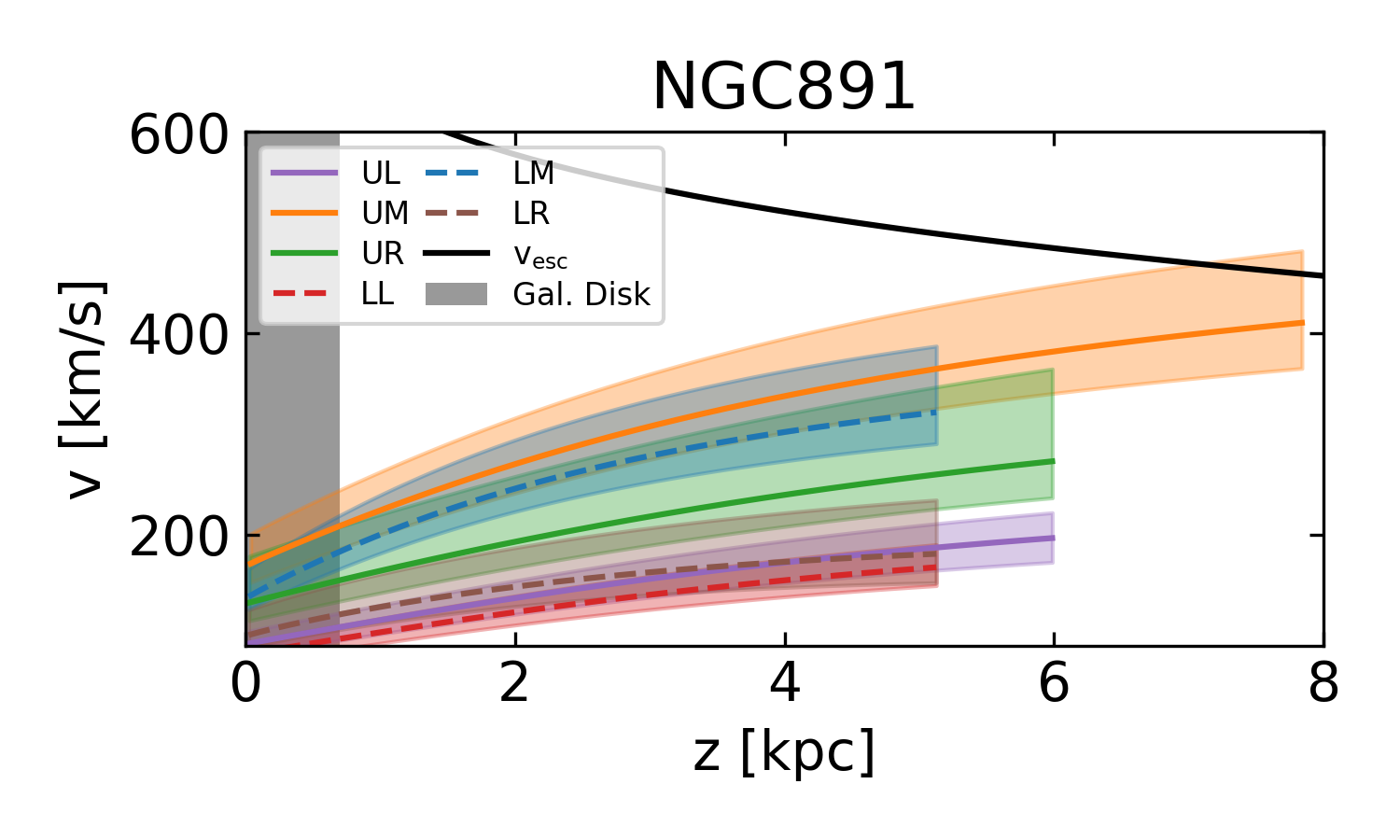}
    \end{subfigure}
   \begin{subfigure}{0.45\linewidth}
       \includegraphics[width=\hsize]{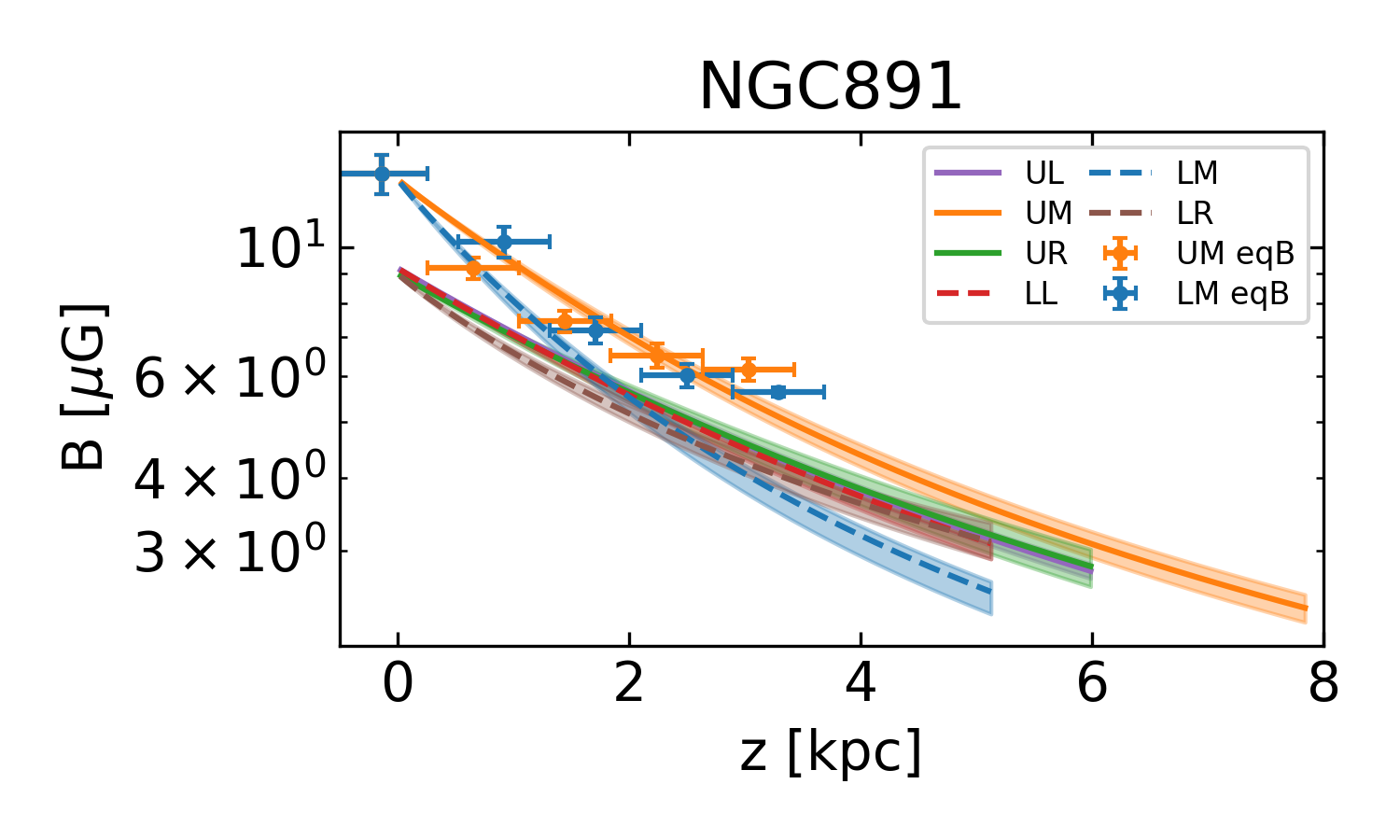}
    \end{subfigure}
     \\
     \begin{subfigure}{0.45\linewidth}
       \includegraphics[width=\hsize]{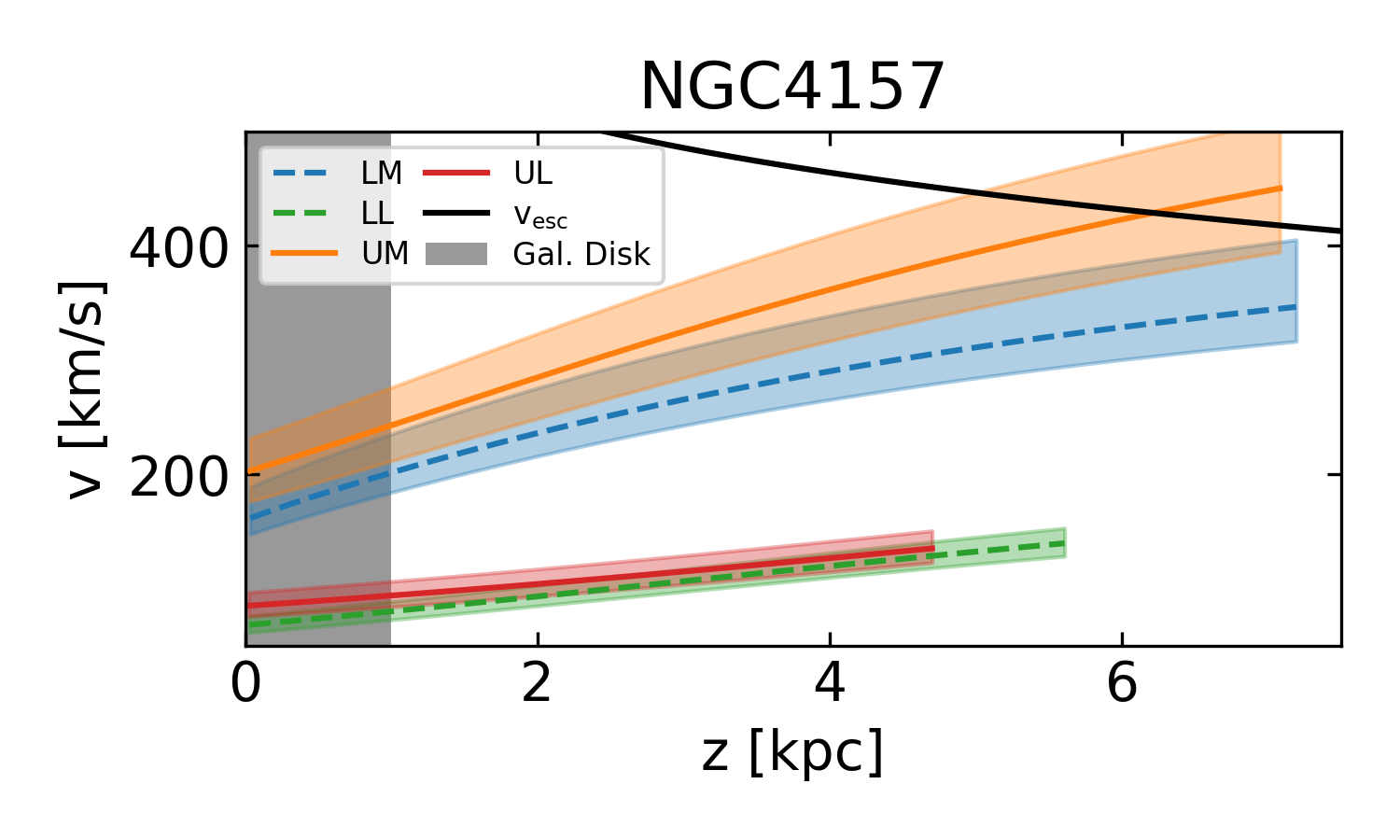}
    \end{subfigure}
   \begin{subfigure}{0.45\linewidth}
       \includegraphics[width=\hsize]{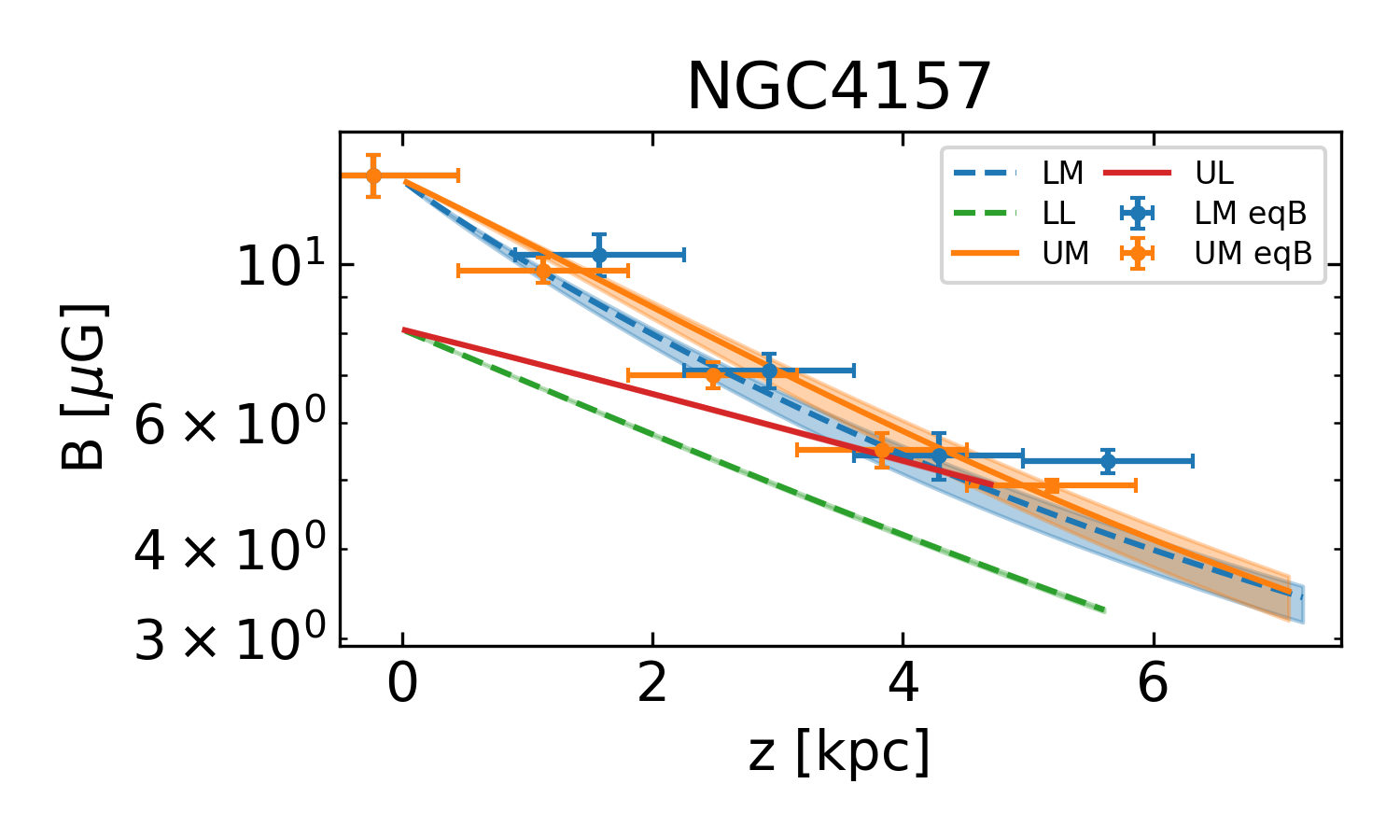}
    \end{subfigure}
    \begin{subfigure}{0.45\linewidth}
       \includegraphics[width=\hsize]{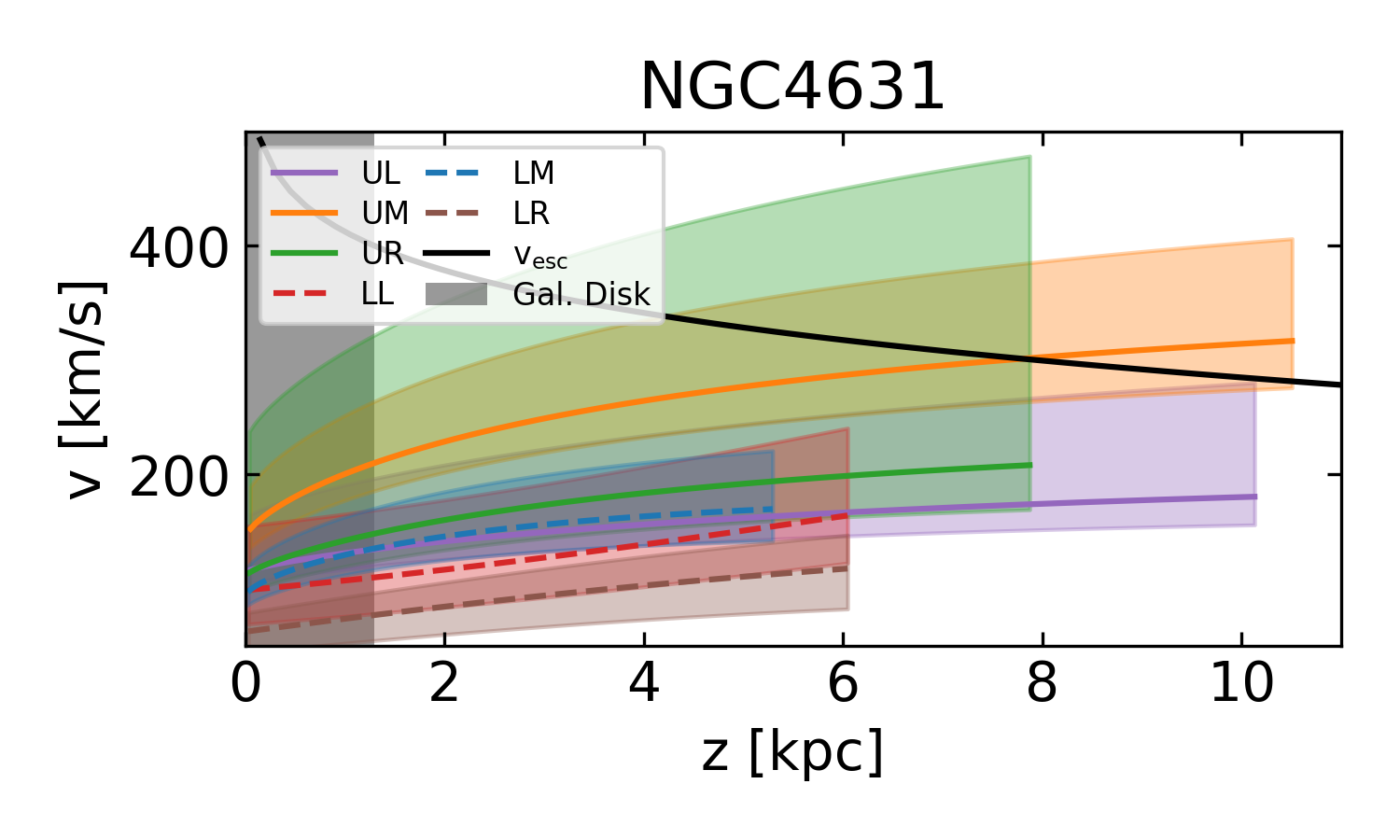}
    \end{subfigure}
   \begin{subfigure}{0.45\linewidth}
       \includegraphics[width=\hsize]{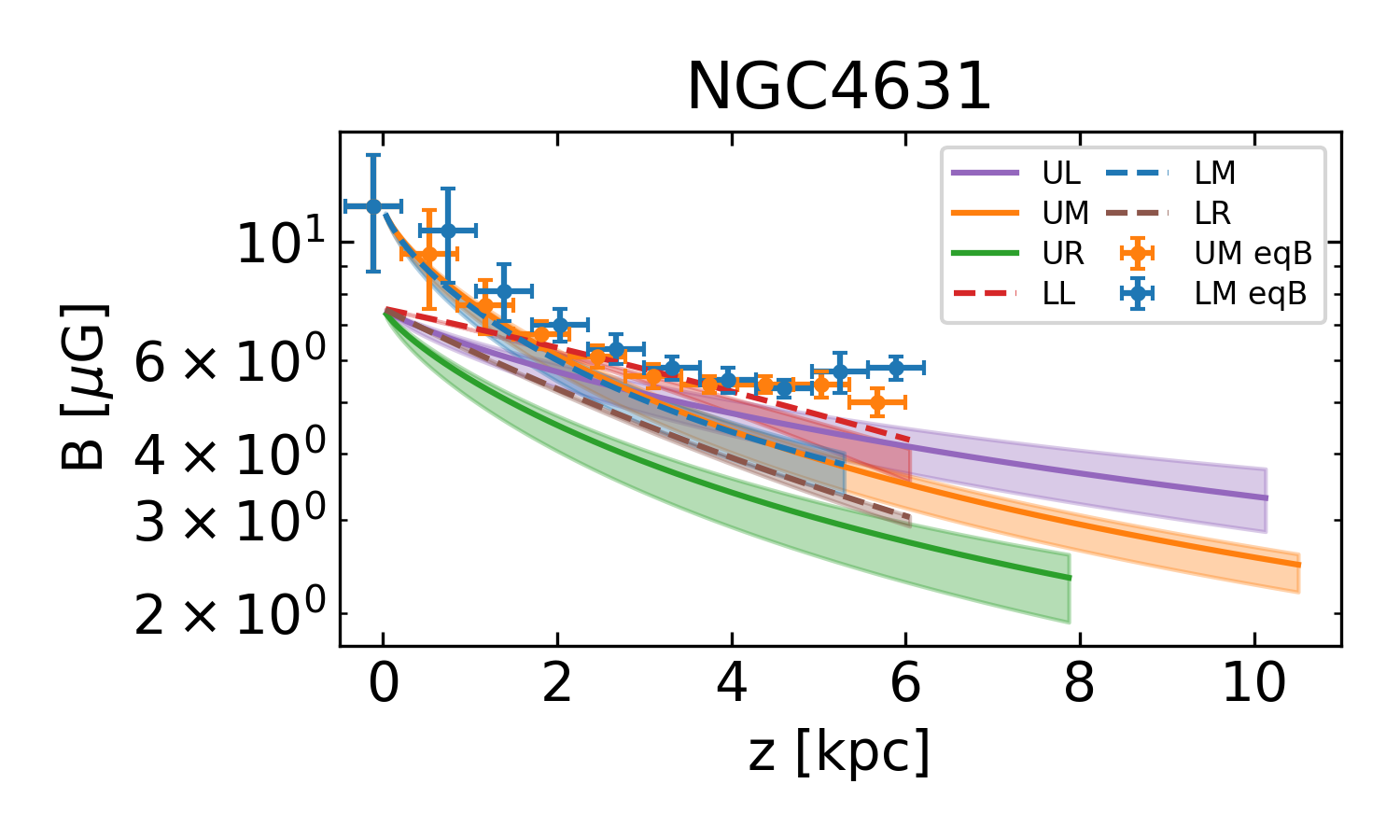}
    \end{subfigure}
   \caption{Velocity (left) and magnetic field (right) profiles for all fitted strips of a galaxy. From top to bottom panel: NGC~891, NGC~4157, NGC~4631. The extent of the galactic disk (estimated via the scale height of the HBA disk components (Table \ref{tab:scale_summary})) is indicated with a black bar in the velocity profile plots. Dashed lines represent strips pointing downward, solid lines represent upward strips. Lines are best fitting models and shaded areas indicate one-$\sigma$ uncertainties. Escape velocities are computed for the central strip of each galaxy via Eq. \ref{eq:escape}, assuming a truncated dark matter halo size of 30\,kpc.}
   \label{fig:spinn_profiles}
\end{figure*}
\subsection{Cosmic Ray Advection}
\label{sec:sub_cr_adv}
For the galaxies that have been overall successfully fitted with the advection flux tube model (NGC~891, NGC~4157, and NGC~4631), we present the velocity and magnetic field profiles in Fig. \ref{fig:spinn_profiles}. In the left column of Fig. \ref{fig:spinn_profiles}, we show the fitted wind velocities. First of all, we note that although the  \spinnaker\ wind profiles start at an altitude of $z=0$\,kpc, the motions within the galactic disk are predominantly turbulent. From there on, the wind starts to evolve and will eventually start to dominate the overall motion. Therefore, we indicate the extent of the galactic disk, approximated from the disk component scale height of the fitted two-component models in each galaxy. At some point within this range, we expect the wind to form and dominate over turbulent motions. Since the wind solutions studied in this paper are accelerated winds, the wind material in this context will eventually leave the gravitational potential. However, in order to better understand the variety of CR-driven winds, we compare the wind velocities to the escape velocity of a spherical, truncated iso-thermal dark matter halo with a assumed size of $R_{\mathrm{max}}=30$\,kpc\footnote{We note that dark matter halo shapes and profiles have shown to be more complex and that the expected halo size strongly depends on the galaxy environment. However, we choose this simplified assumption to be comparable to previous studies. For more details about halo sizes we refer to \citet[e.g.]{2003ApJ...598..260P,2007A&A...461..881L,2018MNRAS.473.2714S} }. Following \citet{2005ARA&A..43..769V} and \citet{2019A&A...622A...9M}, the escape velocity v\textsubscript{esc} at radius r can be computed for known rotation velocities v\textsubscript{rot}:
\begin{equation}
    \varv_{\mathrm{esc}}(r) = \sqrt{2} \cdot \varv_{\mathrm{rot}}\sqrt{1 + \ln\left( \frac{R_{\mathrm{max}}}{r}\right)}.
    \label{eq:escape}
\end{equation}
The escape velocities for the middle strip of each galaxy is shown in black in Fig. \ref{fig:spinn_profiles}, for comparison. Overall, we find that the wind velocities of the central strips are higher than in outer strips. For NGC~4157, the split between inner and outer strips is very distinct. Here, the wind in the upper middle strip reaches the escape velocity at 6\,kpc. The spread of the velocity profiles of NGC~891 is much smaller compared to NGC~4157 and the wind does not reach the escape velocity within the first 8\,kpc of its path. As the rotation velocity of NGC~4631 is much slower compared to NGC~891 and NGC~4157, the computed escape velocity is also lower. Therefore, the galactic wind of NGC~4631 reaches escape velocity in the UM strip at 8\,kpc.   

The right column of Fig. \ref{fig:spinn_profiles} displays the magnetic field strength models for the fitted strips. In addition, we compare these profiles to the magnetic field strength measured on the equipartition maps published by \citet{2022arXiv220811068H}. Here, we can only present equipartition profiles for the central strips as the extent of the maps is not large enough to also properly sample the outer strips. We find good agreement between the equipartition and the modelled magnetic field strength profiles. In the case of NGC~4157 and NGC~4631, one might suspect a flattening of the equipartition profile around a value of approximately 4\,\textmu G but deeper data sets would be needed to confirm such a conjecture. 

\section{Discussion}
\label{sec:discussion}
In the following, we discuss our results of the non-thermal SPIX analysis regarding the influence of thermal absorption, the substructure that was found in the intensity and SPIX profiles, the  CR transport of NGC4013 in particular, as well as the results of the advection flux tube modelling, and we compare our findings to relevant previous works.

\subsection{Non-Thermal Spectral Index}
\label{sec:dis_si}
For the observed frequencies, the power-law index of the CRE injection spectrum is expected to be $\gamma_\mathrm{inj}\approx -2.2$, which translates to an observed radio SPIX of $\alpha_\mathrm{inj} \approx -0.6$\ \citep{2021Ap&SS.366..117H}. In the sample we studied, the observed spectral indices within the galactic disks are flatter than the expected value, with NGC~4631 being the most extreme example. Here, a region within the galactic disk (slightly to the left of the galaxy centre) has a non-thermal SPIX of $\alpha_\mathrm{nth}=-0.4$. In the following, we call this area the region of interest (ROI). We consider two possible rationales for this finding. First of all, the thermal emission within the star forming regions in the galactic disk might have been underestimated. This would lead to a higher flux density in the L-band measurement and therefore to a flatter SPIX. However, to change the measured SPIX from $-0.4$ to $-0.6$, the thermal contamination of the synchrotron emission in the L-band would still have to be of the order of 40\% after thermal correction, which seems very unlikely considering that this is the maximum total fraction of thermal emission found in the sample.

The second mechanism to consider is thermal absorption in the low-frequency measurement. Due to thermal absorption, the low-frequency pure synchrotron flux density measurement would be underestimated and result in a flatter SPIX. Based on measurements within our Galaxy, \citet{2013MNRAS.436.2127O} predict the the peak of thy synchrotron spectrum for the inner Galaxy to be at approximately 100\,MHz, considering that ROI is the region with high SFR, the synchrotron peak will at higher frequencies, which makes a influence of thermal absorption on the low frequency probable.  Whether our measurements are affected by thermal absorption can be checked by comparing our spectral index map to the findings of MP19 and \citet{2022MNRAS.511.3150V}. MP19 show a non-thermal SPIX map based on VLA 1.6 GHz (L-band) and 6 GHz (C-band) data and \citet{2022MNRAS.511.3150V} present a SPIX map based on 315 MHz and 745 MHz data from the upgraded Giant Metrewave Radio Telescope (uGMRT)\footnote{As they use 315\,MHz and 745\,MHz data, the influence of thermal emission can be ruled out.}. For clarity, we summarise the non-thermal SPIX values for the ROI (read from the published SPIX maps) of MP19, our study and \citet{2022MNRAS.511.3150V} here:
\begin{itemize}
    \item MP19 (1.5\,GHz and 6\,GHz):\\ $\alpha_\mathrm{nth}=-0.6$ (expected value for freshly injected CREs).
    \item Our study (144\,MHz and 1.5\,GHz):\\ $\alpha_\mathrm{nth}=-0.4$.
    \item \citet{2022MNRAS.511.3150V} (315\,MHz and 745\,MHz):\\ $-0.2 < \alpha_\mathrm{nth} < -0.4$.
\end{itemize}

\begin{figure}
\centering
\includegraphics[width=\linewidth]{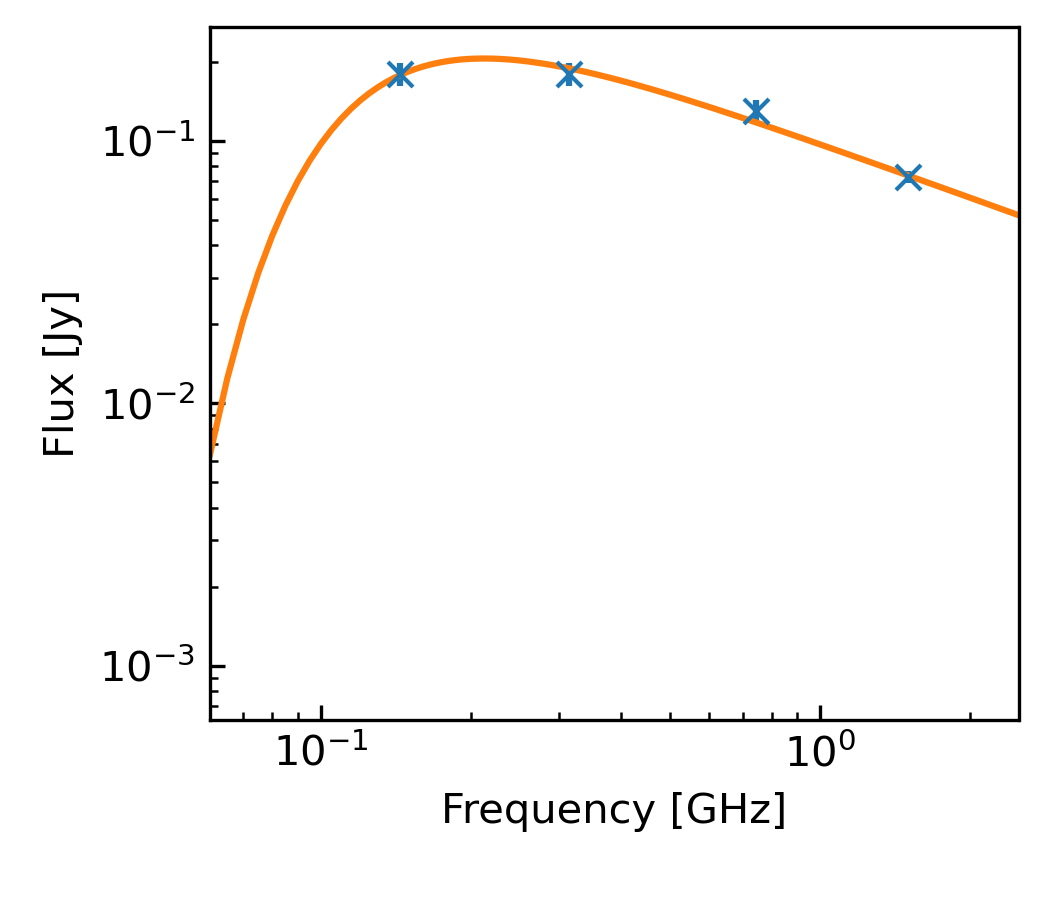}
\caption{Double logarithmic diagram displaying the flux measurements of the region with a very flat spectral index in NGC~4631 at the following frequencies: LOFAR HBA (144\,MHz), uGMRT (315\,MHz and 745\,MHz) \citep[data taken from][]{2022MNRAS.511.3150V}, and VLA L-Band (1.5\,GHz). The orange line displays the synchrotron spectrum, while accounting for thermal absorption with the best fitting emission measure.}
\label{fig:ther_abs}
\end{figure}

  The MP19 finding allows us to conclude that the L-band data are not affected by thermal absorption. As the SPIX for the ROI read from the uGMRT-SPIX map is even flatter than the SPIX value from our analysis, one can suspect that the peak of the synchrotron spectrum for the high star formation regions lies between 315\,MHz and 745\,Mhz, which is the reason for the flat SPIX measurement. We further underline our findings in Fig. \ref{fig:ther_abs}. Here, we fit the emission measure (EM) of the region, as described in \citet[][Eq. 11]{2013A&A...555A..23A}. In the fit we include an opacity $\tau$ that adds an absorption term to the assumed power law of the pure snychrotron emission:
  \begin{equation}
      S = S_0 \left( \frac{\nu}{\nu_0}\right)^\alpha \mathrm{e}^{-\tau}.
  \end{equation}
  Following \citet{1997MNRAS.291..517W} and \citet{2013A&A...555A..23A}, for free-free absorption, the absorption coefficient is given by:
  \begin{equation}
      \tau = \frac{8.2\cdot 10^{-2}\nu^{-2.1} \mathrm{EM}}{T_e^{1.35}}.
  \end{equation}
  Under these assumptions, the peak of the synchrotron spectrum is at $\sim 200$\,MHz. For this region, the best fitting EM is $\sim 3.9\cdot10^5$\,pc\,cm\textsuperscript{-6}, which is comparable to the EM in the core region of M~82 of $\sim 3.2\cdot10^5$\,pc\,cm\textsuperscript{-6} \citep{2013A&A...555A..23A}.
  
  Another possible way to explain the hardening of the observed CR spectrum are re-acceleration processes as described by \citet{2014MNRAS.442.3010T,1980ApJ...239.1089L,1982A&A...107..148L}. \citet{1980ApJ...239.1089L} conclude that re-acceleration processes can change the SPIX between low and high frequency radio data by $\Delta\alpha \le 0.5$.

  Overall, we find thermal absorption to be the more likely process to explain the flat SPIX values in this region. Although \citep{2015MNRAS.449.3879B} conclude that thermal absorption has no significant influence on integrated low-frequency radio measurements of star-forming galaxies, they nevertheless suggest that it may play a role in compact regions close to the galaxy centre. Furthermore,  VLBI measurements of the nuclear starburst of M~82 also show that, for compact regions, thermal absorption influences the low frequency radio data  \citep{1997MNRAS.291..517W,2015A&A...574A.114V}. We therefore deduce that the LOFAR measurements, within the galactic disk, might be affected by thermal absorption.  In order to check if our scale height or CR transport analysis is affected by this phenomenon, we reran the scale height- and \texttt{SPINNAKER}-fitting by removing the $z=0$ flux measurement, but the results did not change significantly.  We therefore do not expect thermal absorption to have a strong influence on our analysis of the radio halo.

\subsection{Substructure in Intensity and Non-Thermal Spectral Index Profiles}
The analysis of the box integration profiles shows that the intensity profiles are preferably fitted with one- or two-component exponential profiles rather than with Gaussian profiles. Still, for some strips of NGC~891 and NGC~4631, the box integration reveals a significant substructure that is imprinted onto the intensity profiles. \citet{2017MNRAS.469.3185P} find in their analysis of 30 high-resolution cosmological zoom simulations of Milky Way-like galaxies that the vertical magnetic field strength profiles in the inner and outer parts of the galaxy each follow different exponential profiles. They report that the break of the profiles typically occur at a height of $5-15$\,kpc. Such a break in the magnetic field strength will certainly affect the observed synchrotron profile, which could be at least part of an explanation for the observed substructure. However, the substructure seems to be more complex than a single break in the magnetic field strength profile could explain. For the UR strip of NGC~891 (see top left panel of Fig \ref{fig:int_prof_891}), this deviation is most probably caused by the residual of an incompletely subtracted background- or foreground source. The substructures imprinted on the profiles of NGC~4631 do not seem to result from point source residuals. A careful manual inspection of the total power image did not reveal any background sources.

The observed substructure of the intensity profiles propagates directly into the observed SPIX profiles. We find plateaus in the SPIX profiles (e. g. Fig. \ref{fig:spin_example} top panel: NGC~891 UM) as well as an actual increase of the SPIX values (e. g. Fig. \ref{fig:spin_example} bottom panel: NGC~4631 UM). While NGC~891 and NGC~4631 show a lot of substructure in their intensity and SPIX profiles (see Appendix \ref{sec:app_spinn}), the fitted profiles of NGC~3432, NGC~4013, and NGC~4157 follow the predicted \texttt{SPINNAKER} models without major deviations. A flattening or an actual increase of the spectral index profile points to a rejuvenation process of the CREs, meaning that there needs to be a process that allows the CREs to loose less energy, or even gain energy. Currently, there is no model implemented in \spinnaker\ that accounts for such a rejuvenation of CREs. In the ideal magneto-hydrodynamic simulations of a rotating Milky Way-type disk galaxy of \citet{2020MNRAS.492.2924V}, a plateau in the vertical synchrotron radiation profiles is found, which becomes more pronounced when the simulation runs with higher SFRs. In a low-frequency uGMRT follow-up study, \citet{2022MNRAS.511.3150V} find evidence for such a plateau in NGC~4631 at a $2-3$\,kpc height and suggest that the plateau might be attributed to CRE re-acceleration in shocks of the galactic outflow. However, we do not find clear evidence for the reported plateau in our data. On the contrary, we find that most of the substructure found in our low-frequency profiles start after 4\,kpc. Another interesting observation is that we detect a much larger radio halo within our LOFAR 144\,MHz data than shown by \citet{2022MNRAS.511.3150V}, even though the spectral bands are comparable (this study: 144\,MHz; \citet{2022MNRAS.511.3150V}: 315\,MHz). Additionally, the simulations by \citet{2020MNRAS.492.2924V} run for about 10\,Myr, which might not be enough to result in an equilibrium state. Additionally, some of the assumptions made in the simulations might not apply.

To conclude this section, we note that the intensity and SPIX profiles analysed in this work show significant substructures at distances larger than $\sim$2\,kpc for NGC~891 and $\sim$4\,kpc for NGC~4631 that are yet not well explained by current model predictions. One possible line of explanation could be that variations in the SFR history change the parameters (gas density, CRE number density, and velocity) of the galactic wind and are therefore imprinted in the radio continuum profile. However, further studies and new models are needed to fully understand the structures found in the galactic halos.

\subsection{Cosmic Ray Diffusion in NGC~4013?}
\label{sec:discussion:N4013}
 Comparing our diffusion fitting results of NGC~4013 to the ST19 analysis, we find a higher diffusion coefficient (this study: $1.4 \cdot 10^{28}$\,cm\,s\textsuperscript{-1}; ST19: $0.55-0.65\cdot 10^{28}$\,cm\,s\textsuperscript{-1}), a slightly lower injection index for the CREs (this study: 2.2; ST19: 2.6) and higher disk scale heights of the assumed exponential profile of the magnetic field (this study: 1.2\,kpc; ST19: 0.1\,kpc). The diffusion coefficients in the analysis of ST19 are based on VLA L- and C-band (6\,GHz) data, which makes a direct comparison to our results more difficult as the CREs probed in this work are much older than the CREs traced by the higher frequency measurements.  Additionally, ST19 find advection to dominate when using 144\,MHz and 1.6\,GHz data for the CR-transport analysis. However, we find much higher wind velocities (this study: $\sim150$\,km\,s\textsuperscript{-1}; ST19: $18-22$\,km\,s\textsuperscript{-1}). Following \citet{2016MNRAS.462.4227R} and \citet{2021Ap&SS.366..117H}, one can estimate the location where the transition from diffusion-dominated to advection-dominated CR-transport happens:
 \begin{equation}
     z_\star \approx 1.2\frac{D/10^{28}\,\mathrm{cm^2 s^{-1}}}{\varv/100\,\mathrm{km s^{-1}}}\,\mathrm{kpc},
     \label{eq:diff_adv}
 \end{equation}
 where $D$ is the diffusion coefficient and $\varv$ the advection wind speed. Using the diffusion coefficient from the UM and UL strips and the wind velocity of the LR and LM strips (taken from Table \ref{tab:spinnaker_param} and \ref{tab:spin_diff}), we find that the transition will happen at $z_\star=1.5$\,kpc. Of course, this is only a zero-order estimate as we get the diffusion coefficients and wind velocities from different locations of the galaxy. The accuracy of the estimate is also affected by the relatively poor sampling of the radio halo, and because diffusion and advection profiles only can be reliably distinguished by analysing extended halo profiles (i.e. distinguishing between the two is difficult for small halos).
 ST19 report the position of this transition to be at $z_\star=1-2$\,kpc\footnote{ST19 use a slightly modified version of Eq. \ref{eq:diff_adv}, with a factor ranging from 0.3 to 0.6, instead of 1.2, which comes from different assumptions of the diffusion process. One-dimensional diffusion would lead to 0.3, while isotropic 3D-diffusion leads to 1.2.}, which is consistent with our estimate. 
 Another galaxy that has been found to be diffusion-dominated is NGC~4565 \citep{2019A&A...628L...3H,2019A&A...632A..12S}. While there is some scatter in the diffusion coefficients of this galaxy, our results fall in the same range. From the diffusion coefficient, one can derive the size of the diffusion halo $L$, if the CRE lifetime is known \citep[adapted from][Eq. 15]{2021Ap&SS.366..117H}:
 \begin{equation}
   L = \sqrt{\frac{D \cdot \tau}{0.75\cdot10^{29}}}\,\mathrm{kpc}  
 \end{equation}
 Here, $D$ is the derived diffusion coefficient in units of $10^{28}\,\mathrm{cm^2 s^{-1}}$ and $\tau$ is the CRE lifetime which is typically about 100\,Myr for the energy range that is traced with the LOFAR HBA data. With this assumption we derive a diffusion halo size for NGC~4013 of $L=4.3$\,kpc (UR stripe excluded). A similar halo size is measured within our own Galaxy \citep{2013MNRAS.436.2127O, 2018MNRAS.475.2724O}. 
 
 To summarise our results about NGC~4013, we note that our analysis remains inconclusive. Overall, the small extent of the galaxy halo makes it very difficult to analyse its CRE transport in great detail. New data reaching farther into the halo would be necessary in order to reliably distinguish the CRE transport mechanisms.  

\subsection{Advection Dominated Galactic Winds}
Of the analysed sample, NGC~891 has been previously modelled with \texttt{SPINNAKER} advection models. \citet{2019A&A...632A..12S} (SC19) analyse the radio halo of NGC~891 with 1.5\,GHz and 6\,GHz VLA \changes\ (+Effelsberg) data. They apply purely advective or diffusive transport models, but restrict the modelling region to a maximal height of 3\, kpc. With the new LoTSS data and the newly cleaned \changes\ L-band data, we can now nearly double the extent of the modelling regions. Furthermore, we analyse six strips (three northern, three southern) instead of two strips (one northern, one southern).  SC19 compare two models, one advection dominated and one diffusion dominated and find that an advection model is a much better fit. With their advection model, they find a mid-plane wind velocity of 150\,km\,s\textsuperscript{-1} for an accelerated galactic wind model. A direct comparison between the model used by SC19 and the flux tube model introduced by \citet{2022MNRAS.509..658H} is difficult because the newer model includes additional processes (e.g. advection losses) and also because the modelling technique has changed. Nevertheless, we find similar velocities in proximity of the galactic disk, while we detect a split between central and outer strips farther out into the halo, where central strips have higher velocities than outer strips. \citet{2019A&A...632A..12S} predict that the galactic wind in NGC~891 reaches the halo escape velocity at a height of $9-17$\,kpc. Our analysis does not allow a confirmation of their results. New, more sensitive data are needed to probe such scales. One might suspect that the UM strip does exceed the escape velocity in this range, as our model uncertainty allow for higher wind velocities compared to the escape velocity at heights larger than 7\,kpc, but this prediction remains uncertain.  

As reported by MP19 and \citet{2022MNRAS.511.3150V}, we expect advection to dominate the CR transport of NGC~4631. Nevertheless, we expect the wind to be much slower than estimated by MP19. MP19 estimate the wind to reach a velocity of 300\,km\,s\textsuperscript{-1} at a height of 3\,kpc, which is much higher than the models in our analysis predict, although the uncertainties are high. Judging from our velocity profile, the galactic wind in NGC~4631 will reach the escape velocity much later (after 8\,kpc in the UM strip, which has the fastest wind), compared to the prediction of MP19. 

Comparing our results from the advection flux tube modelling to the analysis of NGC~5775 by \citet{2022MNRAS.509..658H}, our analysis of NGC~891, NGC~4157, and NGC~4631 points towards lower wind velocities. \cite{2022MNRAS.509..658H} report wind velocities of more than 600\,km/s. However, NGC~5775 also has a much higher integrated SFR (7.5\,$\mathrm{M}_\odot$\,yr\textsuperscript{-1}) and also a higher SFR surface density ($9.4 \times 10^{-3}$\,$\mathrm{M}_\odot$\,yr\textsuperscript{-1}\,kpc\textsuperscript{-2}) than the galaxies that we have analysed, which most likely influences the CR-driven wind since these quantities can be interpreted as the driving source of the wind. Concerning the magnetic field strength profiles, we can confirm the results of \citet{2022MNRAS.509..658H}. As in the case of NGC~5775, the analysed galaxies generally show a good agreement between the  modelled magnetic field strength profiles from \texttt{SPINNAKER} and the equipartition measurements. However, for heights above 6\,kpc, the model profiles of NGC~5775 deviate to lower field strength values compared to the equipartition measurements. We find a similar trend for the LM strip of NGC~891 at heights larger than 2\,kpc, which should be investigated in future studies.

\citet{2008ApJ...674..258E} analyse the galactic wind of the Milky Way by applying a hybrid wind model, where the wind is driven thermally as well as by CR pressure. They find an initial wind velocity of $\varv_{0}=173$\,km\,s\textsuperscript{-1} and the position of the critical point at a height of $z_c=2.4$\,kpc. At the critical point, they predict a sound speed $c_{\star}=251$\,km\,s\textsuperscript{-1}. From their model, \citet{2008ApJ...674..258E} expect the wind velocity to be rather constant until it reaches a height of approximately 1\,kpc and then to accelerate quickly to surpass the sound speed. Therefore the overall shape of the wind profile differs from the wind profiles shown in Fig. \ref{fig:spinn_profiles}, where our velocity increases more gradually, and the predicted wind velocities for the Milky Way seem to lie between  our predictions for NGC~4631 and NGC~891.

\begin{figure*}
    \centering
    \begin{subfigure}{0.3\linewidth}
       \includegraphics[width=\hsize]{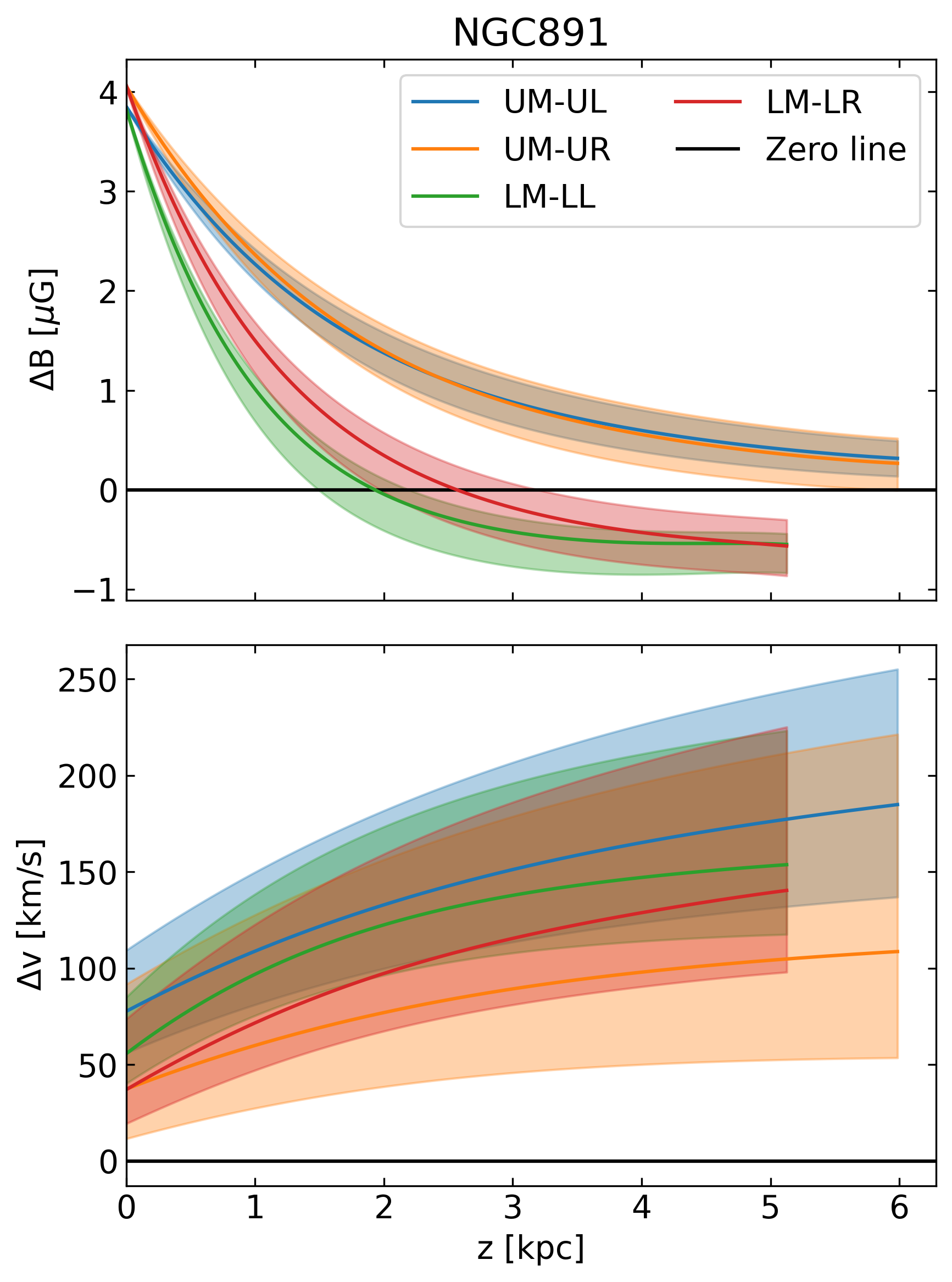}
    \end{subfigure}
    \hfill
   \begin{subfigure}{0.3\linewidth}
       \includegraphics[width=\hsize]{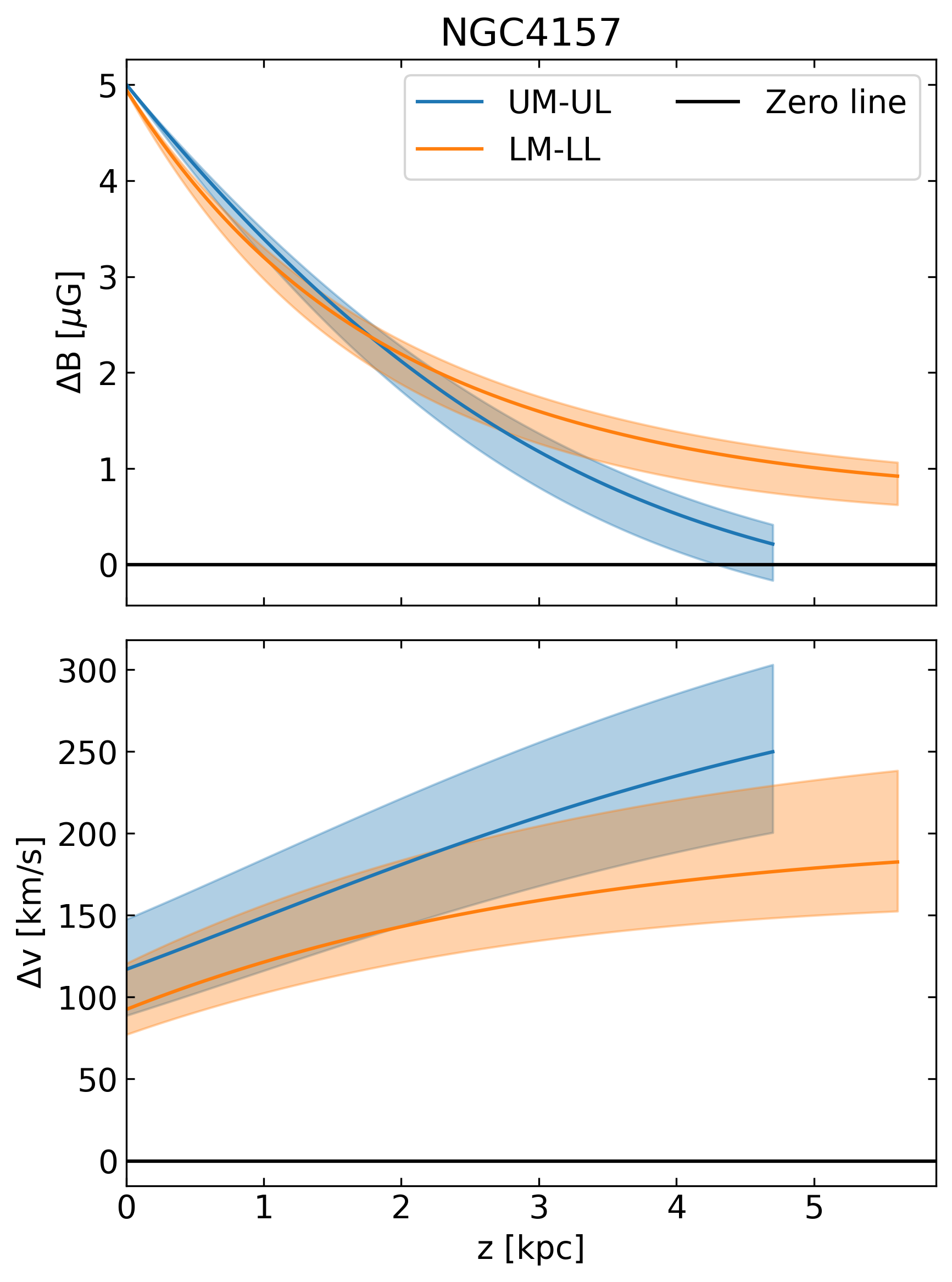}
    \end{subfigure}
    \hfill
    \begin{subfigure}{0.3\linewidth}
    \includegraphics[width=\hsize]{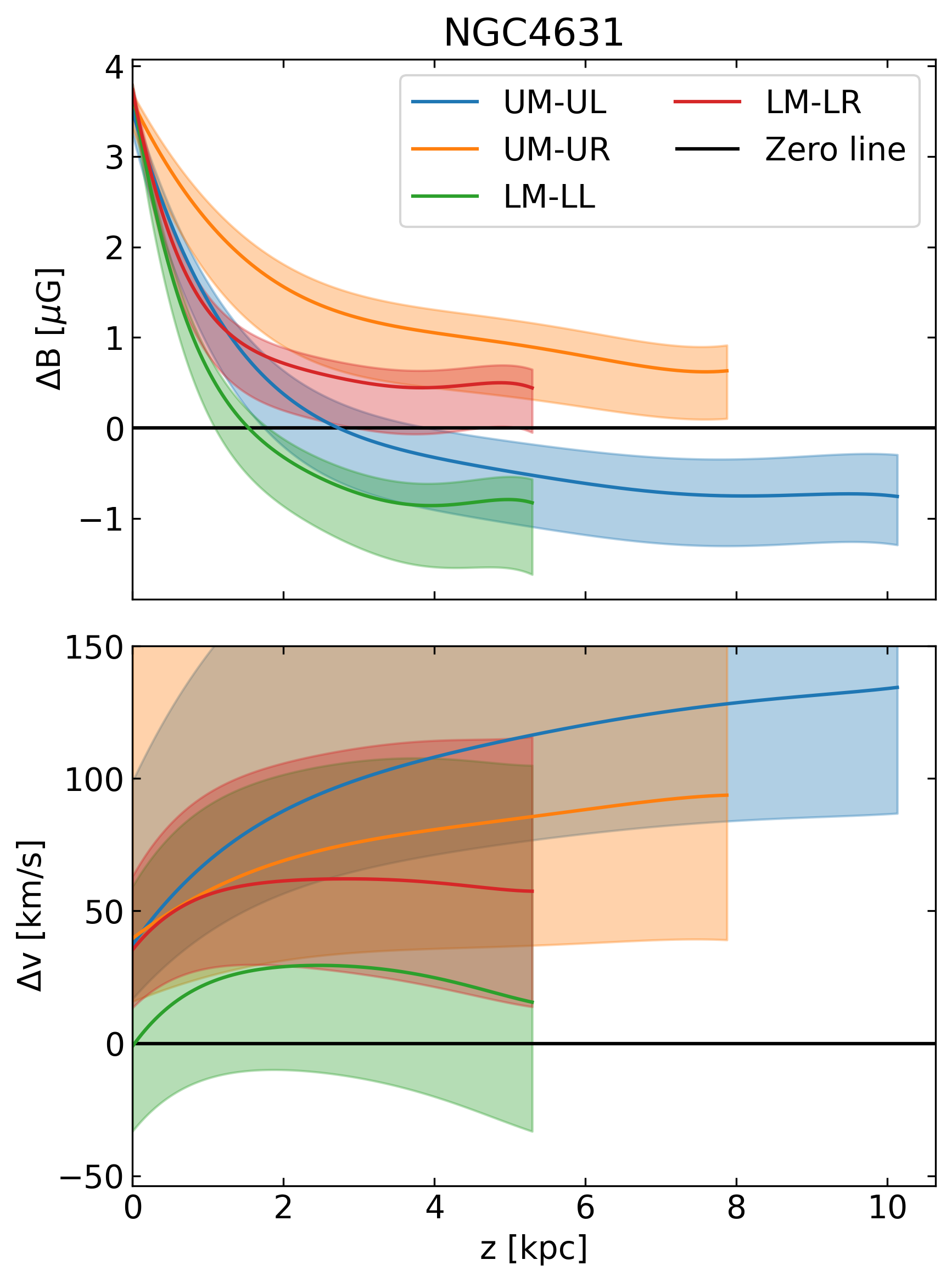}
    \end{subfigure}
    \caption{Difference of magnetic field strength (top row) and wind velocity (bottom row) between central and outer, left and right, strips for NGC~891 (left column), NGC~4157 (middle column, and NGC~4631 (right column). Black lines indicate equal magnetic filed strengths/wind velocities in the central and outer strips.}
    \label{fig:delta_profiles}
\end{figure*}

One of the key benefits of our study, compared to the analysis of NGC~5775 by \citet{2022MNRAS.509..658H}, is the chosen three-strip setup during box integration. This allows us to compare the properties of the galactic wind in the inner and outer part of each galaxy. In Fig. \ref{fig:delta_profiles}, we present the difference of predicted magnetic field strength and wind velocity between the middle and the outer strips
for NGC~891, NGC~4157, and NGC~4631. While we have not found a significant difference between middle and outer strips in the radio scale height analysis, we do see a split in the modelled wind velocity and magnetic field strength. Since the magnetic field strength at $z=0$ is set as the average field strength from the equipartition measurement in each strip, there is an offset between the middle and outer strips. Interestingly, the magnetic field strength in the central regions declines faster than in the outer regions of the galaxy. This results in an approximate equality of the predicted magnetic field strength for high $z$ values in central and outer strips. A similar behaviour is also visible in the cosmological magnetohydrodynamical simulations \citep[][Fig. 5]{2018MNRAS.480.5113M} and simulations of isolated disk galaxies \citep[][Fig. 2]{2019MNRAS.483.1008S}. With regard to wind velocity, the opposite behaviour can be observed. Here the difference in wind velocity between central and outer strips predominately increases (wind velocity increases faster in central strips than in outer ones) in the radio halo. For some strips of NGC~4631, central and outer wind velocities converge, but here the predicted uncertainties are very high.
\section{Summary \& Outlook}
\label{sec:SandO}
In this paper, we have reprocessed VLA L-band data in D-configuration and C-configuration from the CHANG-ES project and combined them with newly published low-frequency data from the LOFAR LoTSS DR2 for an analysis of an edge-on galaxy sample. We investigated resolved non-thermal SPIX maps, intensity profiles perpendicular to the galactic disks, and modelled the CR transport in the galactic halos. We demonstrated that new reduction techniques highly increase data quality, which allowed us to double the extent of the visible radio halo for most of the analysed galaxies. Additionally, we have re-implemented the 'mixture-method' developed by \citet{2018ApJ...853..128V} to estimate the resolved thermal emission of each galaxy and made it publicly available, using common python libraries. We then analysed the resolved non-thermal spectral index and radio scale heights. Lastly we modelled the CR transport using the 1D advection and diffusion models of \texttt{SPINNAKER}. Our main
results can be summarised as follows:
\begin{enumerate}
    \item We find that the LOFAR LoTSS DR2 maps are affected by thermal absorption within the galactic disk, especially in star forming regions, as the non-thermal spectral index maps exceed the expected limit for freshly induced CRE of $\alpha_{\mathrm{nth}}=-0.6$ in these areas.
    
    \item The scale height analysis based on the model comparison with the AICc shows that most of the profiles are equally well fitted with one- as with two-component exponential models. Additionally, we find substructures in the intensity and SPIX profiles that are not well explained by contemporary  CRE transport models.
    
    \item We do not find a strong correlation of the radio halo scale height and physical properties such as galactic diameter, gravitational potential, SFR, SFR surface density, specific SFR, and magnetic field strength measured within the galactic disk. 
    
    \item The modelling of the galactic wind indicates a split of magnetic field strength and wind velocity in the immediate vicinity of the galactic disks for inner and outer strips of the galaxy. While the difference in magnetic field strength evens out in the radio halo, the difference in wind speed increases.
\end{enumerate}

In this paper, we successfully analysed the CRE transport of NGC~891, NGC~4157, and NGC~4631, while the analysis of NGC~3432 and NGC~4013 remained inconclusive. Our results suggest that there are significant differences in CR transport depending on the location within the galactic disk, which should be accounted for in future CRE transport models. To better understand the CRE transport within the radio halo of NGC~4013, a possible approach would be to also include higher frequency data, as done by \citet{2019A&A...632A..13S}, to trace the younger CREs in the halo and then simultaneously fit three frequencies within \texttt{SPINNAKER}. Adding a third frequency to the analysis would lead to higher precision of the model predictions, as it increases the number of profiles that are fitted by \texttt{SPINNAKER} from three (two intensities, one SPIX) to five (three intensities, two SPIXs). However, the relatively small extent of the radio halo of NGC~4013 will most likely always be a limiting factor of the CR transport analysis. Adding information about the magnetic field structure from polarisation measurements and implementing it in new CRE transport models would further help to better understand the role of CRs within galactic feedback processes. More detailed modelling approaches are certainly needed, and will be possible, as the Square Kilometre Array Observatory (SKAO) precursors and pathfinders and eventually the SKAO itself enable us to analyse CR transport within galaxies on an unprecedented scale, both in terms of sensitivity and resolution. 

\begin{acknowledgements}
We thank the anonymous referee for a constructive report that helped to improve our paper. M. S., R.-J. D., B. A., and D. J. B.  acknowledge funding from the German Science Foundation DFG, within the Collaborative Research Center SFB1491 "Cosmic Interacting Matters - From Source to Signal.
M.B. acknowledges funding by the Deutsche Forschungsgemeinschaft (DFG, German Research Foundation) under Germany’s Excellence Strategy – EXC 2121 ‘Quantum Universe’ – 390833306. TW acknowledges financial support from The coordination of the participation in SKA-SPAIN, funded by the Ministry of Science and Innovation (MCIN), from the State Agency for Research of the Spanish Ministry of Science, Innovation and Universities through the "Center of Excellence Severo Ochoa" awarded to the Instituto de Astrofísica de Andalucía (SEV-2017-0709)\\
We thank Tom Jarrett for kindly providing us the resolution enhanced 22 \textmu m WISE maps.\\
This research was supported in part by the National Science Foundation under Grant No. NSF PHY-1748958.\\

LOFAR data products were provided by the LOFAR Surveys Key Science project (LSKSP; \url{https://lofar-surveys.org/}) and were derived from observations with the International LOFAR Telescope (ILT). LOFAR \citep{2013A&A...556A...2V} is the Low Frequency Array designed and constructed by ASTRON. It has observing, data processing, and data storage facilities in several countries, which are owned by various parties (each with their own funding sources), and which are collectively operated by the ILT foundation under a joint scientific policy. The efforts of the LSKSP have benefited from funding from the European Research Council, NOVA, NWO, CNRS-INSU, the SURF Co-operative, the UK Science and Technology Funding Council, the Ministry of Science and Higher Education, Poland, and the Jülich Supercomputing Centre.\\

The following software packages have been used in this work:
Astropy \citep{2013A&A...558A..33A, 2018AJ....156..123A}. 
This research has made use of "Aladin sky atlas" developed at CDS, Strasbourg Observatory, France \citep{2000A&AS..143...33B,2014ASPC..485..277B}; SAOImage DS9 \citep{2003ASPC..295..489J}

\end{acknowledgements}

%
%
\bibliographystyle{aa}
\bibliography{export-bibtex.bib}

\begin{appendix}
\section{Thermal Fraction Maps}
\label{sec:app_tf_maps}
\input{thermal_figs_appendix}
\section{Intensity Profiles}
\label{sec:app_int_prof}
In this section, we present the best fitting model parameters from $\chi^2$-minimisation for one- and two-component exponential profiles. The model comparison was performed with the AICc in Table \ref{tab:profile_param}. Four models were tested (listed from fewest to most free model parameters): one-component exponential without offset in $z-$direction (one\_fo), one-component exponential with offset in $z-$direction (one),
two-component exponential without offset in $z$-direction (two\_fo), and two-component exponential with offset in $z$-direction (two). We also display the intensity profiles from the box-integration technique for LOFAR HBA and VLA L-band as well as the best fitting (based on the AICc model comparison) model as a black dashed line. 
\begin{table*}[h!]
\centering
\caption{Parameters from box integration model fitting: Galaxy name, frequency of fitted data, model identifier, strip identifier, intensity and scale height of the first exponential component, intensity and scale height of the second exponential component (if applicable), offset in $z$-direction (if applicable), reduced $\chi$-square. }
\label{tab:profile_param}
\input{tab_fit_params}
\end{table*}
\input{int_prof_appendix}
\section{SPINNAKER Profiles}
\label{sec:app_spinn}
Best fitting \texttt{SPINNAKER}-models (advection or diffusion) for all galaxies (NGC~891, NGC~3432, NGC~4013, NGC~4157, and NGC~4631) are displayed in this section. The structure of the \texttt{SPINNAKER} output is explained in Fig. \ref{fig:spin_example}.
\input{spinnaker_appendix}
\end{appendix}
\end{document}

%% file: tab_basic_info.tex
\begin{table*}
\caption{Fundamental information about the galaxies analysed in this work: Celestial coordinates at epoch J2000.0\tablefootmark{a}, morphological type\tablefootmark{b}, position angle\tablefootmark{b}, inclination\tablefootmark{c}, distance\tablefootmark{d}, diameter (based on $22\,\upmu$m WISE map)\tablefootmark{d}, rotational velocity\tablefootmark{c},  total baryon mass\tablefootmark{e}. }
\label{tab:basic_info}
\centering
\begin{tabular}{lrrrrrrrrr}
     \hline\hline
     Galaxy & RA & Dec & Morph. Type  & PA & Incl. & Distance & Diam. & $\varv_{\mathrm{rot}}$ & M\textsubscript{baryonic}\\
            &  [H:M:S] &  [D:M:S] &  & [$^\circ$] & [$^\circ$] & [Mpc] & [kpc] & [km\,s\textsuperscript{-1}] & [M$_\odot$]$\cdot$10\textsuperscript{10}\\
    \hline
    NGC~891     & 02h22m33.41s  & +42d20m56.9s  & Sb    & 22.0 & 90.0 & 9.1 & 25.1 & 212 & 8.63\\
    NGC~3432    & 10h52m31.13s  & +36d37m07.6s  & SBm   & 33.0 & 85.0 & 9.4 & 9.96 & 110 & 0.86\\
    NGC~4013    & 11h58m31.380s & +43d56m47.70s & SABb  & 58.6 & 88.0 & 16 & 16.1 & 182 &  4.99\\
    NGC~4157    & 12h11m04.37s  & +50d29m04.8s  & SABb  & 64.7 & 83.0 & 15.6 &  16.6 & 189 & 5.75 \\
    NGC~4631    & 12h42m08.01s  & +32d32m29.4s  & SBcd  & 85.7 & 85.0 &  7.4 & 23.4 & 139 & 1.93\\
    \hline
\end{tabular}

\tablefoot{\\
\tablefoottext{a}{Taken from the NASA NED (\url{https://ned.ipac.caltech.edu/}).\\}
\tablefoottext{b}{Taken from the HyperLeda Database (\url{http://leda.univ-lyon1.fr/}).\\}
\tablefoottext{c}{Taken from \citet{2022arXiv220400635H}\\}
\tablefoottext{d}{Taken from \citet{2015AJ....150...81W}\\}
\tablefoottext{e}{Taken from \citet{2016MNRAS.456.1723L}}
}
\end{table*}

%% file: total_int_figs.tex
\begin{figure*}
\centering

\begin{subfigure}{0.45\linewidth}
\centering
\includegraphics[width=1\linewidth]{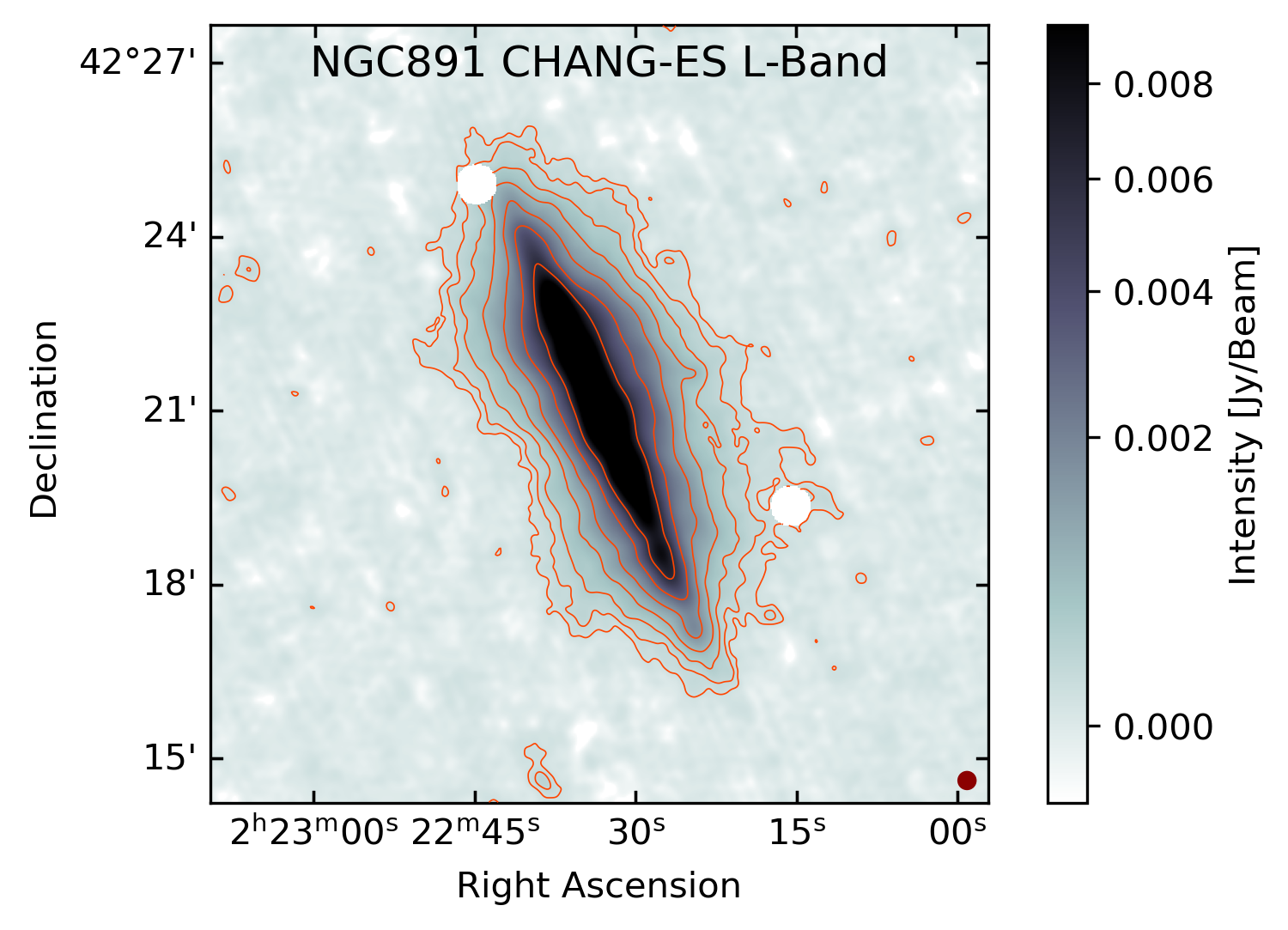}
\end{subfigure}
\hfill
\begin{subfigure}{0.45\linewidth}
\centering
\includegraphics[width=1\linewidth]{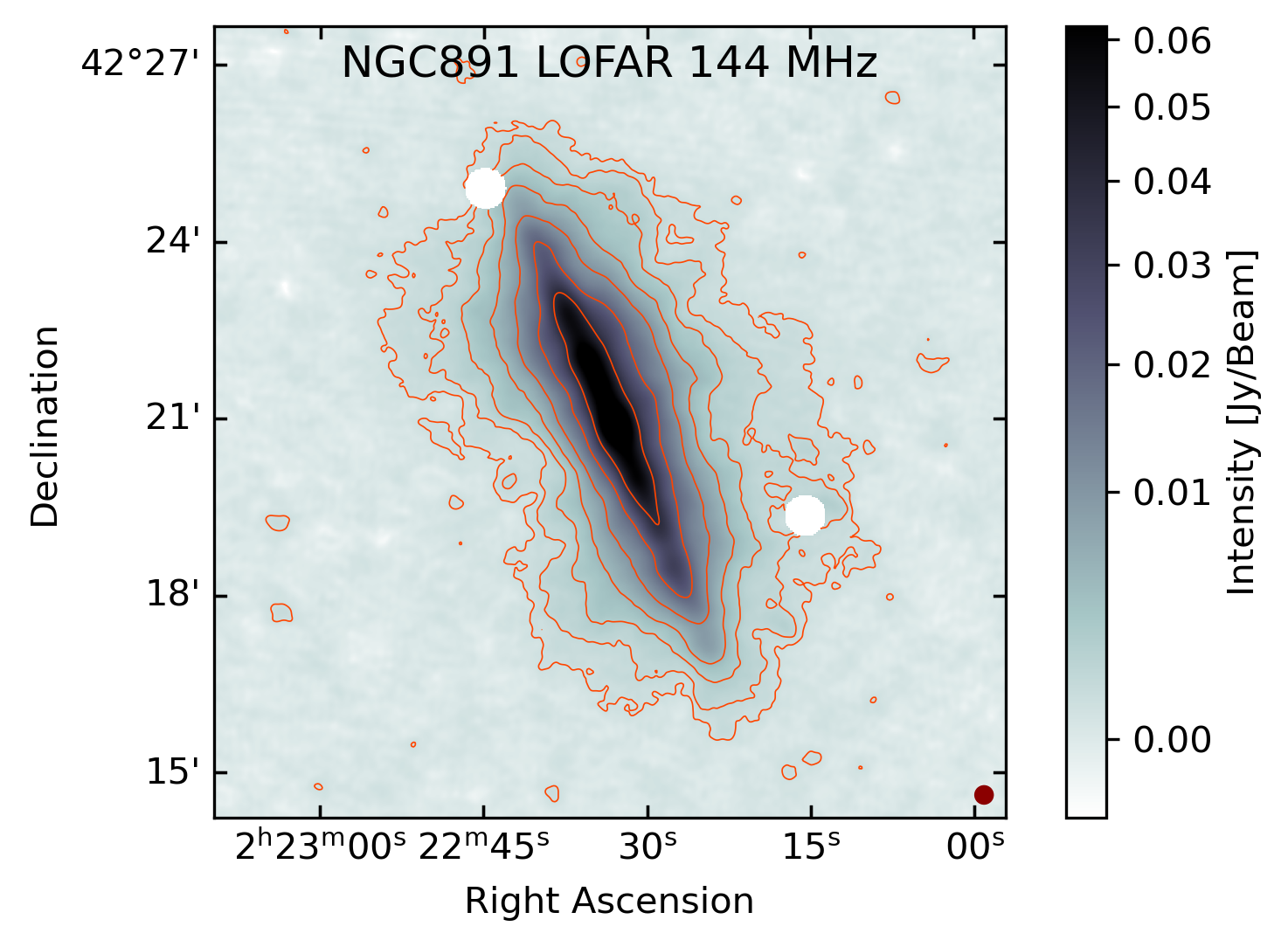}
\end{subfigure}
\\
\begin{subfigure}{0.45\linewidth}
\includegraphics[width=1\linewidth]{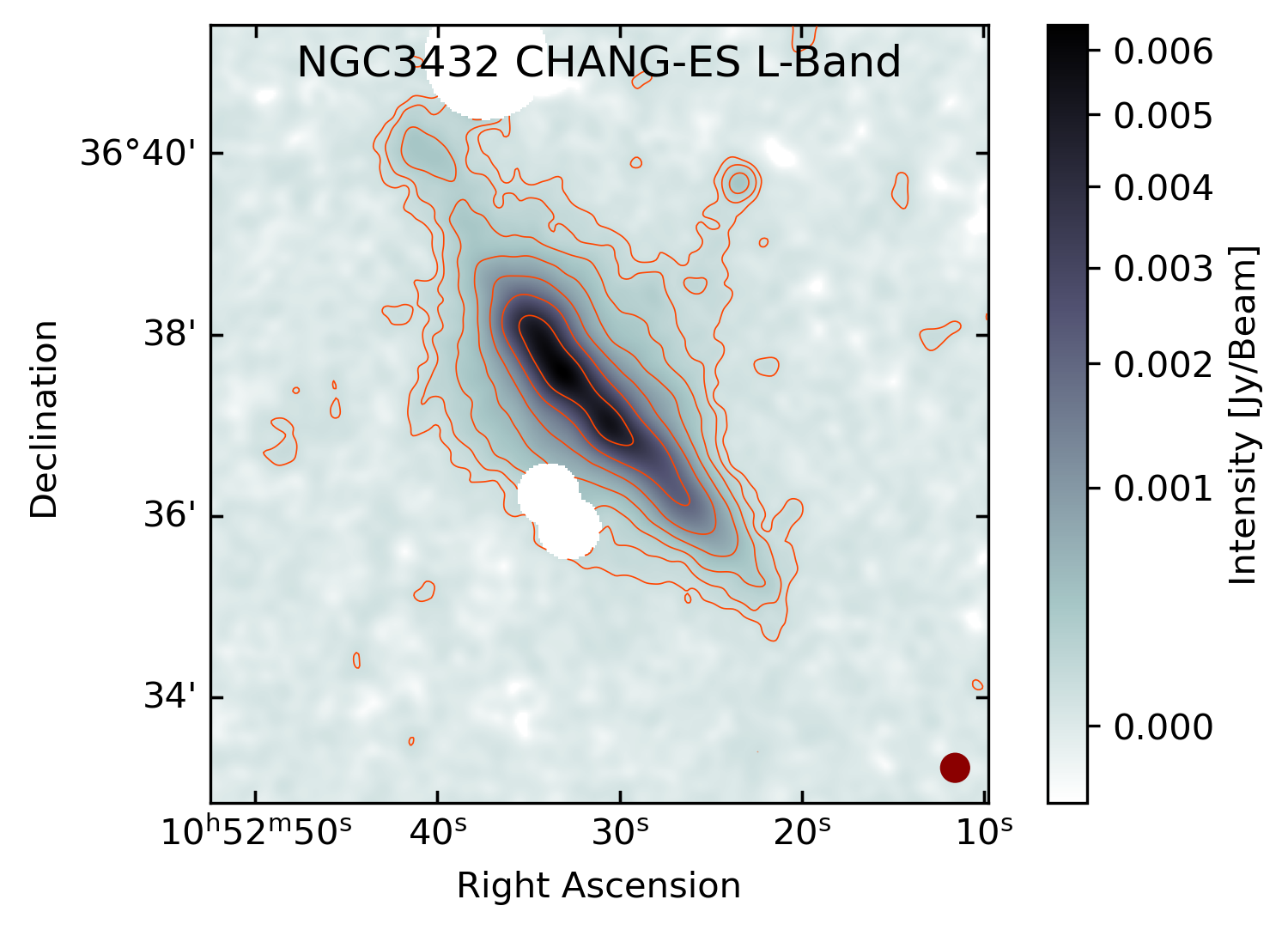}
\end{subfigure}
\hfill
\begin{subfigure}{0.45\linewidth}
\includegraphics[width=1\linewidth]{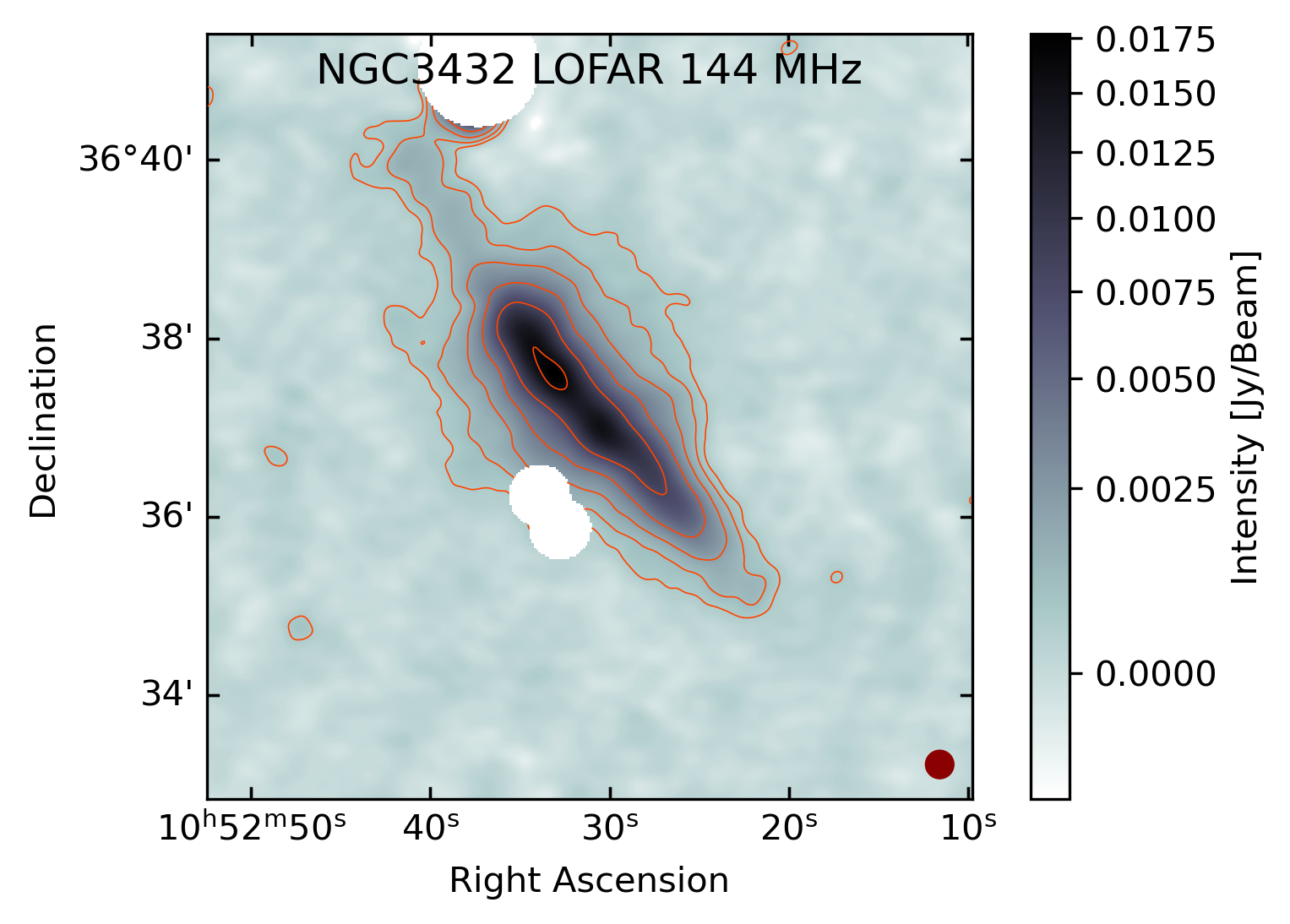}
\end{subfigure}
\\
\begin{subfigure}{0.45\linewidth}
\includegraphics[width=1\linewidth]{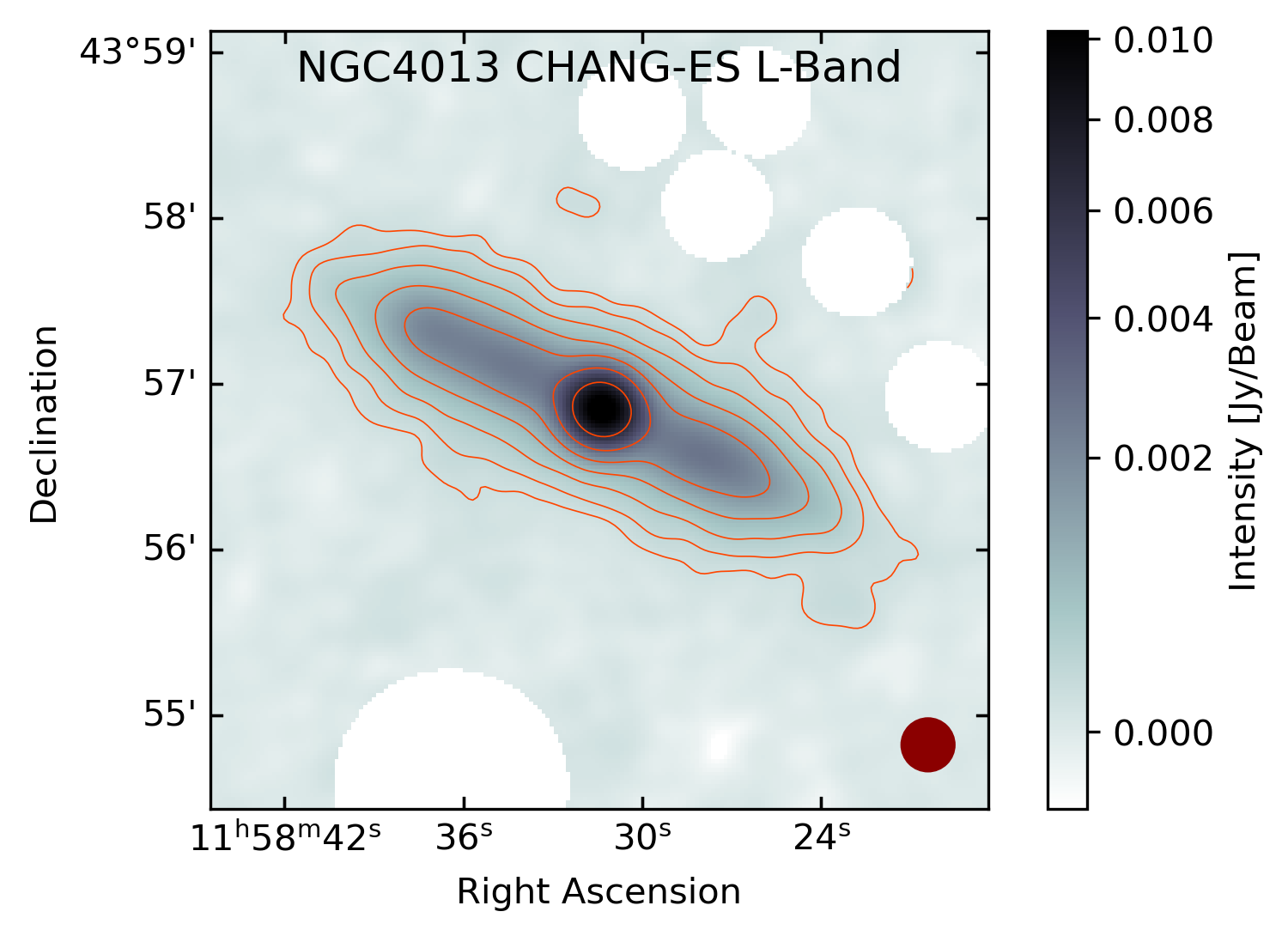}
\end{subfigure}
\hfill
\begin{subfigure}{0.45\linewidth}
\includegraphics[width=1\linewidth]{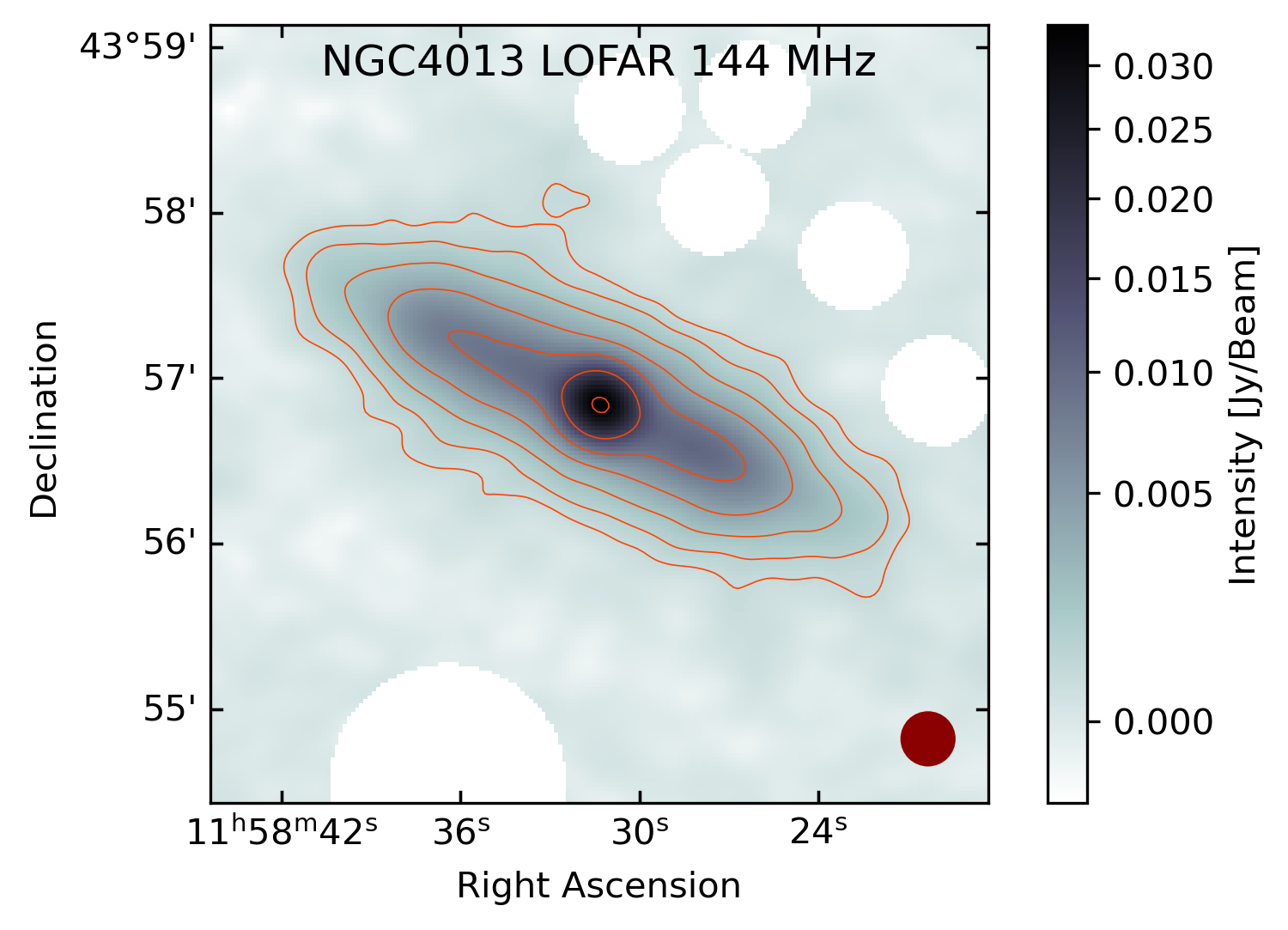}
\end{subfigure}
\caption{Total Intensity maps + contours from CHANG-ES L-band (left column) and LOFAR HBA (right column) after subtracting/masking point sources in the vicinity of the galaxy for NGC~891 (top row), NGC~3432 (middle row), and NGC~4013 (bottom row). Contours start at 3$\sigma$ above the background noise with an increment of 2. The beam is displayed in the bottom right corner of each map as a dark red circle.}
\label{fig:int_maps_1}
\end{figure*}

\begin{figure*}
\centering

\begin{subfigure}{0.45\linewidth}
\includegraphics[width=1\linewidth]{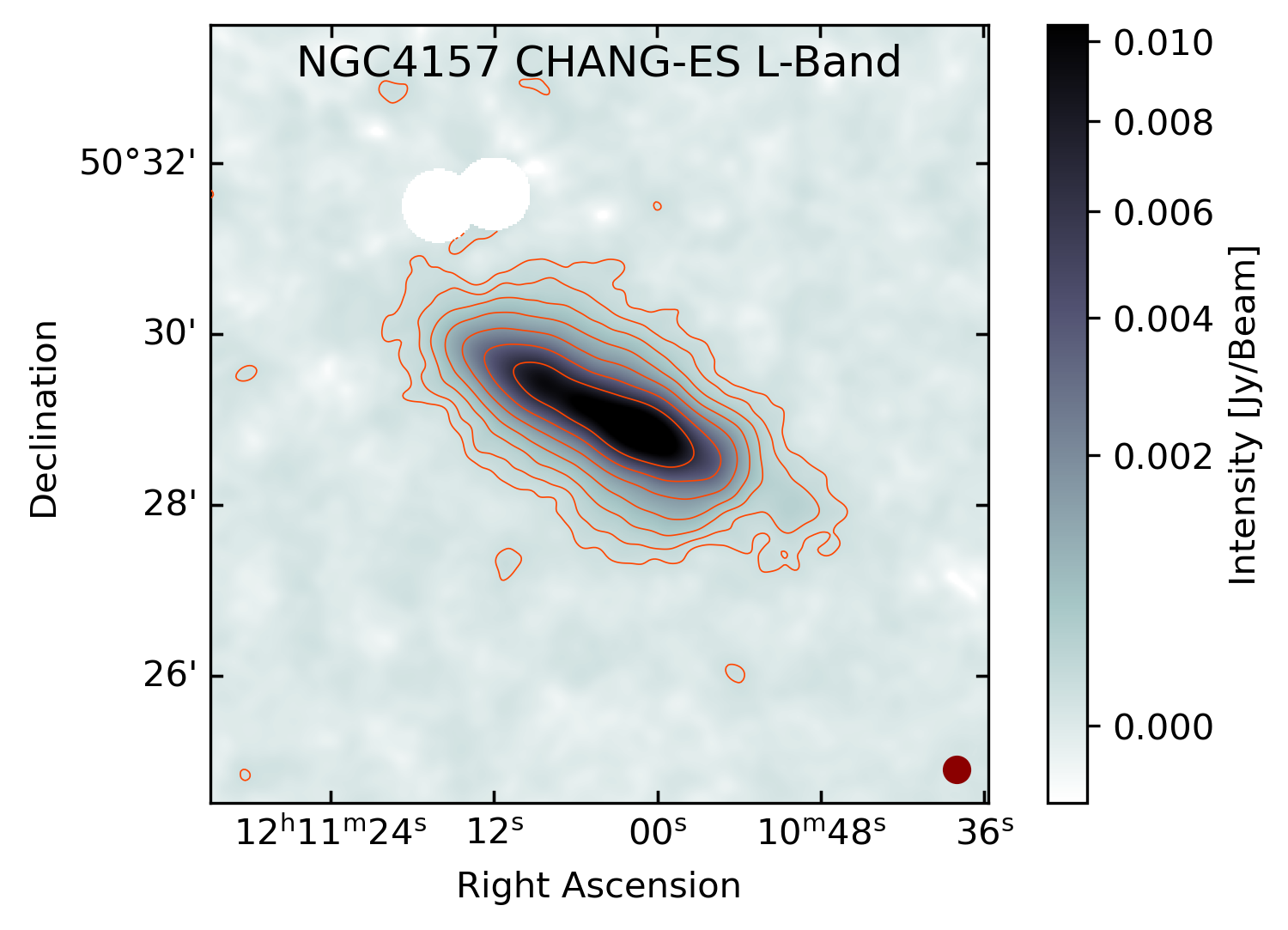}
\end{subfigure}
\hfill
\begin{subfigure}{0.45\linewidth}
\includegraphics[width=1\linewidth]{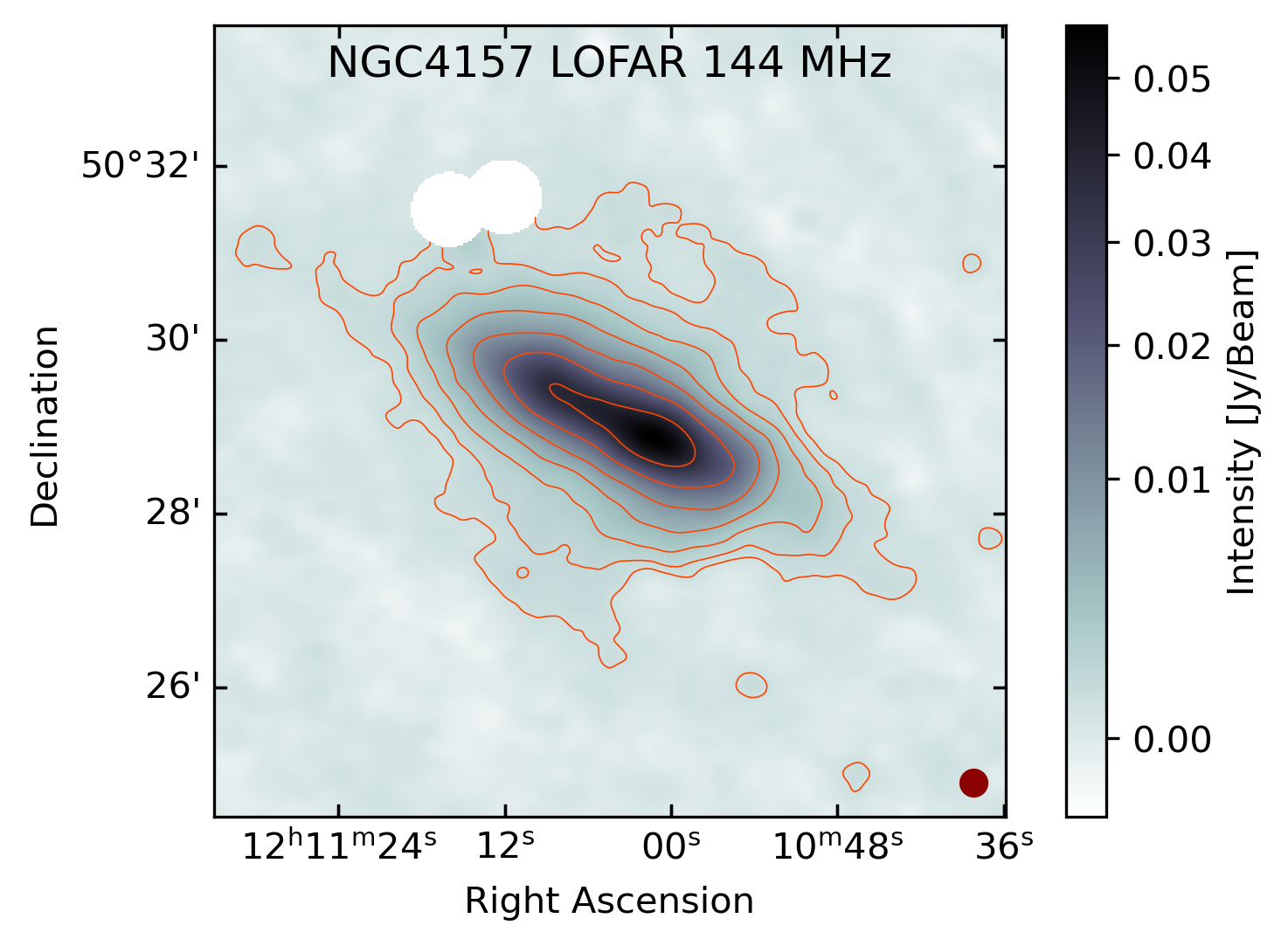}
\end{subfigure}
\\
\begin{subfigure}{0.45\linewidth}
\includegraphics[width=1\linewidth]{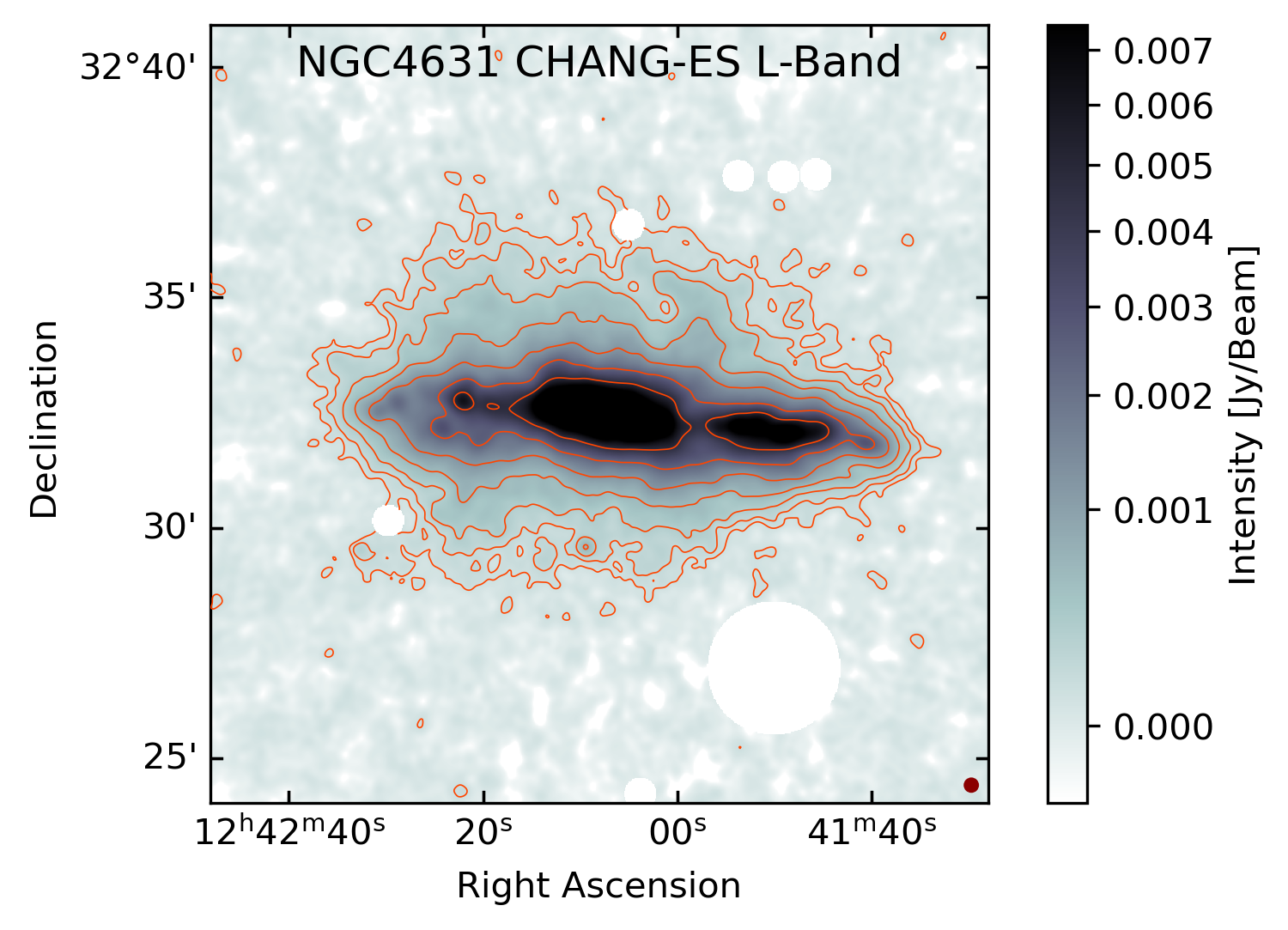}
\end{subfigure}
\hfill
\begin{subfigure}{0.45\linewidth}
\includegraphics[width=1\linewidth]{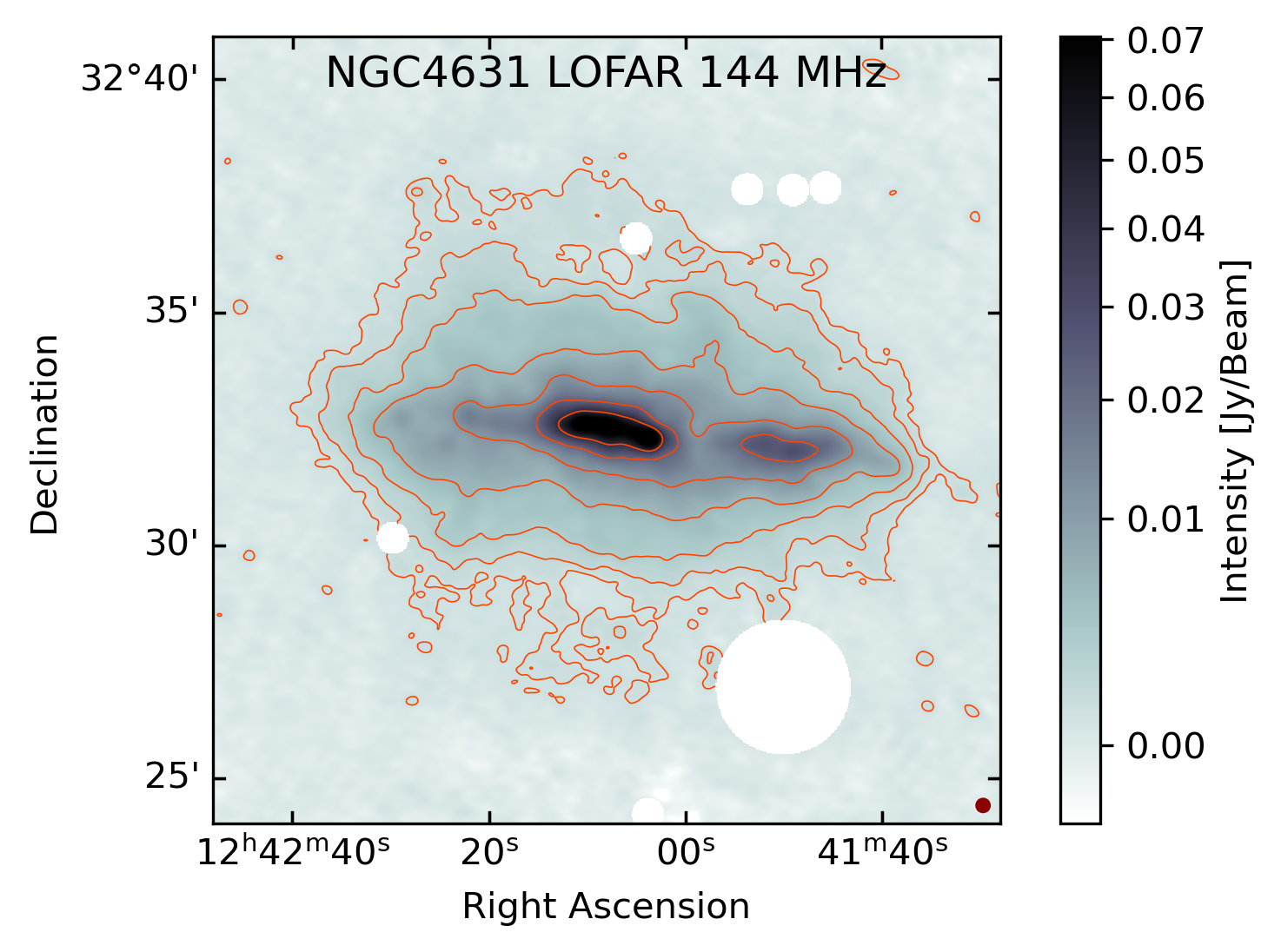}
\end{subfigure}
\caption{Continuation of Figure \ref{fig:int_maps_1}. Total Intensity maps from CHANG-ES L-band (left column) and LOFAR HBA (right column) after subtracting/masking point sources in the vicinity of the galaxy for NGC~4157 (top row), NGC~4631(bottom row).}
\label{fig:int_maps_2}
\end{figure*}

%% file: tab_obs_info.tex
\begin{table*}
\caption{Basic information about the galaxies EVLA C-Array observations.}
\label{tab:obs_info}
\centering
\begin{tabular}{l r r  r r r}
    \hline
    \hline                        
    Galaxy  & Observing Date & Integration Time & Flux Cal. & Phase Cal. & SB ID \\
            & [M-D-Y]       & [M:S]             &   &\\
    \hline
    NGC~891  & 02-11-2012, 04-01-2012    & 46:40 & 3C48  & J0314+4314 & 8237609, 8182311\\
    NGC~3432 & 03-25-2012                & 43:20 & 3C286 & J1006+3454 & 8256114\\
    NGC~4013 & 03-31-2012                & 44:00 & 3C286 & J1219+4829 & 8318174\\
    NGC~4157 & 03-31-2012                & 41:00 & 3C286 & J1219+4829 & 8318174\\
    NGC~4631 & 02-04-2012                & 77:00 & 3C286 & J1221+2813 & 8312467\\

\hline                                   
\end{tabular}

\end{table*}

%% file: si_fig.tex
\begin{figure*}
\centering
\begin{subfigure}{0.45\linewidth}
\centering
\includegraphics[width=1\linewidth]{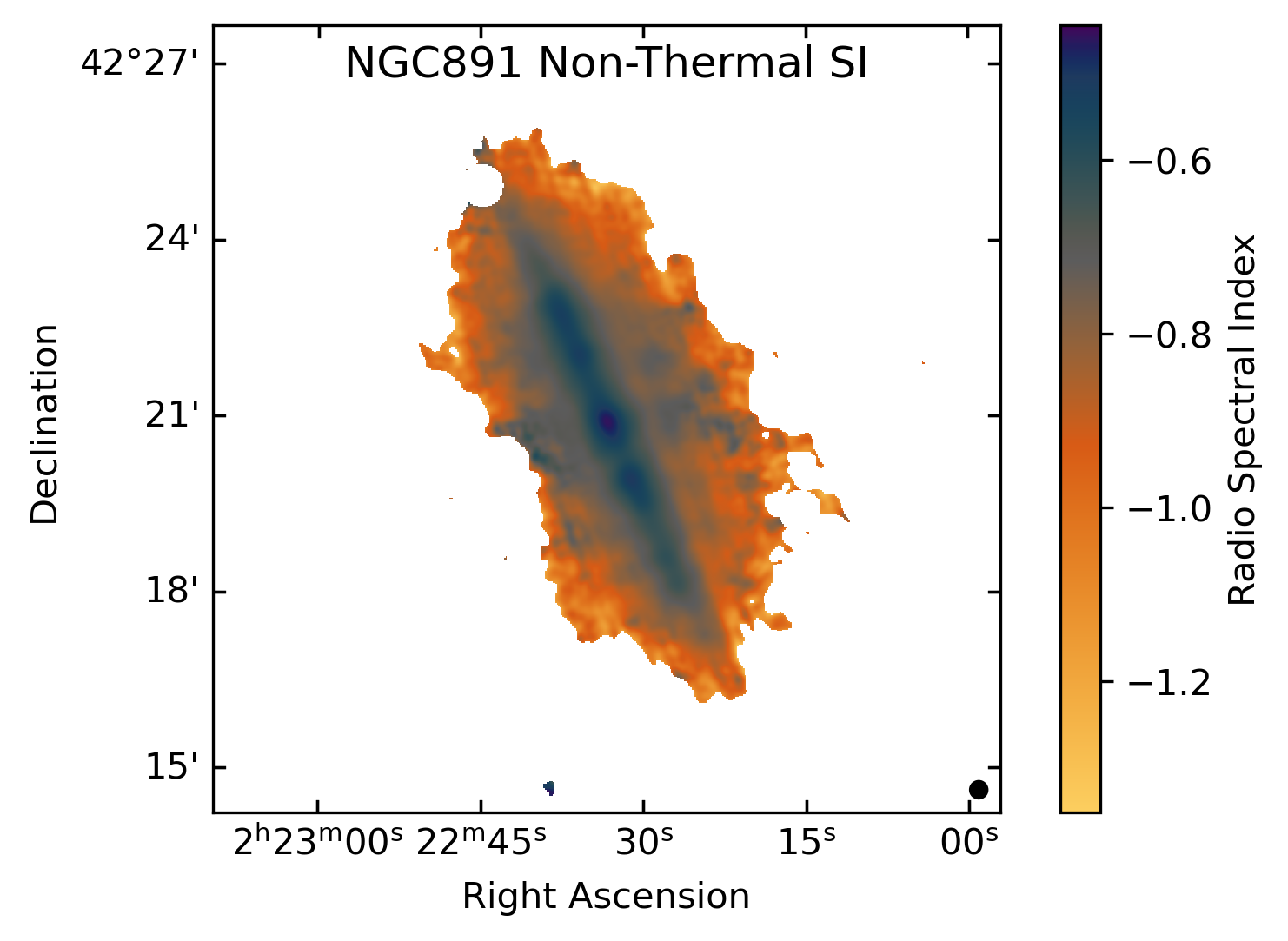}
\end{subfigure}
\hfill
\begin{subfigure}{0.45\linewidth}
\centering
\includegraphics[width=1\linewidth]{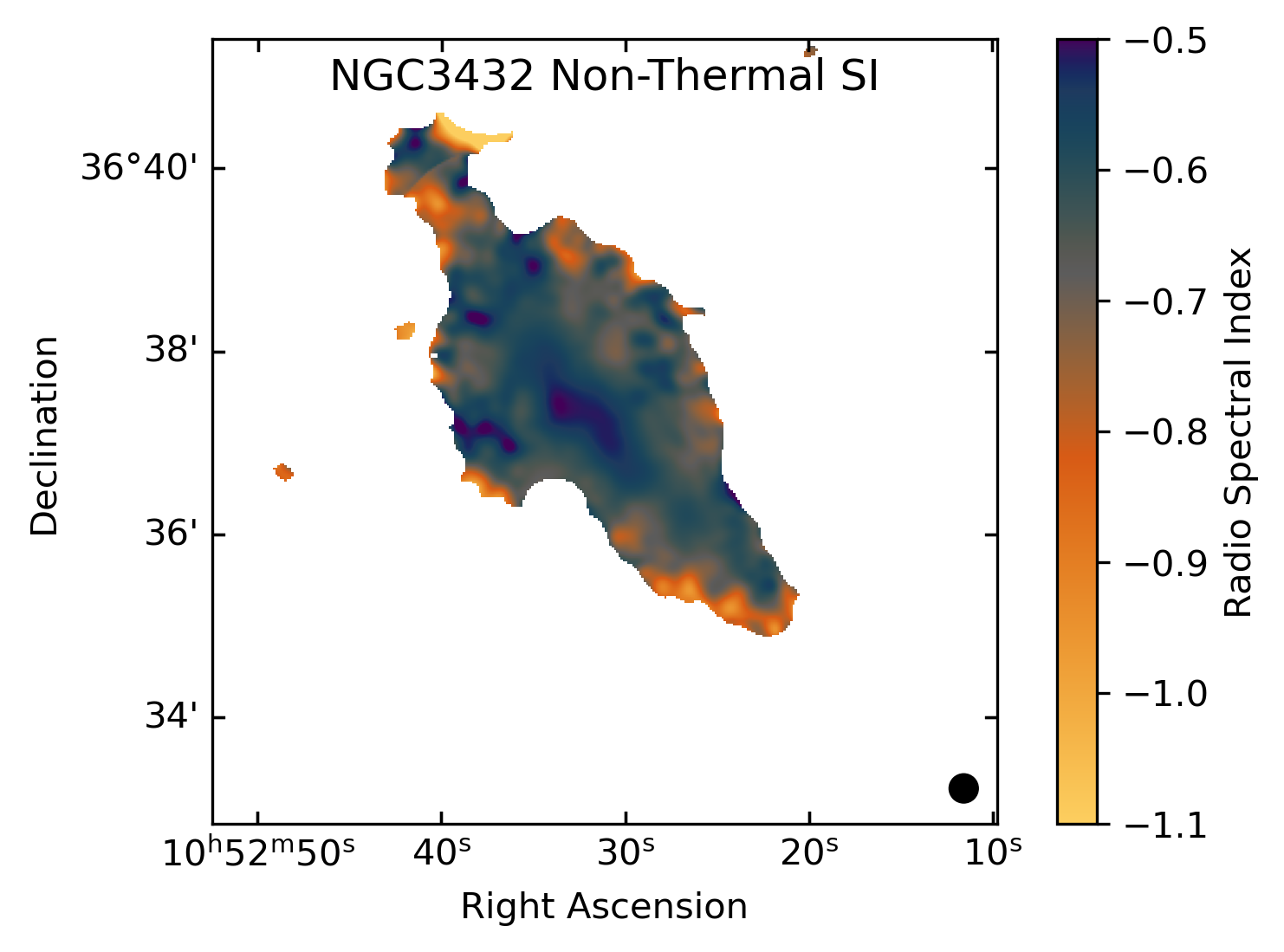}
\end{subfigure}
\\
\begin{subfigure}{0.45\linewidth}
\includegraphics[width=1\linewidth]{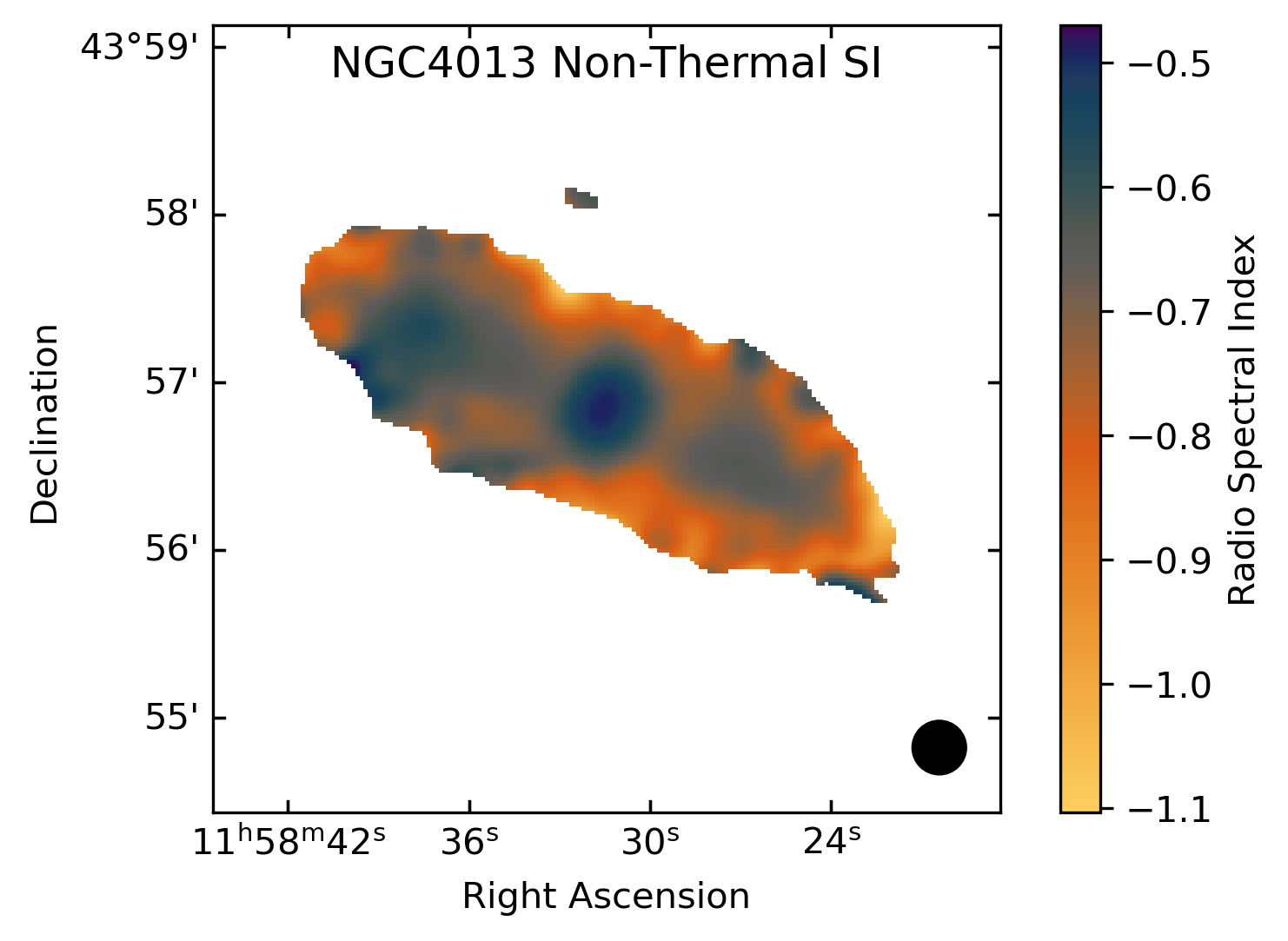}
\end{subfigure}
\hfill
\begin{subfigure}{0.45\linewidth}
\includegraphics[width=1\linewidth]{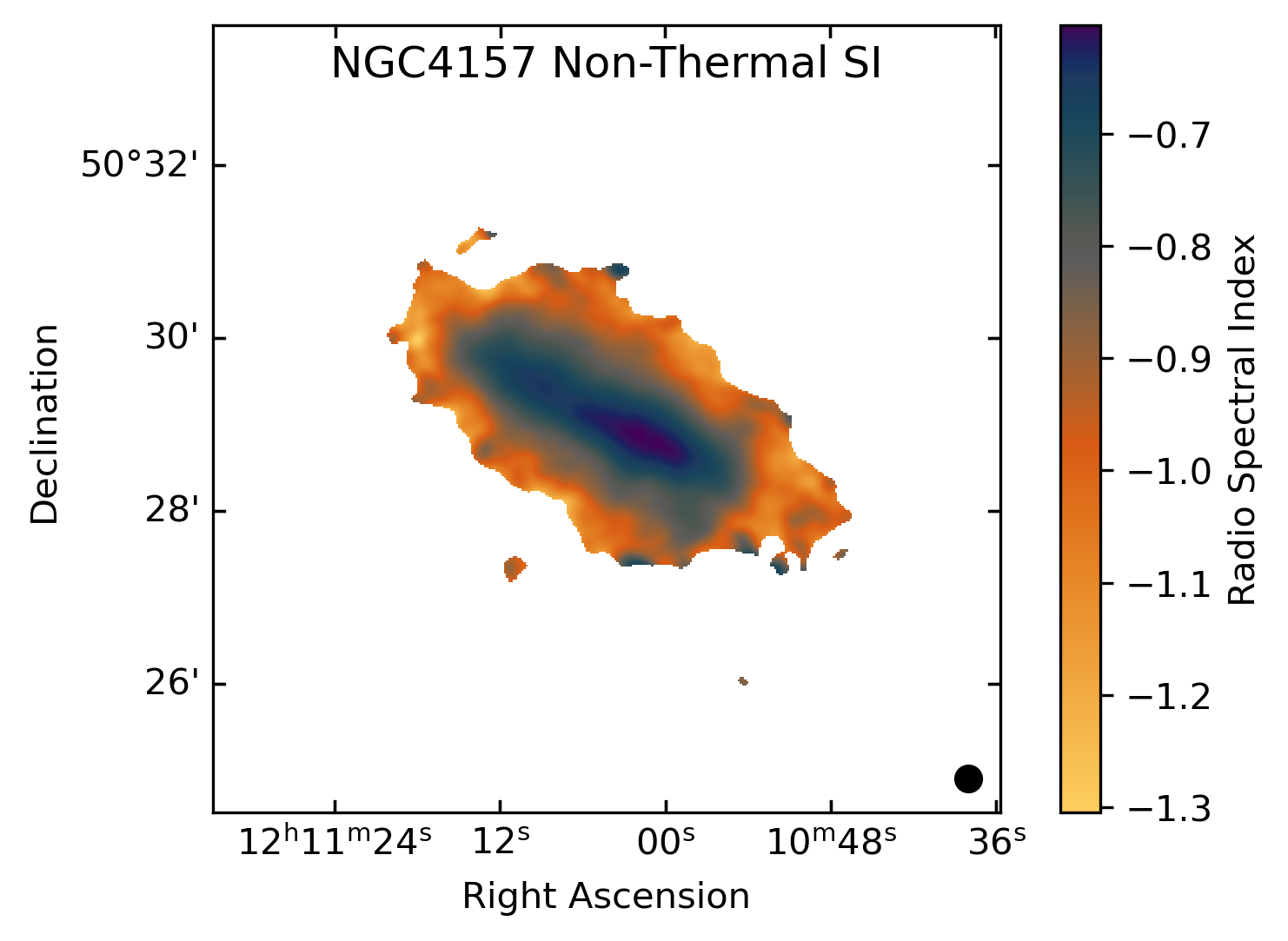}
\end{subfigure}
\\
\begin{subfigure}{0.45\linewidth}
\includegraphics[width=1\linewidth]{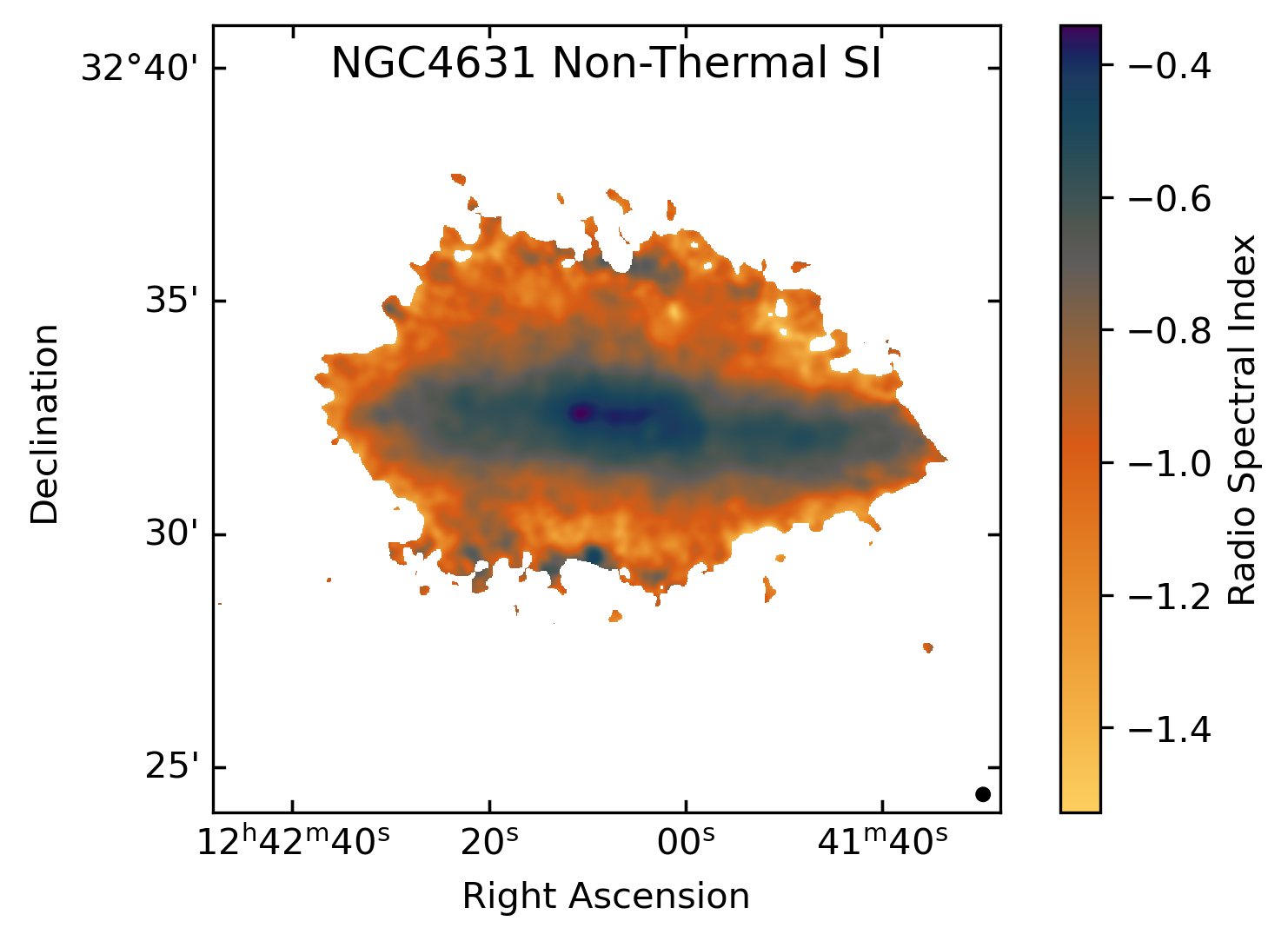}
\end{subfigure}
\caption{Non-thermal spectral index (L-band corrected for thermal emission \& uncorrected HBA) maps. We use a custom colormap, where a flat SI is coloured blue to plum and a steep SI orange to yellow. The beam is displayed in the bottom right corner of each map as black circle.}
\label{fig:si_maps}
\end{figure*}


%% file: tab_radio_param.tex
\begin{table*}
	\caption{Flux densities and map properties of the analysed sample. We list the integrated flux density measurements in HBA (S\textsubscript{HBA} and L-Band (S\textsubscript{L-Band}), based on the described aperture, accounting for background noise ($\sigma$\textsubscript{HBA} \& $\sigma$\textsubscript{L-Band}) as well as calibration errors. Physical diameters, based on the $3\sigma$-contours in HBA, parallel  ($d_{\mathrm{HBA}}^{\mathrm{disk}}$) and orthogonal  ($d_{\mathrm{HBA}}^{\mathrm{halo}}$) to the galactic disk are also listed.
	}
	\label{tab:flux}
\centering          
\begin{tabular}{l l r r r r r r}     
\hline\hline       
Galaxy & Ellipse $ a \times b$\tablefootmark{a} & $S$\textsubscript{HBA} & $\sigma$\textsubscript{HBA}\, & $S$\textsubscript{L-Band} & $\sigma$\textsubscript{Lband}& $d_{\mathrm{HBA}}^{\mathrm{disk}}$ & $d_{\mathrm{HBA}}^{\mathrm{halo}}$\\
& [arcmin$^2$] & [Jy] & [\muJyBeam] & [Jy] & [\muJyBeam] & [kpc] & [kpc]\\
\hline 

NGC~891  &$5.9\times 4.55 $ &3.3$\pm$0.3 &207 &0.68$\pm$0.03 &31 & 28.4 & 15.6\\
NGC~3432 &$4.35\times 2.2 $ &0.30$\pm$0.03 &165 &0.086$\pm$0.004 &22 & 14.4 & 8.5\\
NGC~4013 &$2.36\times 1.45$ &0.151$\pm$0.015 &167 &0.0381$\pm$0.0019 &34 & 16.1 & 5.4\\
NGC~4157 &$4.25\times 2.08$ &1.01$\pm$0.10 &196 &0.180$\pm$0.009 &35 & 34.8 & 21.7 \\
NGC~4631 &$8.8 \times 7.5$ &4.46$\pm$0.45 &235& 1.03$\pm$0.05 &25  & 30.0 & 24.6\\
\hline                  
\end{tabular}
\tablefoot{
\tablefoottext{a}{Ellipses are defined in \citet{2022arXiv220400635H} and encompass the $3\sigma$-contours of the re-calibrated LoTSS map.}
}
\end{table*}

%% file: tab_thermal_emission.tex
\begin{table}
	\caption{Integrated thermal flux densities and non-thermal fractions in the analysed sample.}
	\label{tab:therm flux}
\centering          
\begin{tabular}{l r r r r}     
\hline\hline       
Galaxy & $S_\mathrm{{HBA}^{th}}$ & $S_\mathrm{{L-Band}^{th}}$ & NTF\textsubscript{HBA} & NTF\textsubscript{L-Band}\\
& [Jy] & [Jy] & & \\
\hline 

NGC~891  &0.056 &0.044 &0.98 &0.94 \\
NGC~3432 &0.013 &0.010 &0.95 &0.86 \\
NGC~4013 &0.006 &0.005 &0.96 &0.86 \\
NGC~4157 &0.018 &0.014 &0.98 &0.93 \\
NGC~4631 &0.117 &0.092 &0.97 &0.91 \\
\hline                  
\end{tabular}

\end{table}

%% file: tab_box_setup.tex
\begin{table*}
\centering
\caption{Box widths (W) used for box integration, radio beam FWHM converted to physical scales, assumed flux tube radius (R\textsubscript{0}) used for cosmic ray transport modelling for each galaxy; SFR, SFR surface density, and average equipartition magnetic field strength in the galactic disk for individual stripes; gravitational pull for central and outer stripes (as we assume the gravitational potential to be symmetric, there is no distinction for left and right stripes).}             
\label{tab:box_setup}      
\begin{tabular}{l r r r | r r r | r r r | r r r | r r}        
    \hline\hline                 
    Galaxy  & W         & Beam  & R\textsubscript{0}& \multicolumn{3}{c}{SFR}                           & \multicolumn{3}{|c}{$\Sigma$\textsubscript{SFR}} & \multicolumn{3}{|c}{B-Field Strength} & \multicolumn{2}{|c}{Gravitational Pull}\\
            &[\arcsec]  &[kpc]  & [kpc]             &\multicolumn{3}{|c}{[M\textsubscript{\(\odot\)}\,yr\textsuperscript{-1}]} & \multicolumn{3}{|c}{[M\textsubscript{\(\odot\)}\,yr\textsuperscript{-1}\,kpc\textsuperscript{-2}]$\cdot10^{-3}$}& \multicolumn{3}{|c}{[\textmu G]} & \multicolumn{2}{|c}{[N]$\cdot10^{-10}$}\\
            &           &       &                   & left & middle & right & left & middle & right & left & middle & right & middle & outer\\ 

    \hline                        
    NGC~891  & 177   & 0.88 & 7.8    & 0.5 & 1.0 & 0.4   & 3.9 & 5.6 & 2.8   & 9.2 & 13.1 & 9.0 & 3.7 & 0.6\\       
    NGC~3432 & 102   & 0.91 & 4.6      & 0.2 & 0.3 & 0.1   & 3.7 & 4.2 & 1.1   & 8.6 & 10.2 & 6.7 & 1.6 & 0.3\\
    NGC~4013 & 63    & 1.36 & 4.3      & 0.2 & 0.3 & 0.1   & 4.6 & 5.6 & 3.8   & 7.5 &  9.1 & 7.3 & 5.0 & 0.8\\  
    NGC~4157 & 120   & 1.51 & 9.0   & 0.2 & 1.3 & 0.2   & 1.3 & 5.5 & 1.2   & 8.1 & 13.1 & 7.4 & 2.5 & 0.4\\ 
    NGC~4631 & 266   & 0.72 & 9.6    & 0.3 & 1.6 & 0.5   & 1.8 & 6.1 & 2.8   & 7.5 & 11.7 & 8.9 & 1.3 & 0.2\\ 
\hline                                   
\end{tabular}\\
\end{table*}

%% file: tab_scale_summary.tex
\begingroup
\renewcommand*{\arraystretch}{1.2}
\begin{table}
\centering
\caption{Fitted radio scale heights per galaxy. One and two component model are listed separately. For the two component models, we distinguish disk and halo scale heights. The Number of stripes (N) that have been fitted with each model is also indicated. If only one stripe has been fitted with a certain model, the uncertainty of the scale heights is computed within $\chi^2$-minimisation. If multiple stripes have been been fitted with a model, the uncertainty is computed via the standard deviation of the sample. }             
\label{tab:scale_summary}      
\begin{tabular}{l r r r| r r}        
    \hline\hline                 
    Band  & \multicolumn{3}{c}{Two-Component} & \multicolumn{2}{|c}{One-Component} \\
            & Disk & Halo & N  & Overall & N \\
            &[kpc] &[kpc] &     &[kpc]\\
    \hline
    \multicolumn{4}{l|}{\textbf{NGC~891}}   \\
    L-Band  & -             &           - & 0 & 1.09$\pm$0.16 & 6 \\
    HBA     & 0.76$\pm$0.25 & 2.9$\pm$0.5 & 4 & 1.76$\pm$0.09 & 2    \\
    \hline
    \multicolumn{4}{l|}{\textbf{NGC~3432}}  \\
    L-Band  & - & - & 0 & 0.86$\pm$0.21 & 6   \\
    HBA     & - & - & 0 & 0.7$\pm$0.4 & 6    \\
    \hline
    \multicolumn{4}{l|}{\textbf{NGC~4013}} \\
    L-Band  & - & - & 0 & 0.60$\pm$0.08 & 6    \\
    HBA     & - & - & 0 & 0.75$\pm$0.14 & 6\\
    \hline
    \multicolumn{4}{l|}{\textbf{NGC~4157}}  \\
    L-Band  & 0.7$\pm$0.4 & 1.2$\pm$0.2 & 1 & 1.16$\pm$0.19 & 5  \\
    HBA     & 1.05$\pm$0.16 & 7$\pm$3 & 2 & 2.2$\pm$0.6 & 4\\
    \hline
    \multicolumn{4}{l|}{\textbf{NGC~4631}}  \\
    L-Band  & 0.50$\pm$0.08 & 2.1$\pm$0.3 & 3 & 1.7$\pm$0.5 & 3    \\
    HBA     & 1.45$\pm$0.01 & 4.9$\pm$0.4 & 2 & 2.8$\pm$0.6 & 4    \\
\hline                                   
\end{tabular}\\
\end{table}
\endgroup

%% file: tab_spin_param.tex
\centering
\begingroup
\renewcommand{\arraystretch}{1.3} 
\begin{tabular}{l l r r r r r r r}        
    \hline\hline                 
    Stripe & $z_{\mathrm{max}}$ & $\gamma$ & $\varv_c$ & $z_0$ & $\beta$ & z\textsubscript{c} & $\chi^2_\nu$\\ 
        &       [kpc]            &                                      & [km\,s\textsuperscript{-1}] & [kpc] &  & [kpc] & \\ 
    \hline                        
    \multicolumn{8}{c}{NGC~891}\\
    \hline
    LR  & 5.2      & 2.40$^{+0.08}_{-0.06}$ & 128$^{+18}_{-8}$ & 3.9$^{+0.5}_{-0.7}$ & 1.53$^{+0.04}_{-0.1}$ & 1.0$^{+0.6}_{-0.6}$ & 3.1	\\
    LM     & 5.2      & 2.16$^{+0.07}_{-0.05}$ & 152$^{+24}_{-11}$ & 2.4$^{+0.2}_{-0.3}$ & 1.73$^{+0.05}_{-0.08}$ & 0.2$^{+0.1}_{-0.1}$ & 3.4\\
    LL   & 5.2    & 2.20$^{+0.08}_{-0.07}$ & 129$^{+8}_{-6}$ & 4.9$^{+0.6}_{-0.6}$ & 1.85$^{+0.02}_{-0.05}$ & 2.3$^{+0.9}_{-0.7}$ & 3.6\\
    UR  & 6.0   & 2.32$^{+0.13}_{-0.07}$ & 167$^{+38}_{-14}$ & 5$^{+0.7}_{-0.9}$ & 1.79$^{+0.13}_{-0.18}$ & 1.1$^{+0.6}_{-0.6}$ & 2.2\\
    UM     & 7.8                         & 2.11$^{+0.08}_{-0.05}$ & 189$^{+26}_{-16}$ & 3.55$^{+0.3}_{-0.4}$ & 1.78$^{+0.06}_{-0.07}$ & 0.34$^{+0.12}_{-0.13}$ & 3.4\\ 
    UL   & 6.0   & 2.27$^{+0.08}_{-0.08}$ & 137$^{+9}_{-8}$ & 5.0$^{+0.8}_{-0.5}$ & 1.85$^{+0.03}_{-0.04}$ & 2.0$^{+0.9}_{-0.5}$ & 3.6\\
    \hline
    \multicolumn{8}{c}{NGC~3432}\\
    \hline
    LM & 3.9 & 2.13$^{+0.09}_{-0.16}$ & 240$^{+420}_{-135}$ & 3.0$^{+0.4}_{-0.2}$ & 2.10$^{+0.30}_{-0.35}$ & <0.9 & 1.5\\
    UM & 3.8 & 2.17$^{+0.12}_{-0.12}$ & 233$^{+760}_{-100}$ & 2.7$^{+0.2}_{-0.4}$ & 2.2$^{+0.20}_{-0.4}$ & <0.9 & 0.55	\\			\hline
    \multicolumn{8}{c}{NGC~4013}\\
    \hline
    LR & 3.6 & 2.4$^{+0.2}_{-0.2}$ & 122$^{+85}_{-26}$ & 2.5$^{+0.4}_{-0.3}$ & 2.0 & 0.9$^{+0.9}_{-0.6}$ & 0.4\\	
    LM & 2.9 & 2.30$^{+0.08}_{-0.04}$ & 100$^{+55}_{-14}$ & 1.1$^{+0.1}_{-0.1}$ & 1.6 & <1.4 & 1.7\\
    LL & 3.6 & 2.25$^{+0.15}_{-0.10}$ & 180$^{+610}_{-60}$ & 2.1$^{+0.4}_{-0.3}$ & 2.0 & <1.4 & 1	\\							
    \hline
    \multicolumn{8}{c}{NGC~4157}\\
    \hline
    LM     & 7.2   & 2.21$^{+0.09}_{-0.06}$ & 177$^{+24}_{-12}$ & 4.7$^{+0.4}_{-0.7}$ & 1.72$^{+0.04}_{-0.10}$ & 0.4$^{+0.1}_{-0.2}$ & 1.7										 \\
    LL   & 6.0    & 2.39$^{+0.09}_{-0.10}$ & 128$^{+6}_{-5}$ & 8.2$^{+0.8}_{-1.1}$ & 2.21$^{+0.02}_{-0.02}$ & 4.7$^{+1.0}_{-1.2}$ & 0.5										\\
    UM     & 7.1   & 2.26$^{+0.08}_{-0.08}$ & 226$^{+26}_{-24}$ & 5.3$^{+0.4}_{-0.5}$ & 2.05$^{+0.14}_{-0.12}$ & 0.6$^{+0.2}_{-0.2}$ & 2.81										\\
    UL   & 4.7  & 2.34$^{+0.10}_{-0.08}$ & 164$^{+10}_{-8}$ & 13.0$^{+2.0}_{-2.5}$ & 2.66$^{+0.02}_{-0.04}$ & 7.0$^{+0.5}_{-2.0}$ & 2.9																				\\
    \hline
    \multicolumn{8}{c}{NGC~4631}\\
    \hline
    LR  & 6.0    & 2.2$^{+0.16}_{-0.20}$ & 95$^{+12}_{-12}$ & 7.3$^{+1.7}_{-2.1}$ & 1.80$^{+0.10}_{-0.14}$ & 3.1$^{+3.2}_{-2.0}$ & 0.58																			\\
    LM     & 5.3     & 2.1$^{+0.10}_{-0.04}$ & 95$^{+20}_{-10}$ & 3.1$^{+0.5}_{-0.7}$ & 1.20$^{+0.08}_{-0.08}$ & <0.72 & 1.0																				\\
    LL   & 6.0    & 2.3$^{+0.30}_{-0.26}$ & 150$^{+55}_{-26}$ & 13$^{+7}_{-4}$ & 2.80$^{+0.10}_{-1.30}$ & 4.9$^{+0.9}_{-4.3}$ & 0.33																				\\
    UR  & 7.9   & 2.45$^{+0.24}_{-0.09}$ & 115$^{+114}_{-22}$ & 4.7$^{+1.0}_{-2.3}$ & 1.30$^{+0.15}_{-0.36}$ & <0.72 & 1.9																				\\
    UM     & 10.5     & 2.20$^{+0.14}_{-0.05}$ & 145$^{+34}_{-15}$ & 3.7$^{+0.4}_{-1.0}$ & 1.25$^{+0.10}_{-0.10}$ & <0.72 & 3.0																				\\
    UL   & 10.1     &2.45$^{+0.28}_{-0.12}$ & 125$^{+40}_{-15}$ & 9.4$^{+3.2}_{-3.1}$ & 1.25$^{+0.12}_{-0.14}$ & 0.5$^{+3.2}_{-0.5}$ & 0.5																				\\
\hline                                   
\end{tabular}

\endgroup

%% file: thermal_figs_appendix.tex
 The thermal emission maps in for the VLA L-band and LOFAR HBA data as well as the corresponding thermal fraction maps are displayed in this section. In the thermal fraction maps, we plot contours at factors of 0.25, 0.5, and 0.75 of the peak thermal fraction that was measured for each galaxy.
\begin{figure*}
\centering
\begin{subfigure}{0.45\linewidth}
\centering
\includegraphics[width=1\linewidth]{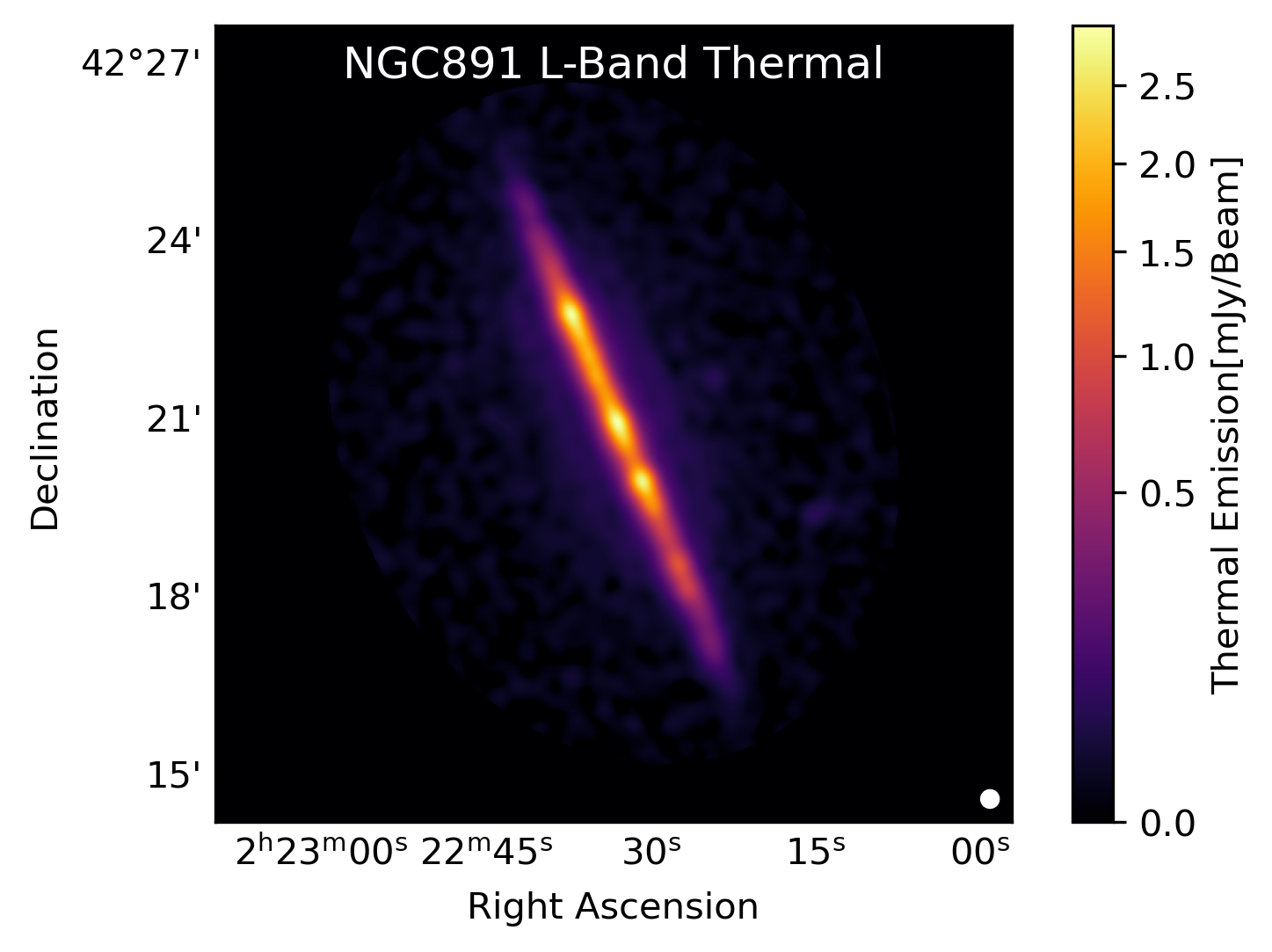}
\end{subfigure}
\begin{subfigure}{0.45\linewidth}
\centering
\includegraphics[width=1\linewidth]{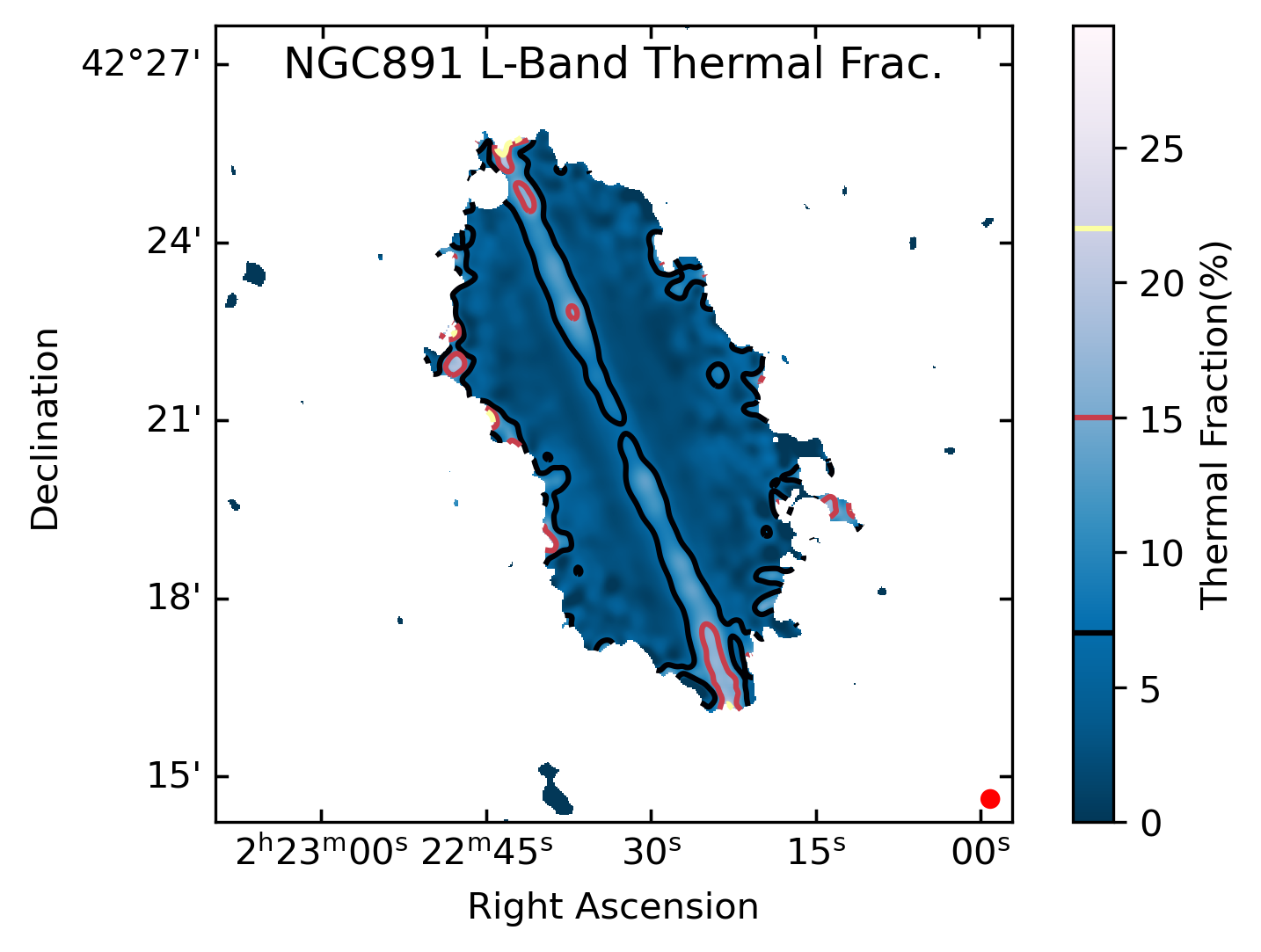}
\end{subfigure}
\\
\begin{subfigure}{0.45\linewidth}
\includegraphics[width=1\linewidth]{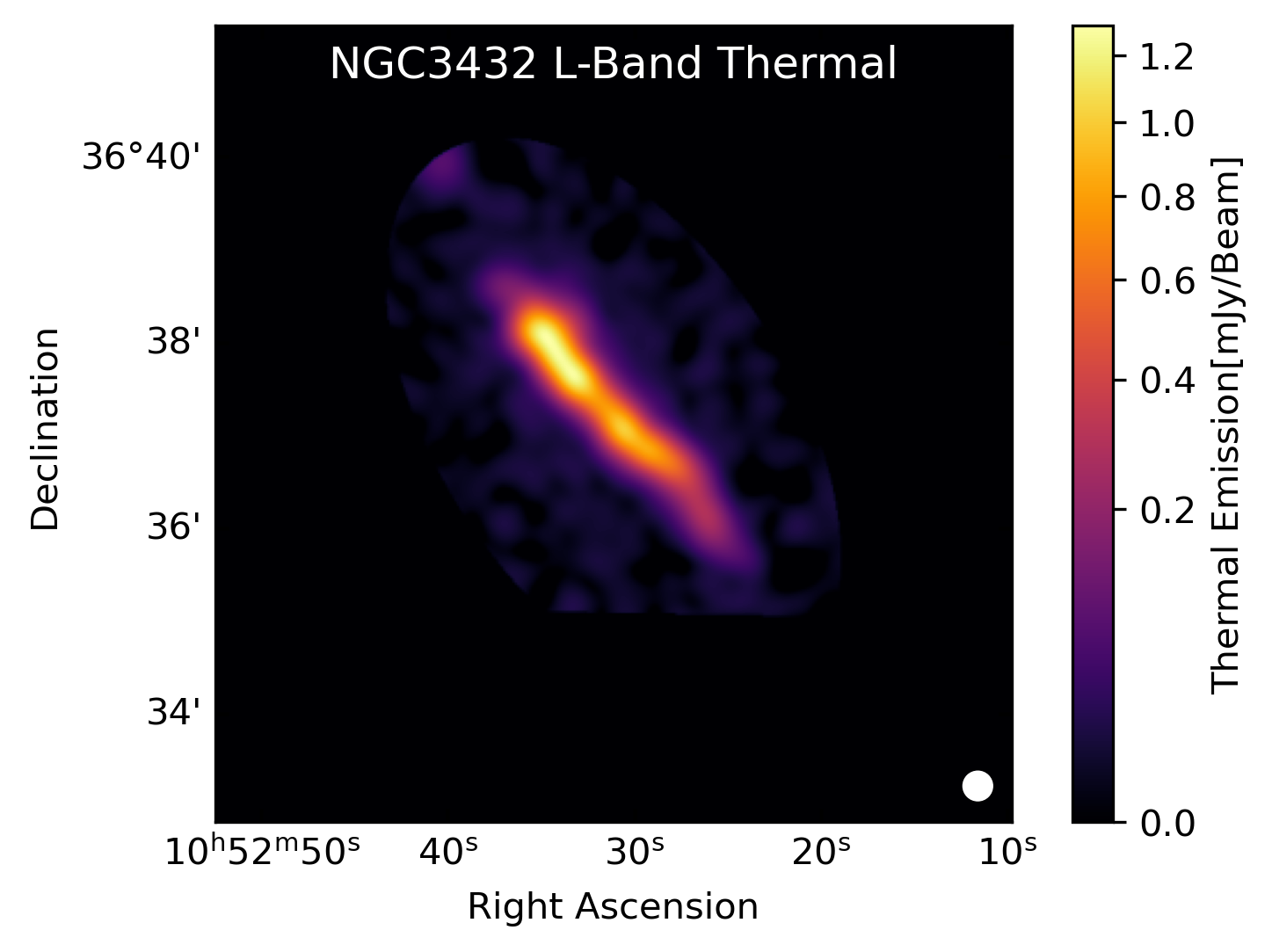}
\end{subfigure}
\begin{subfigure}{0.45\linewidth}
\includegraphics[width=1\linewidth]{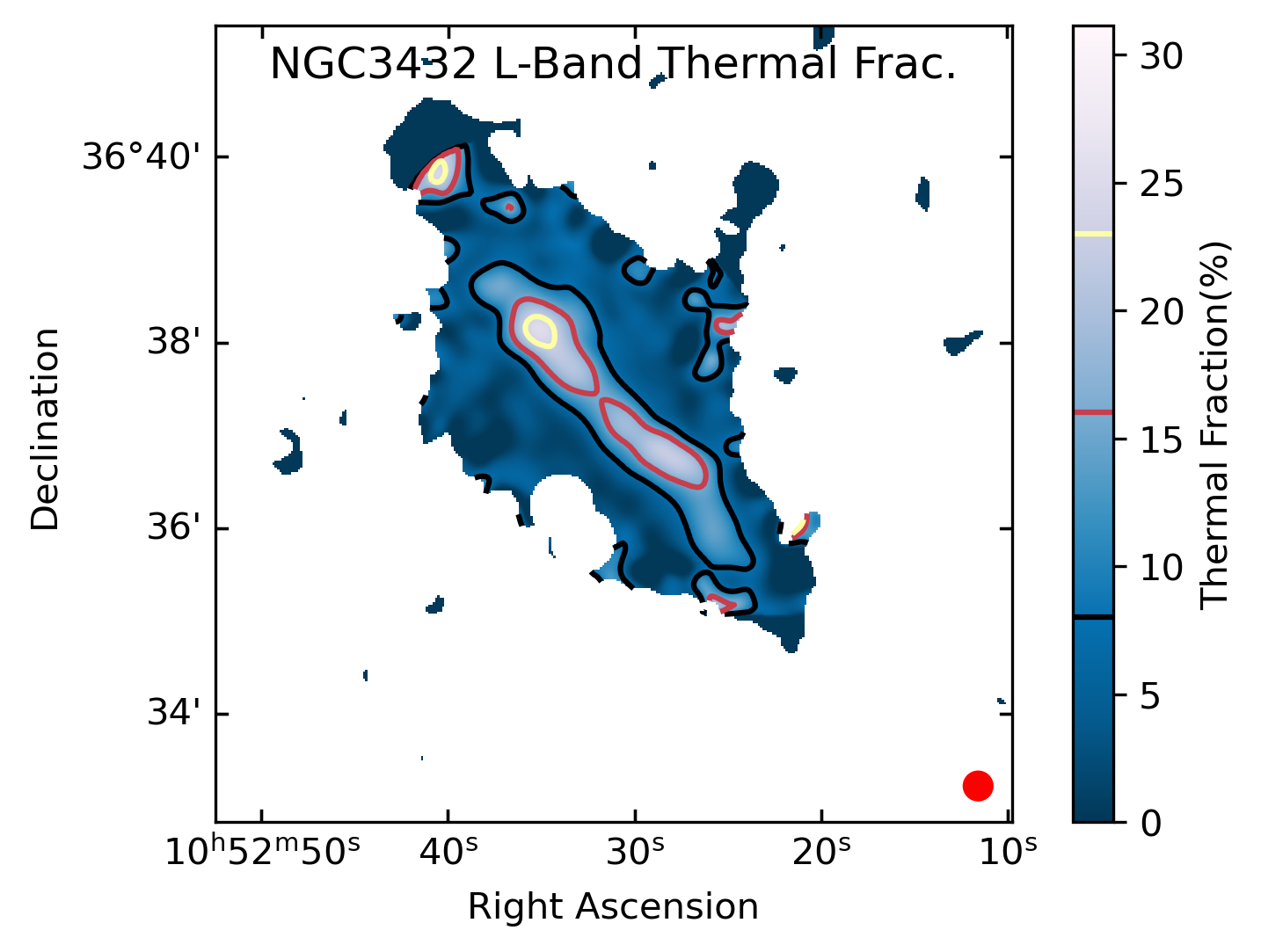}
\end{subfigure}
\\
\begin{subfigure}{0.45\linewidth}
\includegraphics[width=1\linewidth]{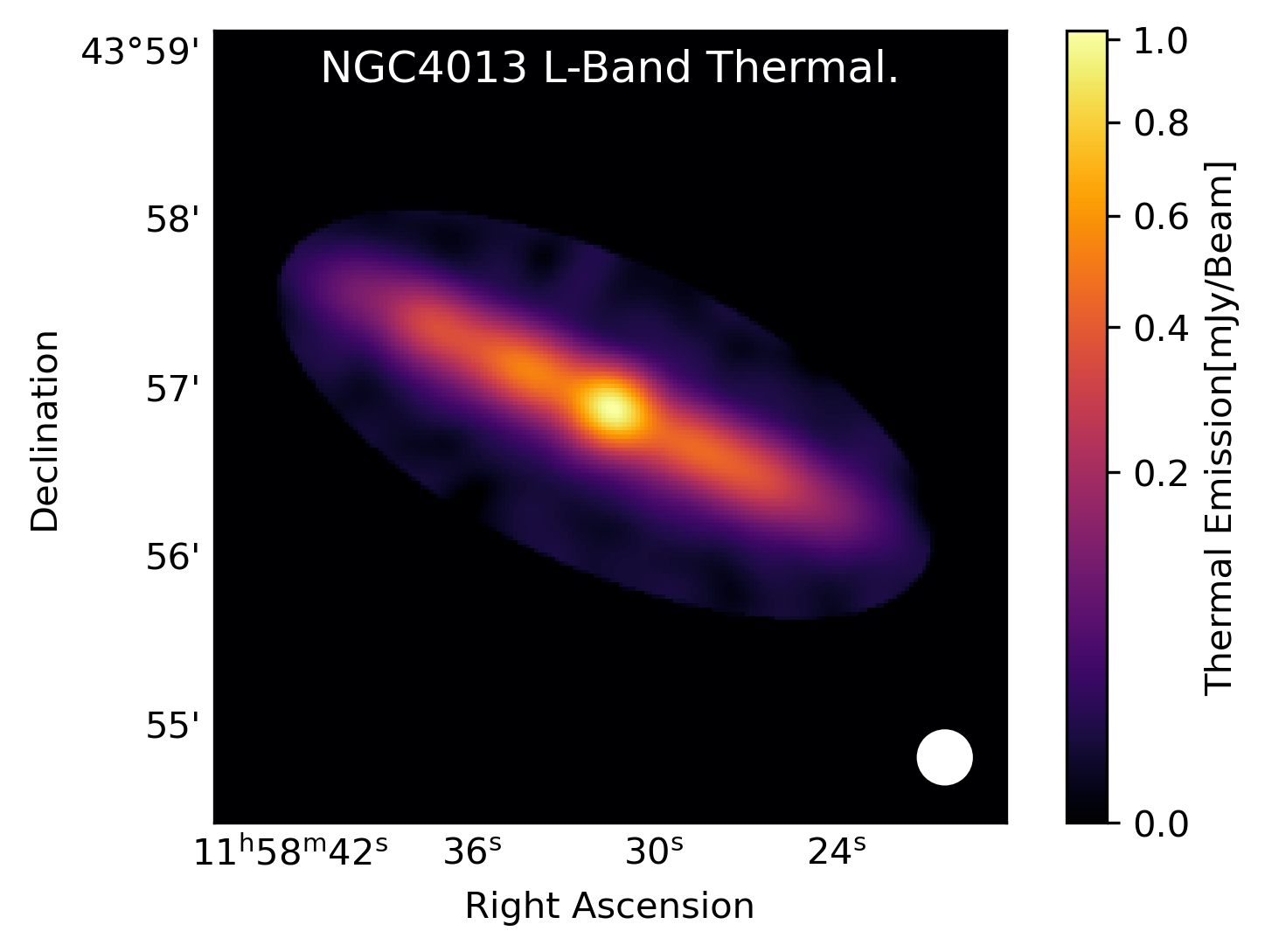}
\end{subfigure}
\begin{subfigure}{0.45\linewidth}
\includegraphics[width=1\linewidth]{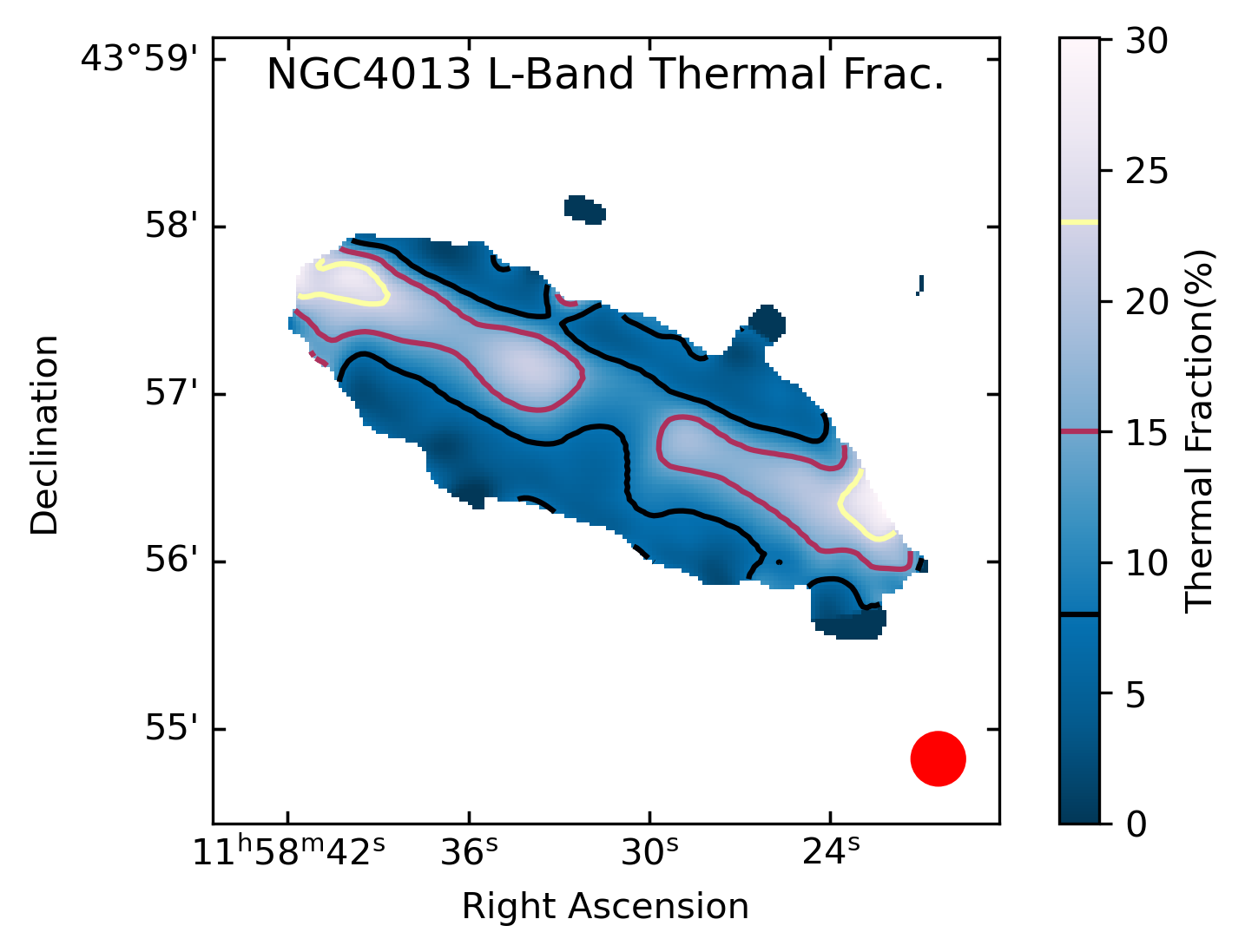}
\end{subfigure}
\caption{Thermal emission at 1.5\,GHz (CHANG-ES L-band) (left column) and thermal fraction maps of the L-band data (right column) of NGC~891 (top row), NGC~3432 (middle row), and NGC~4013 (bottom row). Thermal emission maps use a power law scaling with a power law index of 0.5. Thermal fraction maps are clipped below 3$\sigma$ above the background noise. The beam is displayed in the bottom right corner of each map as a white (thermal emission maps) or red (thermal fraction maps) circle.}
\label{fig:thermal_maps_lband1}
\end{figure*}

\begin{figure*}
\centering
\begin{subfigure}{0.45\linewidth}
\includegraphics[width=1\linewidth]{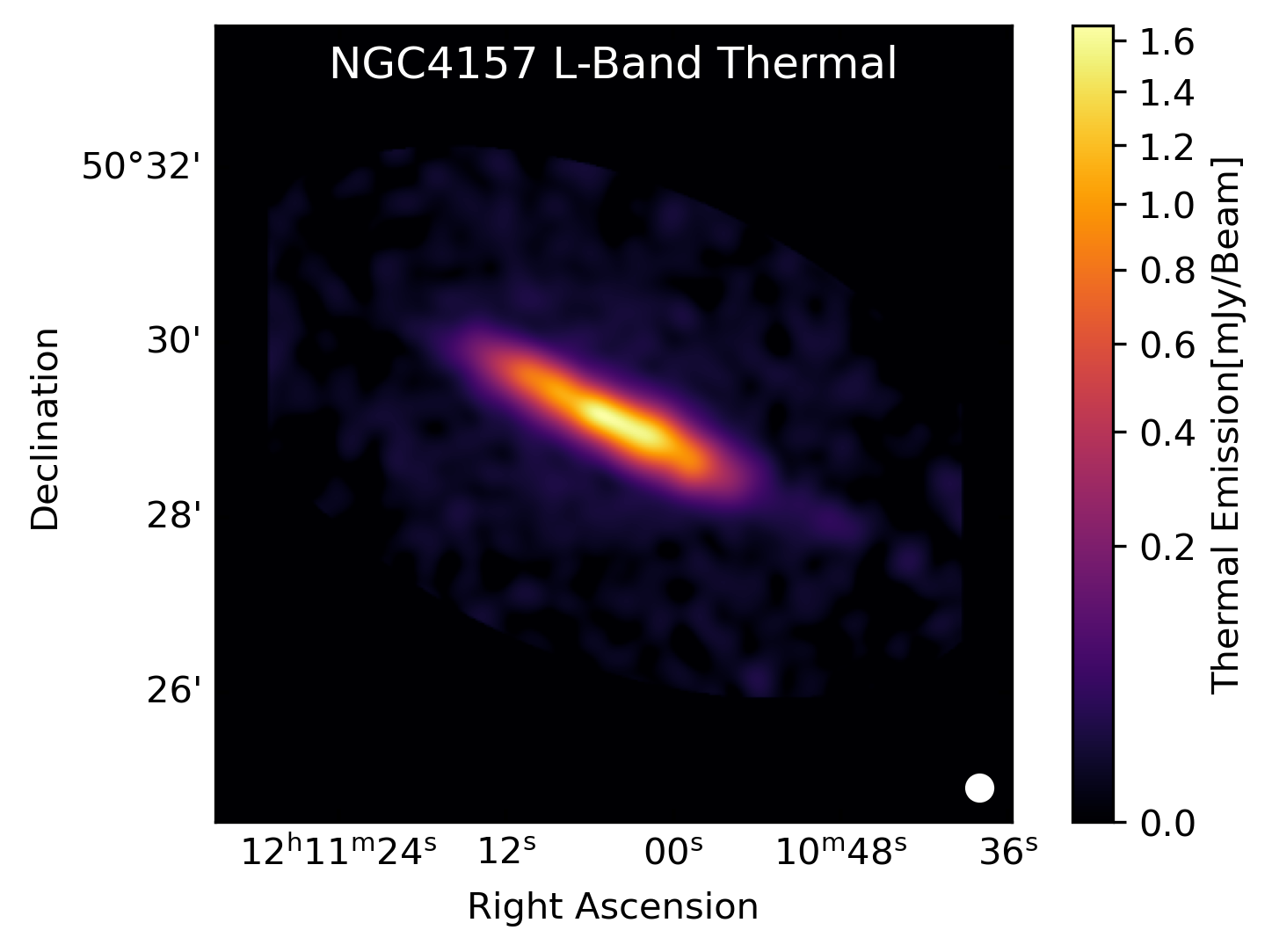}
\end{subfigure}
\begin{subfigure}{0.45\linewidth}
\includegraphics[width=1\linewidth]{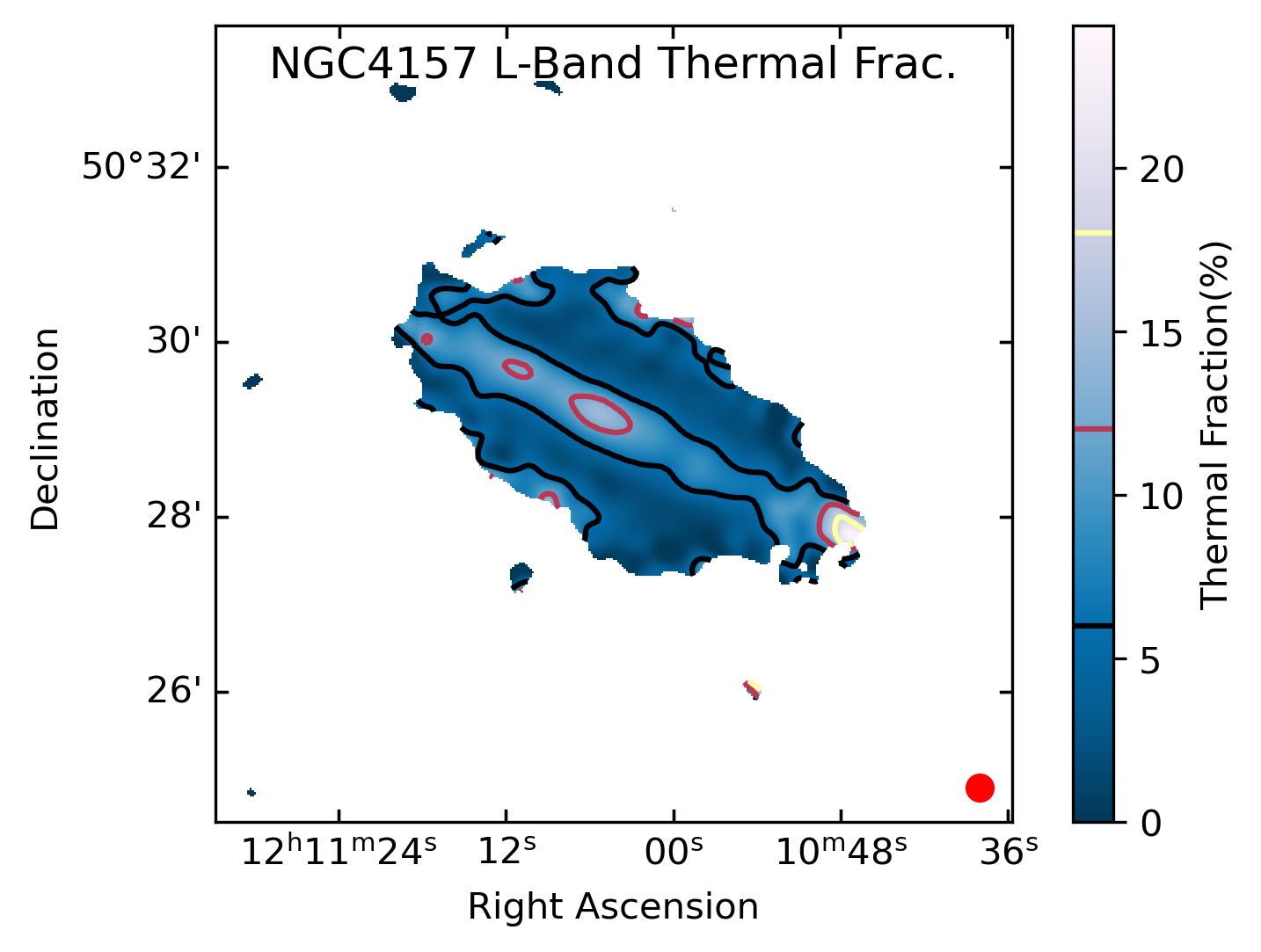}
\end{subfigure}
\\
\begin{subfigure}{0.45\linewidth}
\includegraphics[width=1\linewidth]{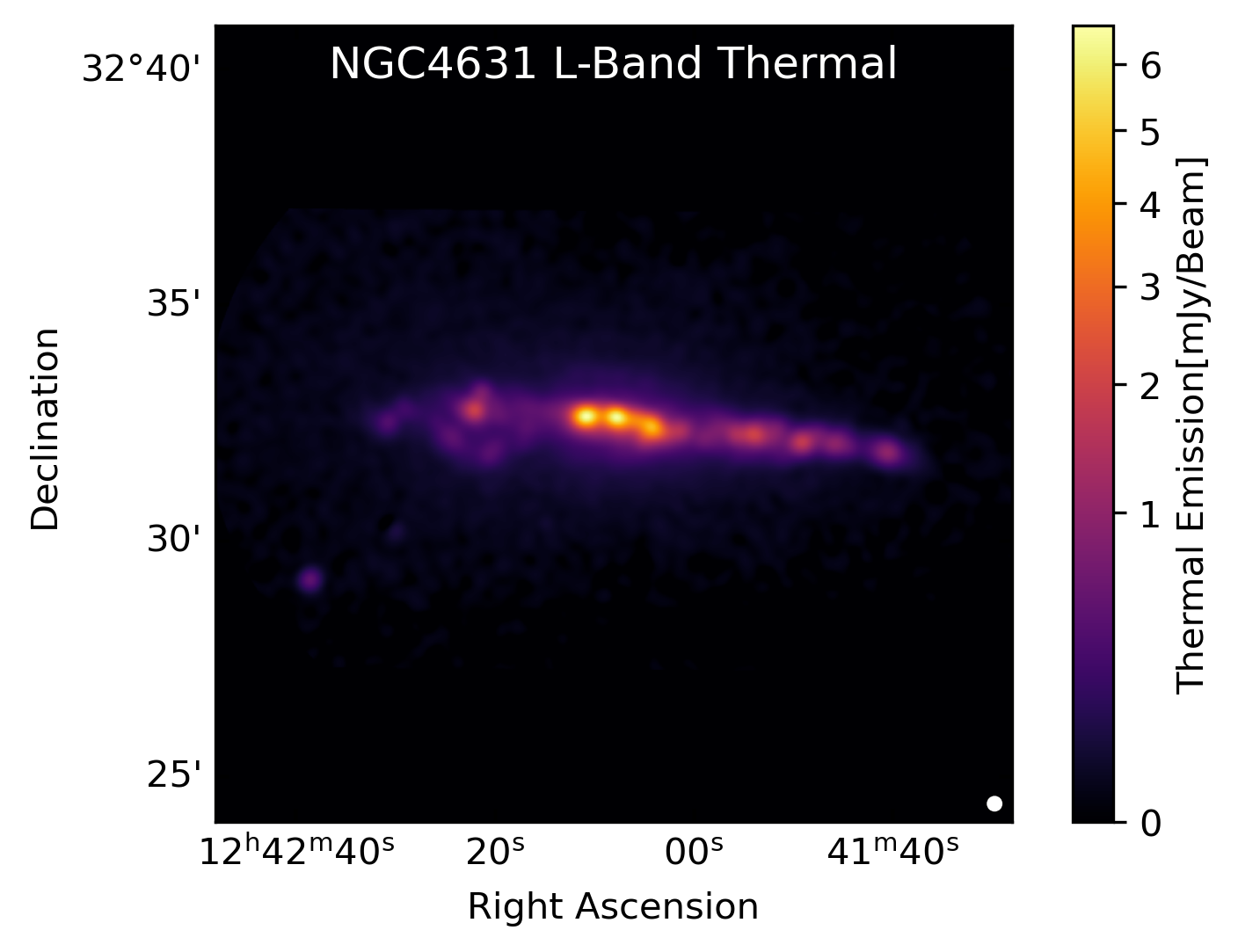}
\end{subfigure}
\begin{subfigure}{0.45\linewidth}
\includegraphics[width=1\linewidth]{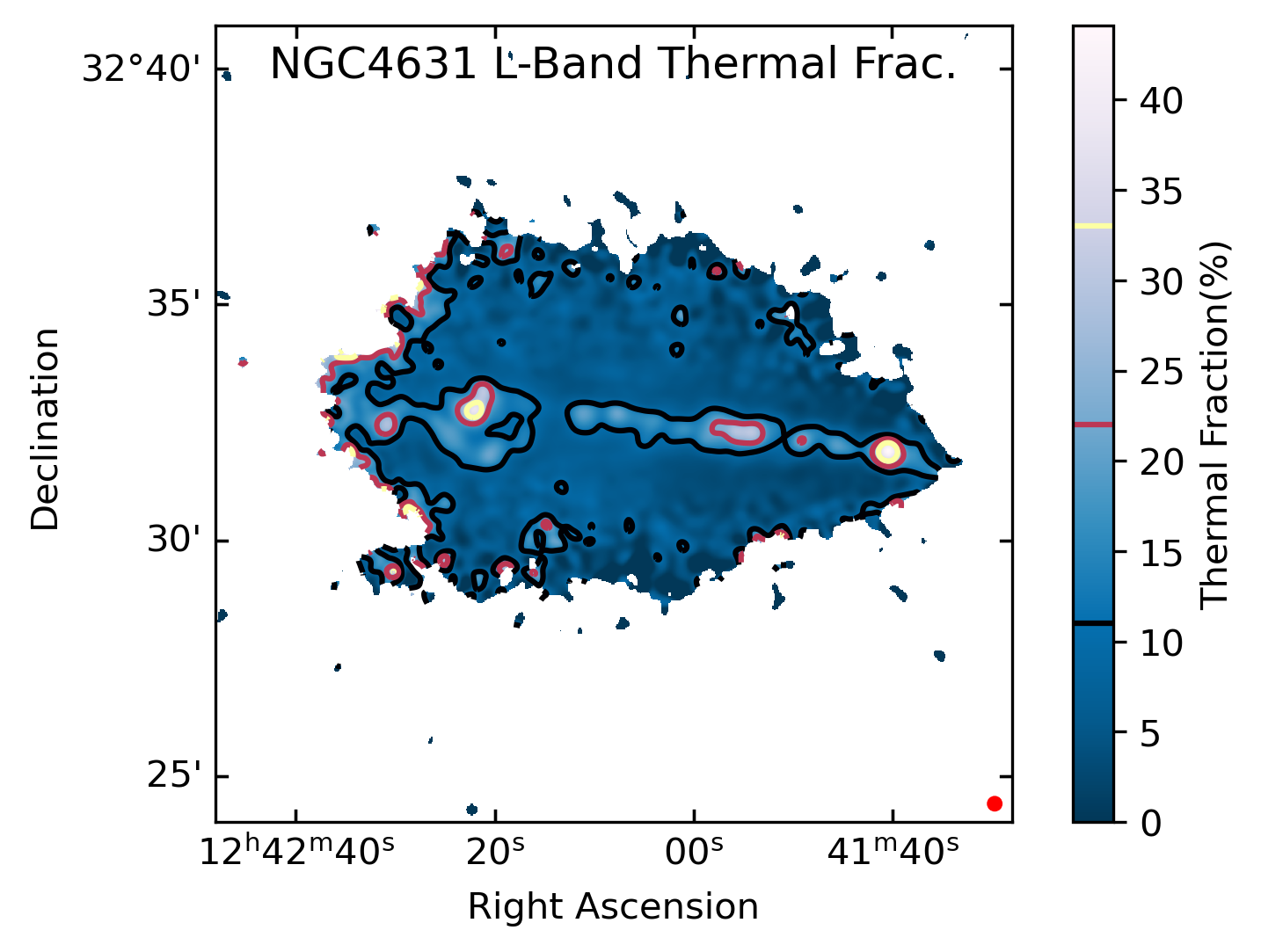}
\end{subfigure}
\caption{Fig. \ref{fig:thermal_maps_lband1} continued: Thermal emission at 1.5\,GHz (CHANG-ES L-band) (left column) and thermal fraction maps of the L-band data (right column) of NGC~4157 (top row) and NGC~4631 (bottom row).}
\label{fig:thermal_maps_n3432_l}
\end{figure*}

\begin{figure*}
\centering

\begin{subfigure}{0.45\linewidth}
\centering
\includegraphics[width=1\linewidth]{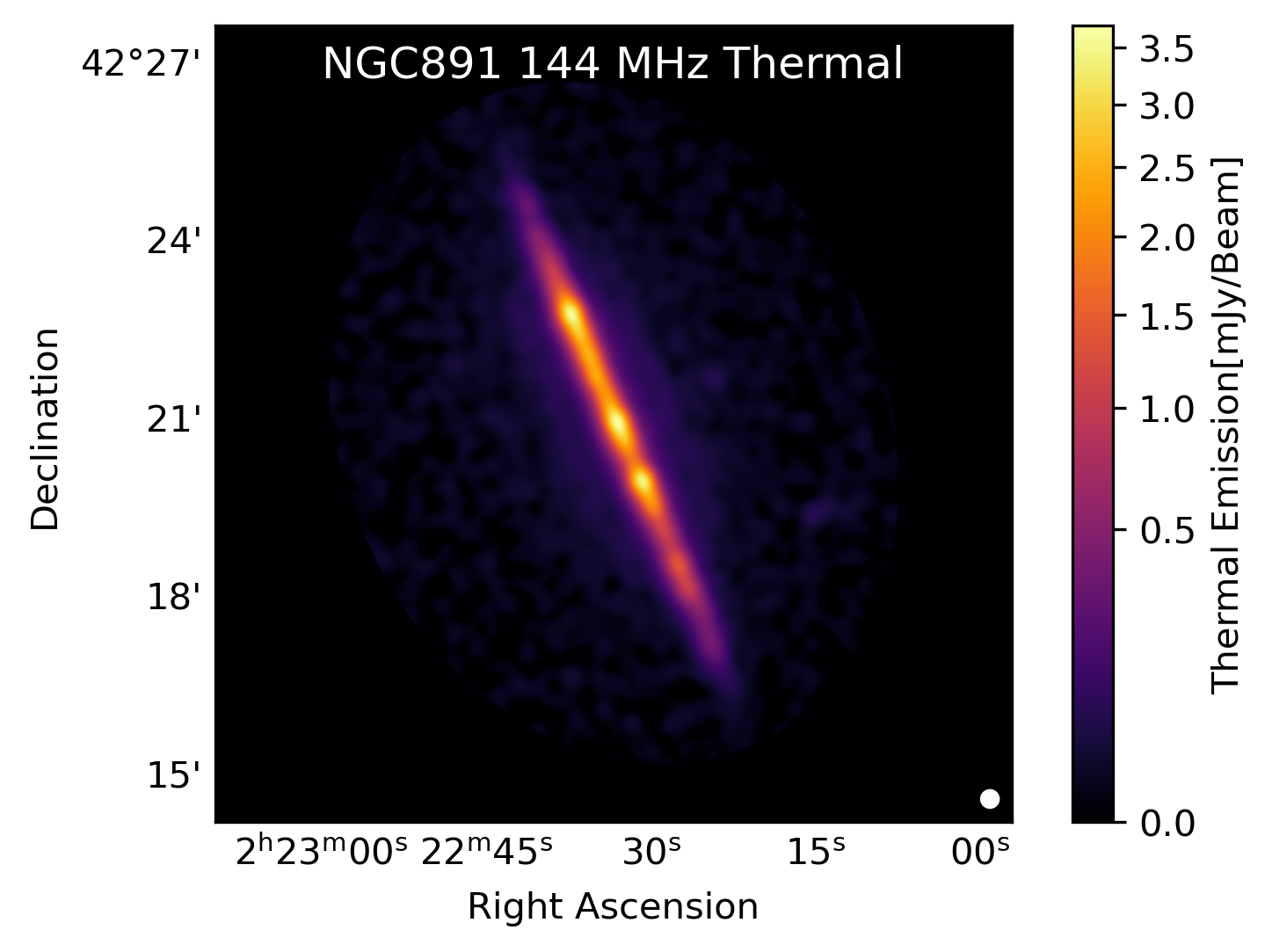}
\end{subfigure}
\begin{subfigure}{0.45\linewidth}
\centering
\includegraphics[width=1\linewidth]{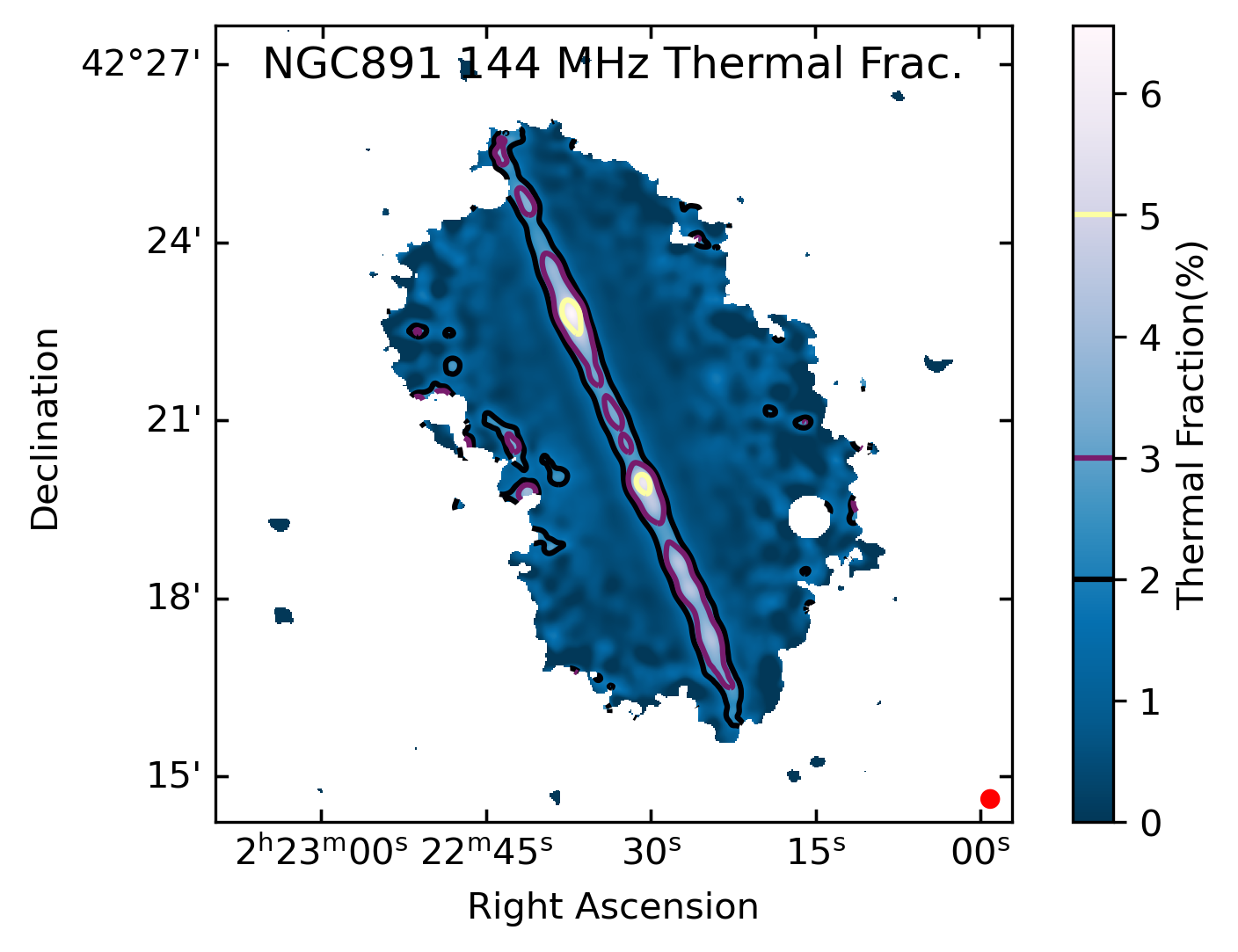}
\end{subfigure}
\\
\begin{subfigure}{0.45\linewidth}
\includegraphics[width=1\linewidth]{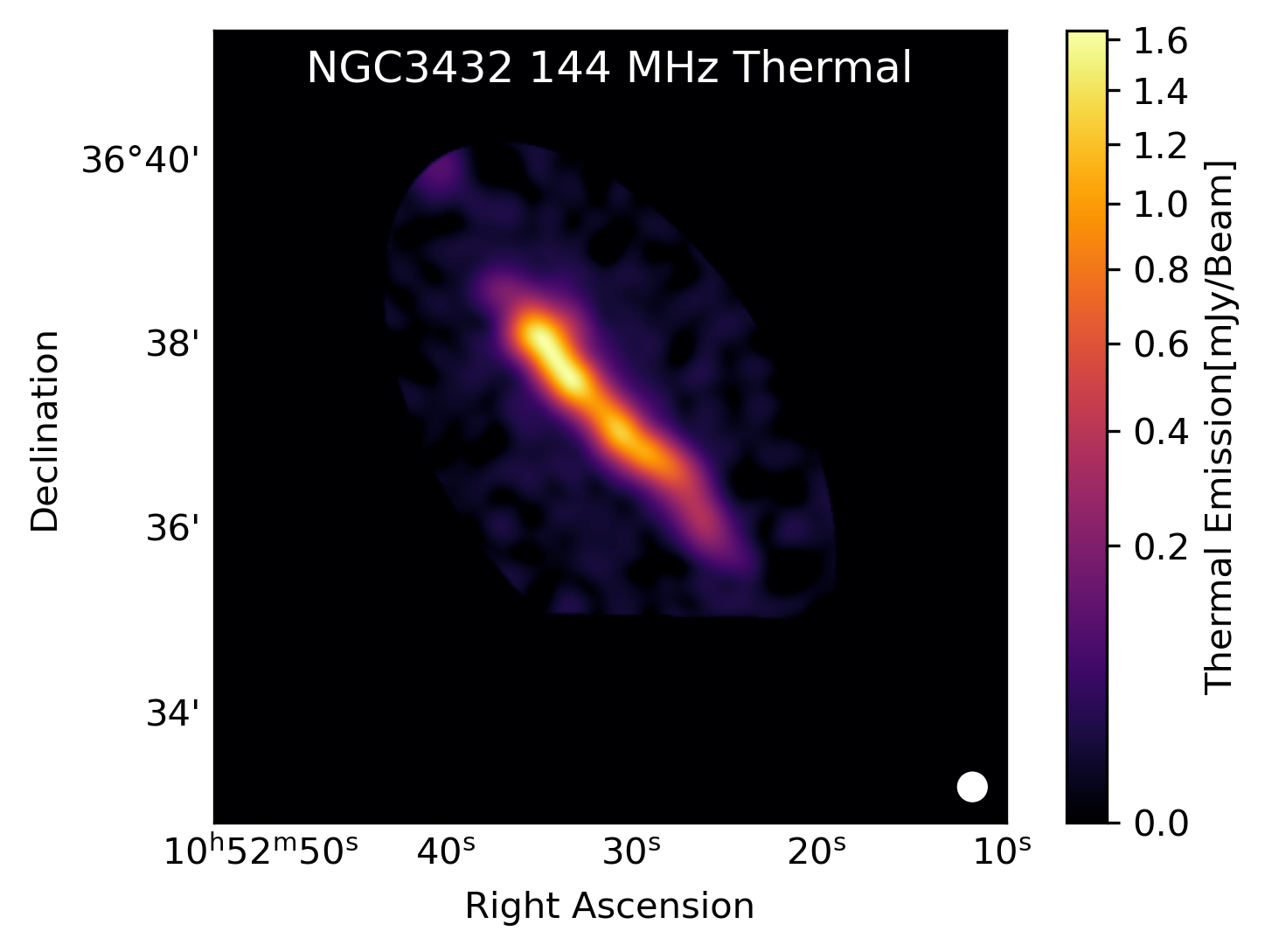}
\end{subfigure}
\begin{subfigure}{0.45\linewidth}
\includegraphics[width=1\linewidth]{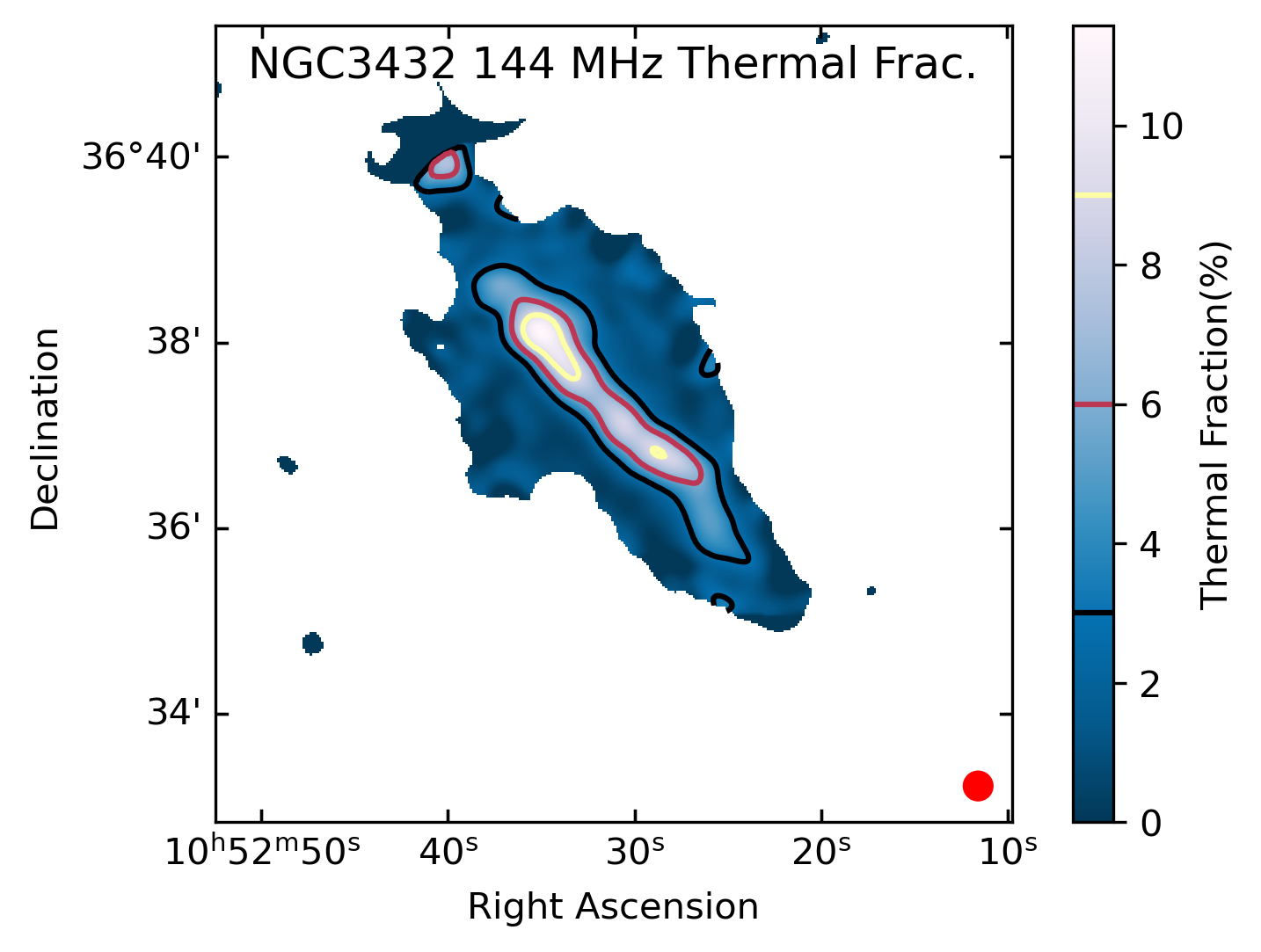}
\end{subfigure}
\\
\begin{subfigure}{0.45\linewidth}
\includegraphics[width=1\linewidth]{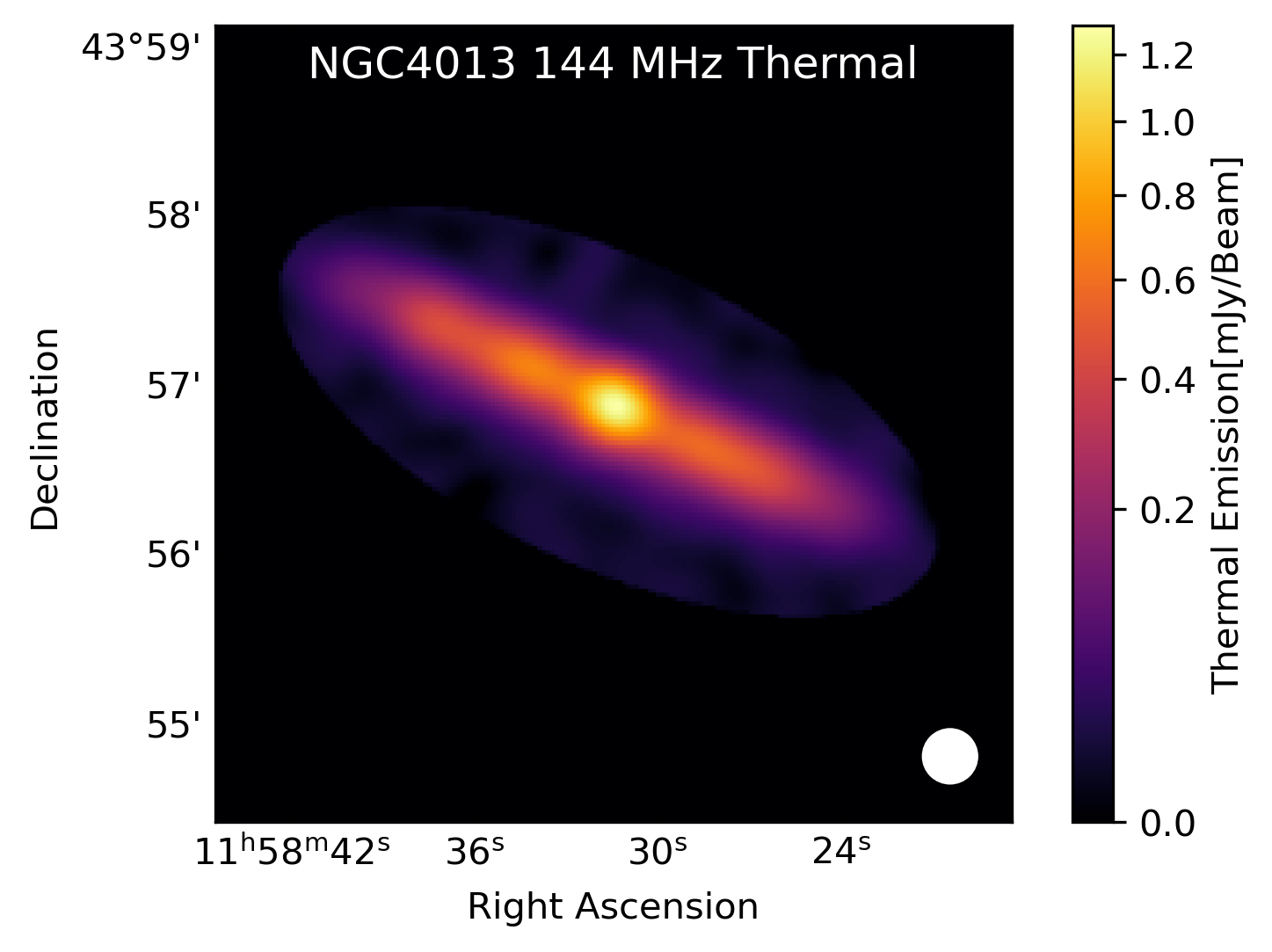}
\end{subfigure}
\begin{subfigure}{0.45\linewidth}
\includegraphics[width=1\linewidth]{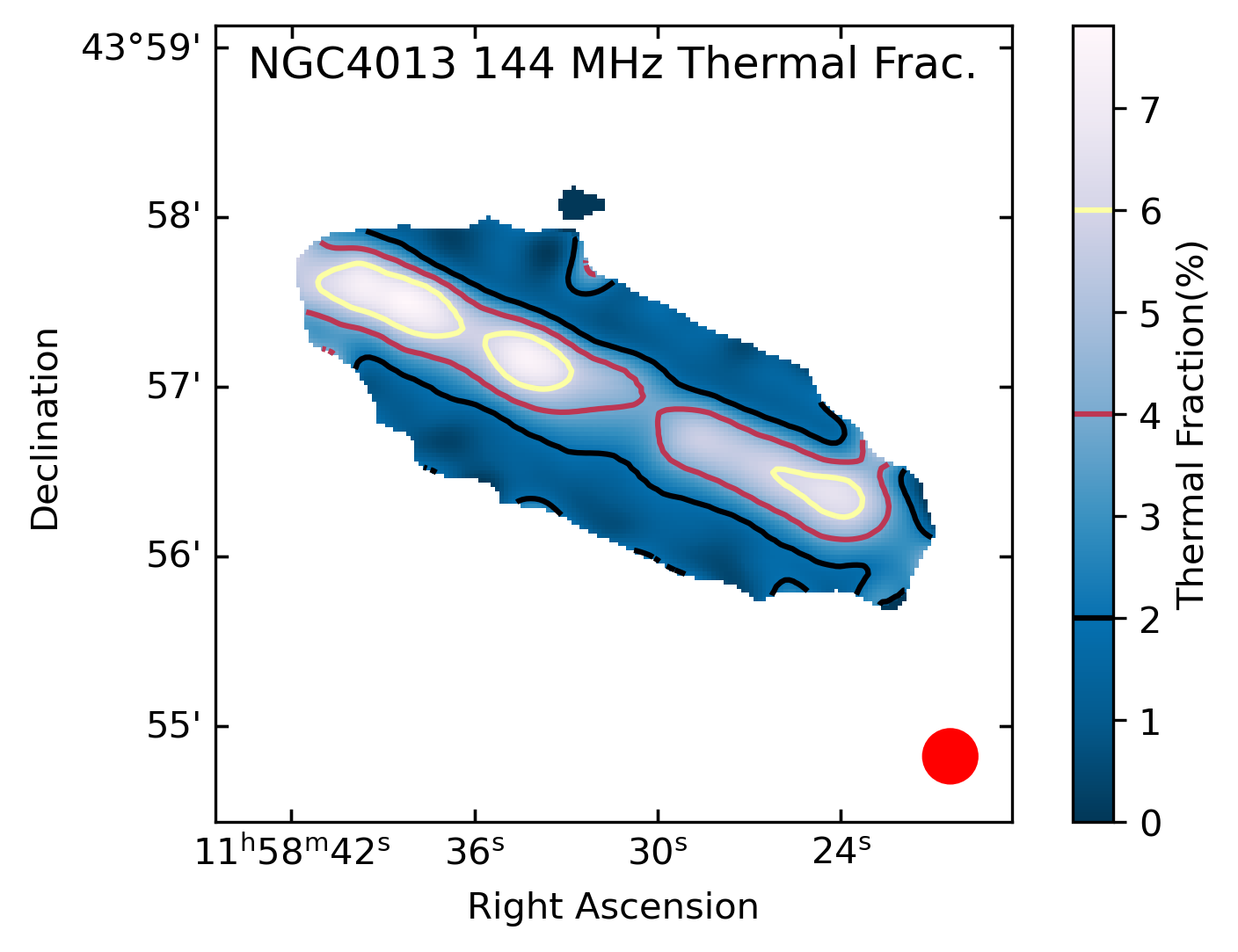}
\end{subfigure}
\caption{Thermal emission at 144\,MHz (LOFAR HBA) (left column) and thermal fraction maps of the HBA data (right column) of NGC~891 (top row), NGC~3432 (middle row), and NGC~4013 (bottom row). Thermal emission maps use a power law scaling with a power law index of 0.5. Thermal fraction maps are clipped below 3$\sigma$ above the background noise. The beam is displayed in the bottom right corner of each map as a white (thermal emission maps) or red (thermal fraction maps) circle.}
\label{fig:thermal_maps_hba1}
\end{figure*}

\begin{figure*}
\centering
\begin{subfigure}{0.45\linewidth}
\includegraphics[width=1\linewidth]{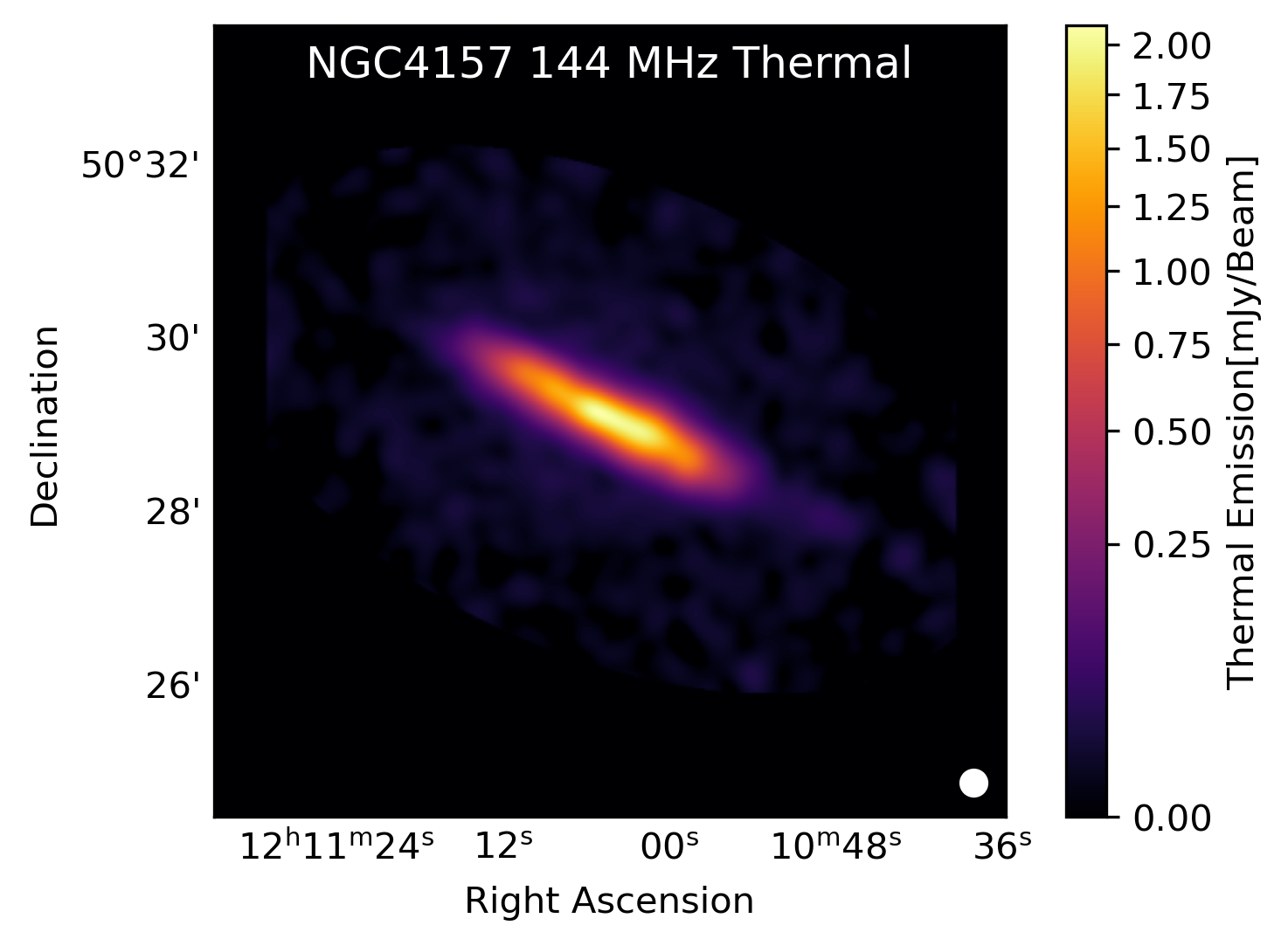}
\end{subfigure}
\begin{subfigure}{0.45\linewidth}
\includegraphics[width=1\linewidth]{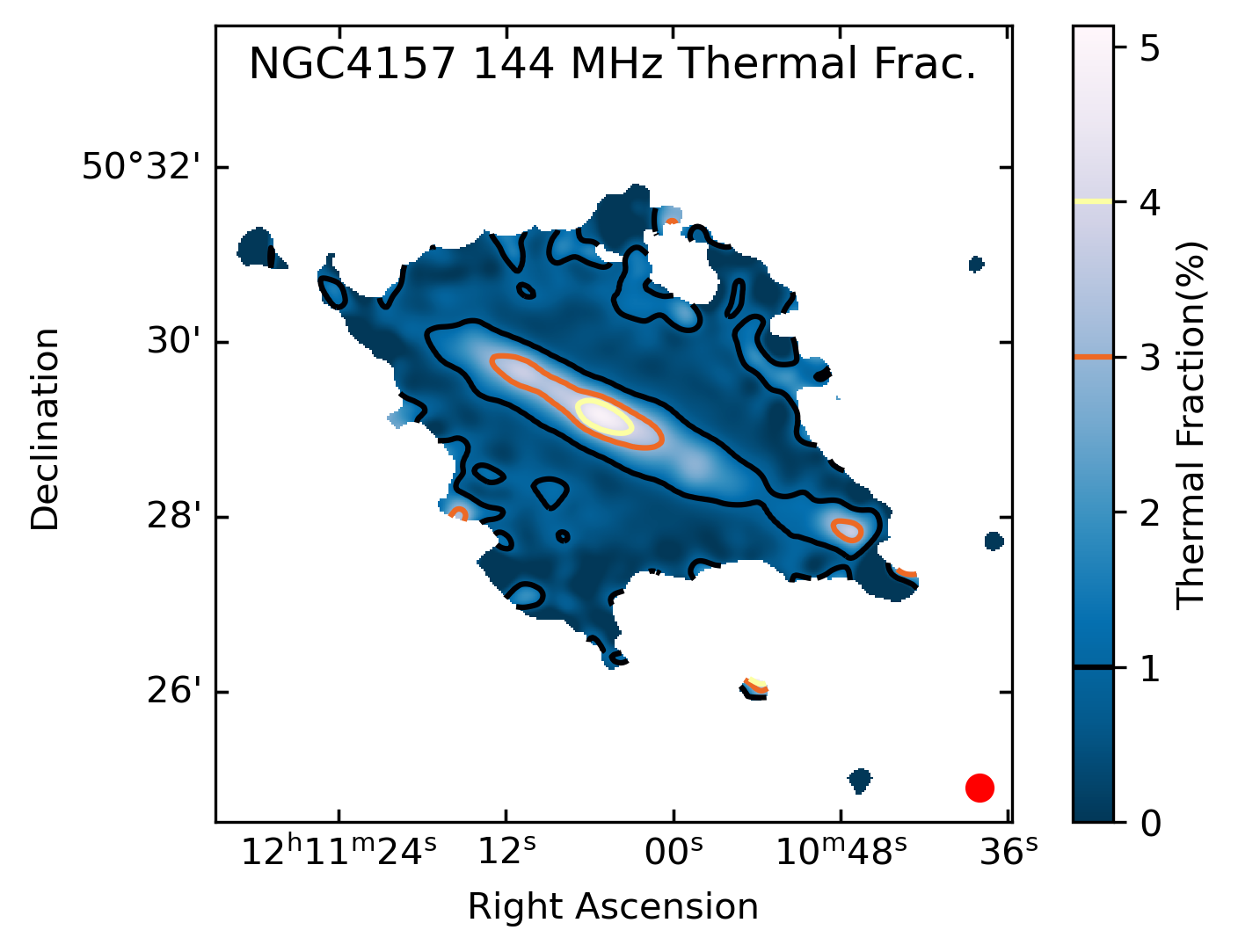}
\end{subfigure}
\\
\begin{subfigure}{0.45\linewidth}
\includegraphics[width=1\linewidth]{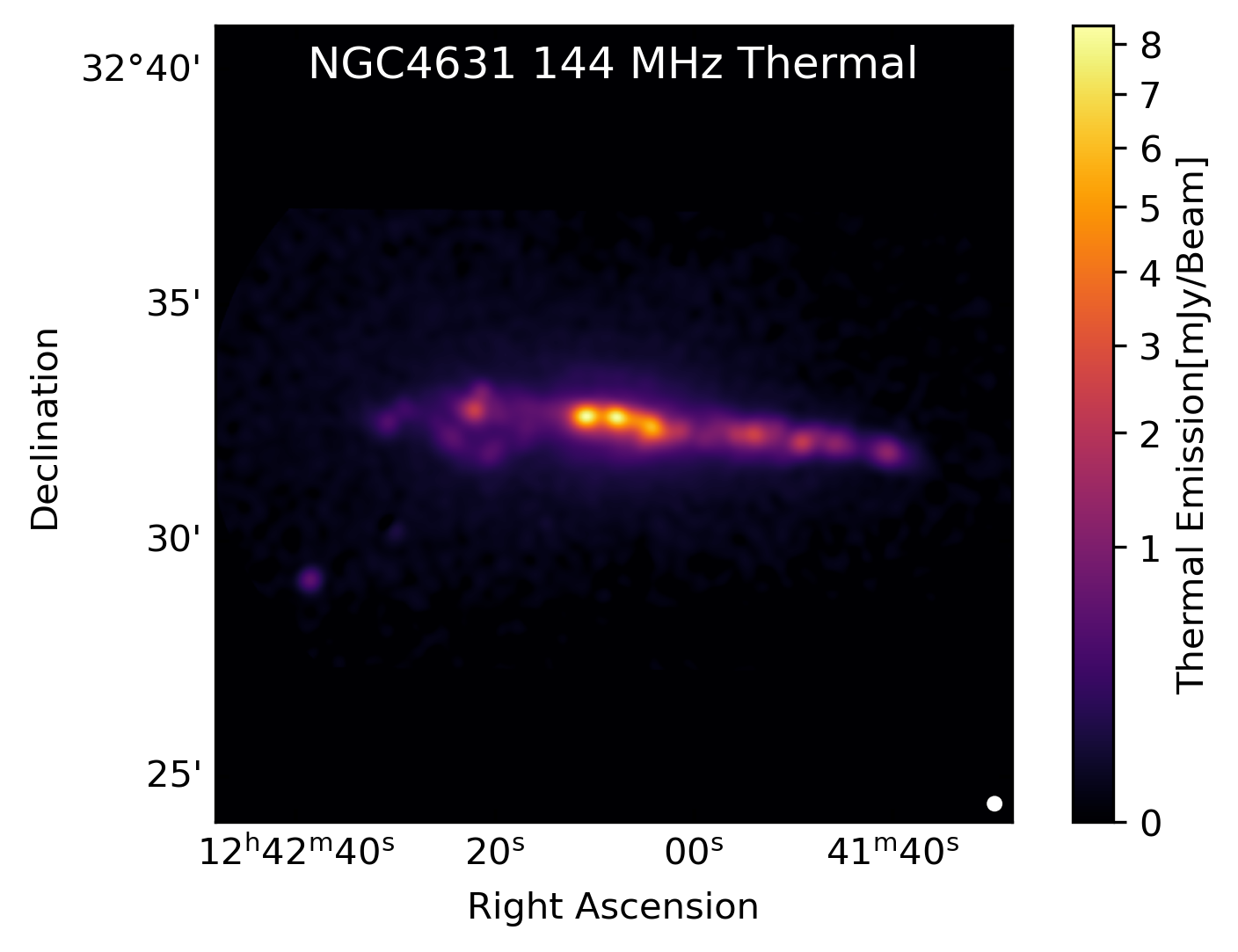}
\end{subfigure}
\begin{subfigure}{0.45\linewidth}
\includegraphics[width=1\linewidth]{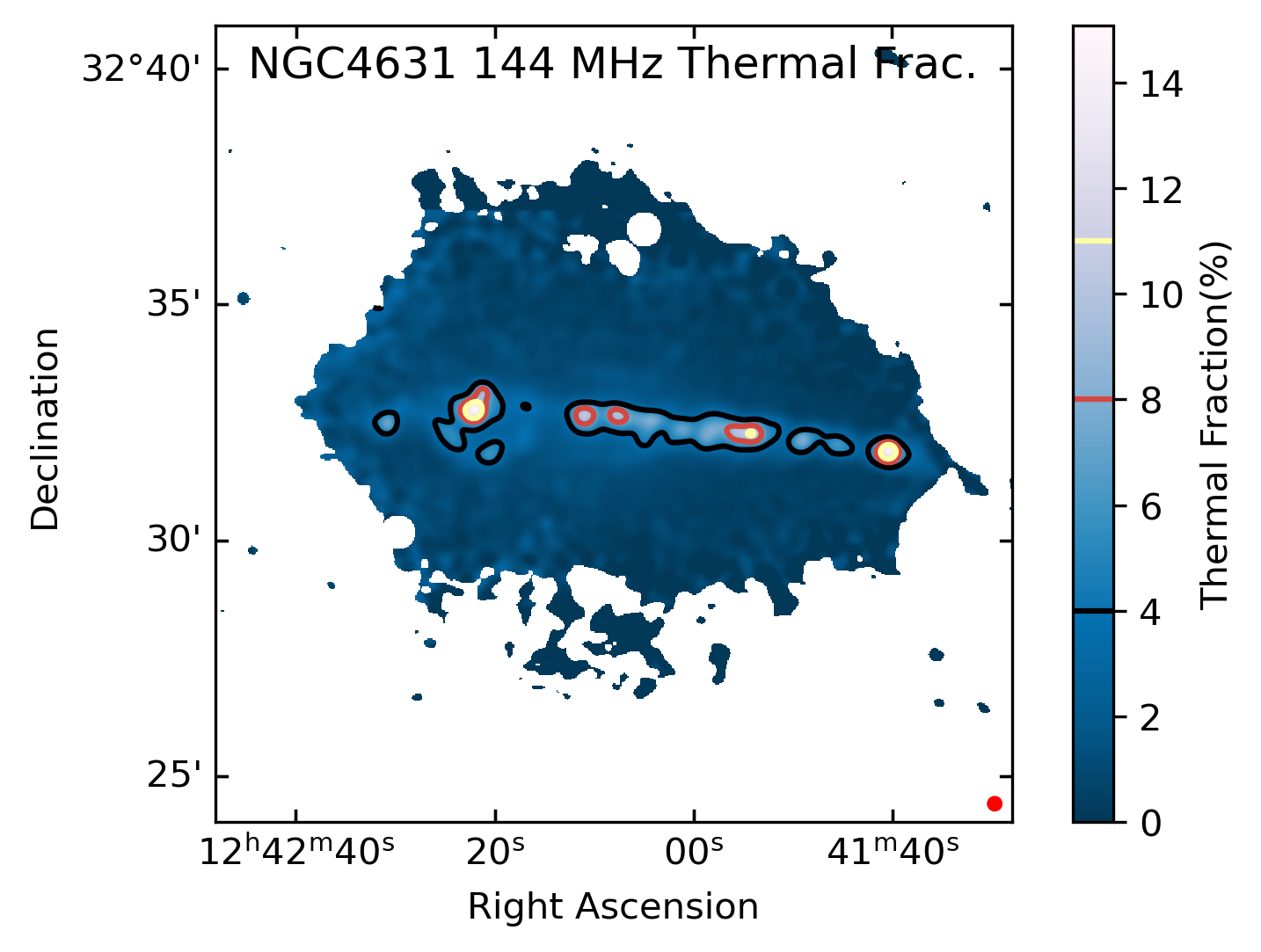}
\end{subfigure}
\caption{Fig. \ref{fig:thermal_maps_hba1} continued: Thermal emission at 144\,MHz (LOFAR HBA) (left column) and thermal fraction maps of the HBA data (right column) of NGC~4157 (top row), and NGC~4631 (bottom row).}
\label{fig:thermal_maps_app_2}
\end{figure*}

%% file: tab_fit_params.tex
\begingroup
\renewcommand{\arraystretch}{0.8} 
\begin{tabular}{lrlllllllr}
\hline
\hline
Galaxy & $\nu$ & Model & ID & $w_0$ & $z_0$ & $w_1$ & $z_1$ & $Z_0$ & $\chi^2_\nu$\\
 & [MHz] & & & [\textmu Jy\,Beam\textsuperscript{-1}] & [kpc] & [\textmu Jy\,Beam\textsuperscript{-1}] & [kpc] & [kpc] &\\
\toprule
N~3432 & 1500 & one\_fo & LR & 0.32$\pm$0.04 &   1.16$\pm$0.09 &             - &             - &              - &  3.72 \\
N~3432 &  144 & one\_fo & LR &   7.5$\pm$2.3 &   0.28$\pm$0.07 &             - &             - &              - &  2.41 \\
N~3432 & 1500 &     one & UR & 2.20$\pm$0.20 &   0.45$\pm$0.03 &             - &             - &  0.54$\pm$0.07 &  4.27 \\
N~3432 &  144 &     one & UR &     31$\pm$21 &   0.13$\pm$0.09 &             - &             - &  0.72$\pm$0.06 &  1.39 \\
N~3432 & 1500 & one\_fo & LM & 4.68$\pm$0.27 & 0.940$\pm$0.028 &             - &             - &              - &  5.23 \\
N~3432 &  144 & one\_fo & LM &  16.7$\pm$1.3 &   1.01$\pm$0.04 &             - &             - &              - &  1.11 \\
N~3432 & 1500 & one\_fo & UM & 4.43$\pm$0.21 & 0.789$\pm$0.019 &             - &             - &              - & 10.07 \\
N~3432 &  144 & one\_fo & UM &  16.7$\pm$1.2 &   0.87$\pm$0.03 &             - &             - &              - &  2.26 \\
N~3432 & 1500 & one\_fo & LL & 2.47$\pm$0.29 &   0.91$\pm$0.05 &             - &             - &              - &  1.44 \\
N~3432 &  144 & one\_fo & LL &   8.0$\pm$1.0 &   1.14$\pm$0.07 &             - &             - &              - &  1.77 \\
N~3432 & 1500 & one\_fo & UL & 1.55$\pm$0.16 &   0.91$\pm$0.05 &             - &             - &              - &  1.97 \\
N~3432 &  144 & one\_fo & UL &   6.2$\pm$0.8 &   1.03$\pm$0.07 &             - &             - &              - &  1.50 \\
N~4013 & 1500 & one\_fo & LR & 2.09$\pm$0.20 & 0.607$\pm$0.027 &             - &             - &              - &  1.77 \\
N~4013 &  144 & one\_fo & LR &   9.6$\pm$1.0 &   0.75$\pm$0.04 &             - &             - &              - &  0.55 \\
N~4013 & 1500 & one\_fo & UR & 2.32$\pm$0.22 &   0.73$\pm$0.04 &             - &             - &              - &  0.83 \\
N~4013 &  144 & one\_fo & UR &  12.4$\pm$1.4 &   0.69$\pm$0.04 &             - &             - &              - &  0.12 \\
N~4013 & 1500 & one\_fo & LM &   4.7$\pm$0.5 & 0.526$\pm$0.025 &             - &             - &              - &  8.92 \\
N~4013 &  144 & one\_fo & LM &  20.6$\pm$2.1 & 0.639$\pm$0.026 &             - &             - &              - &  3.15 \\
N~4013 & 1500 & one\_fo & UM &   9.4$\pm$0.9 & 0.476$\pm$0.018 &             - &             - &              - &  0.97 \\
N~4013 &  144 & one\_fo & UM &      33$\pm$3 & 0.614$\pm$0.023 &             - &             - &              - &  0.30 \\
N~4013 & 1500 & one\_fo & LL & 2.45$\pm$0.18 & 0.670$\pm$0.030 &             - &             - &              - &  2.89 \\
N~4013 &  144 & one\_fo & LL &  10.0$\pm$1.0 &   0.74$\pm$0.04 &             - &             - &              - &  1.41 \\
N~4013 & 1500 & one\_fo & UL & 3.41$\pm$0.22 & 0.578$\pm$0.020 &             - &             - &              - &  1.11 \\
N~4013 &  144 & one\_fo & UL &   8.8$\pm$0.8 &   1.05$\pm$0.05 &             - &             - &              - &  5.63 \\
N~4157 & 1500 & one\_fo & LR &   3.2$\pm$0.4 &   1.32$\pm$0.07 &             - &             - &              - &  0.68 \\
N~4157 &  144 &     one & LR &  15.7$\pm$2.3 &   1.25$\pm$0.07 &             - &             - &  0.55$\pm$0.26 &  5.26 \\
N~4157 & 1500 & one\_fo & UR &   3.7$\pm$0.6 &   0.96$\pm$0.06 &             - &             - &              - &  0.17 \\
N~4157 &  144 & one\_fo & UR &   6.9$\pm$1.2 &   2.31$\pm$0.19 &             - &             - &              - &  6.10 \\
N~4157 & 1500 & one\_fo & LM &  13.6$\pm$0.6 & 1.121$\pm$0.021 &             - &             - &              - &  4.79 \\
N~4157 &  144 & two\_fo & LM &      66$\pm$7 &   0.88$\pm$0.11 &   8.6$\pm$2.0 &   3.5$\pm$0.3 &              - &  2.34 \\
N~4157 & 1500 & two\_fo & UM &       8$\pm$7 &     0.7$\pm$0.4 &       9$\pm$8 & 1.25$\pm$0.20 &              - &  0.65 \\
N~4157 &  144 &     two & UM &      64$\pm$5 &   1.21$\pm$0.10 &   1.3$\pm$0.6 &      10$\pm$5 &  0.24$\pm$0.17 &  2.88 \\
N~4157 & 1500 & one\_fo & LL &   2.6$\pm$0.4 &   0.96$\pm$0.08 &             - &             - &              - &  0.03 \\
N~4157 &  144 & one\_fo & LL &   5.0$\pm$0.8 &   2.63$\pm$0.19 &             - &             - &              - &  4.17 \\
N~4157 & 1500 & one\_fo & UL &   2.4$\pm$0.3 &   1.45$\pm$0.07 &             - &             - &              - &  2.24 \\
N~4157 &  144 & one\_fo & UL &  10.0$\pm$1.1 &   2.59$\pm$0.13 &             - &             - &              - &  0.40 \\
N~4631 & 1500 & one\_fo & LR &   6.8$\pm$0.4 & 1.053$\pm$0.029 &             - &             - &              - &  1.11 \\
N~4631 &  144 & one\_fo & LR &  19.4$\pm$1.1 &   1.89$\pm$0.05 &             - &             - &              - &  0.74 \\
N~4631 & 1500 &     two & UR &  10.5$\pm$2.8 &   0.52$\pm$0.16 & 0.89$\pm$0.22 &   2.6$\pm$0.3 &   -0.6$\pm$0.4 &  0.39 \\
N~4631 &  144 & two\_fo & UR &  11.6$\pm$1.3 &   1.44$\pm$0.27 &   2.7$\pm$1.2 &   4.5$\pm$0.8 &              - &  1.07 \\
N~4631 & 1500 &     two & LM &     42$\pm$16 &   0.59$\pm$0.10 &   6.3$\pm$1.2 & 1.80$\pm$0.07 &   -0.7$\pm$0.4 &  1.88 \\
N~4631 &  144 & two\_fo & LM &  37.0$\pm$2.9 &   1.46$\pm$0.12 &   3.0$\pm$1.0 &   5.3$\pm$0.7 &              - &  5.03 \\
N~4631 & 1500 &     two & UM &     45$\pm$10 &   0.39$\pm$0.11 &   5.0$\pm$0.6 & 2.01$\pm$0.05 & -0.44$\pm$0.27 &  1.68 \\
N~4631 &  144 & one\_fo & UM &  19.1$\pm$0.8 &   2.90$\pm$0.05 &             - &             - &              - &  4.45 \\
N~4631 & 1500 & one\_fo & LL & 2.00$\pm$0.14 &   1.93$\pm$0.06 &             - &             - &              - &  1.12 \\
N~4631 &  144 & one\_fo & LL &   8.6$\pm$0.5 &   2.84$\pm$0.07 &             - &             - &              - &  0.86 \\
N~4631 & 1500 & one\_fo & UL & 1.15$\pm$0.09 &   2.15$\pm$0.09 &             - &             - &              - &  1.07 \\
N~4631 &  144 & one\_fo & UL &   6.1$\pm$0.3 &   3.71$\pm$0.10 &             - &             - &              - &  0.75 \\
 N~891 & 1500 & one\_fo & LR &   4.3$\pm$0.3 &   1.09$\pm$0.04 &             - &             - &              - &  2.55 \\
 N~891 &  144 & one\_fo & LR &  15.2$\pm$1.0 &   1.85$\pm$0.04 &             - &             - &              - &  4.02 \\
 N~891 & 1500 & one\_fo & UR &   4.5$\pm$0.3 &   1.35$\pm$0.04 &             - &             - &              - &  3.50 \\
 N~891 &  144 & two\_fo & UR &      32$\pm$5 &   0.77$\pm$0.15 &   6.7$\pm$1.4 & 3.04$\pm$0.26 &              - &  3.41 \\
 N~891 & 1500 & one\_fo & LM &  17.2$\pm$0.8 & 0.822$\pm$0.016 &             - &             - &              - &  7.41 \\
 N~891 &  144 &     two & LM &   188$\pm$113 &   0.87$\pm$0.08 &   5.8$\pm$2.1 &   2.9$\pm$0.4 &   -1.0$\pm$0.7 &  1.01 \\
 N~891 & 1500 & one\_fo & UM &  16.9$\pm$0.6 & 1.176$\pm$0.014 &             - &             - &              - &  5.88 \\
 N~891 &  144 &     two & UM &    111$\pm$18 &   1.05$\pm$0.10 &   6.6$\pm$2.3 &   3.6$\pm$0.5 & -0.33$\pm$0.27 &  0.50 \\
 N~891 & 1500 & one\_fo & LL &   7.4$\pm$0.6 &   1.05$\pm$0.03 &             - &             - &              - &  1.21 \\
 N~891 &  144 & two\_fo & LL &     43$\pm$22 &   0.37$\pm$0.21 &  13.9$\pm$1.6 & 2.18$\pm$0.08 &              - &  1.22 \\
 N~891 & 1500 & one\_fo & UL &   6.9$\pm$0.6 &   1.07$\pm$0.04 &             - &             - &              - &  1.05 \\
 N~891 &  144 & one\_fo & UL &  25.4$\pm$1.9 &   1.67$\pm$0.05 &             - &             - &              - &  1.05 \\
\bottomrule
\end{tabular}
\endgroup

%% file: int_prof_appendix.tex
\begin{figure*}
\centering
\begin{subfigure}{0.49\linewidth}
\centering
\includegraphics[width=1\linewidth]{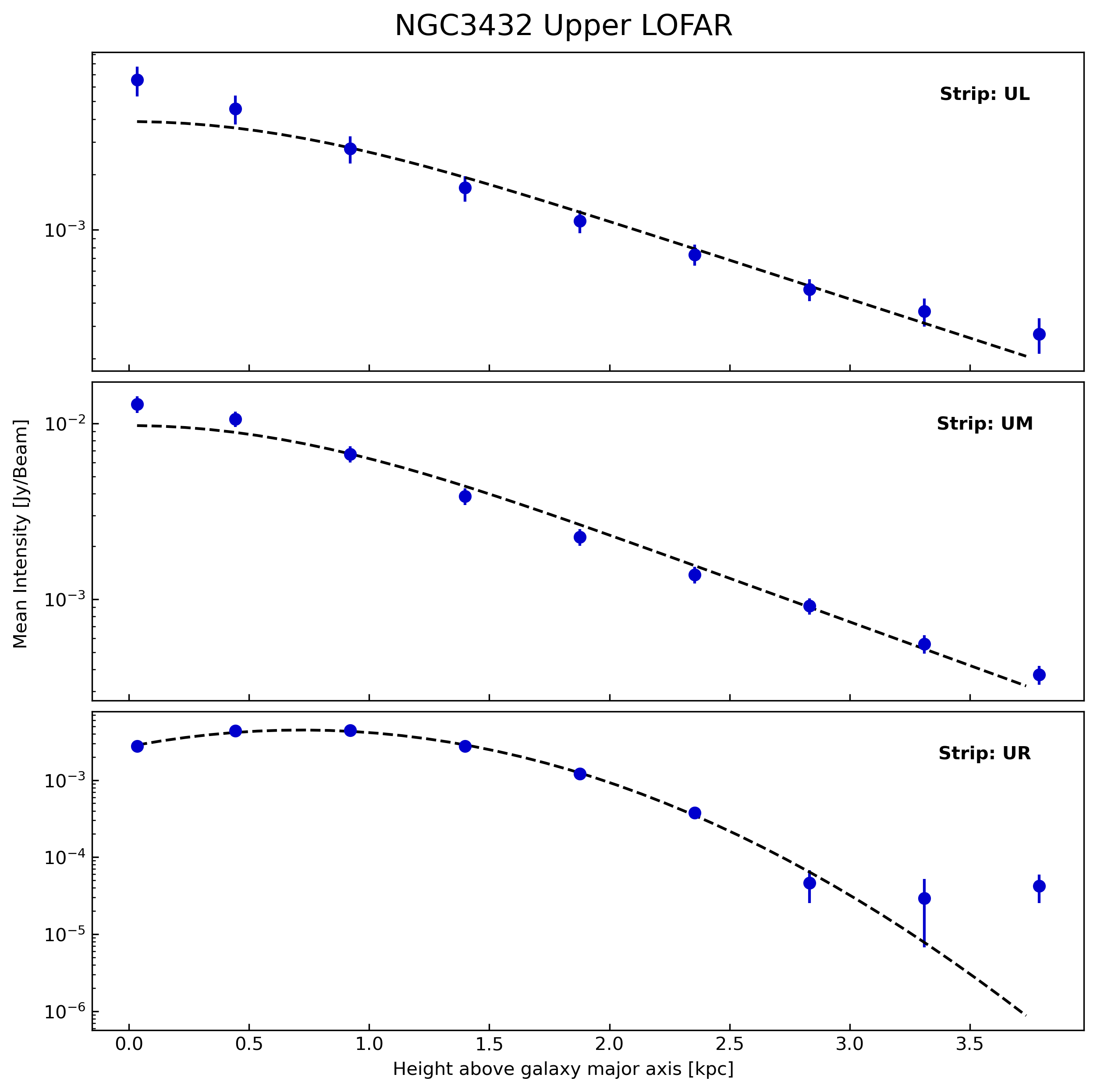}
\end{subfigure}
\hfill
\begin{subfigure}{0.49\linewidth}
\centering
\includegraphics[width=1\linewidth]{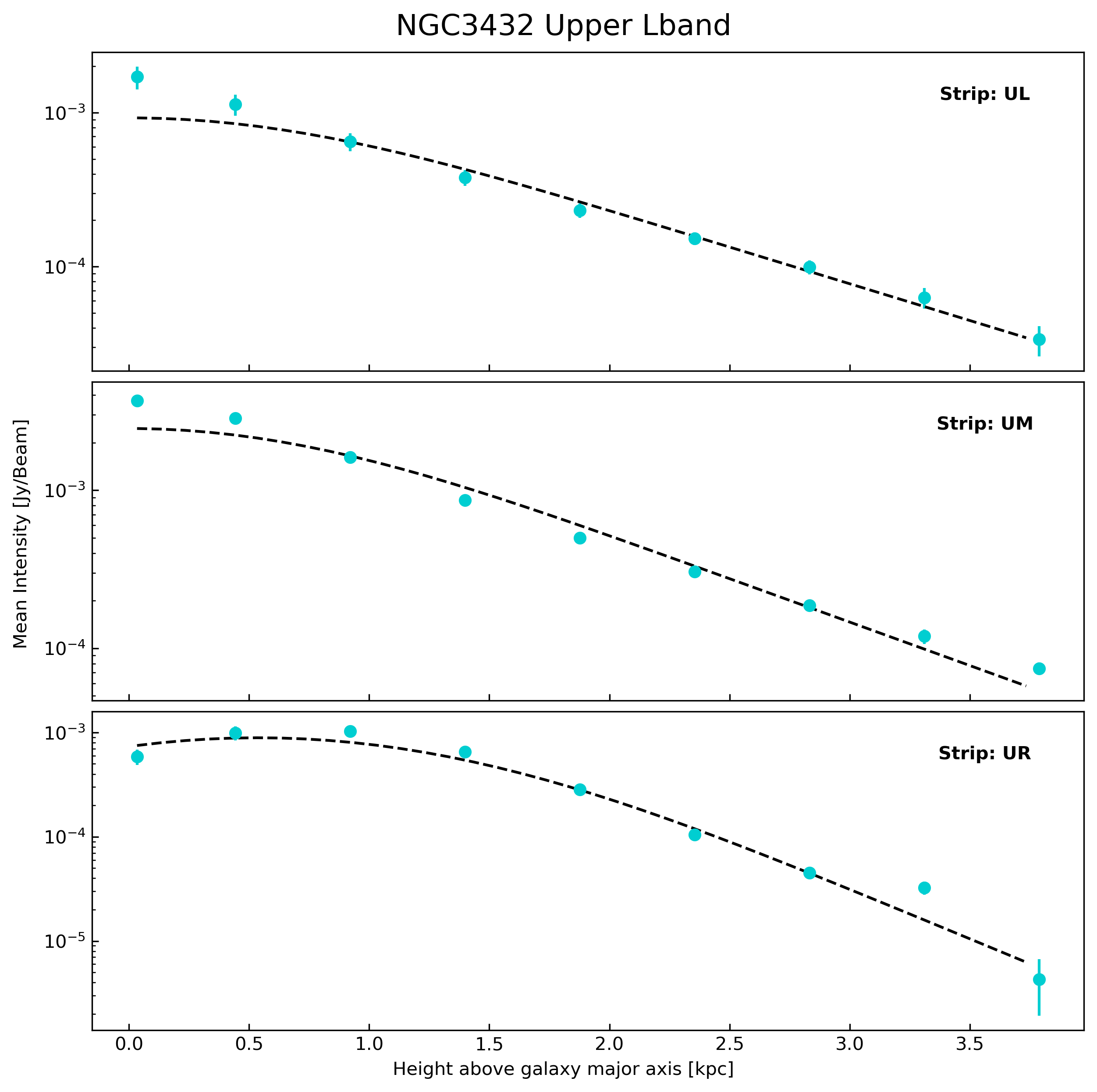}
\end{subfigure}
\\
\begin{subfigure}{0.49\linewidth}
\centering
\includegraphics[width=1\linewidth]{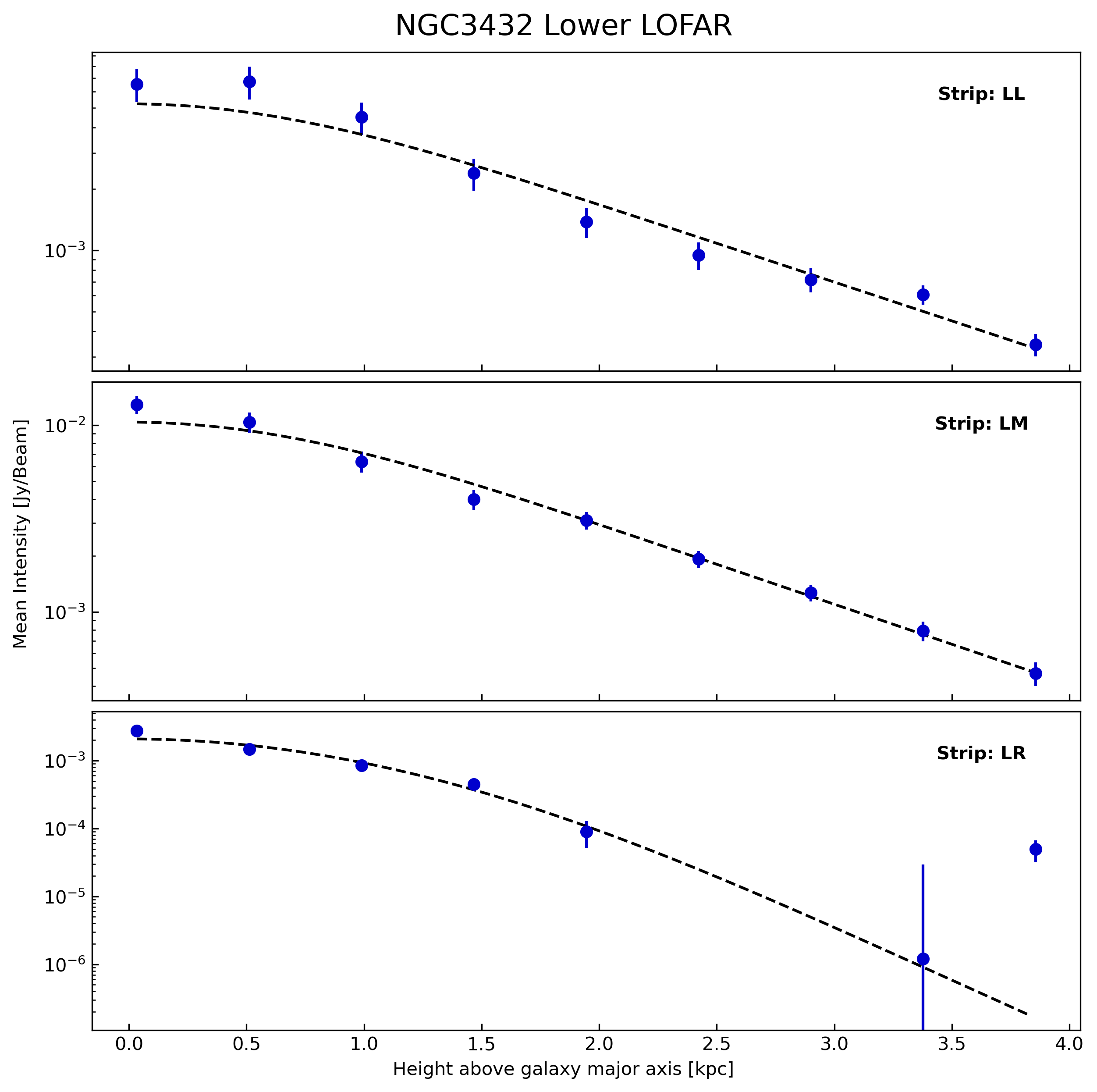}
\end{subfigure}
\hfill
\begin{subfigure}{0.49\linewidth}
\centering
\includegraphics[width=1\linewidth]{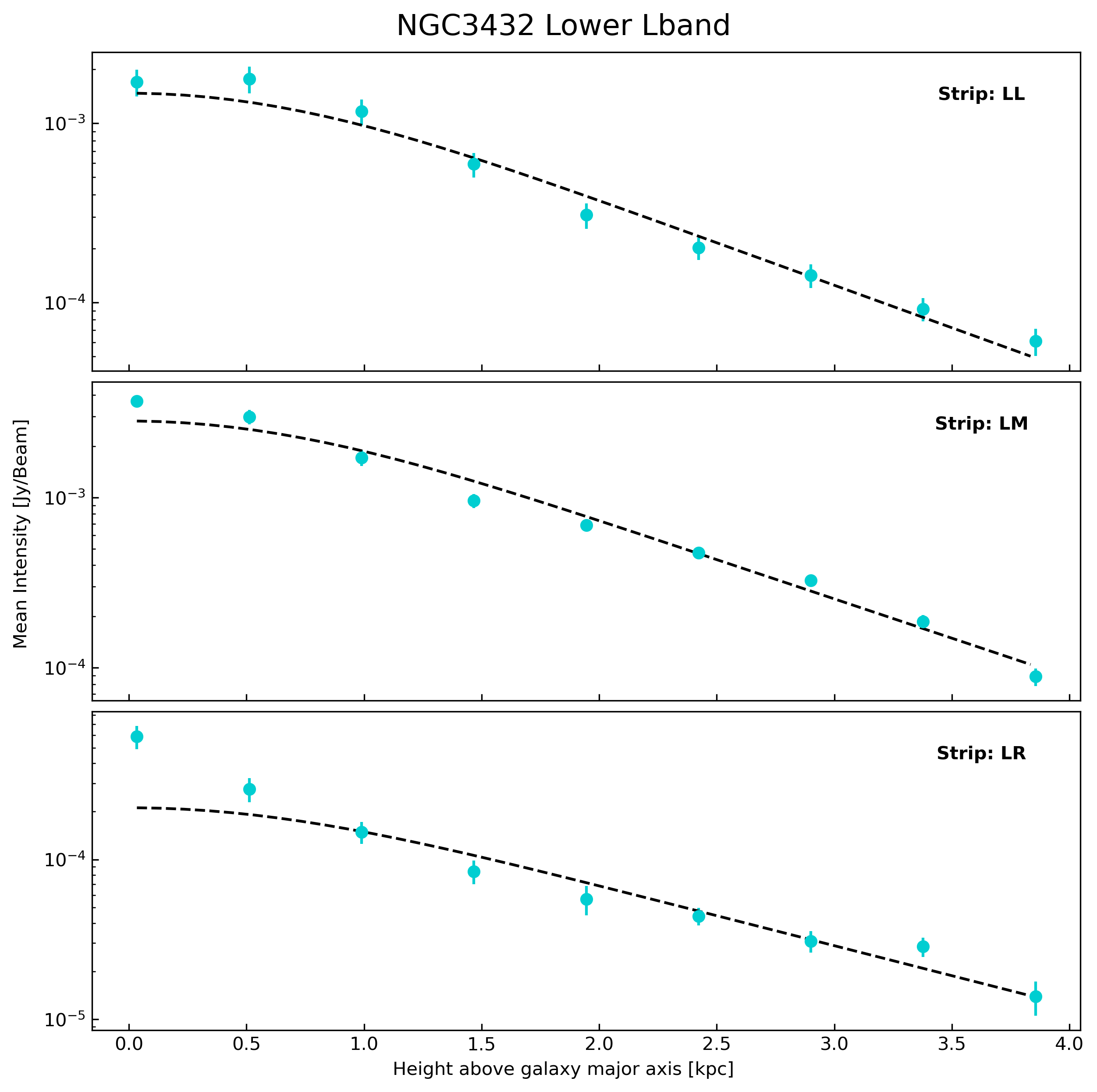}
\end{subfigure}

\caption{Intensity Profiles of NGC~3432}
\label{fig:int_prof_app_3432}
\end{figure*}

\begin{figure*}
\centering
\begin{subfigure}{0.49\linewidth}
\centering
\includegraphics[width=1\linewidth]{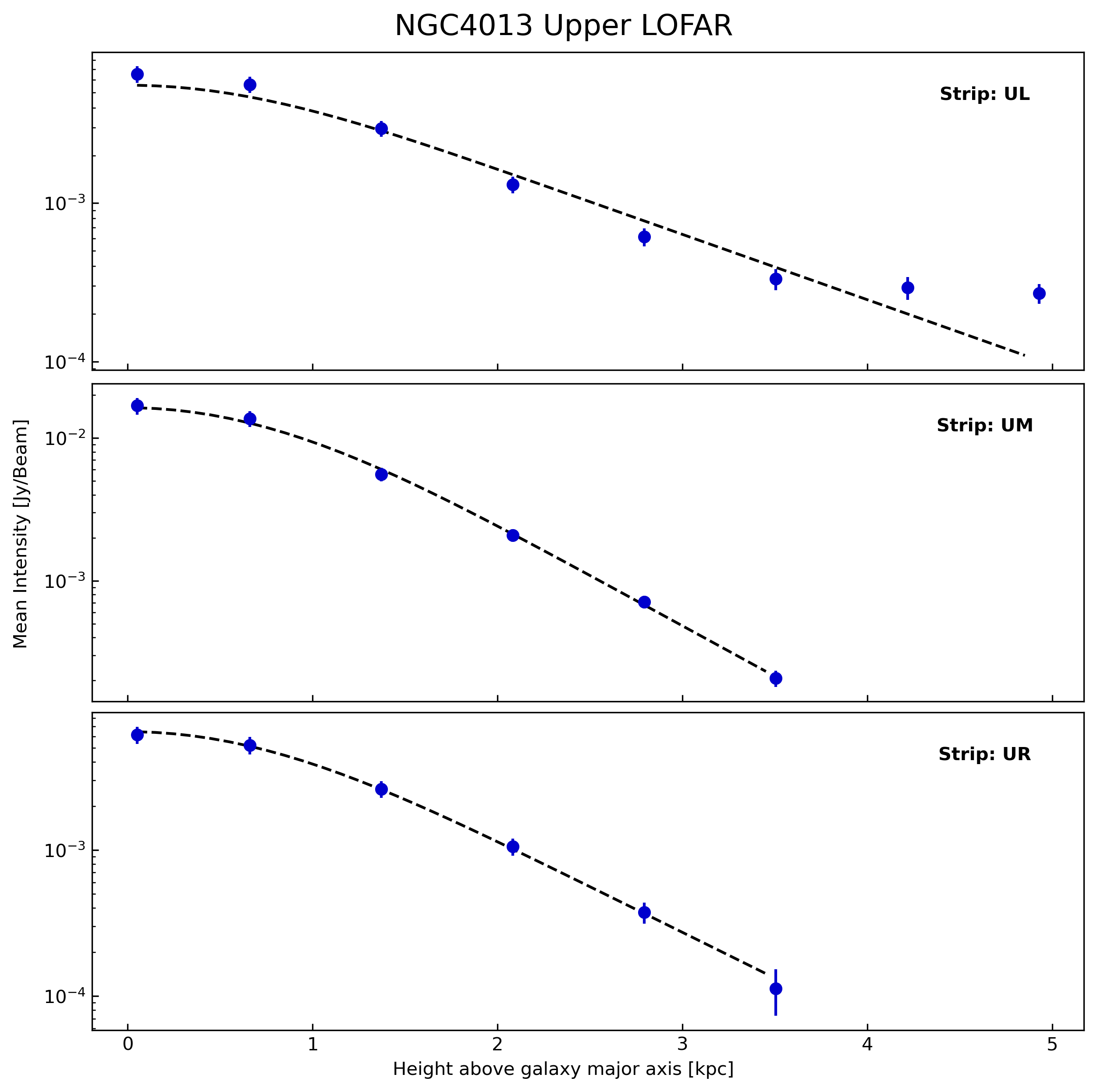}
\end{subfigure}
\hfill
\begin{subfigure}{0.49\linewidth}
\centering
\includegraphics[width=1\linewidth]{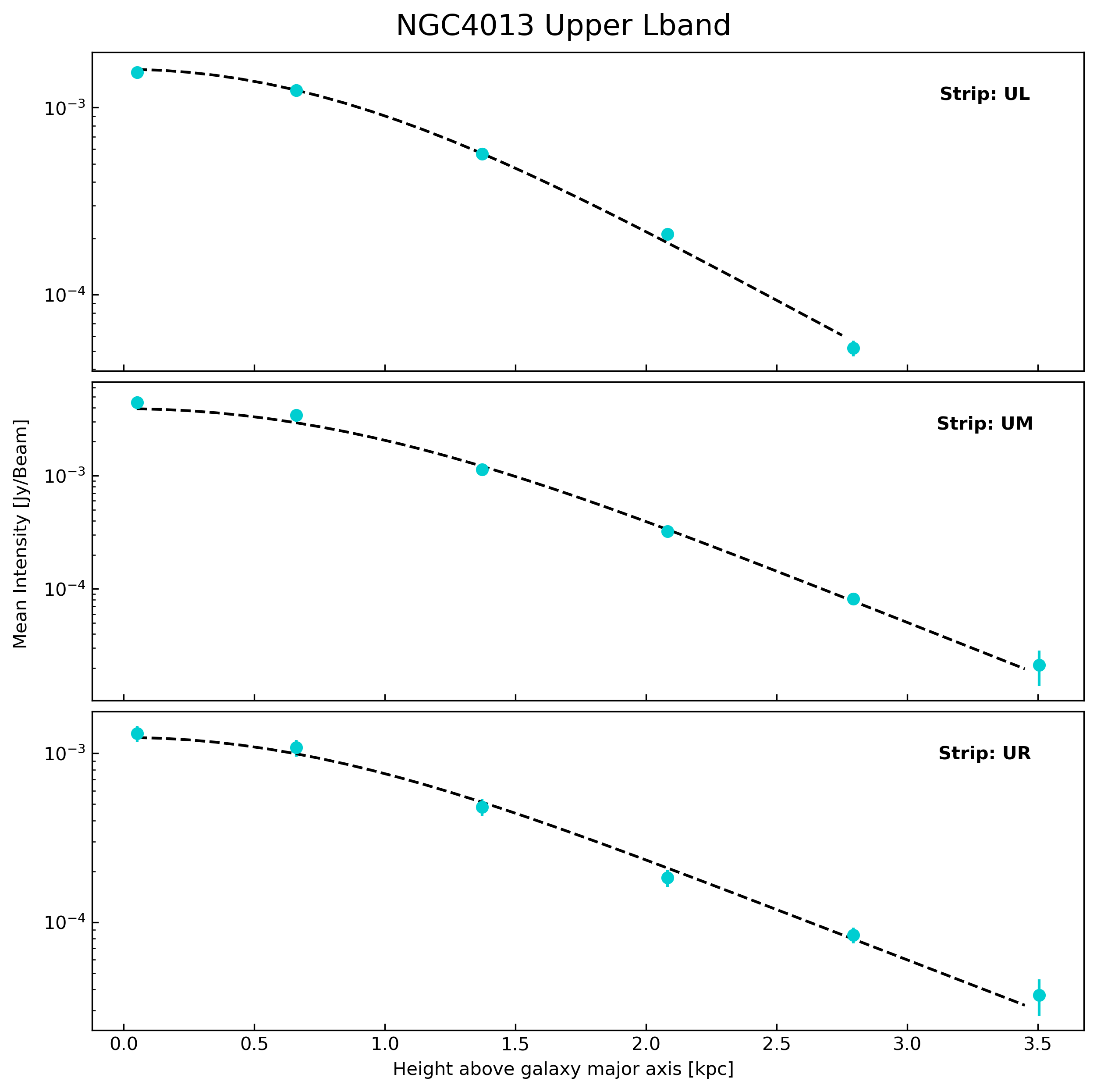}
\end{subfigure}
\\
\begin{subfigure}{0.49\linewidth}
\centering
\includegraphics[width=1\linewidth]{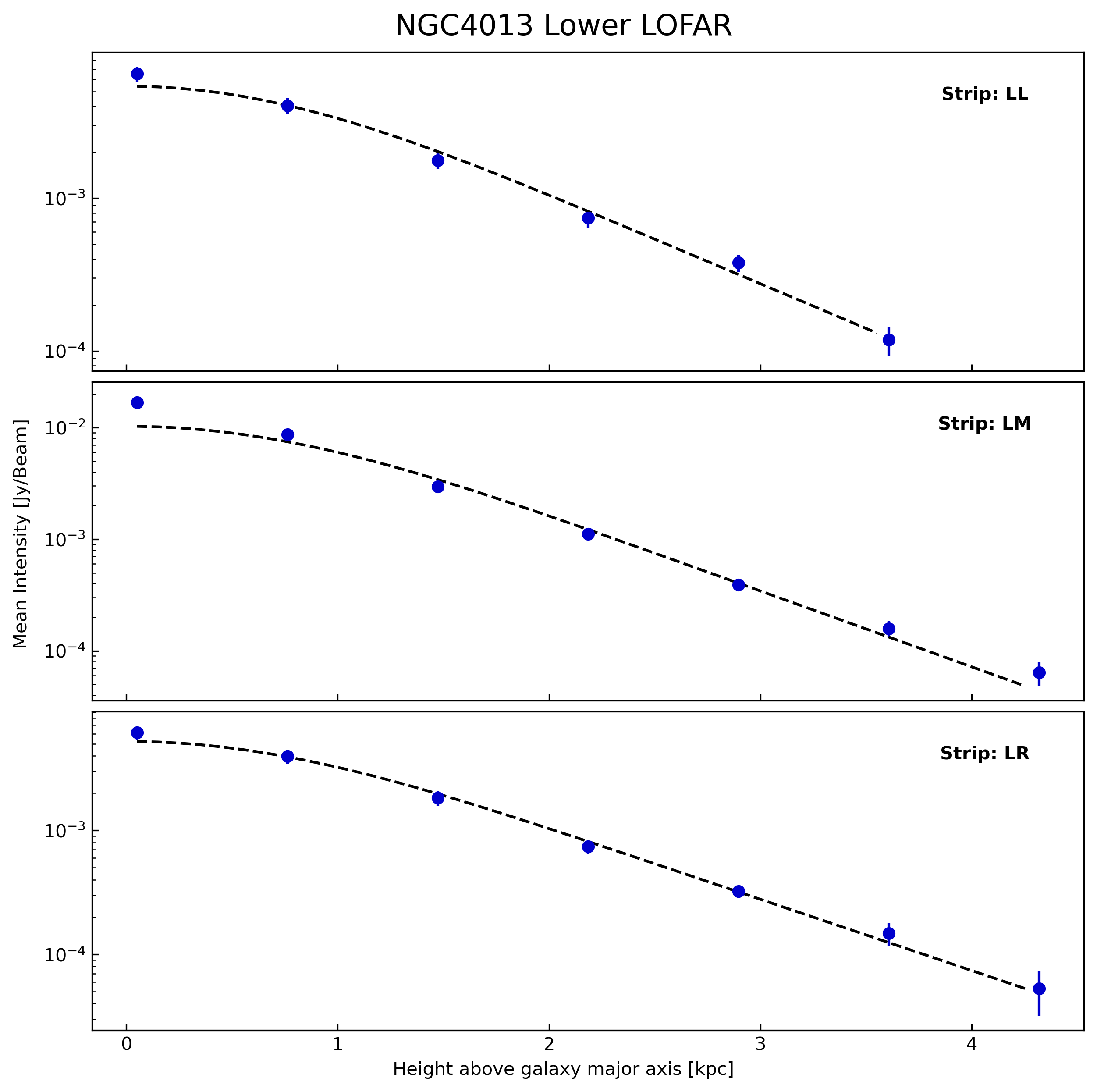}
\end{subfigure}
\hfill
\begin{subfigure}{0.49\linewidth}
\centering
\includegraphics[width=1\linewidth]{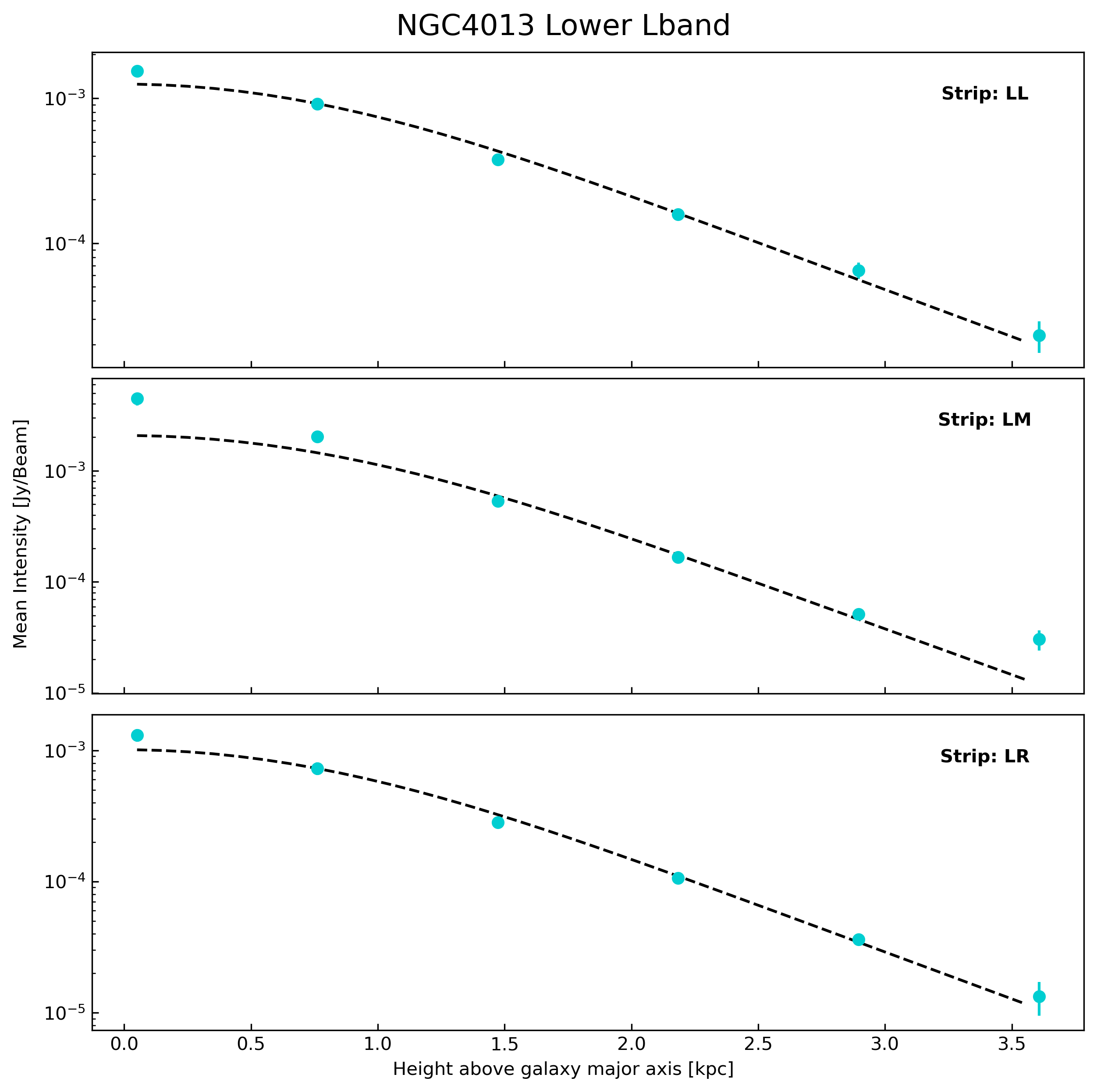}
\end{subfigure}

\caption{Intensity Profiles of NGC~4013}
\label{fig:int_prof_app_4013}
\end{figure*}

\begin{figure*}
\centering
\begin{subfigure}{0.49\linewidth}
\centering
\includegraphics[width=1\linewidth]{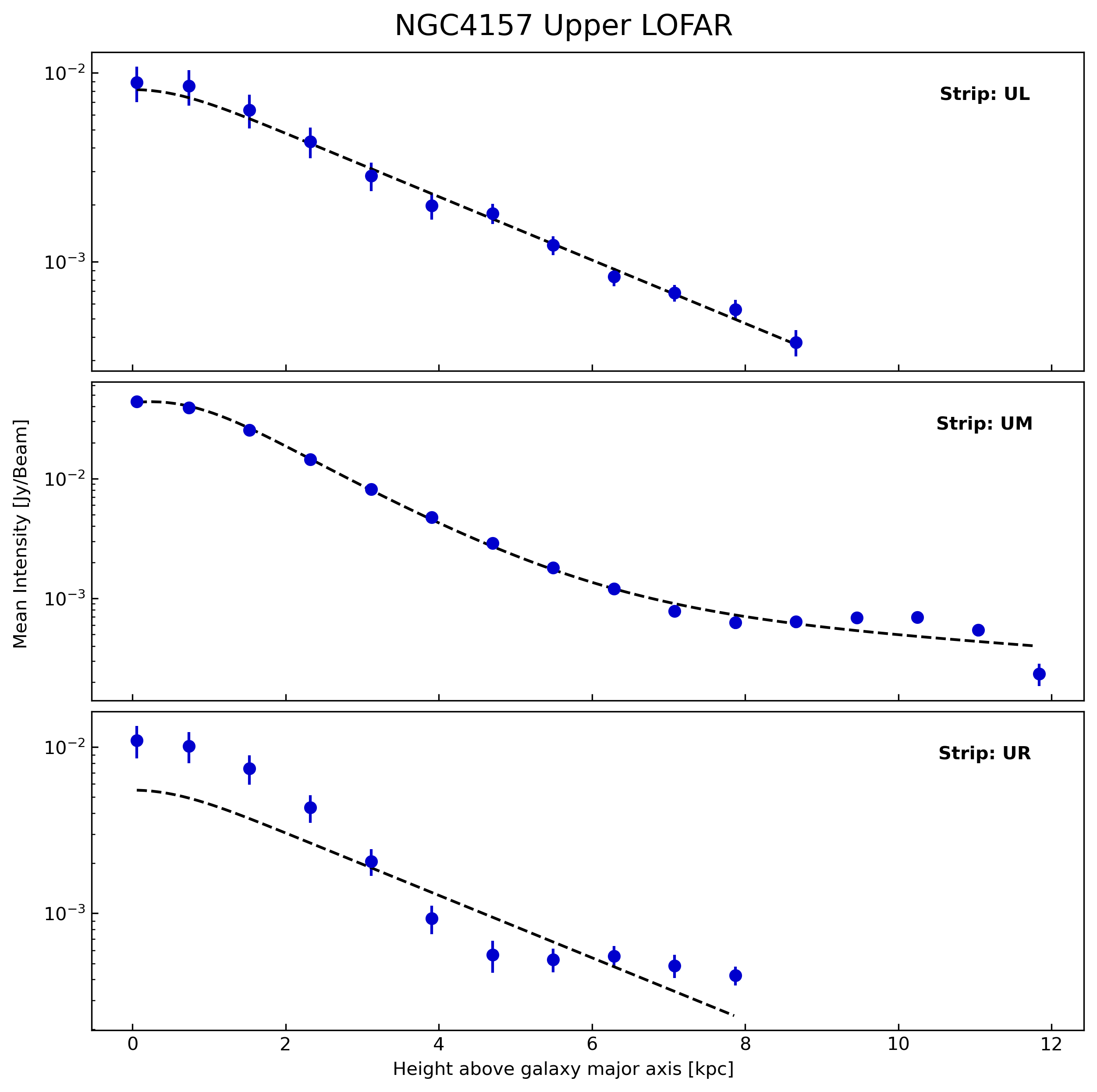}
\end{subfigure}
\hfill
\begin{subfigure}{0.49\linewidth}
\centering
\includegraphics[width=1\linewidth]{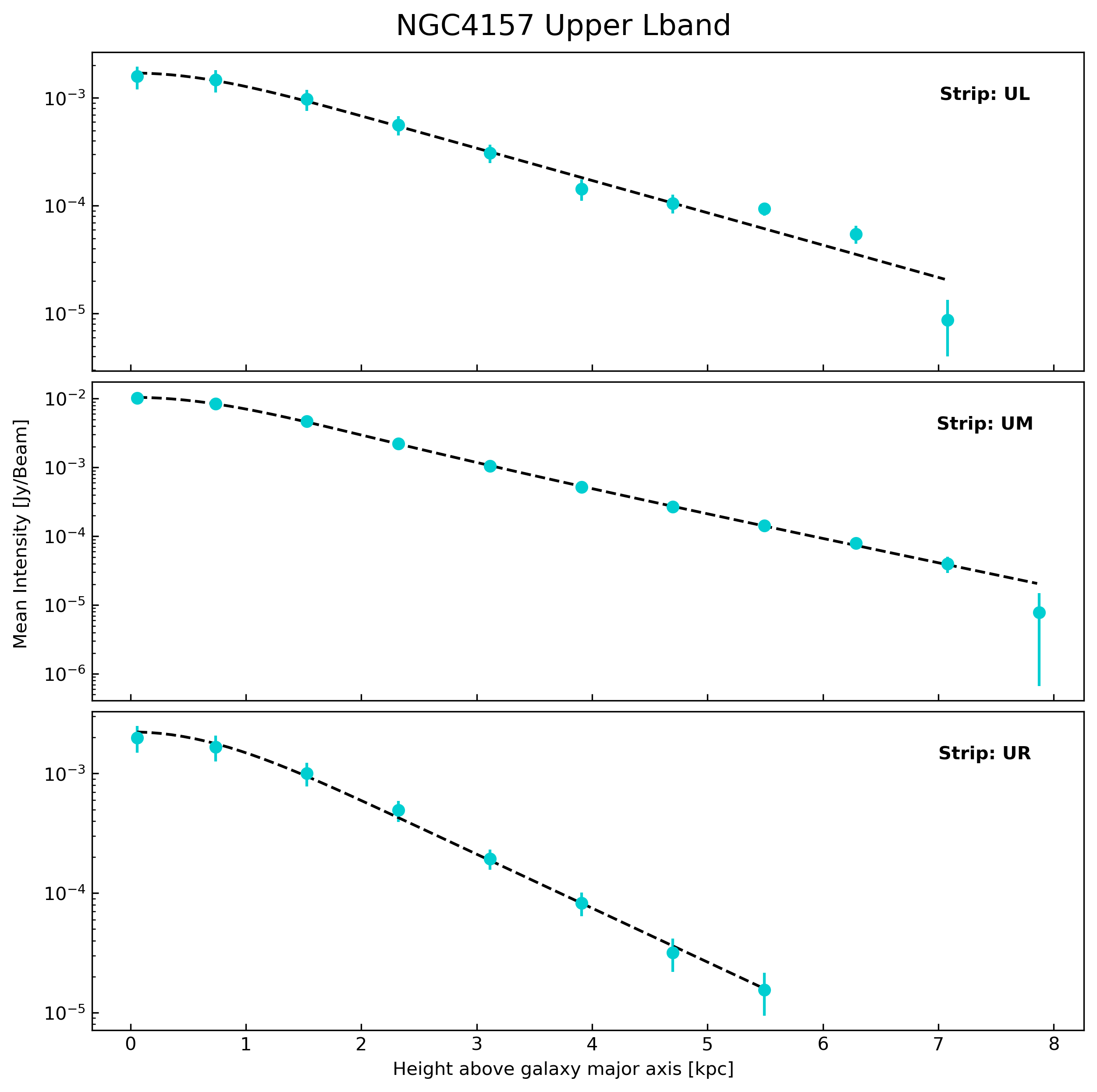}
\end{subfigure}
\\
\begin{subfigure}{0.49\linewidth}
\centering
\includegraphics[width=1\linewidth]{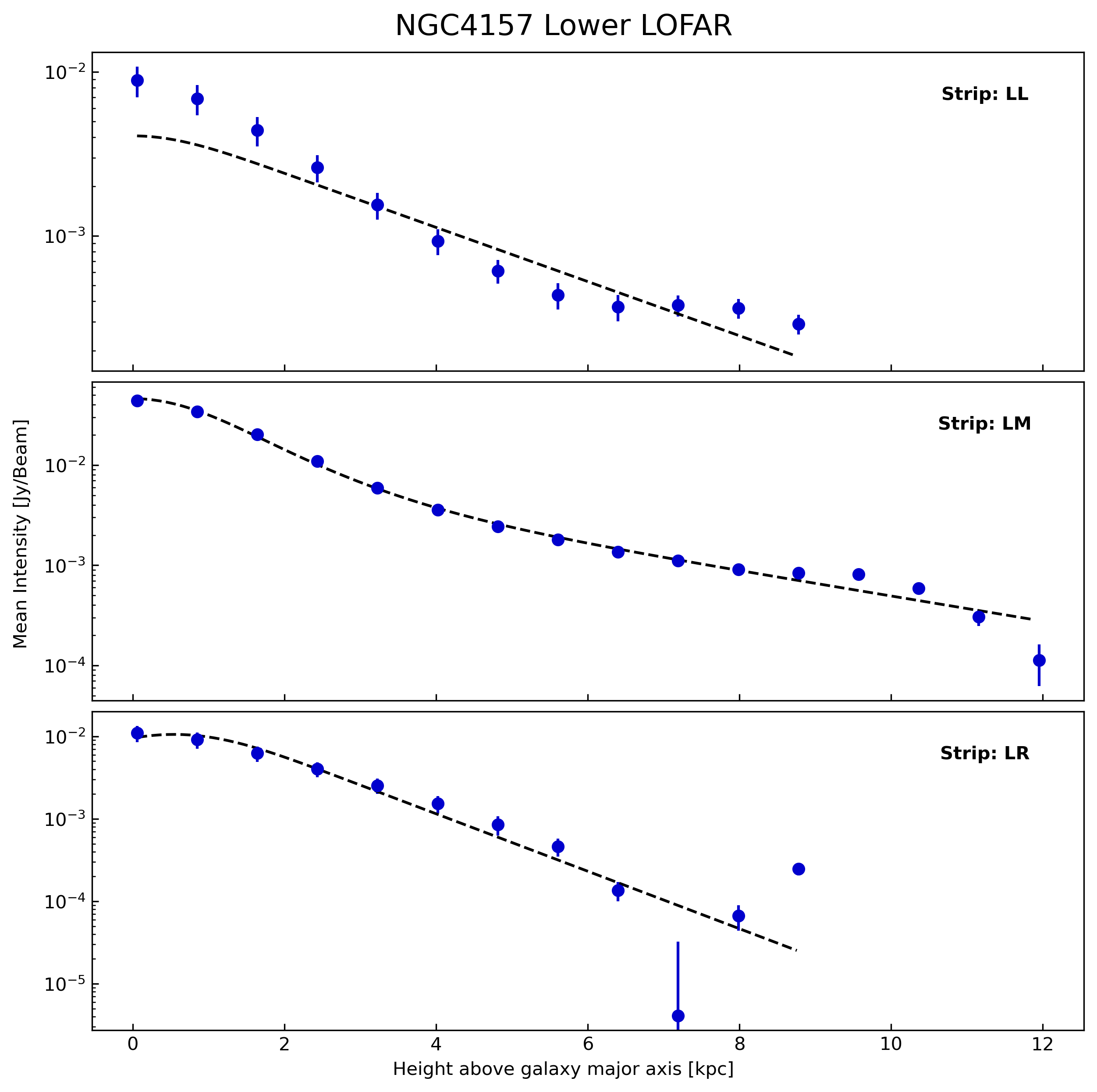}
\end{subfigure}
\hfill
\begin{subfigure}{0.49\linewidth}
\centering
\includegraphics[width=1\linewidth]{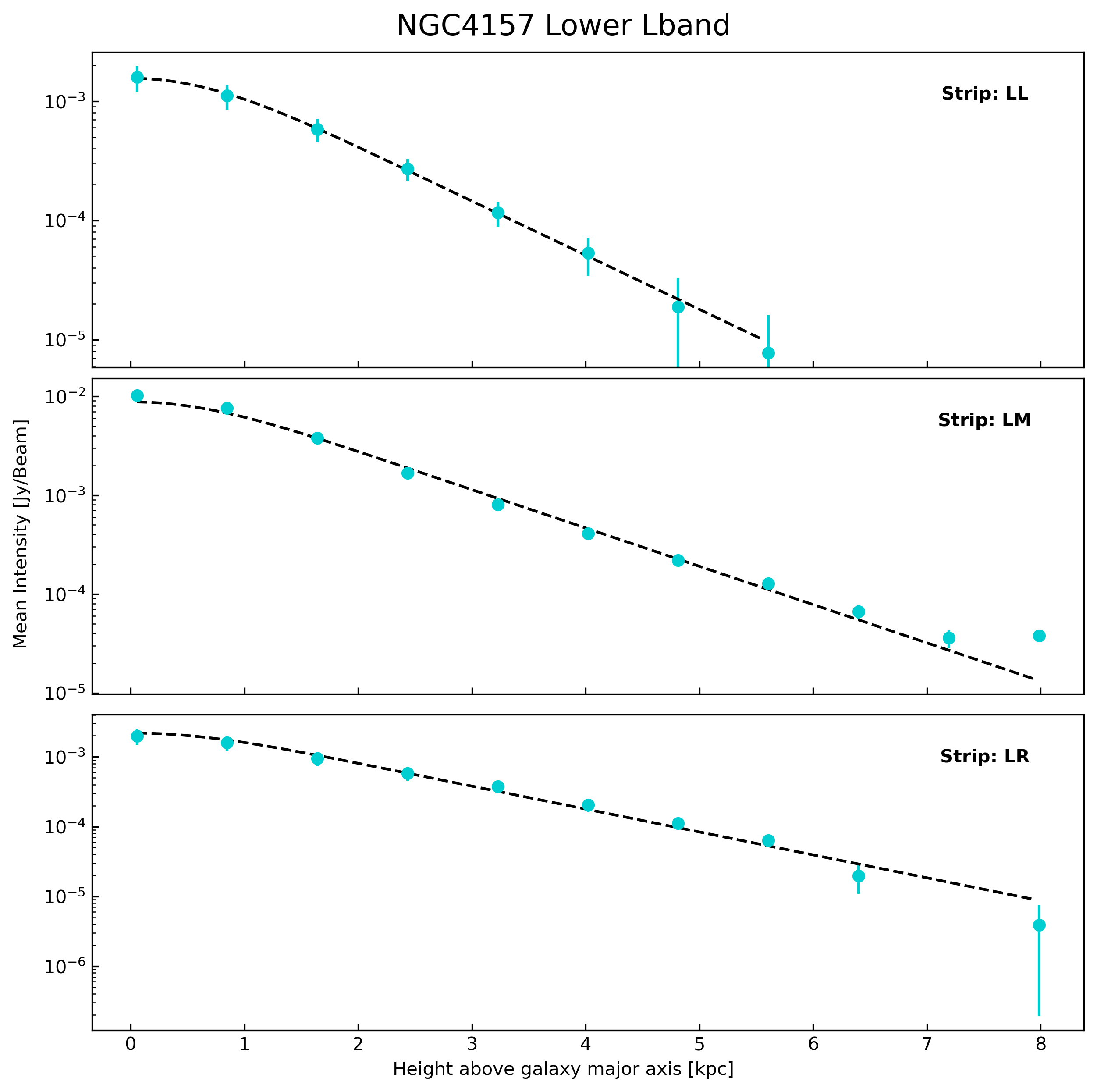}
\end{subfigure}

\caption{Intensity Profiles of NGC~4157}
\label{fig:int_prof_app_4157}
\end{figure*}

\begin{figure*}
\centering
\begin{subfigure}{0.49\linewidth}
\centering
\includegraphics[width=1\linewidth]{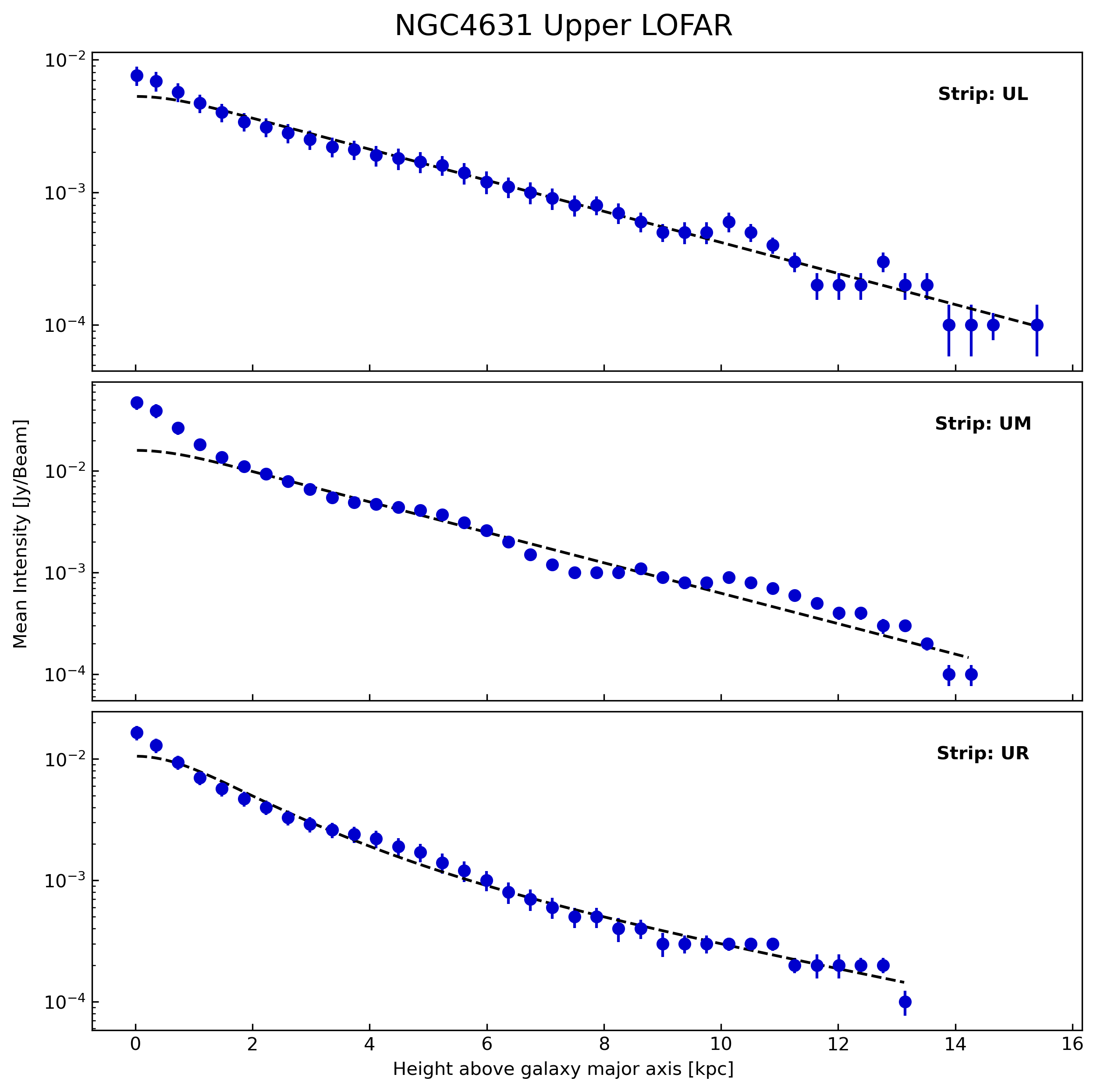}
\end{subfigure}
\hfill
\begin{subfigure}{0.49\linewidth}
\centering
\includegraphics[width=1\linewidth]{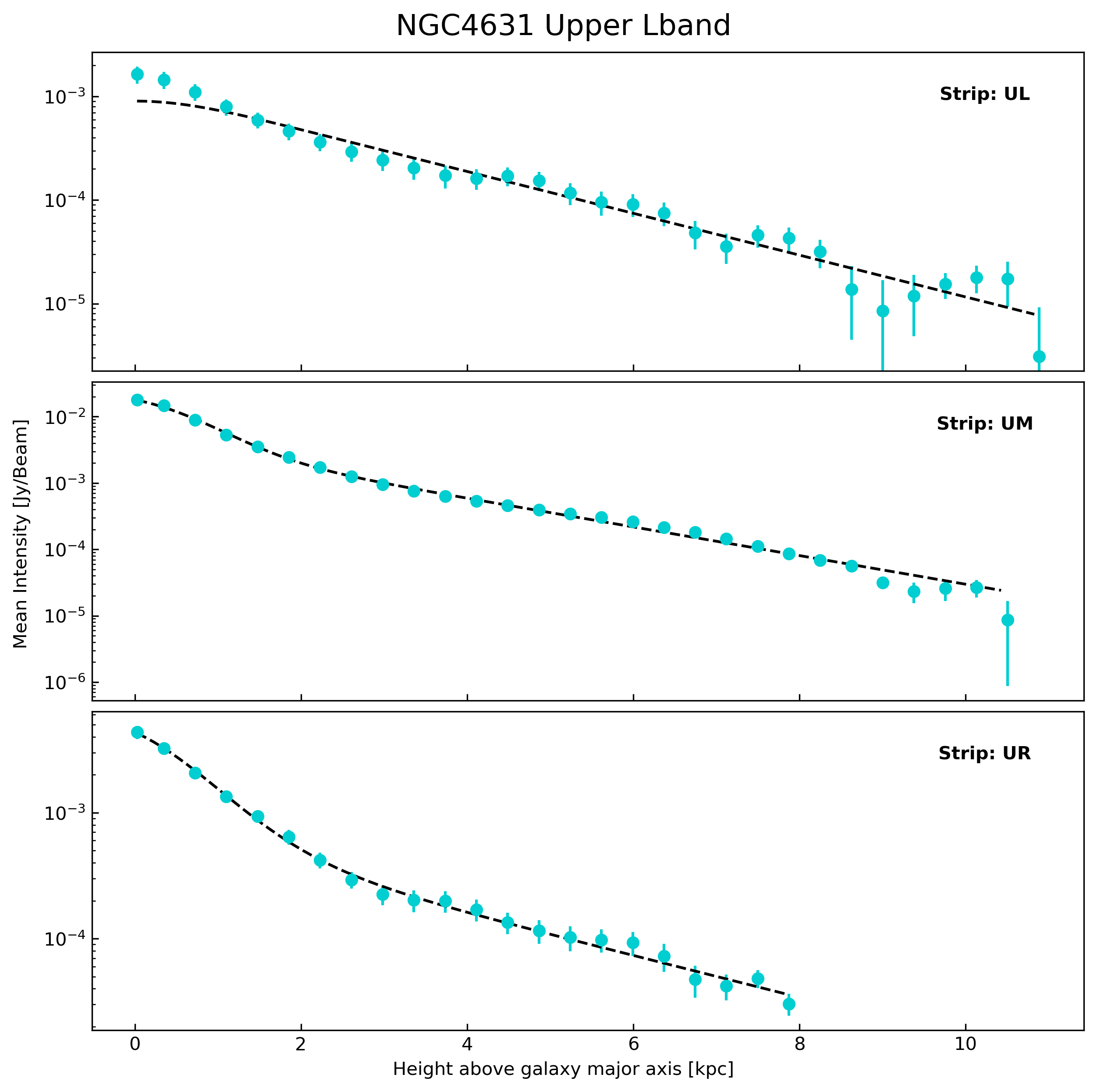}
\end{subfigure}
\\
\begin{subfigure}{0.49\linewidth}
\centering
\includegraphics[width=1\linewidth]{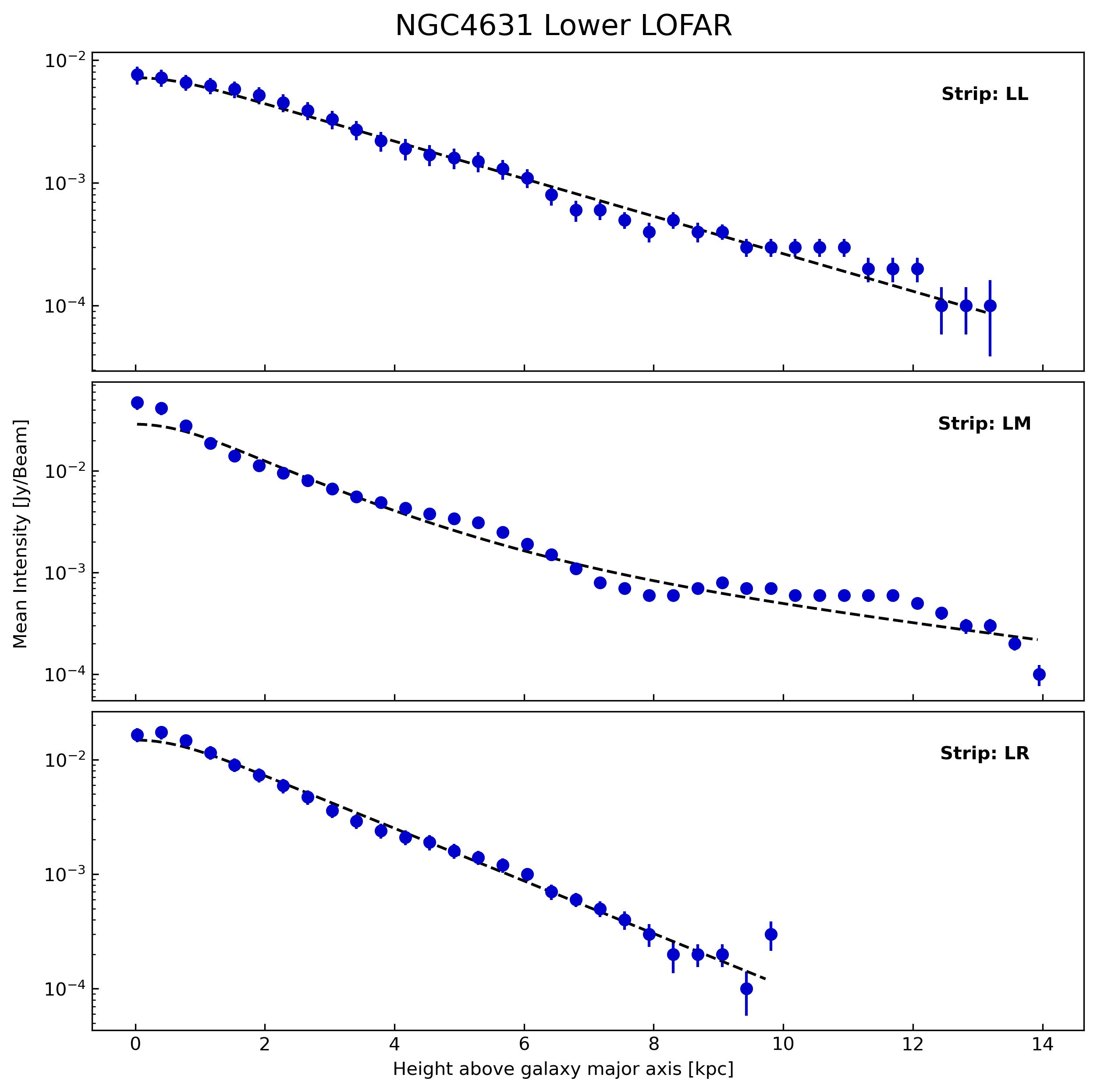}
\end{subfigure}
\hfill
\begin{subfigure}{0.49\linewidth}
\centering
\includegraphics[width=1\linewidth]{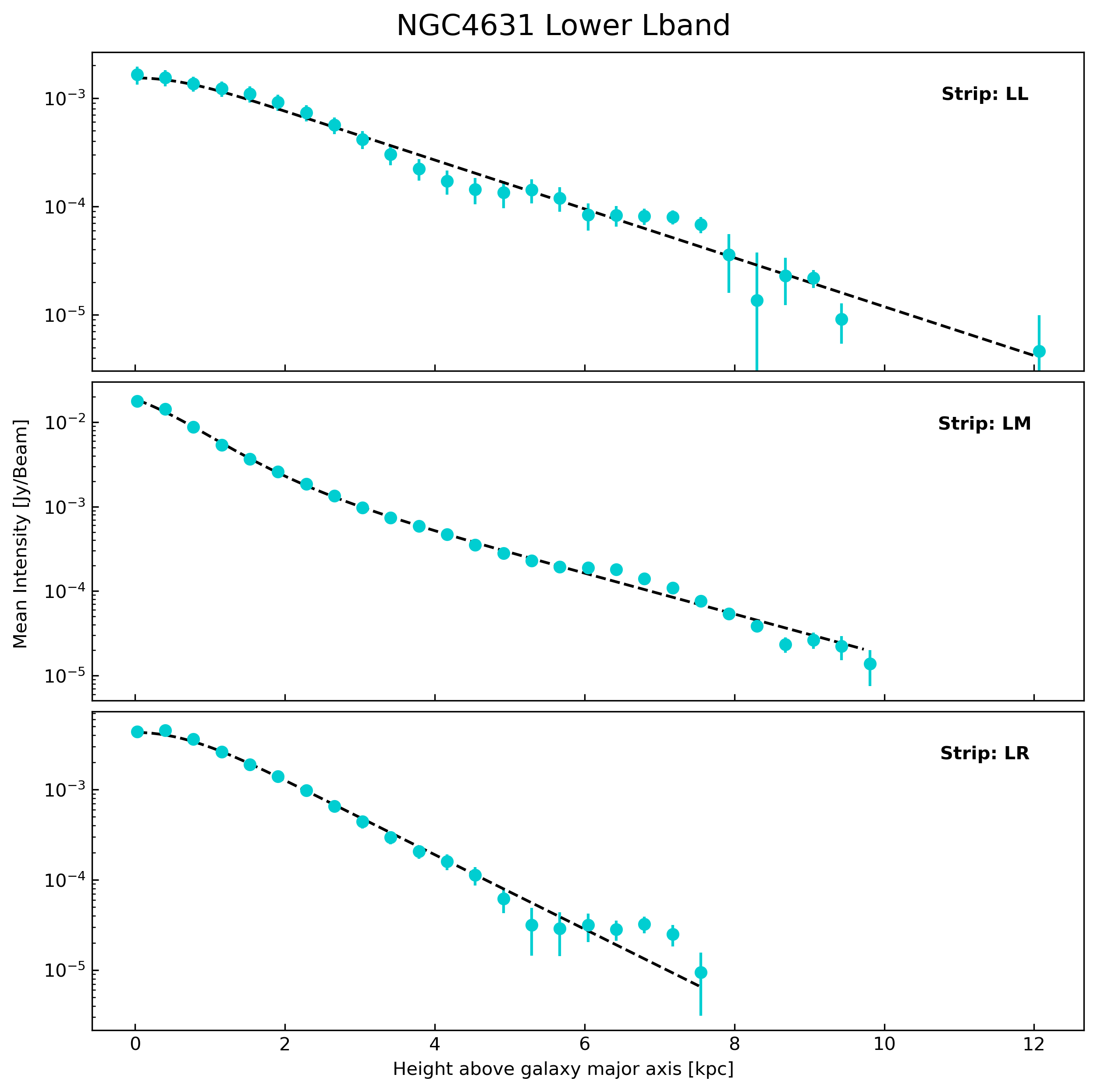}
\end{subfigure}

\caption{Intensity Profiles of NGC~4631}
\label{fig:int_prof_app_4631}
\end{figure*}

%% file: spinnaker_appendix.tex
\begin{figure*}
\centering
\begin{subfigure}{0.49\linewidth}
\centering
\includegraphics[width=1\linewidth]{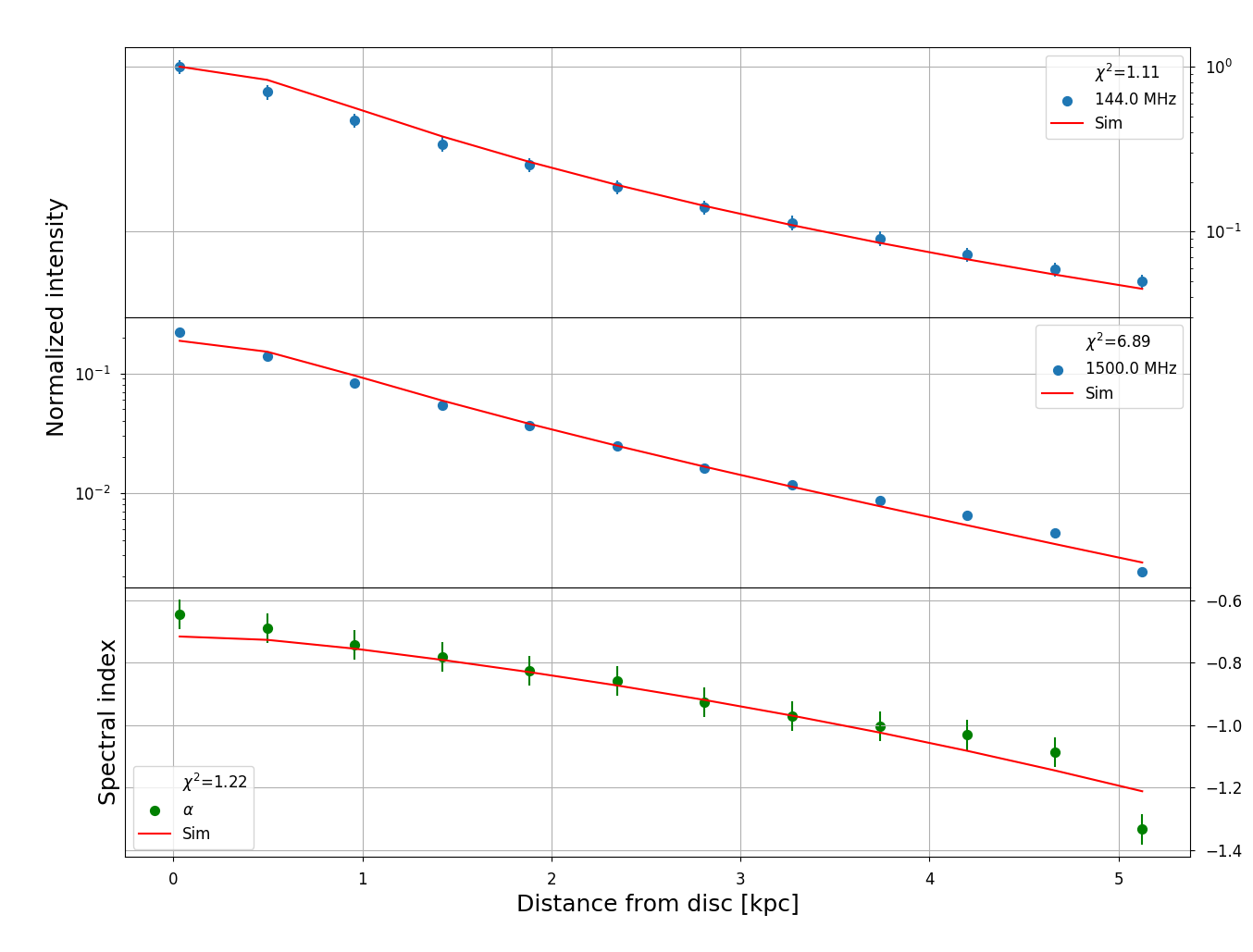}
\end{subfigure}
\hfill
\begin{subfigure}{0.49\linewidth}
\centering
\includegraphics[width=1\linewidth]{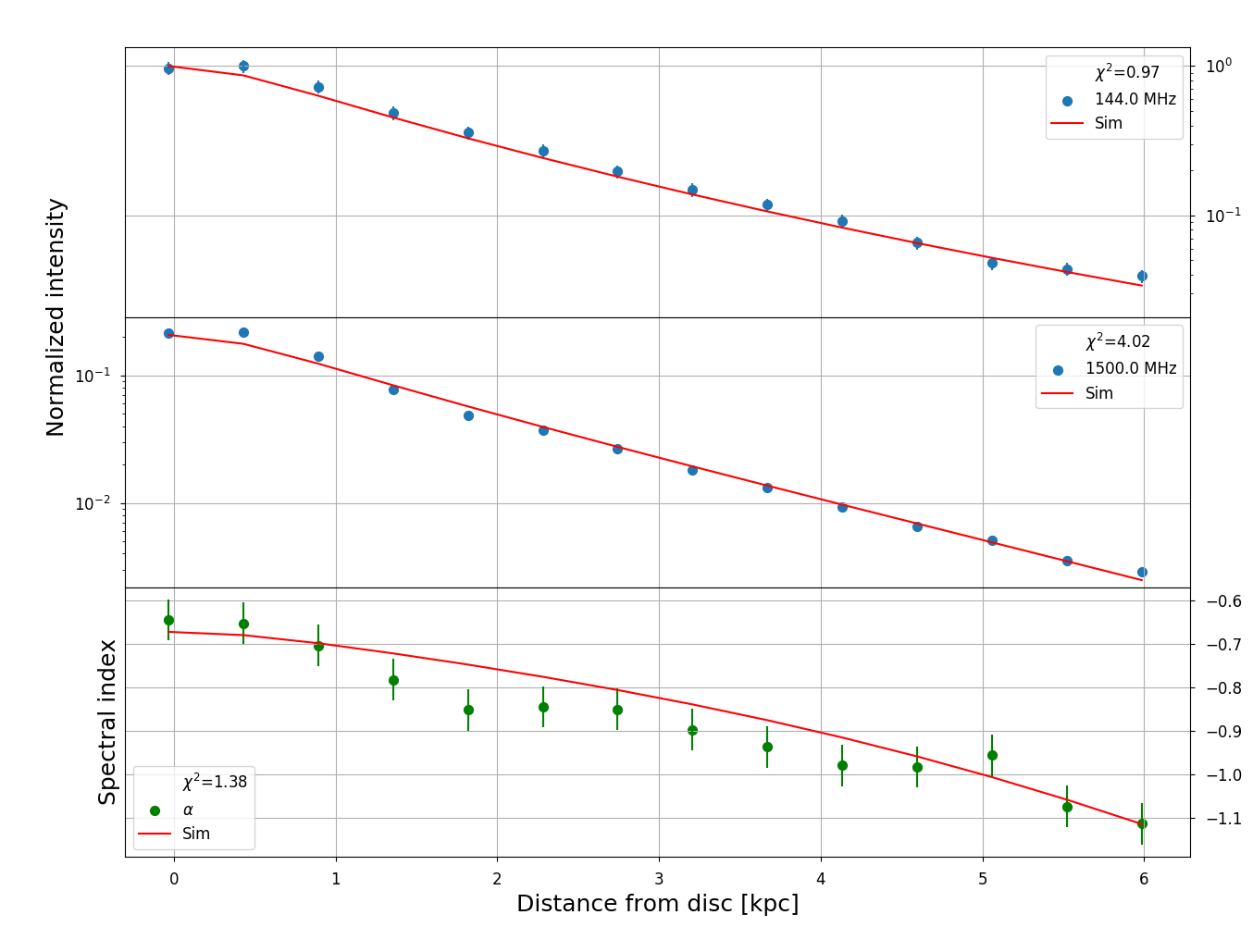}
\end{subfigure}
\\
\begin{subfigure}{0.49\linewidth}
\centering
\includegraphics[width=1\linewidth]{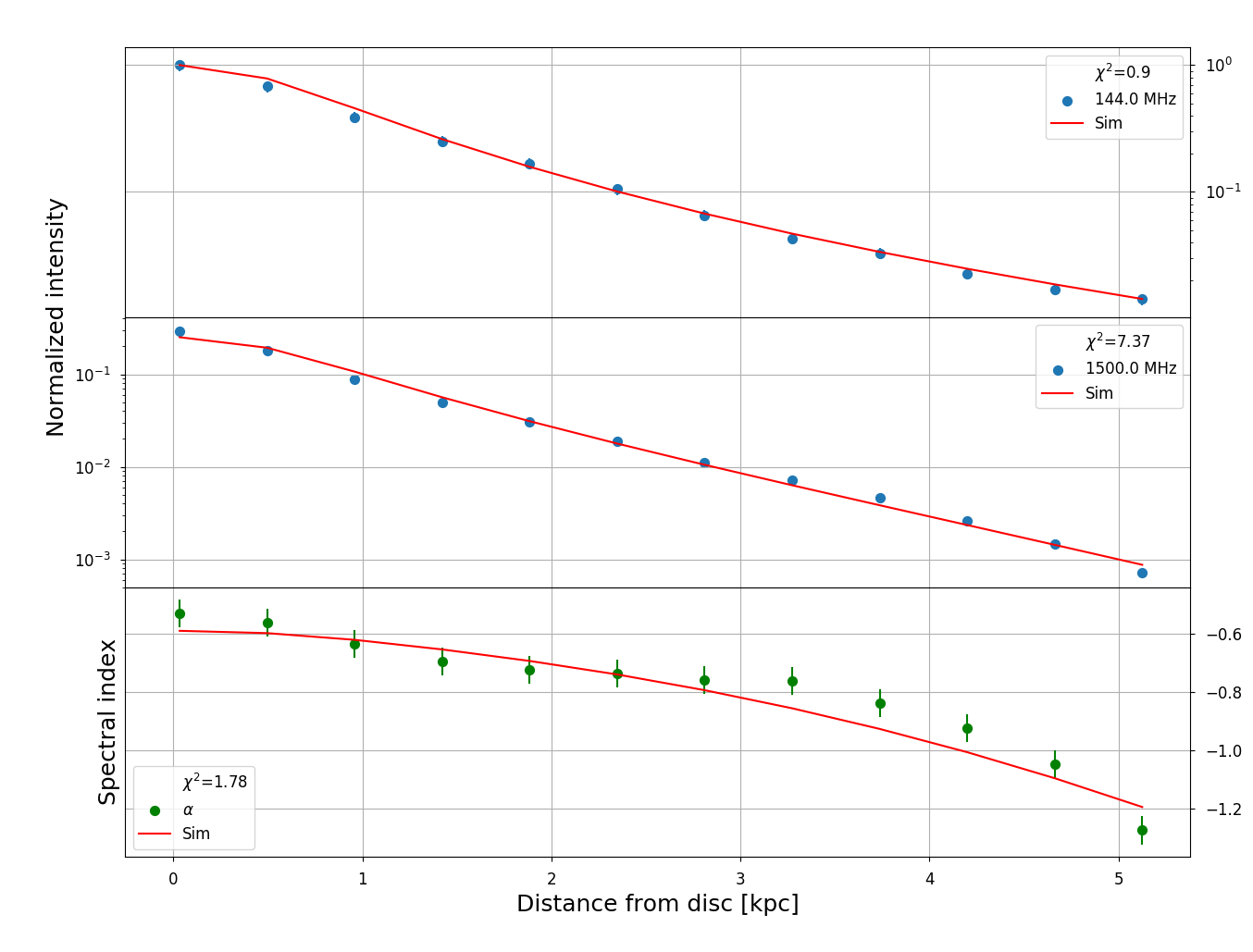}
\end{subfigure}
\hfill
\begin{subfigure}{0.49\linewidth}
\centering
\includegraphics[width=1\linewidth]{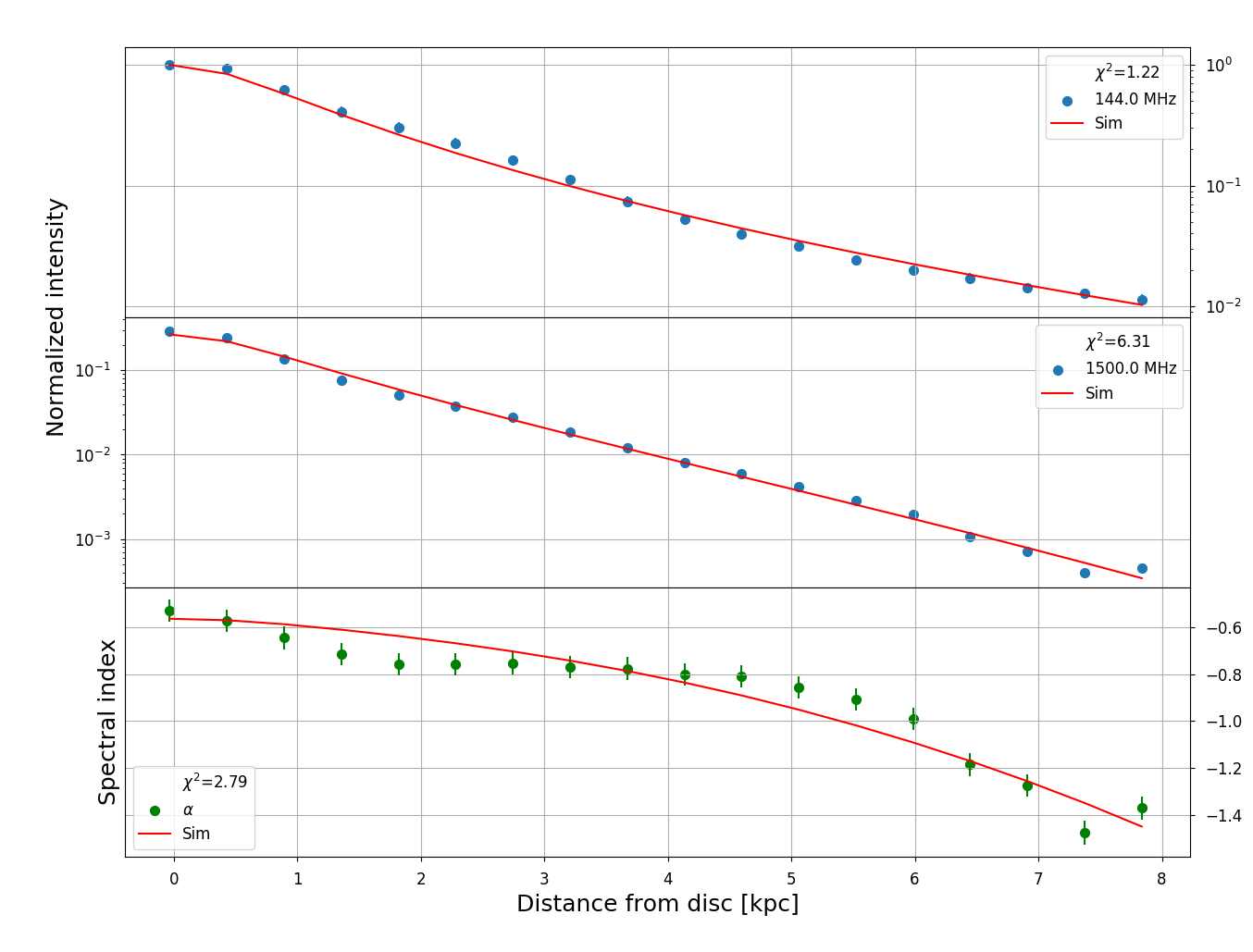}
\end{subfigure}
\\
\begin{subfigure}{0.49\linewidth}
\centering
\includegraphics[width=1\linewidth]{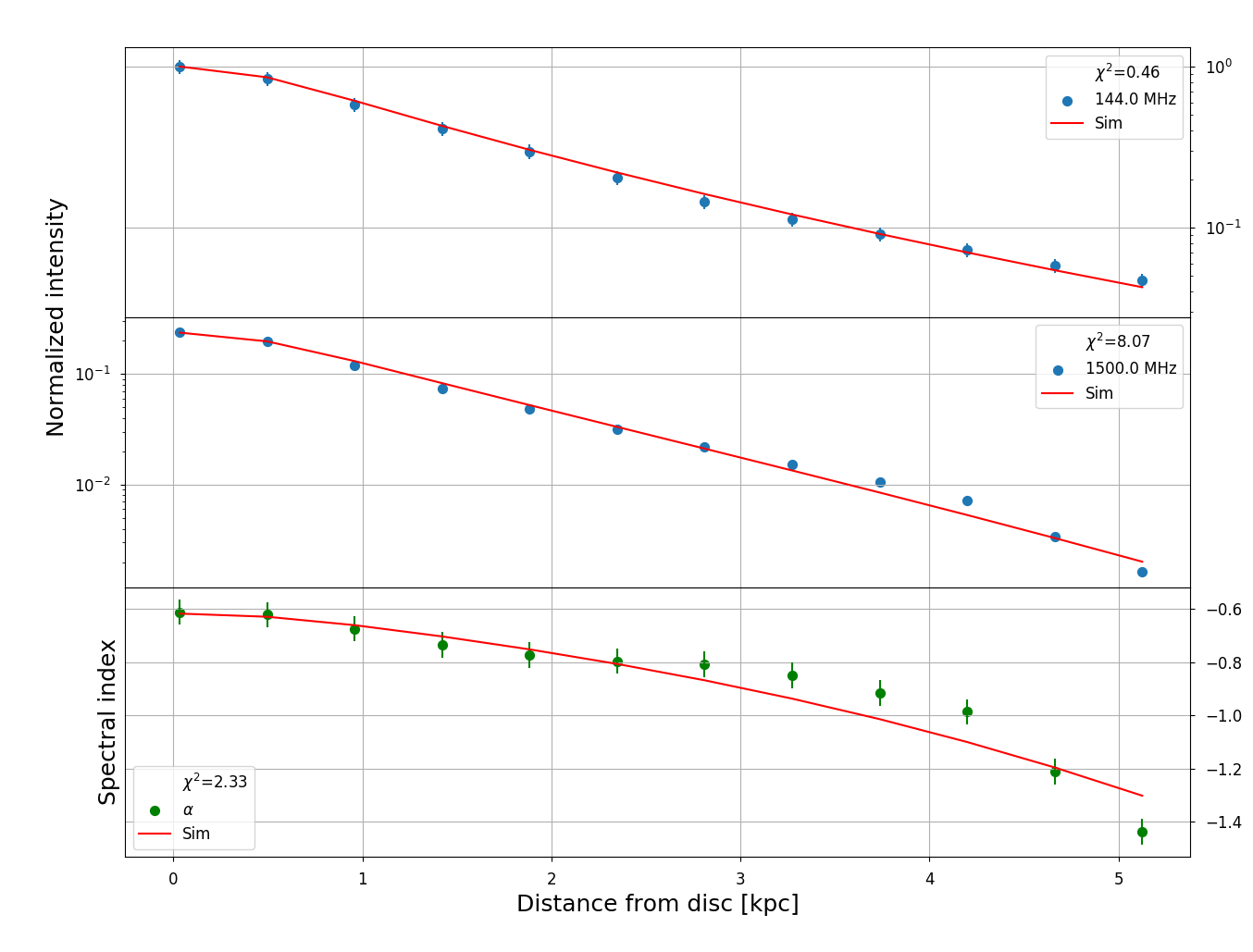}
\end{subfigure}
\hfill
\begin{subfigure}{0.49\linewidth}
\centering
\includegraphics[width=1\linewidth]{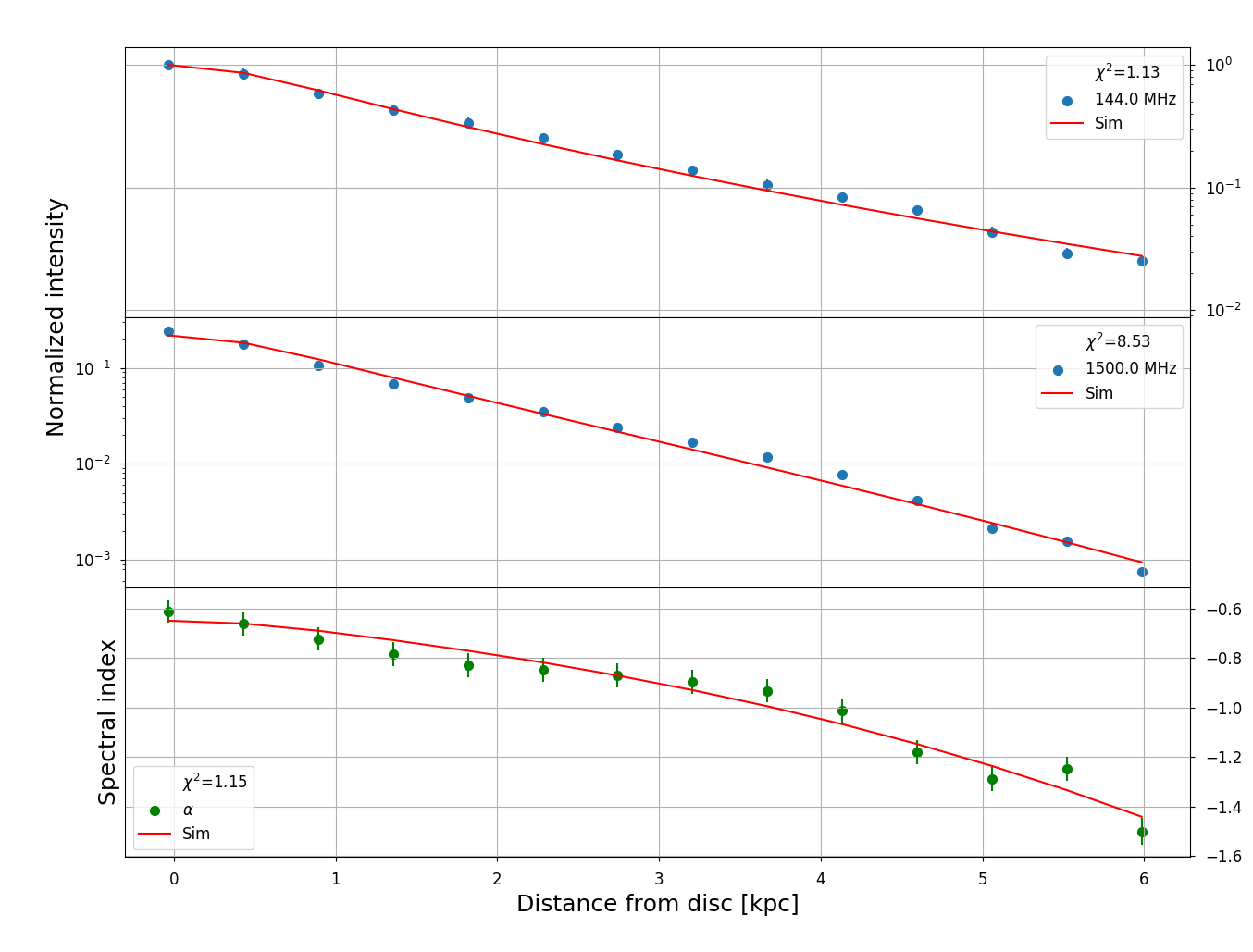}
\end{subfigure}
\caption{\texttt{SPINNAKER} profiles of NGC~891 (all advection). Strips are presented as follows: top left: LR, top right: UR, middle left: LM, middle right: UM, bottom left: LL, and bottom right: UL.}
\end{figure*}
\begin{figure*}
\centering
\begin{subfigure}{0.49\linewidth}
\centering
\includegraphics[width=1\linewidth]{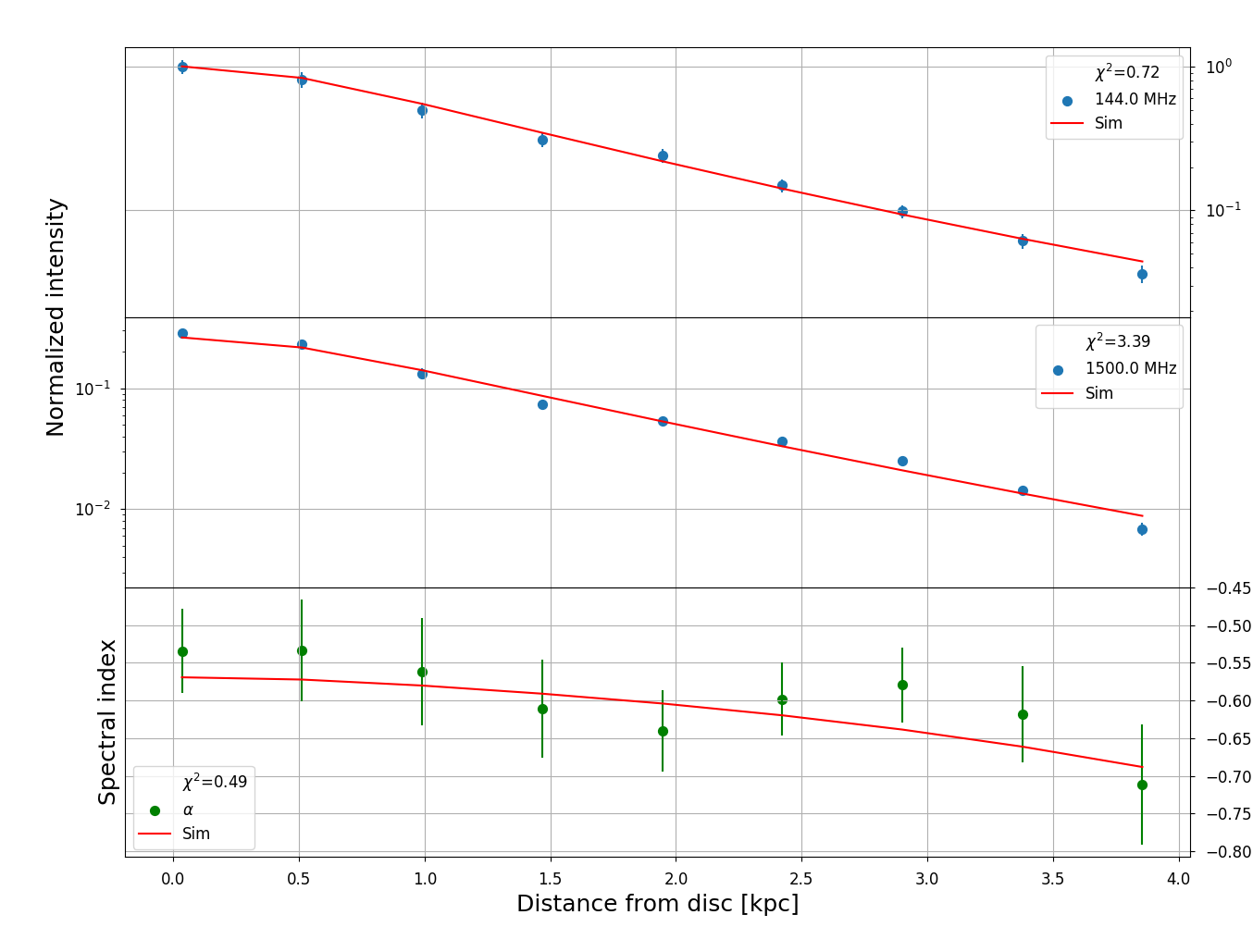}
\end{subfigure}
\hfill
\begin{subfigure}{0.49\linewidth}
\centering
\includegraphics[width=1\linewidth]{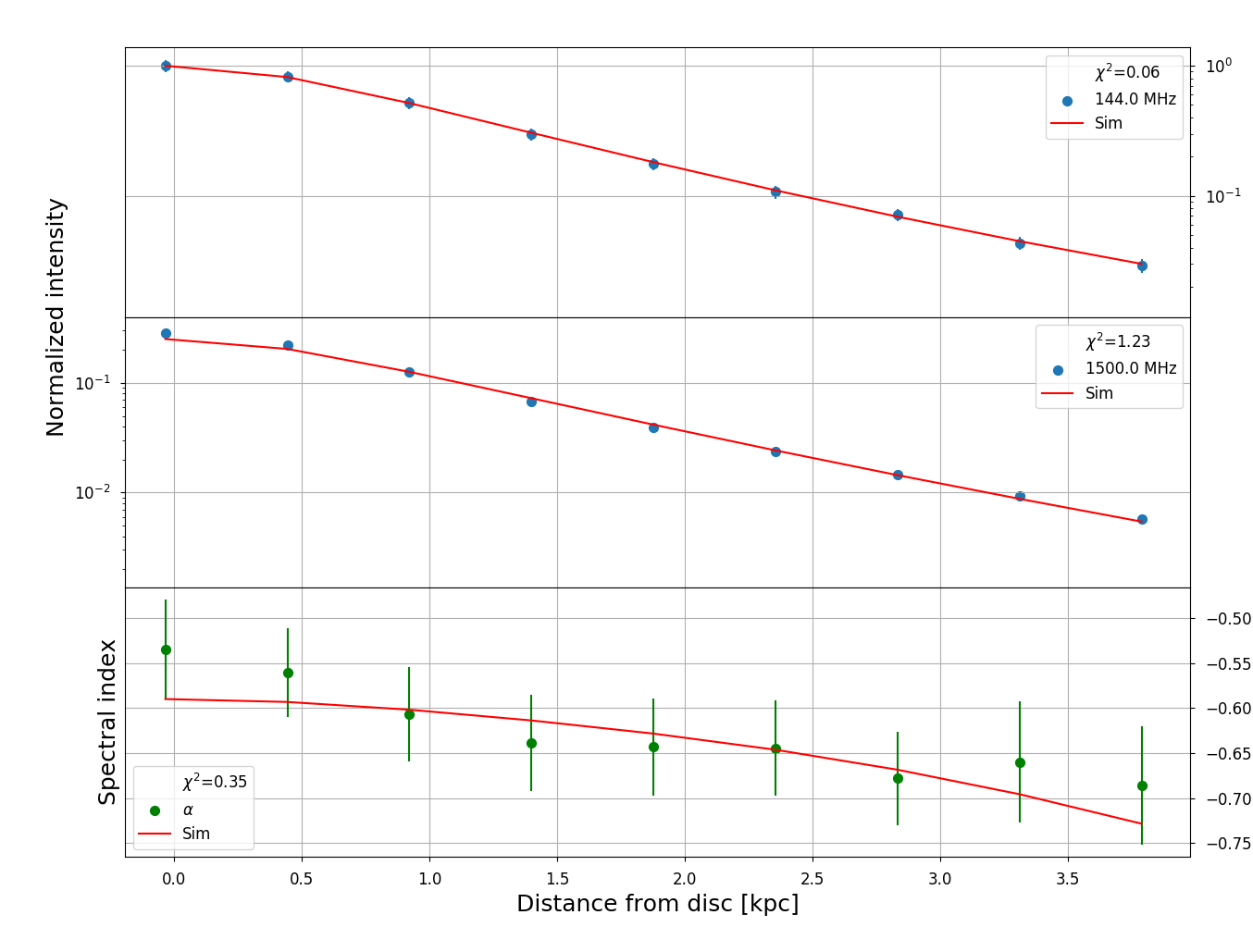}
\end{subfigure}
\caption{\texttt{SPINNAKER} profiles of NGC~3432 (all advection). Strips are presented as follows: left: LM and right: UM.}
\end{figure*}

\begin{figure*}
\centering
\begin{subfigure}{0.49\linewidth}
\centering
\includegraphics[width=1\linewidth]{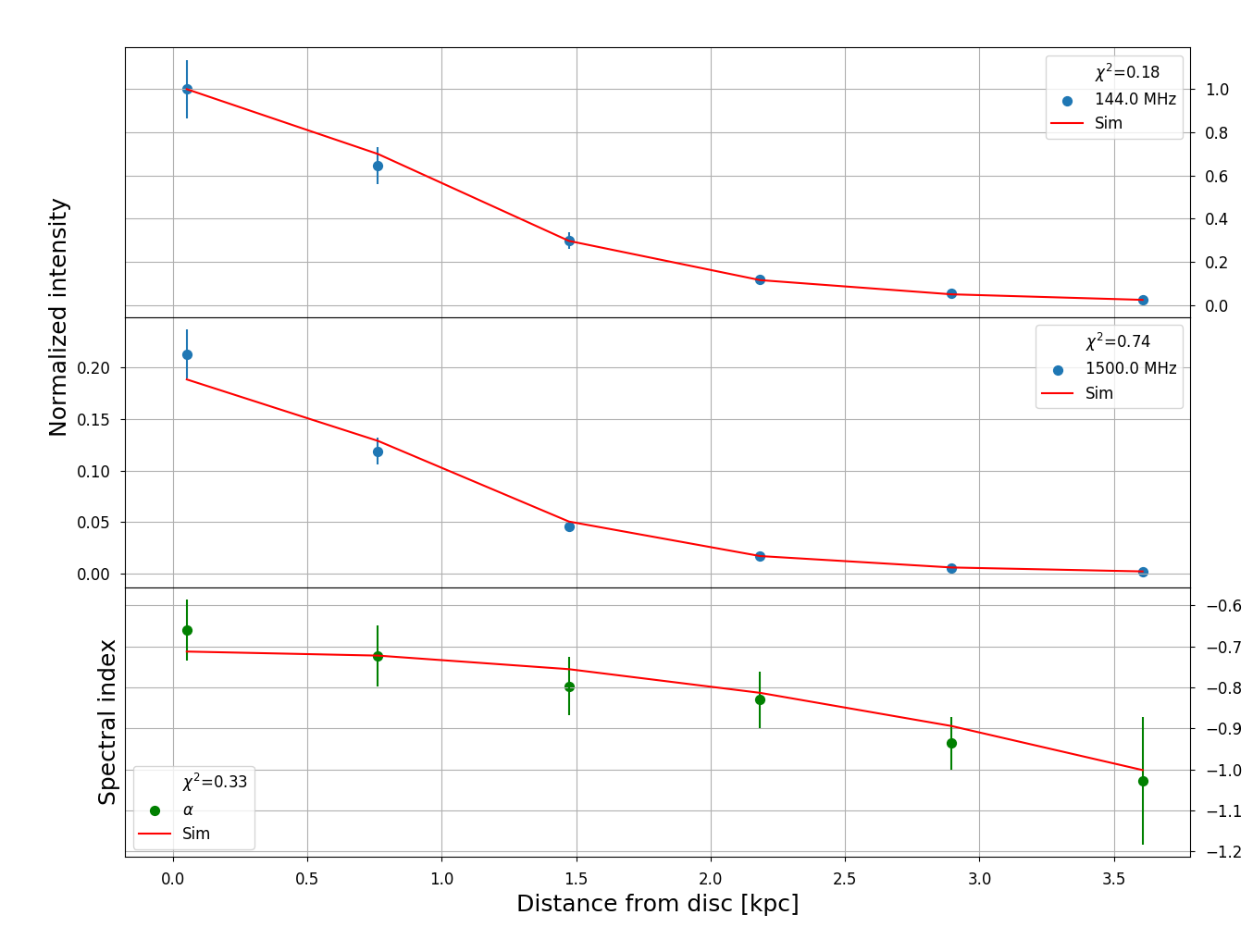}
\end{subfigure}
\hfill
\begin{subfigure}{0.49\linewidth}
\centering
\includegraphics[width=1\linewidth]{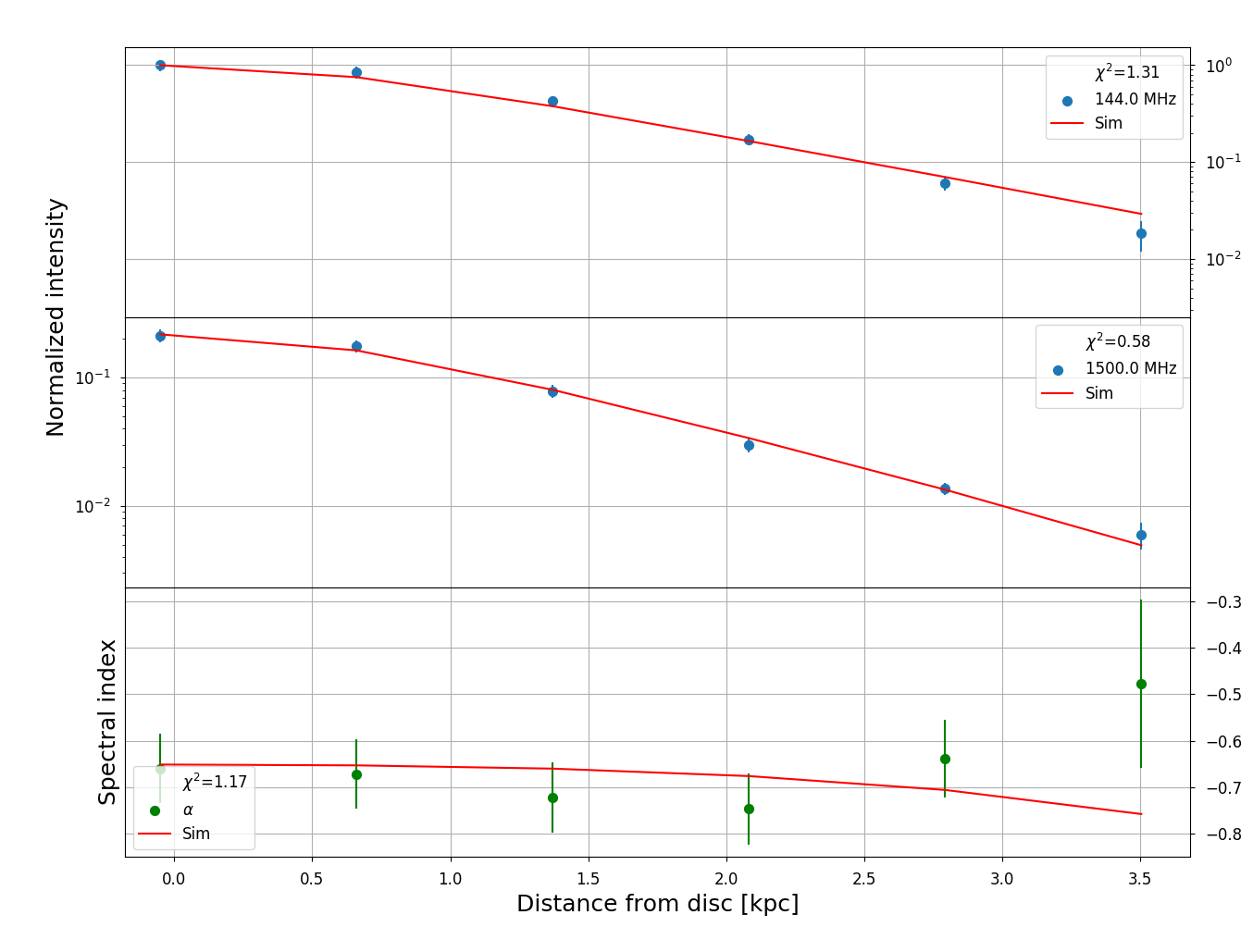}
\end{subfigure}
\\
\begin{subfigure}{0.49\linewidth}
\centering
\includegraphics[width=1\linewidth]{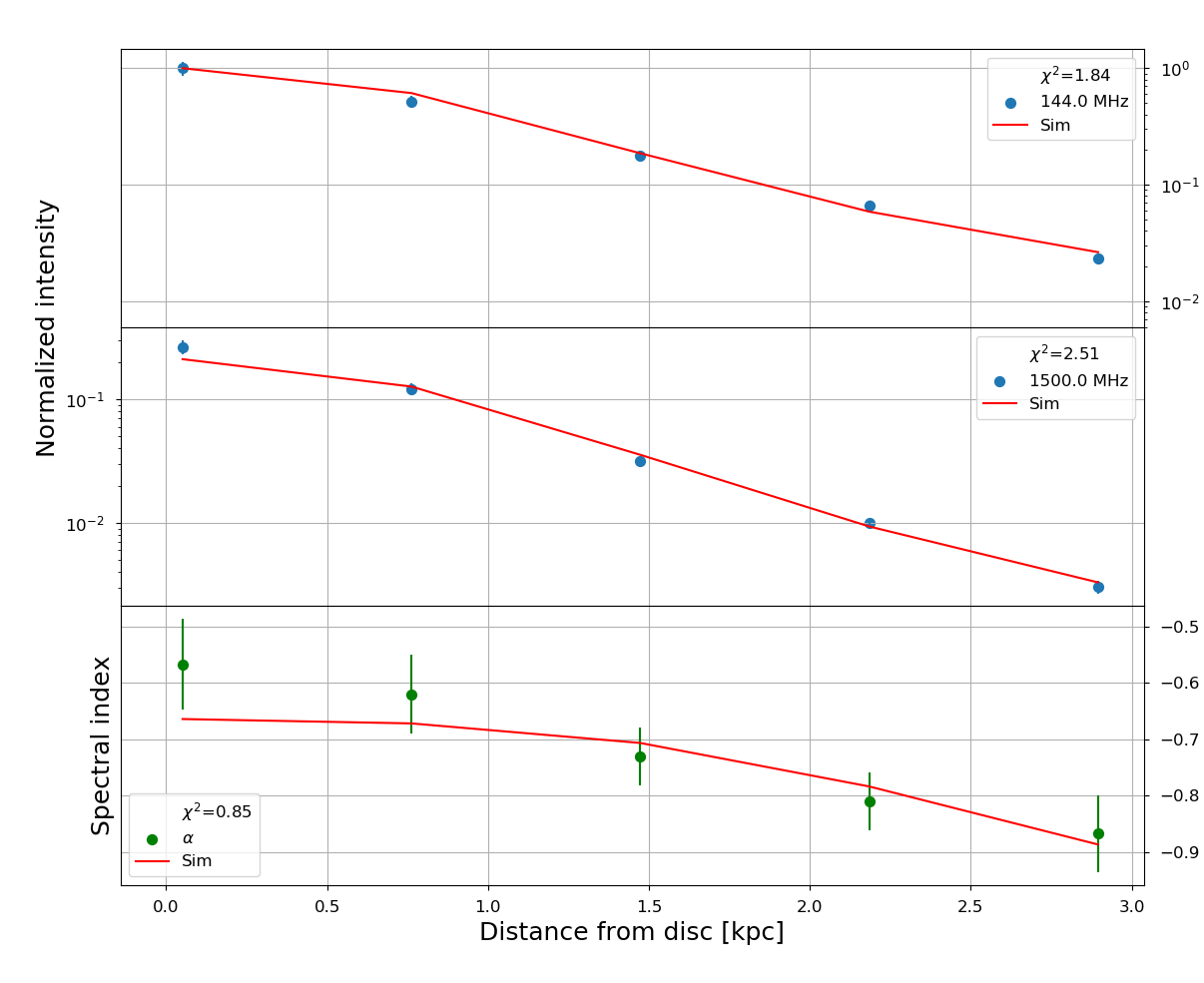}
\end{subfigure}
\hfill
\begin{subfigure}{0.49\linewidth}
\centering
\includegraphics[width=1\linewidth]{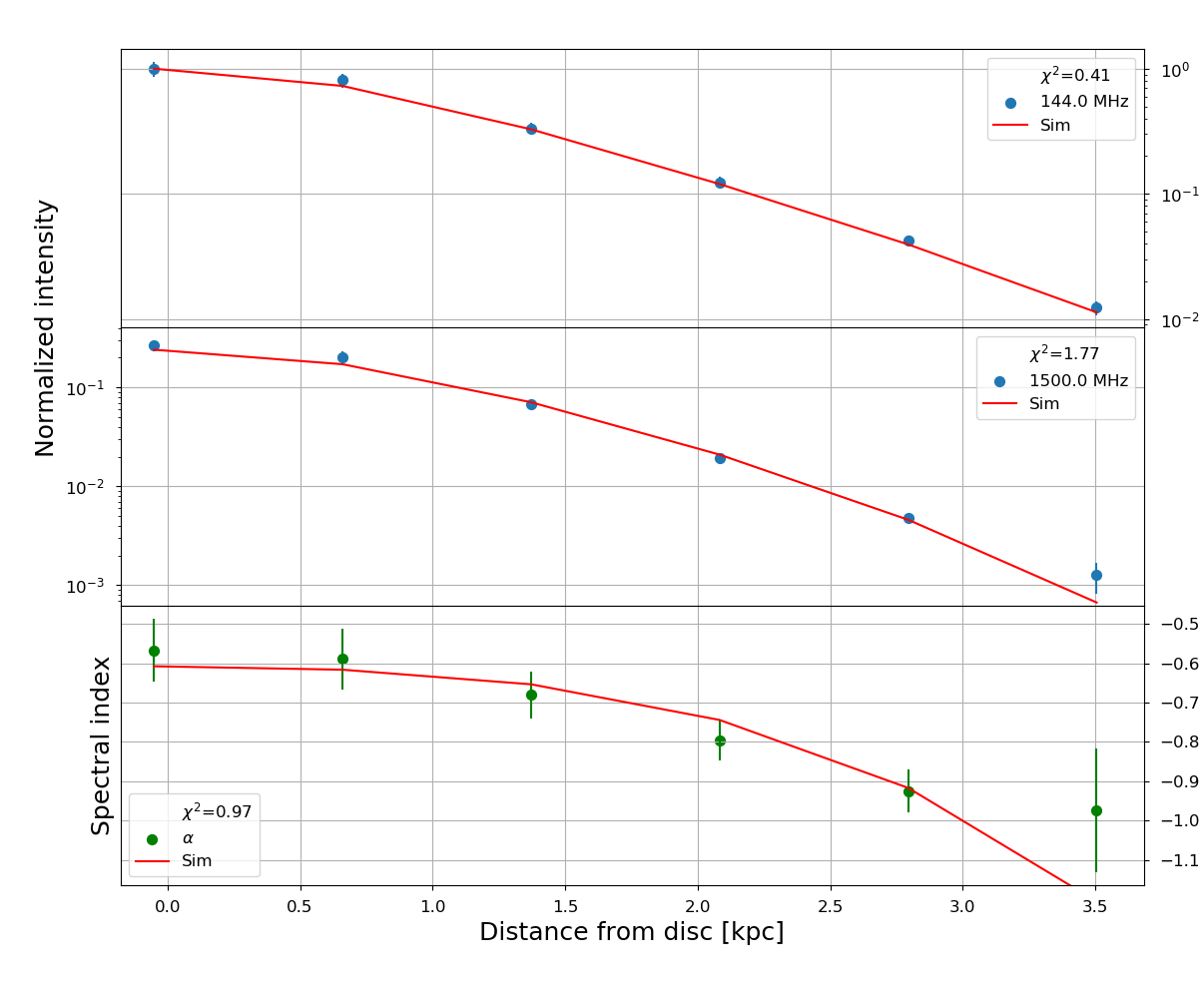}
\end{subfigure}
\\
\begin{subfigure}{0.49\linewidth}
\centering
\includegraphics[width=1\linewidth]{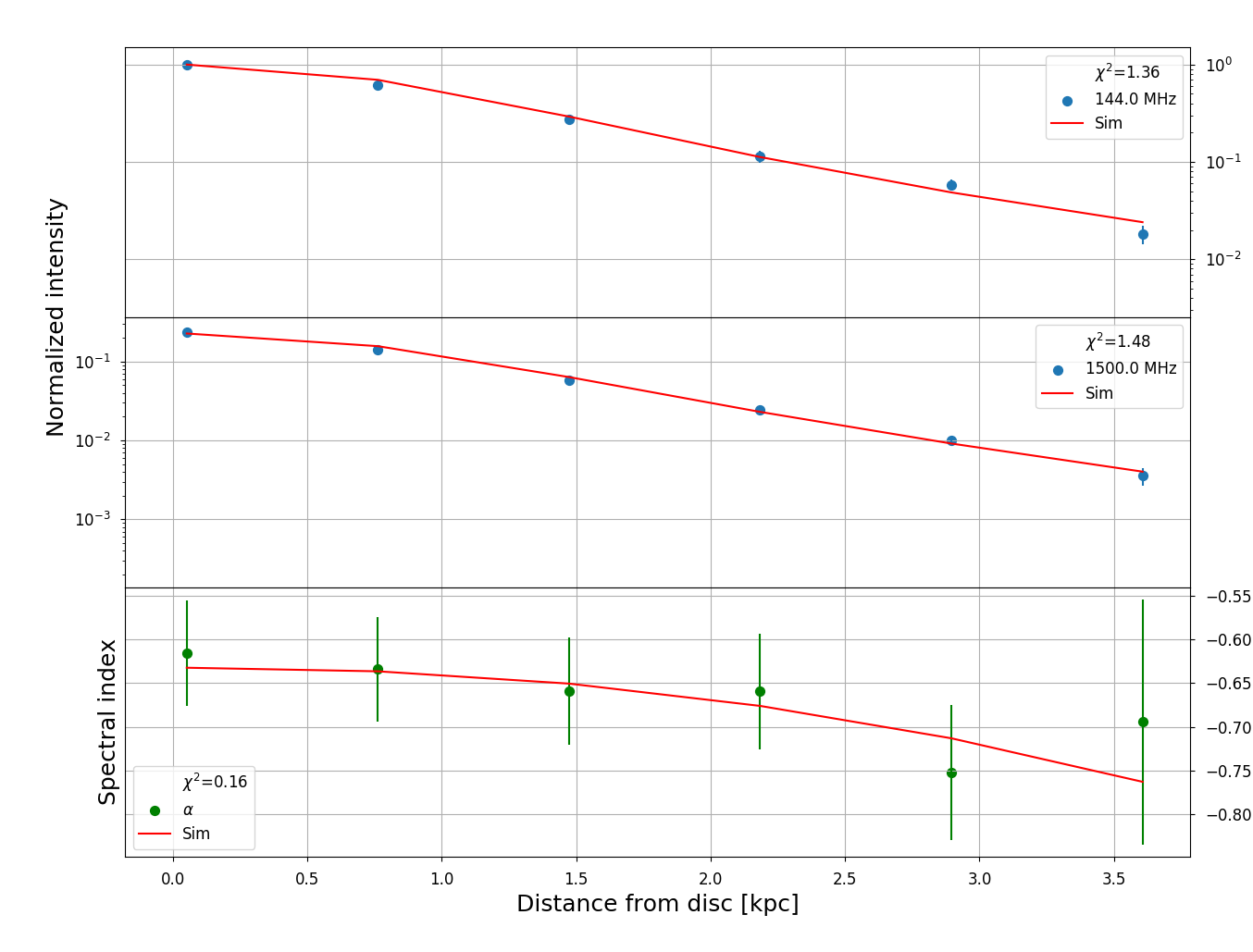}
\end{subfigure}
\hfill
\begin{subfigure}{0.49\linewidth}
\centering
\includegraphics[width=1\linewidth]{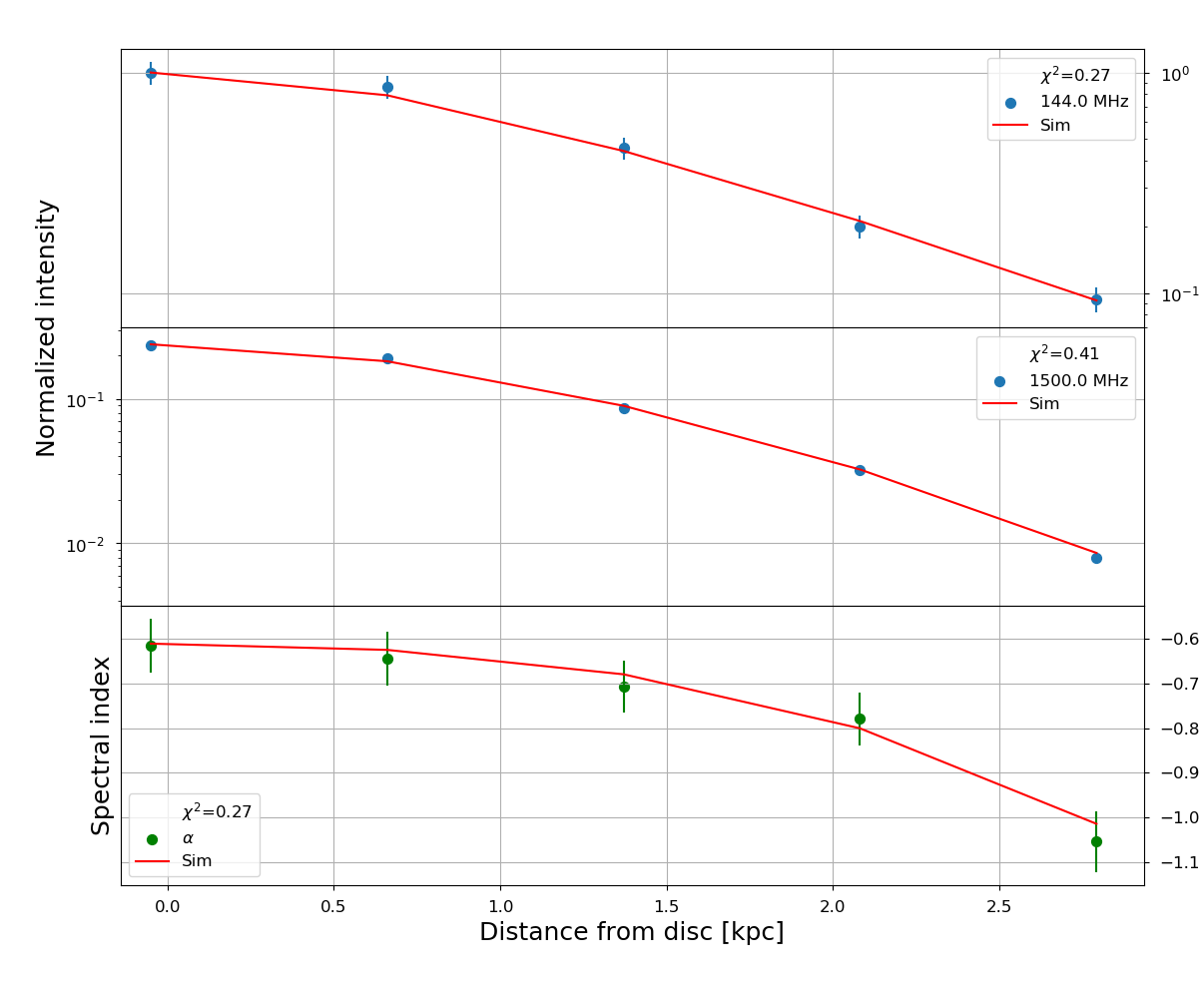}
\end{subfigure}
\caption{\texttt{SPINNAKER} profiles of NGC~4013. Strips are presented as follows: top left: LR (advection), top right: UR (diffusion), middle left: LM (advection), middle right: UM (diffusion), bottom left: LL (advection), and bottom right: UL (diffusion).}
\end{figure*}

\begin{figure*}
\centering
\begin{subfigure}{0.49\linewidth}
\centering
\includegraphics[width=1\linewidth]{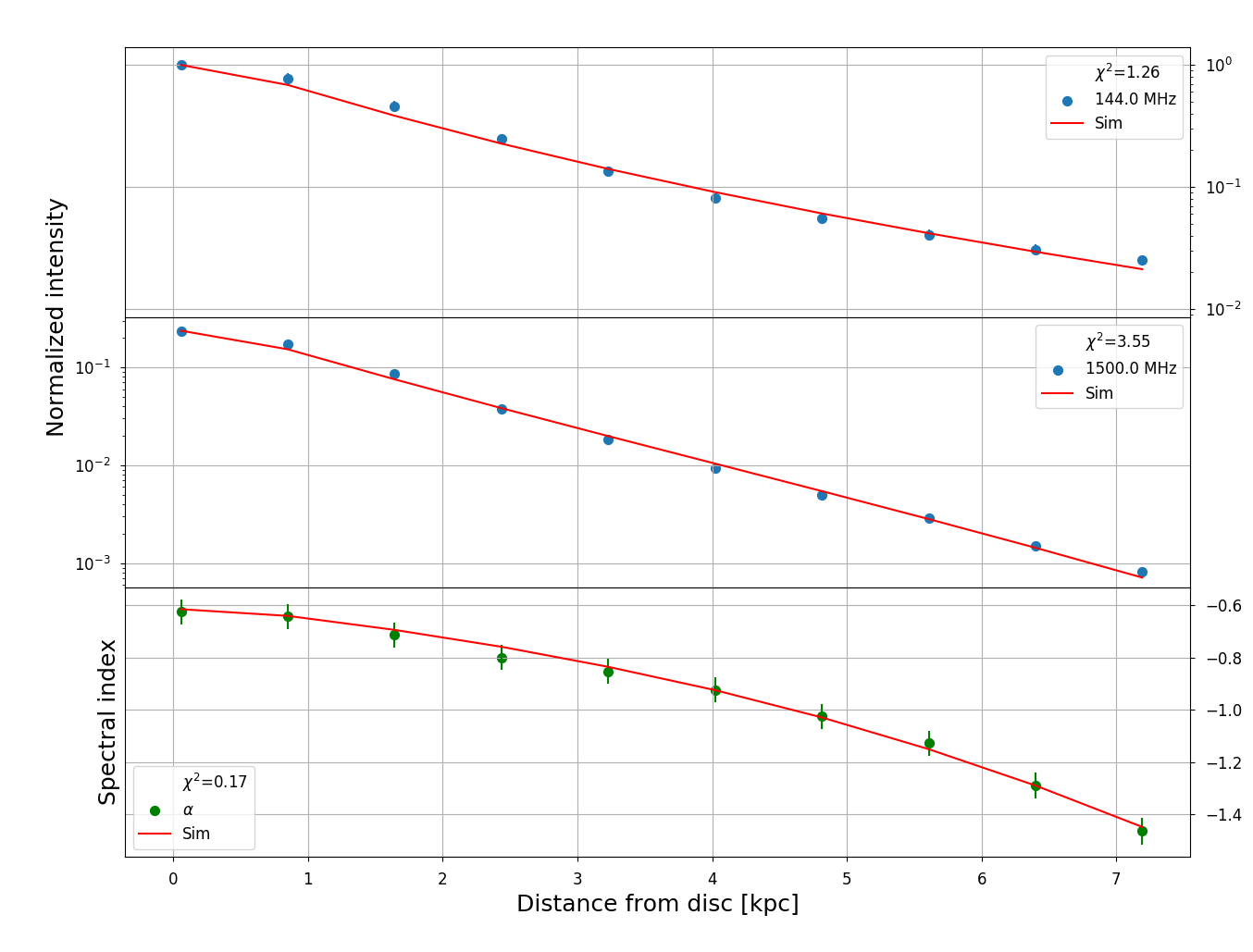}
\end{subfigure}
\hfill
\begin{subfigure}{0.49\linewidth}
\centering
\includegraphics[width=1\linewidth]{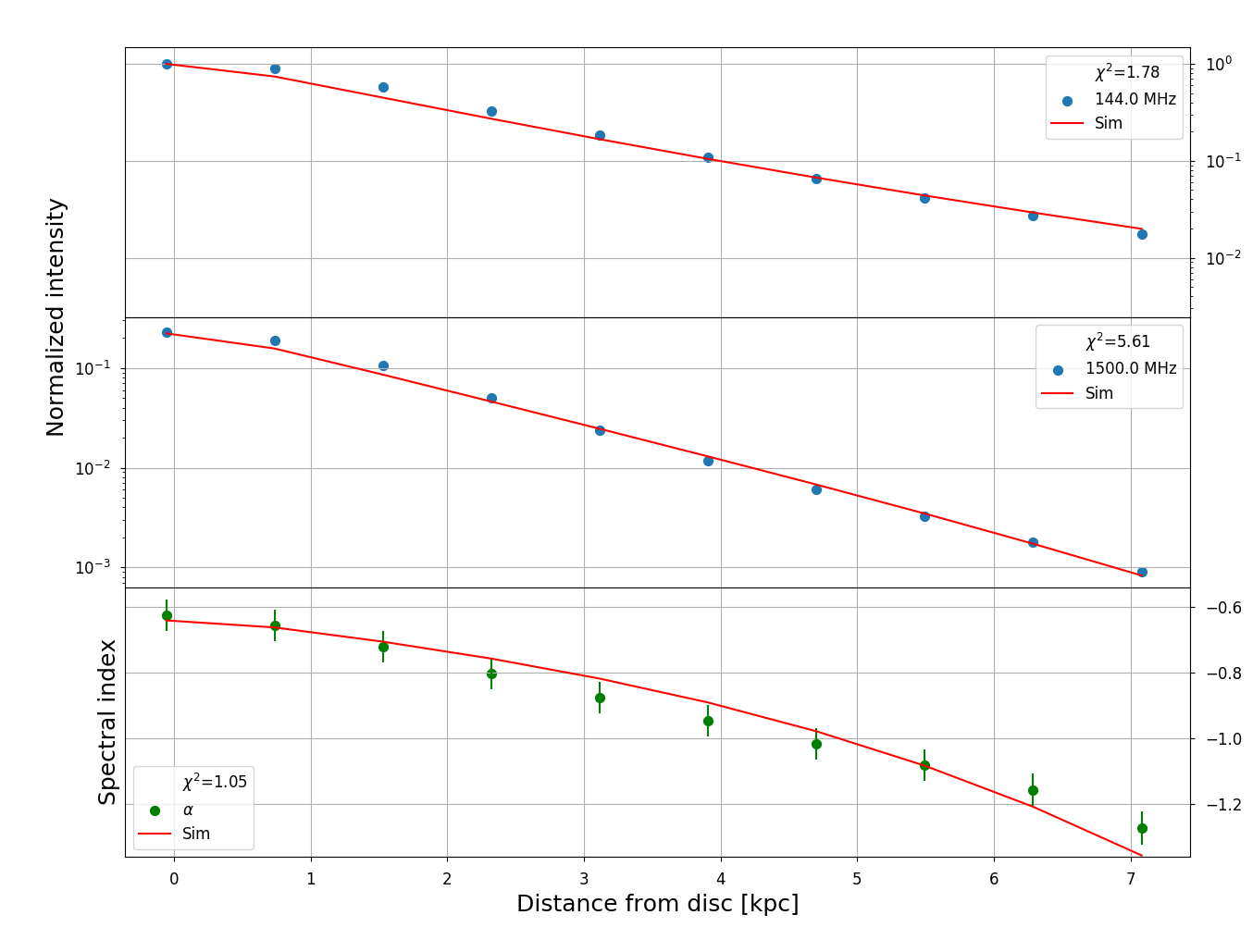}
\end{subfigure}
\\
\begin{subfigure}{0.49\linewidth}
\centering
\includegraphics[width=1\linewidth]{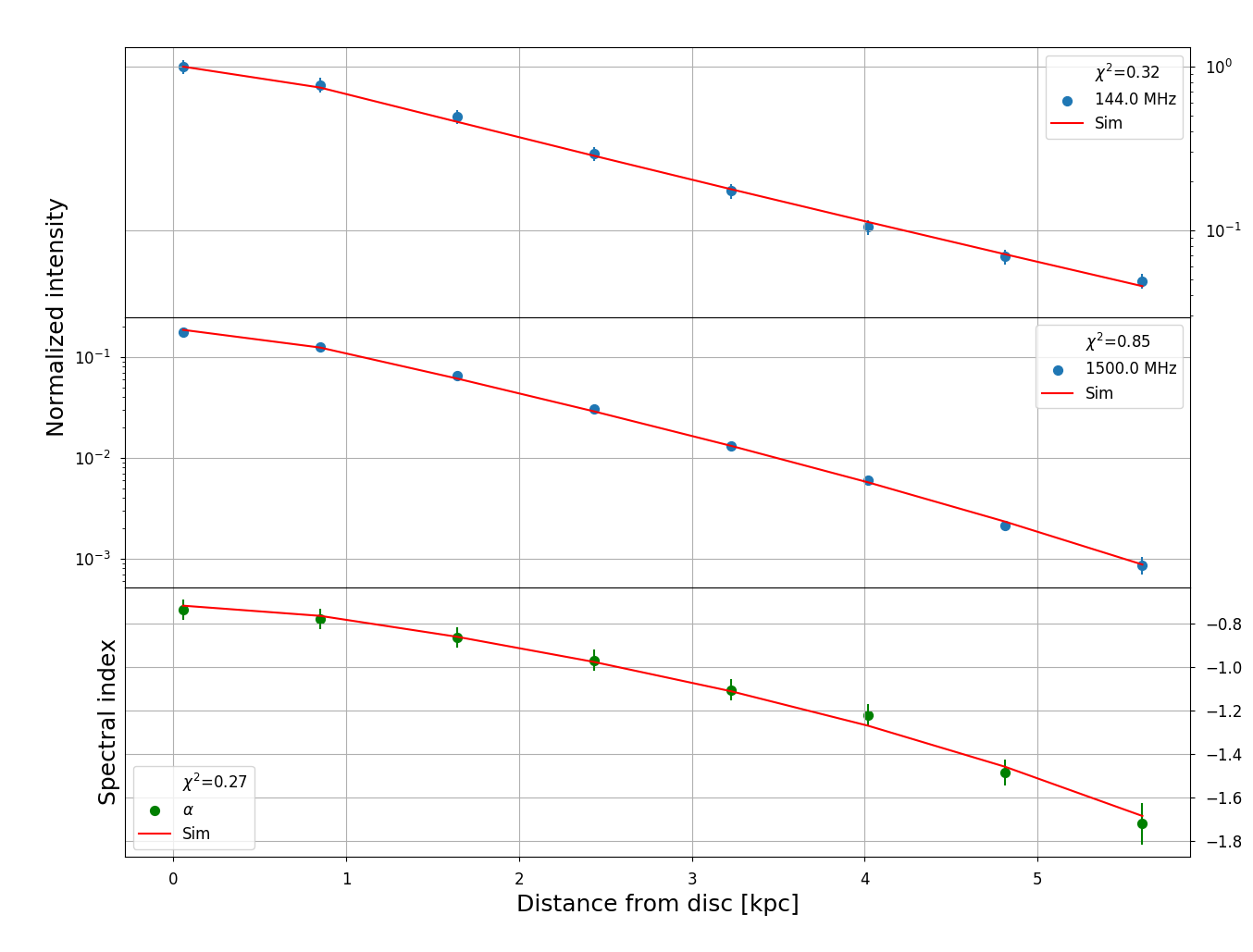}
\end{subfigure}
\hfill
\begin{subfigure}{0.49\linewidth}
\centering
\includegraphics[width=1\linewidth]{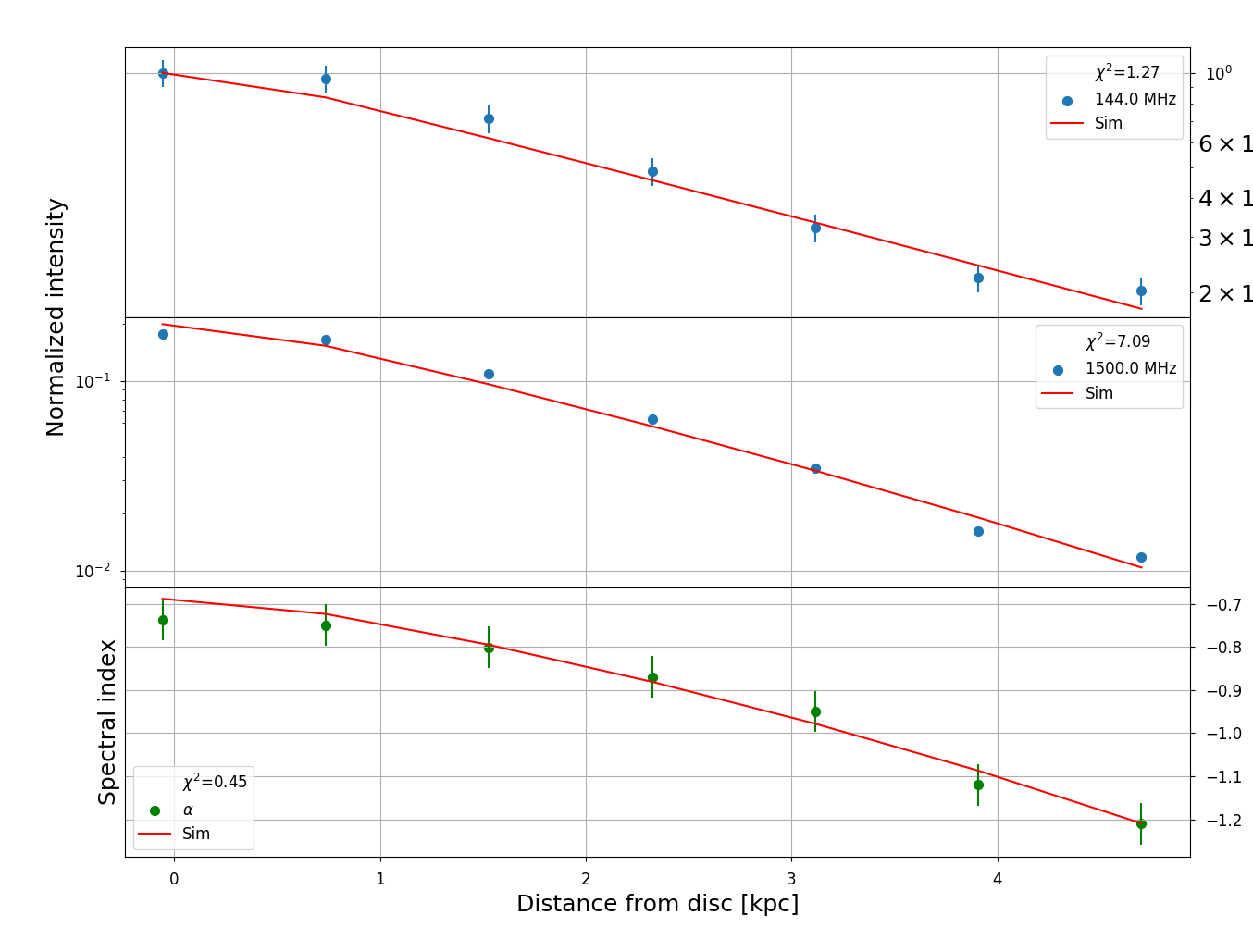}
\end{subfigure}
\caption{\texttt{SPINNAKER} profiles of NGC~4157 (all advection). Stripes are present as follows: top left: LM, top right: UM, bottom left: LL, and bottom right: UL.}
\end{figure*}

\begin{figure*}
\centering
\begin{subfigure}{0.49\linewidth}
\centering
\includegraphics[width=1\linewidth]{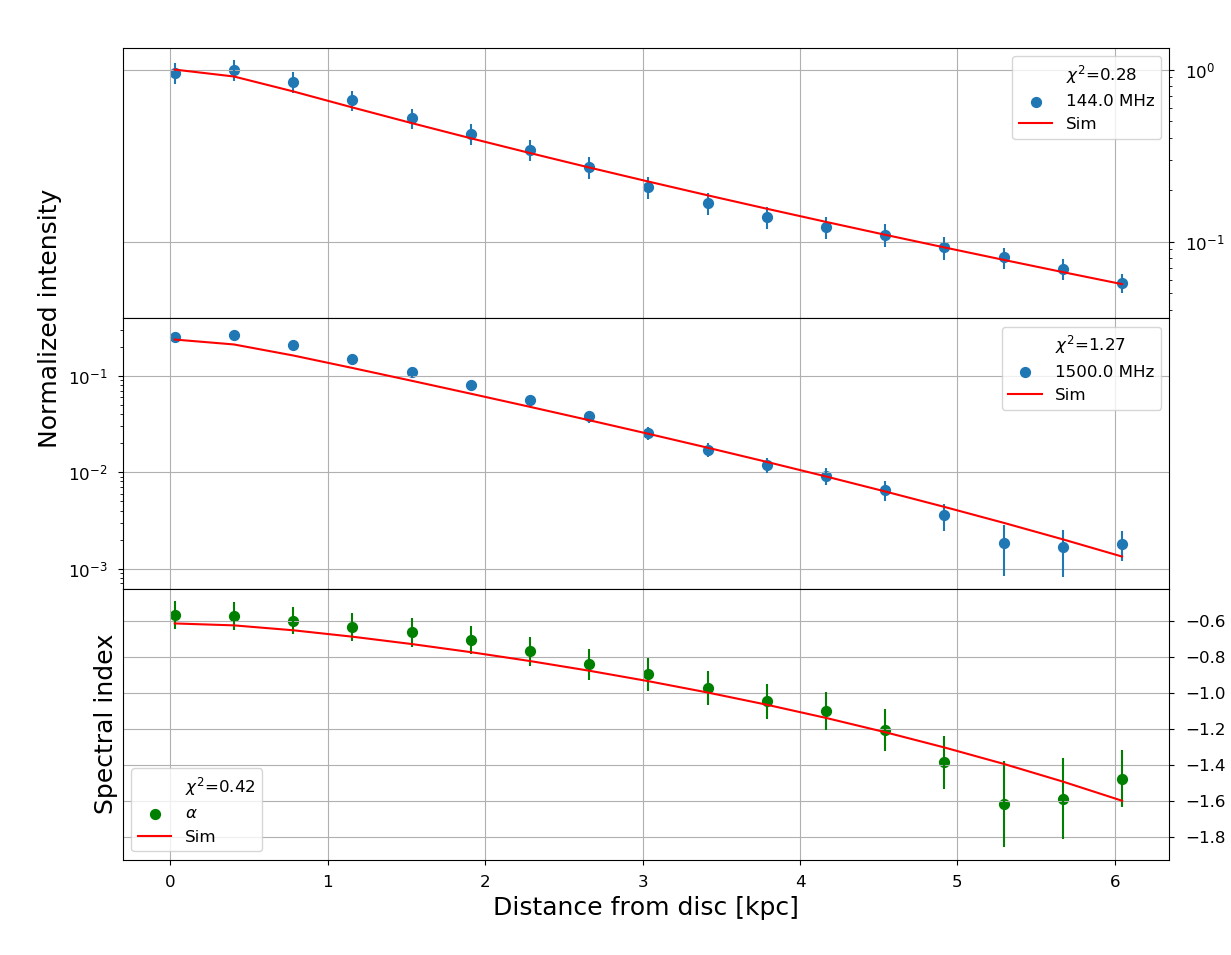}
\end{subfigure}
\hfill
\begin{subfigure}{0.49\linewidth}
\centering
\includegraphics[width=1\linewidth]{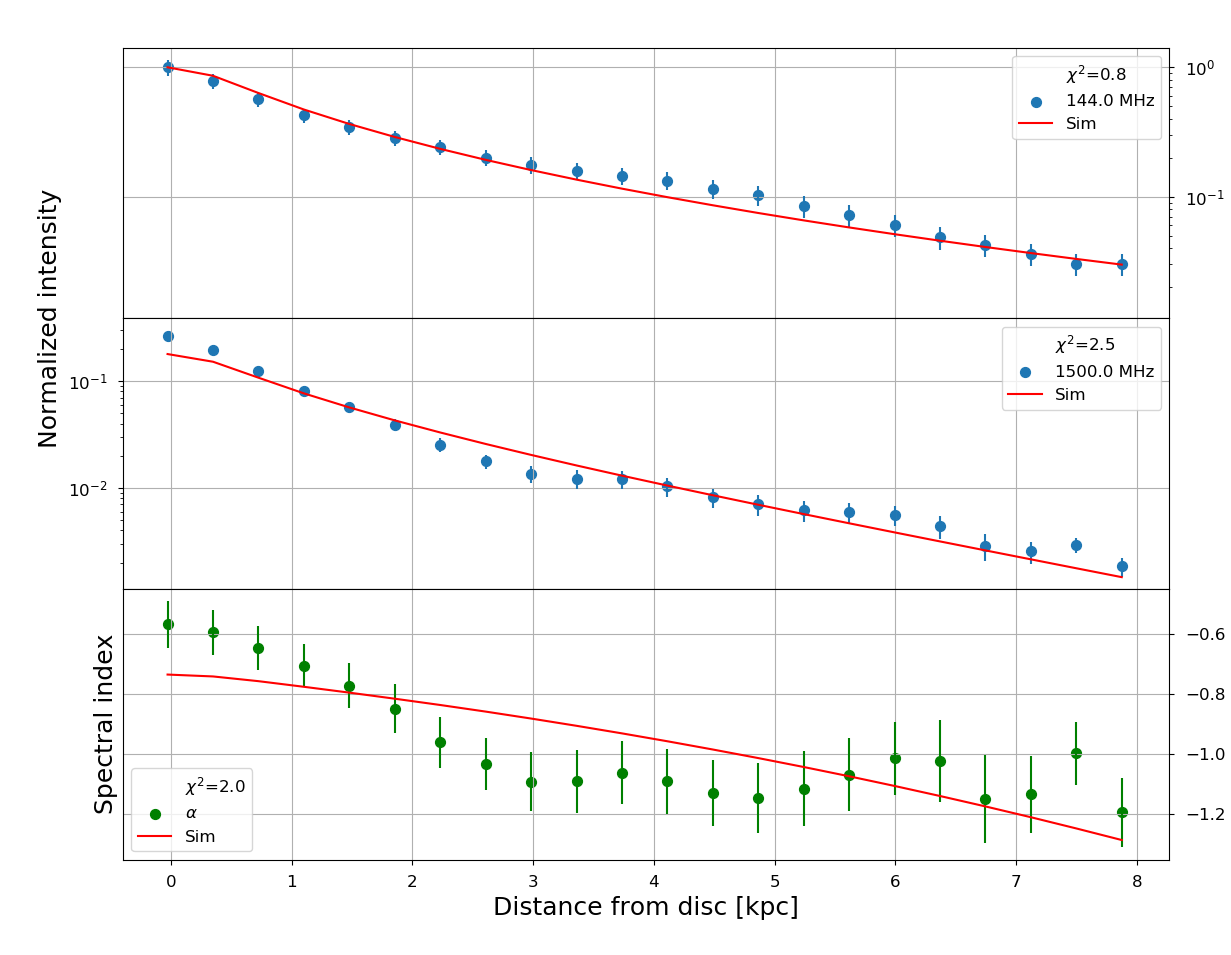}
\end{subfigure}
\\
\begin{subfigure}{0.49\linewidth}
\centering
\includegraphics[width=1\linewidth]{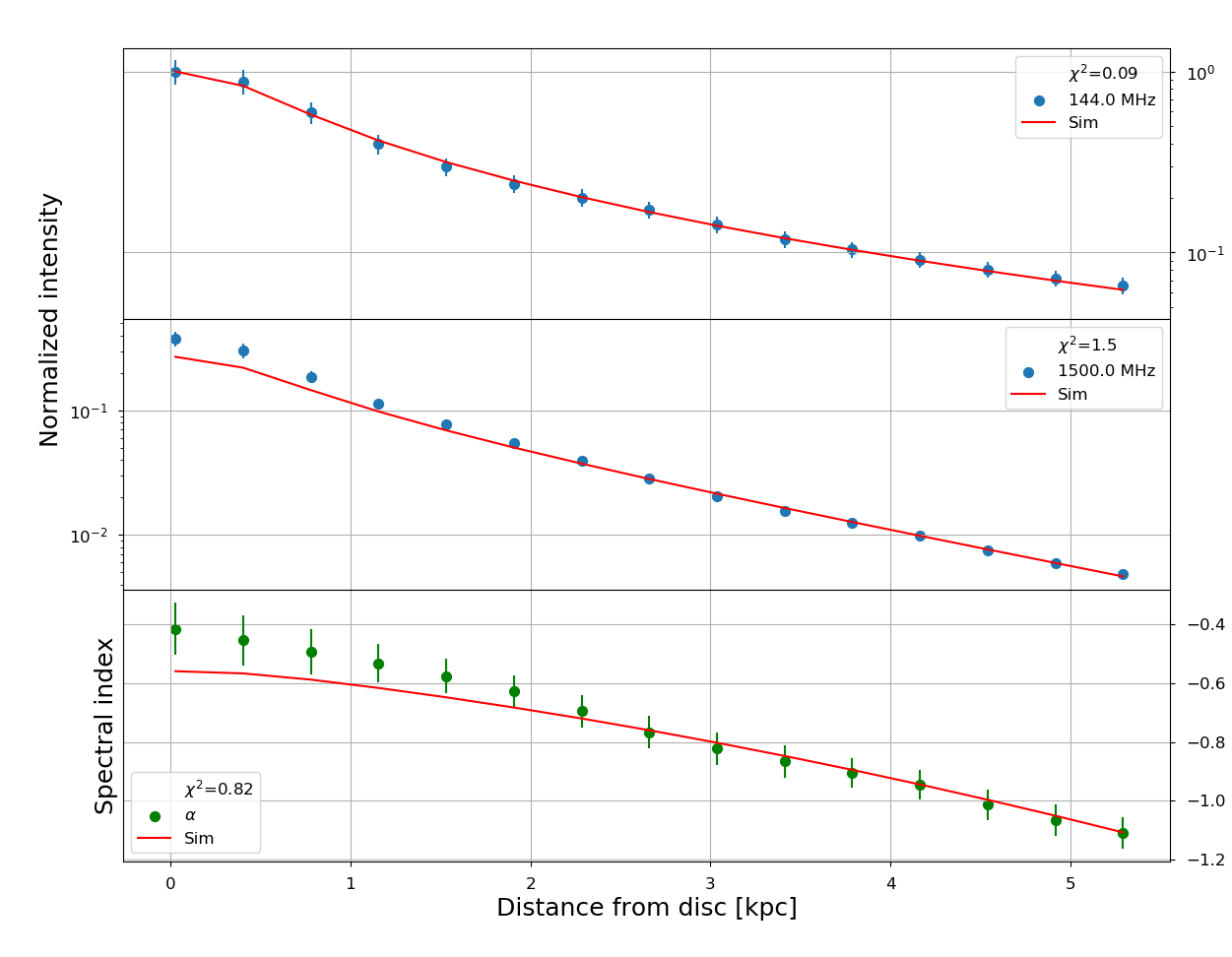}
\end{subfigure}
\hfill
\begin{subfigure}{0.49\linewidth}
\centering
\includegraphics[width=1\linewidth]{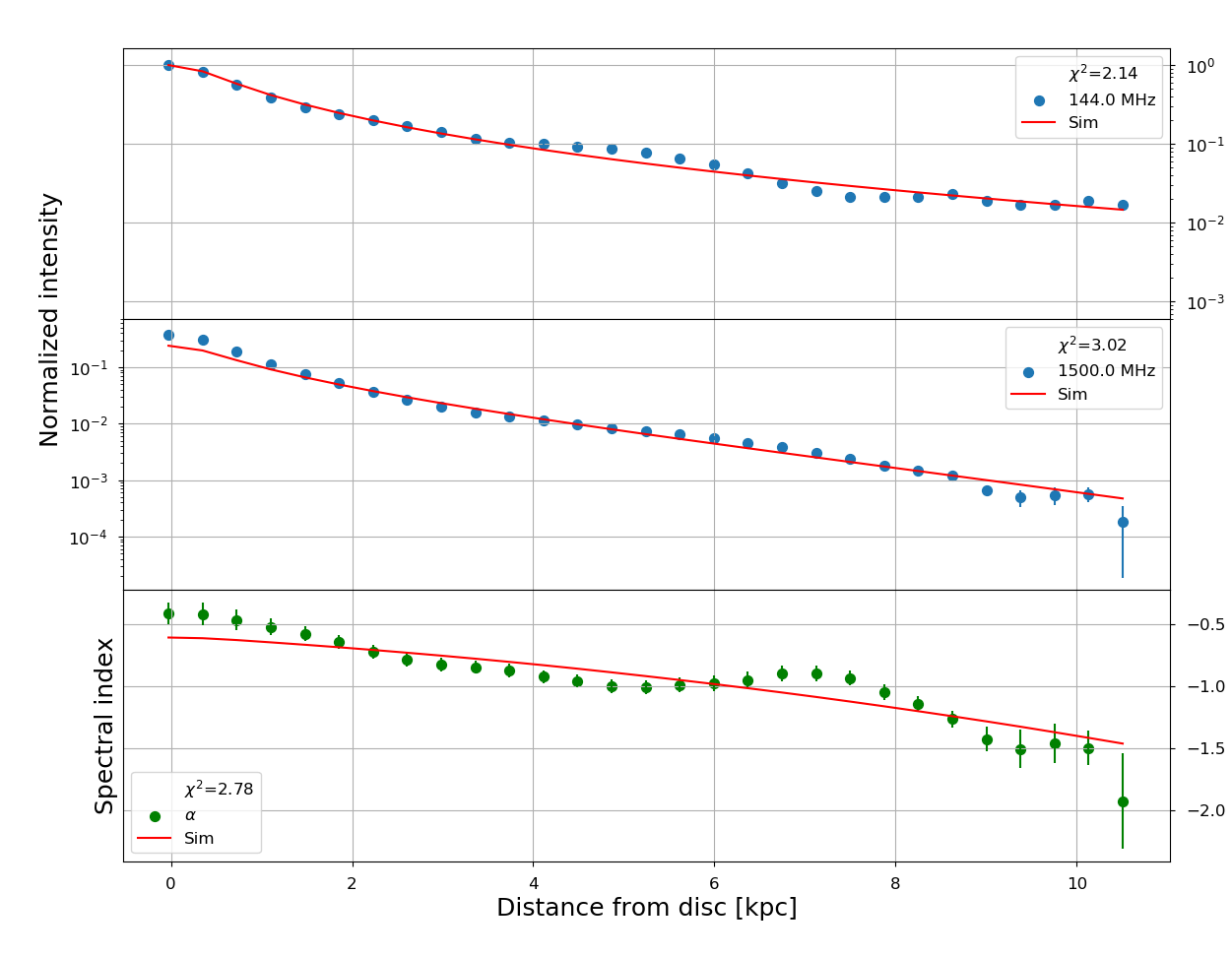}
\end{subfigure}
\\
\begin{subfigure}{0.49\linewidth}
\centering
\includegraphics[width=1\linewidth]{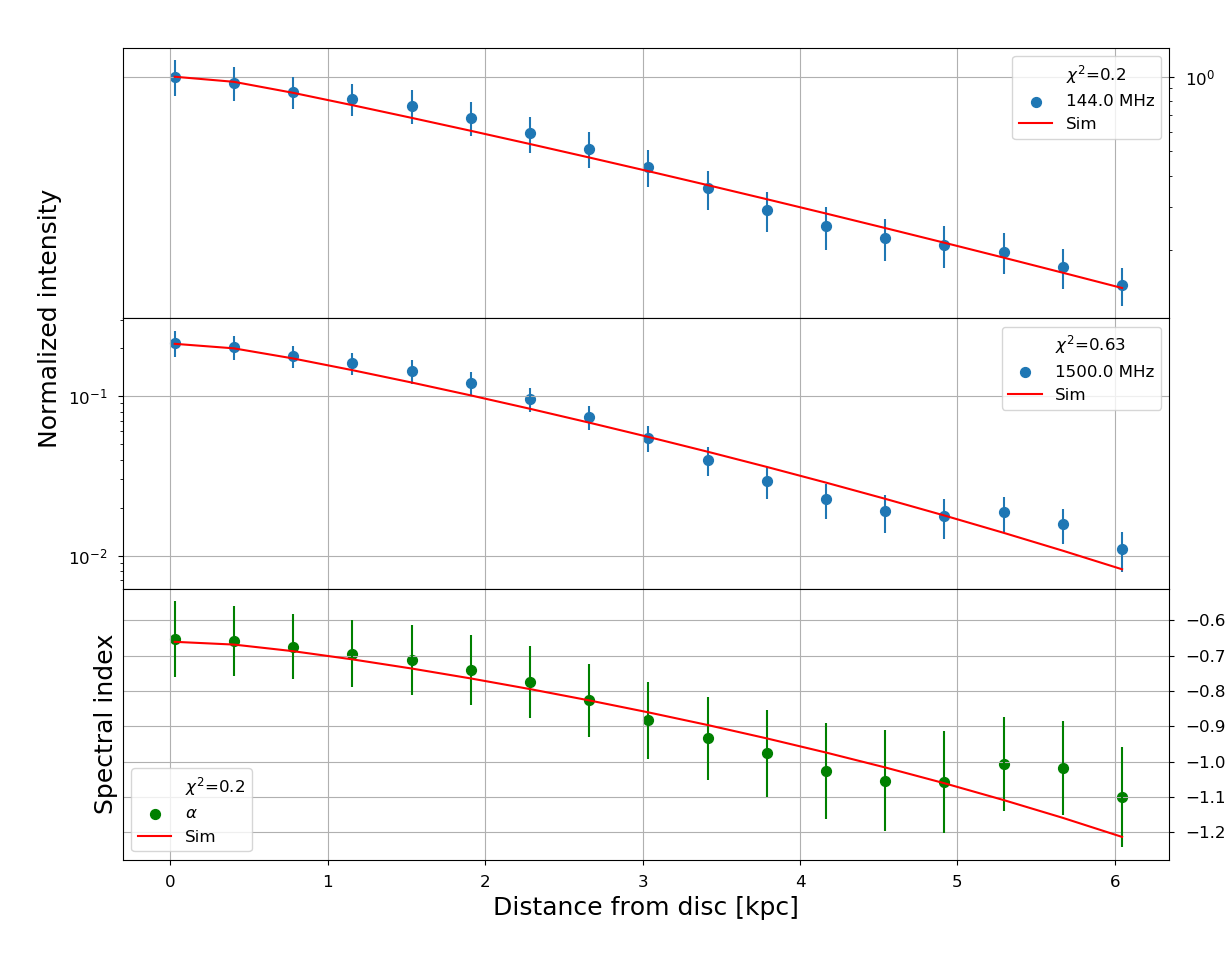}
\end{subfigure}
\hfill
\begin{subfigure}{0.49\linewidth}
\centering
\includegraphics[width=1\linewidth]{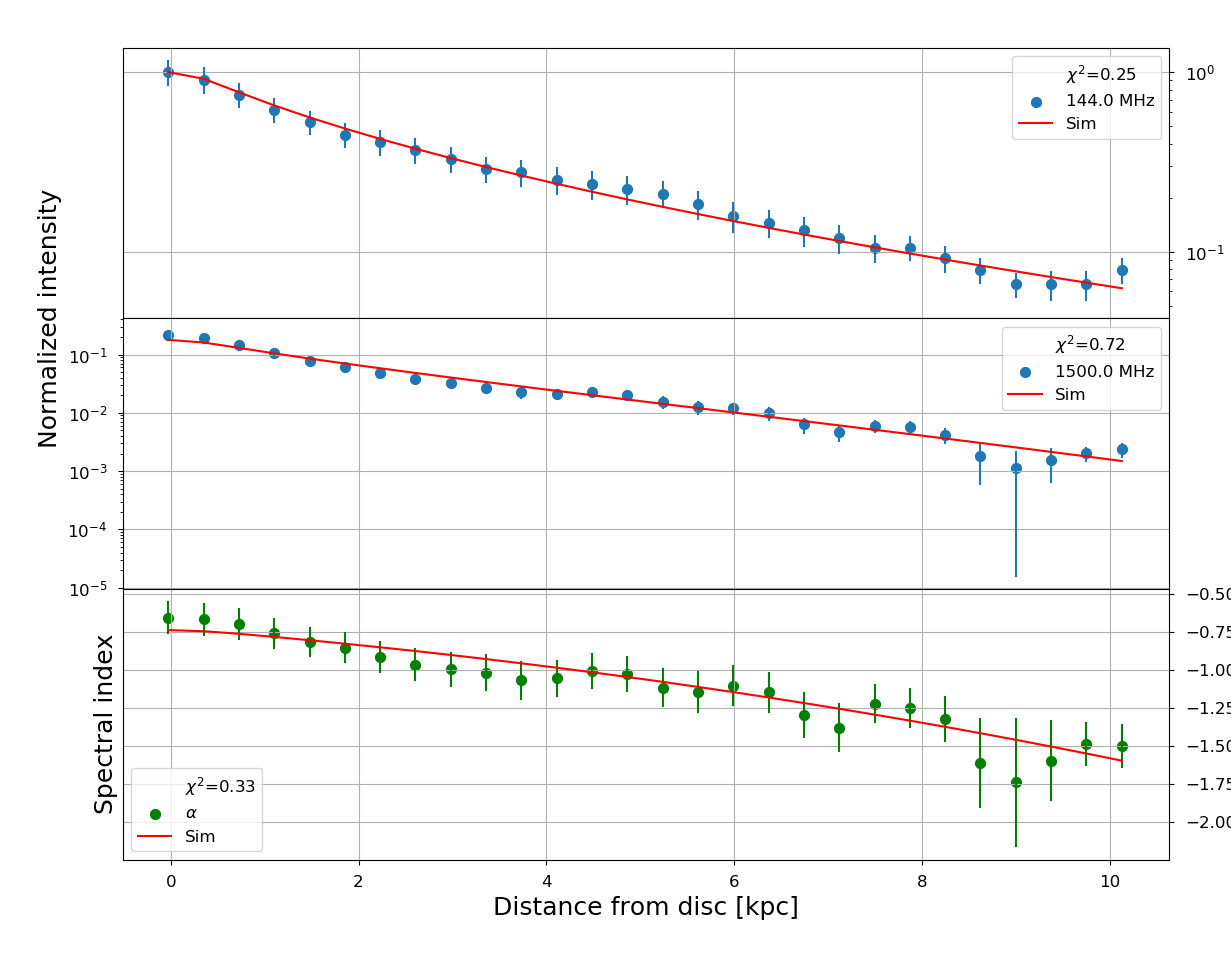}
\end{subfigure}
\caption{\texttt{SPINNAKER} profiles of NGC~4631 (all advection). Strips are presented as follows: top left: LR, top right: UR, middle left: LM, middle right: UM, bottom left: LL, and bottom right: UL.}
\end{figure*}